\documentclass[12pt]{report}
\usepackage[a4paper]{geometry}
\usepackage{german,alpheqn,epsfig,amsmath,amssymb,mathrsfs,alpheqn}

\newcommand{\stern}[1]{\stackrel{\ast}{#1}}
\newcommand{\halb}{\underline{\cap}}
\newcommand{\rf}[1]{(\ref{#1})}
\newcommand{\plmi}{\begin{array}{c}\scr+\\\scr-\end{array}}

\newcommand{\scrc}{{_\prime}\!c}
\newcommand{\scr}{\scriptstyle}
\newcommand{\scrscr}{\scriptscriptstyle}
\newcommand{\dis}{\displaystyle}
\newcommand{\bm}[1]{\boldmath $#1$ \unboldmath $\!\!$}
\newcommand{\calL}{{\cal L}}
\newcommand{\calM}{{\cal M}}
\newcommand{\calS}{{\cal S}}
\newcommand{\calR}{{\cal R}}
\newcommand{\calP}{{\cal P}}
\newcommand{\calT}{{\cal T}}
\newcommand{\calU}{{\cal U}}
\newcommand{\calD}{{\cal D}}
\newcommand{\calDhalb}{\stackrel{\halb}{\calD}}

\newcommand{\calDquer}{\bar{\calD}}
\newcommand{\calDquerF}{\bar{\mbox{\bm{\calD}}}}
\newcommand{\calDhalbF}{\stackrel{\halb}{\mbox{\bm{\calD}}}}
\newcommand{\calDasympF}{\stackrel{\asymp}{\mbox{\bm{\calD}}}}
\newcommand{\calDasymp}{\stackrel{\asymp}{\calD}}
\newcommand{\calB}{{\cal B}}
\newcommand{\calF}{{\cal F}}
\newcommand{\calH}{{\cal H}}
\newcommand{\calE}{{\cal E}}
\newcommand{\hF}{\hat{\cal F}{}}
\newcommand{\hE}{\hat{\cal E}{}}

\newcommand{\pdach}{\hat{p}{}}

\newcommand{\tdach}{\hat{t}{}}

\newcommand{\rdach}{\hat{r}{}}
\newcommand{\Pdach}{\hat{P}{}}
\newcommand{\ndach}{\hat{n}{}}
\newcommand{\hdach}{\hat{h}{}}
\newcommand{\Gammaquer}{\bar{\Gamma}{}}
\newcommand{\Gquer}{\bar{G}}
\newcommand{\Gammaschl}{\tilde{\Gamma}{}}

\newcommand{\Bschl}{\tilde{B}{}}
\newcommand{\tauschl}{\tilde{\tau}_4{}}
\newcommand{\Gammastern}{\stackrel{\ast}{\Gamma}{}\!\!}
\newcommand{\calBstern}{\stackrel{\ast}{\calB}{}\!\!}
\newcommand{\Pcap}{\stackrel{\cap}{P}{}\!\!}
\newcommand{\Gammadach}{\stackrel{\cap}{\Gamma}{}\!}
\newcommand{\gammadach}{\stackrel{\cap}{\gamma}{}\!}

\newcommand{\Gammaasymp}{\stackrel{\asymp}{\Gamma}{}\!}
\newcommand{\nablaasymp}{\stackrel{\asymp}{\nabla}{}\!}
\newcommand{\tauasymp}{\stackrel{\asymp}{\tau}{}\!}

\newcommand{\nablastern}{\stackrel{\ast}{\nabla}}
\newcommand{\Gstern}{\stackrel{\ast}{G}}
\newcommand{\Kstern}{\stackrel{\ast}{K}{}\!\!}
\newcommand{\Rstern}{\stackrel{\ast}{R}{}\!\!}
\newcommand{\Dstern}{\stackrel{\ast}{D}{}\!\!}
\newcommand{\hplus}{\!\stackrel{\scrscr +}{h}{}\!\!}
\newcommand{\hminus}{\!\stackrel{\scrscr -}{h}{}\!\!}
\newcommand{\Gdach}{\stackrel{\cap}{G}}
\newcommand{\taudach}{{\stackrel{\cap}{\tau}_4}}
\newcommand{\tauquer}{\overline{\tau}_4}
\newcommand{\lambdaquer}{\overline{\lambda}_4}
\newcommand{\taukr}{\stackrel{\circ}{\tau}_4}
\newcommand{\Gammakr}{\stackrel{\circ}{\Gamma}}
\newcommand{\lambdakr}{\stackrel{\circ}{\lambda}_4}

\newcommand{\omegakr}{\stackrel{\circ}{\omega}{}\!\!}
\newcommand{\Omegakr}{\stackrel{\circ}{\Omega}}
\newcommand{\Deltadach}{\stackrel{\cap}{\Delta}}
\newcommand{\nabladach}{\stackrel{\cap}{\nabla}{}\!\!}
\newcommand{\Deltaquer}{\overline{\Delta}}
\newcommand{\Basymp}{\stackrel{\asymp}{B}{}\!}
\newcommand{\Kasymp}{\stackrel{\asymp}{K}{}\!}
\newcommand{\Zasymp}{\stackrel{\asymp}{Z}{}\!}
\newcommand{\Bdach}{\stackrel{\cap}{B}}
\newcommand{\Zdach}{\stackrel{\cap}{Z}{}\!}
\newcommand{\Kdach}{\stackrel{\cap}{K}{}\!}
\newcommand{\Rdach}{\stackrel{\cap}{R}{}\!}
\newcommand{\Kcup}{\stackrel{\cup}{K}{}\!}

\newcommand{\Zcup}{\stackrel{\cup}{Z}{}\!}
\newcommand{\Aquer}{\bar{A}{}}
\newcommand{\Bquer}{\bar{B}{}}
\newcommand{\Dquer}{\bar{D}{}}
\newcommand{\Fquer}{\bar{F}{}}
\newcommand{\Squer}{\bar{S}{}}
\newcommand{\Qquer}{\bar{Q}{}}
\newcommand{\ndp}{{\rm ^{(||)}}{}\!\!\nabladach}         
\newcommand{\nds}{{\rm ^{(\perp)}}{}\!\!\nabladach}      
\newcommand{\Kdp}{{\rm ^{(||)}}{}\!\!\Kdach}             
\newcommand{\Kds}{{\rm ^{(\perp)}}{}\!\!\Kdach}          
\newcommand{\Zdp}{{\rm ^{(||)}}{}\!\!\Zdach}             
\newcommand{\Zds}{{\rm ^{(\perp)}}{}\!\!\Zdach}          
\newcommand{\cDhs}{{\rm ^{(\perp)}}{}\!\!\calDhalb}      
\newcommand{\cDhp}{{\rm ^{(||)}}{}\!\!\calDhalb}         
\newcommand{\SigmaB}{{\rm ^{(B)}}\Sigma}
\newcommand{\SigmaW}{{\rm ^{(W)}}\Sigma}
\newcommand{\gsigma}{{\it ^{(\sigma)}}g}
\newcommand{\lambdasigma}{{\it ^{(\sigma)}}\hat{\lambda}_4}
\newcommand{\SigmacalB}{{\rm ^{(\calB)}}\Sigma}
\newcommand{\SigmacalBquer}{{\rm ^{(\calB)}}\bar{\Sigma}}
\newcommand{\Sigmap}{{\it ^{(p)}}\Sigma}

\newcommand{\calTF}{{\rm ^{(F)}}{\cal T}}
\newcommand{\calTW}{{\rm ^{(F)}}{\cal T}}
\newcommand{\TB}{{\rm ^{(B)}}T}
\newcommand{\calTB}{{\rm ^{(B)}}{\cal T}}
\newcommand{\Rpara}{{}^{||} R}
\newcommand{\Rperp}{{}^{\perp} R}
\newcommand{\tB}{{\rm ^{(B)}}t}
\newcommand{\tM}{{\rm ^{(m)}}t}
\newcommand{\te}{{\rm ^{(e)}}t}
\newcommand{\TW}{{\rm ^{(W)}}T}
\newcommand{\tW}{{\rm ^{(W)}}t}
\newcommand{\TF}{{\rm ^{(F)}}T}
\newcommand{\TM}{{\rm ^{(m)}}T}

\newcommand{\Te}{{\rm ^{(e)}}T}
\newcommand{\Tf}{{\rm ^{(f)}}T}
\newcommand{\Tp}{{\rm ^{(p)}}T}
\newcommand{\TcalB}{{\rm ^{(\calB)}}T}
\newcommand{\TcalF}{{\rm ^{(\calF)}}T}
\newcommand{\omkrsig}{{\it ^{(\sigma)}}\!\!\omegakr}
\newcommand{\Omkrsig}{{\it ^{(\sigma)}}\!\!\Omegakr}
\newcommand{\Tid}{{\rm ^{(id)}}T}

\newcommand{\calHnull}{{\rm ^{(0)}}\calH}
\newcommand{\calPnull}{{\rm ^{(0)}}\calP}
\newcommand{\calFnull}{{\rm ^{(0)}}\calF}
\newcommand{\calUnull}{{\rm ^{(0)}}\calU}
\newcommand{\calMf}{{\rm ^{(f)}}{\scr\calM}}
\newcommand{\calPf}{{\rm ^{(f)}}{\scr\calP}}
\newcommand{\Unull}{{\rm ^{(0)}}\!U}
\newcommand{\Tnull}{{\rm ^{(0)}}\!T}
\newcommand{\Bnull}{{\rm ^{(0)}}\!B}
\newcommand{\Hnull}{{\rm ^{(0)}}\!H}
\newcommand{\knull}{{\rm ^{(0)}}\!k}
\newcommand{\Rnull}{{\rm ^{(0)}}\!R}
\newcommand{\Cnull}{{\rm ^{(0)}}\!C}
\newcommand{\Enull}{{\rm ^{(0)}}\!E}
\newcommand{\Mex}{{\rm ^{(ex)}}\!M}
\newcommand{\Min}{{\rm ^{(in)}}\!M}

\newcommand{\pdual}{{^* p}}

\begin{document}
\title{Die Allgemeine Relativit"atstheorie als SO(3)-Eichtheorie}
\author{\ }
\date{\normalsize Von der Fakult"at Physik der Universit"at Stuttgart zur
Erlangung der W"urde eines Doktors der Naturwissenschaften (Dr.~rer.~nat.) genehmigte
Abhandlung\\[3cm]
Vorgelegt von Michael Mattes aus Stuttgart\\[2cm]
Hauptberichter: Prof.~Dr.~Dr.~h.c.~W.Weidlich\\
Mitberichter: Prof.~Dr.~H.R.Trebin\\
Tag der m"undlichen Pr"ufung: 18.~Juli 1990\\[2cm]
Institut f"ur theoretische Physik der Universit"at Stuttgart\\1990}
\maketitle
\tableofcontents
\listoffigures
\pagebreak
\renewcommand{\thepage}{\arabic{page}}
\setcounter{page}{1}

\chapter*{Kurzfassung}
\addcontentsline{toc}{chapter}{Kurzfassung}
\setcounter{page}{1}
Die Einsteinsche Theorie der Gravitation ("`Allgemeine Relativit"atstheorie"') beruht
auf der Annahme, da\3 die geometrischen Eigenschaften des vierdimensionalen
Raumzeit-Kontinuums im wesentlichen von der darin befindlichen Materie bestimmt
werden. Im Gegensatz hierzu mi\3t die Newtonsche Mechanik dem Raum und der Zeit eine
absolute, materieunabh"angige Bedeutung bei. Die vorliegende Arbeit stellt einen
Kompromi\3 zwischen diesen beiden Vorstellungen dar: Es wird zwar einerseits die
spontane Aufspaltung der Raum-Zeit in eine "`universale Zeit"' und einen "`absoluten
Raum"' im Sinne Newtons akzeptiert, andrerseits aber erh"alt diese (1+3)-Aufspaltung
den Status eines dynamischen Objektes im Sinne Einsteins.

Hierbei wird die (gro\3r"aumige) Eigendynamik der (1+3)-Zerlegung nur schwach an die
(lokalen) Schwankungen der Materiedichte mit Hilfe der Einstein-Gleichungen
angekoppelt, soda\3 diese nach wie vor ihre G"ultigkeit f"ur die
Gravitationsph"anomene in einem begrenzten Raumbereich behalten (Planeten, Sterne
Galaxien). Auf einer kosmischen Skala jedoch werden die Eigenschaften des Universums
als Ganzes im wesentlichen von der Eigendynamik der Raumzeit-Aufspaltung bestimmt.

Die Geometrie der (1+3)-Zerlegung und die Expansion des leeren Universums werden im
Detail untersucht. Die neue Gravitationstheorie enth"alt im Vergleich zur
Einsteinschen Theorie einen zus"atzlichen Feldfreiheitsgrad, der als Tr"ager f"ur den
Energie-Impuls-Inhalt der Gravitationswechselwirkung in Frage kommt. Der
"`Grundzustand"' des leeren Universums ($\leadsto$ maximale Symmetrie) wird durch
eine de Sitter-Geometrie beschrieben, wobei sich nur das {\em expandierende}
Universum als stabil erweist. Ein Teil der vorliegenden Doktorarbeit (Kap.V) wurde bei der
Zeitschrift f"ur Naturforschung zur Ver"offentlichung eingereicht.\cite{MaSo89_a}


\chapter{Erfolge und Probleme der Allgemeinen Relativit"atstheorie}
\indent

Die popul"arwissenschaftliche Literatur reagiert bekanntlich mit einiger Verz"ogerung
auf die neuen bahnbrechenden Erkenntnisse, die von den akademischen Spezialisten in
ihren Labors und Denkzellen an den universit"aren Forschungseinrichtungen erarbeitet
werden. Diesem Retardierungseffekt kann man sicherlich auch einige positive Seiten
abgewinnen wenn man bedenkt, da\3 so manche "`wissenschaftliche Revolution"' wieder
zerplatzt, bevor sie "uberhaupt ans Licht der "Offentlichkeit gelangt (und so dem
Ansehen ihres Sch"opfers Schaden zuf"ugen k"onnte). Gerade wegen dieser
Selektionseigenschaft k"onnen aber die Reaktionen in der popul"arwissenschaftlichen
Presse manchmal einen durchaus seri"osen Hinweis darauf  abgeben, da\3 die
normalerweise nur f"ur Eingeweihte zug"angliche Forschungsfront wieder einmal in
Bewegung gekommen ist.

Diese Annahme scheint umso mehr berechtigt, wenn sich auch noch die eigentlichen
K"onner darum bem"uhen, die neuesten Forschungsergebnisse ihres engeren Fachgebietes
unter die Leute zu bringen. Die gegenw"artige Welle von popul"arwissenschaftlichen
Darstellungen der Relativit"atstheorie, die von renommierten theoretischen Physikern
verfa\3t werden (z.B. \cite{Haw}-\cite{BaSi86}), deutet jedenfalls daraufhin, da\3 70
Jahre nach der ersten Formulierung dieser Theorie durch Einstein eine
aufsehenerregende Neuentwicklung eingetreten sein mu\3. Was sind die Ursachen f"ur
diese Renaissance der (Allgemeinen) Relativit"atstheorie?

\section{Kurzer, historischer R"uckblick}
\indent

Um diese Frage zu beantworten, mu\3 man nochmals in die Anfangsphase der Allgemeinen
Relativit"atstheorie zur"uckgehen. Sofort nach Aufstellung seiner Gravitationstheorie
im Jahr 1915 hat Einstein selbst diese einem ersten "`H"artetest"' unterzogen, indem
er mit ihrer Hilfe die Periheldrehung der Merkurbahn (43 Bogensekunden pro
Jahrhundert) f"ur das damalige Verst"andnis "`richtig"' erkl"arte. Damit schien
zun"achst einmal der Siegeszug der Allgemeinen Relativit"atstheorie gesichert zu
sein. Paradoxerweise ist vom heutigen Standpunkt aus gesehen gerade dieser erste
Erfolg der Einsteinschen Theorie h"ochst zweifelhaft, da Einstein bei seiner
Berechnung der Periheldrehung von einer exakten Kugelgestalt der Sonne ausging,
w"ahrend man heutzutage wei\3, da\3 aufgrund der inhomogenen Rotation der Sonne ein
Quadrupolmoment auftritt, welches Periheldrehungen in der Gr"o\3enordnung der
allgemein-relativistischen Korrektur bewirkt (Dicke und Goldenberg). Als Folge davon
scheint heute niemand so richtig zu verstehen, warum die Vorhersagen der
Einsteinschen Theorie so gut mit dem beobachteten Wert "ubereinstimmen.

Zun"achst aber kam noch ein weiterer gl"ucklicher Umstand der Relativit"atstheorie
in ihrer fr"uhen Phase zu Hilfe: die totale Sonnenfinsternis von 1919. Es wurden zwei
Expeditionen (Eddington) zur Beobachtung der Ablenkung des Sternenlichtes durch die
verfinsterte Sonne durchgef"uhrt, welche die Einsteinschen Vorausagen bis auf eine
Genauigkeit von 20\% (bzw. 7\%) best"atigten. Dies wurde als eine hinreichende
Verifizierung der Einsteinschen Theorie akzeptiert, vor allem auch deshalb, weil die
damalige Experimentiertechnik einen gr"o\3eren Genauigkeitsgrad nicht zulie\3. Mit
dieser etwas d"urftigen experimentellen "`Untermauerung"' hat die Allgemeine
Relativit"atstheorie eine 40-j"ahrige Stagnationsphase "uberstanden, wobei die
weitere {\em theoretische} Durcharbeitung eine gewisse "Uberlebenshilfe geboten hat
(Wheeler, Chandrasekhar, Schild, Zel'dovich).

Allerdings erfolgte nun aber in den 60er Jahren eine intensive Wiederbelebung der
Relativit"atstheorie sowohl von der theoretischen wie auch der experimentellen Seite
her. Der Beitrag der Theoretiker bestand einerseits in der Aufstellung einer
Konkurrenz-Theorie (Skalar-Tensor-Theorie von Brans und Dicke), andrerseits in der
Anwendung neuer Rechenmethoden  (Penrose) und der Aufdeckung wichtiger theoretischer
Ph"anomene der Theorie (Hawking, Thorne, Novikov, Bardeen). Aber auch die
Experimentalphysiker konnten mit aufsehenerregenden Entdeckungen aufwarten: Pulsare,
Quasare, Hintergrund-Strahlung, Gravitationslinsen, relativistische
Doppelstern-Systeme. Dar"uberhinaus hatten sich die experimentellen Methoden so
verfeinert, da\3 man nicht mehr auf blo\3e Beobachtung der allgemein-relativistischen
Effekte auf einer kosmologischen Skala angewiesen war, sondern auch Experimente
innerhalb des planetaren Ma\3stabs durchf"uhren konnte. So wurde z.B. die
Gravitations-Rotverschiebung nachgewiesen mit Hilfe eines Raketenexperimentes
(Vessot), durch Erdumrundung an Bord eines Passagierflugzeuges (Hafele und Keating)
und schlie\3lich sogar innerhalb eines Geb"audes (Jefferson-Turm der
Harvard-University, Pound-Rebka-Snider). Ferner wurde ein Doppelstern-Pulsar
entdeckt (Hulse und Taylor), der aufgrund seiner g"unstigen Systemparameter die
allgemein-relativistischen Effekte im Sinne Einsteins wie aus dem Bilderbuch
produziert (z.B. ist hier die Periastron-Drehung 36000 mal so gro\3 wie bei der
Merkur-Bahn; Bahnradius $\approx$ Sonnenradius, Bahnperiode 7,75\,h, Pulsperiode
0,059\,sec).

Auf dem Hintergrund dieser erdr"uckenden Beweislast gegen die Konkurrenten der
Allgemeinen Relativit"atstheorie und zugunsten der Einsteinschen Ideen mag die in
j"ungster Zeit beobachtbare Euphorie der Relativit"atstheoretiker verst"andlich
erscheinen. Man sollte dar"uber aber nicht vergessen, da\3 die logisch-konzeptionelle
Geschlossenheit der Einsteinschen Theorie keineswegs so einstimmig akzeptiert wird
wie ihre experimentellen Erfolge. Die Uneinigkeit in der theoretischen Bewertung der
Einsteinschen Theorie wird allein schon an ihrer unterschiedlichen Benennung durch
verschiedene Autoren deutlich: "`Chronogeometry"' (Fock, Fokker), "`geochronometrical
gravity"' (Treder), "`geometrodynamics"' (Wheeler), "`gravi\-dy\-na\-mics"' (deWitt,
Ivanenko), "`gravitodynamics"' (Mercier). Abgesehen von diesen "au\3erlichen
Meinungsverschiedenheiten, gibt es auch inhaltliche Differenzen: Hatte Einstein noch
durch seine Namensgebung "`Allgemeine Relativit"atstheorie"' zum Ausdruck bringen
wollen, da\3 es sich hier um eine Verallgemeinerung des Relativit"atsprinzips seiner
Speziellen Theorie handeln soll, so bestreiten z.B. Fock und Fokker, da\3 die
Allgemeine Theorie (im Gegensatz zur Speziellen) "uberhaupt ein Relativit"ats-Prinzip
enth"alt. Die in Einsteins Theorie vorhandene Kovarianz wird von manchen Autoren als
eine glatte Trivialit"at angesehen (Kretschmann, Havas).

In der vorliegenden Arbeit wird von der bestehenden Kritik an der Einsteinschen
Theorie ausgegangen und versucht eine L"osung f"ur gewisse Schwachpunkte der
Allgemeinen Relativit"atstheorie zu finden. Diese unbefriedigenden und
verbesserungsw"urdigen Aspekte sollen zun"achst einmal pr"azisiert werden.

\section{Problematische Aspekte der Einsteinschen Theorie}
\centerline{{\large\bf 1.}}\indent

Das hervorstechende Merkmal der Allgemeinen Relativit"atstheorie ist sicherlich die
Eigenschaft der allgemeinen Kovarianz. Als Folge dieser Eigenschaft verlieren die
Koordinaten der Raum-Zeit ihre unmittelbare physikalische Bedeutung, die sie in der
Speziellen Theorie noch besitzen (globale Inertialsysteme). An die Stelle der
globalen Inertialsysteme tritt in der Allgemeinen Theorie der Begriff des {\em
Bezugssystems} (reference frame), ohne den weder die theoretische Seite (Spinoren,
etc.) noch die experimentelle Seite (Verifikation) auskommen kann. Anschaulich
gesehen kann man sich ein Bezugssystem als ein Tetradenfeld "uber der Raum-Zeit
vorstellen, das nur bis auf eine SO(3)-Umeichung eindeutig bestimmbar ist. Der
zeitartige Tetradenvektor ist dabei ein SO(3)-Skalar und definiert eine
3-dimensionale raumartige Distribution, die im Falle ihrer Integrabilit"at zu einer
(1+3)-Foliation der 4-dimensionalen Raum-Zeit f"uhrt. F"ur die konkrete Bearbeitung
der allermeisten relativistischen Probleme ist die {\em Wahl} einer solchen Foliation
unvermeidbar \cite{Nester}! Auf diese Weise erh"alt --- trotz der allgemeinen
Kovarianz --- die Wahl eines Bezugssystems eine so wichtige Bedeutung, da\3 wir uns
versucht f"uhlen werden, die (1+3)-Foliation von ihrem untergeordneten Status einer
rechentechnischen Hilfsgr"o\3e zu befreien und zu einem ordentlichen, dynamischen
Objekt der Theorie zu machen. Obwohl dies die Bevorzugung eines bestimmten
Bezugssystems bedeutet, mu\3 es nicht unbedingt die Verletzung der allgemeinen
Kovarianz der Theorie nach sich ziehen (siehe sp"ater). Ein solches Vorgehen w"urde
lediglich zur Folge haben, da\3 die Zahl der (kovarianten!) Variablen der
Gravitationstheorie um diejenigen vergr"o\3ert wird, welche die (1+3)-Foliation
beschreiben ($\leadsto$ "`"Atherfelder"' und SO(3)-Eichfeld).

\pagebreak
\vspace{1cm}\centerline{{\large\bf 2.}}

Eng verwandt mit der Problematik der Bezugssysteme ist das {\em "Aquivalenzprinzip}.
Ent\-sprechend diesem Prinzip mu\3 es m"oglich sein, f"ur jeden Punkt der Raum-Zeit
ein (lokales) Bezugsystem einzurichten, bez"uglich dem die physikalischen Gesetze
ihre speziell-relativistische Form annehmen. Dadurch wird die Gravitationskraft zu
einer Scheinkraft degradiert, die durch "Ubergang zu einem geeigneten Bezugssystem
(lokal) wegtransformiert werden kann. Dies erzeugt nun aber f"ur die Theorie ein
gewisses Problem, denn wenn die Gravitationskraft lokal zum Verschwinden gebracht
werden kann, dann gilt dies sicher auch f"ur ihren Energie/Impulsinhalt,falls man
sich diesen aus der Metrik und ihren ersten kovarianten Ableitungen aufgebaut denken
darf. Man k"onnte darauf antworten, da\3 das "Aqui\-valenz\-prinzip gar nicht f"ur
alle physikalischen Gesetze gilt, da manche Beziehungen (z.B. die
Einstein-Gleichungen) "uberhaupt kein speziell-relativistisches Analogon haben. Diese
Art der Argumentation hat dazu gef"uhrt, eine "`schwache"' Form des
"Aquivalenzprinzips einzuf"uhren, die nur f"ur Gesetze der ersten
Differentiations-Ordnung gelten soll, also z.B. nicht f"ur die Einstein-Gleichungen,
die zweite Ableitungen der Metrik enthalten. Dem mu\3 aber entgegengehalten werden,
da\3 der Energie/Impulsinhalt eines Feldes, das einer Bewegungsgleichung zweiter
Ordnung gen"ugt, aus den ersten Ableitungen der Feldvariablen aufgebaut sein sollte
und demnach auch dem (schwachen) "Aquivalenzprinzip gen"ugen sollte; dies f"uhrt
jedoch im speziellen Fall des metrischen Feldes zu einer Trivialit"at, denn die
ersten kovarianten Ableitungen der Metrik verschwinden identisch und die Verwendung
einer nicht-kovarianten Ableitung der Metrik kommt wegen der Kovarianzforderung an
den entsprechenden Energie-Impuls-Tensor nicht in Frage.

In diesem Zusammenhang mu\3 nun auf folgenden Sachverhalt hingewiesen werden: Wenn
die Gravitation mit anderen Formen der Energie und Materie (z.B. mit einem
Gravitationswellen-Detektor) Energie-Impuls nach den "ublichen Erhaltungsgesetzen
austauschen kann, dann sollte auf der rechten Seite der Einstein-Gleichungen u.a. auch
ein Term erscheinen, der den Energie-Impulsinhalt der Gravitation beschreibt. Dieser
Umstand f"uhrt sofort zu weitreichenden Folgerungen, von denen wir hier nur eine
anf"uhren wollen: Wenn man alle Energie und Materie aus dem Universum entfernt,
stehen zur Beschreibung dieses leeren Universums die Einstein-Gleichungen mit dem
allein verbleibenden Gravitationsterm auf der rechten Seite zur Verf"ugung. F"ur
dieses inhomogene Gleichungssystem ist aber eine ganz andere L"osungsmannigfaltigkeit
im Vergleich zu den homogenen Einstein-Gleichungen zu erwarten. Es ergibt sich hier
die Frage, ob die experimentell feststellbare Expansion des Universums wirklich von
der sichtbaren Materiedichte
wesentlich beeinflu\3t wird, wie es das Standardmodell der Kosmologie verlangt, oder
ob die Expansion nicht eine Folge der "`Vakuum-Dynamik"' des {\em leeren} Universums
ist?

Wir glauben, da\3 wir zu diesem ungel"osten Problem der Einsteinschen Theorie eine
neue L"osung anbieten k"onnen. Da n"amlich das metrische Feld seine fundamentale
Bedeutung in unserer Theorie verliert und diese an die neuen Variablen abtritt,
welche die (1+3)-Aufspaltung beschreiben, k"onnen wir diese neuen Variablen
ben"utzen, um f"ur das Gravitationsfeld eine Energie-Impulsdichte zu definieren.
Damit kann die Dynamik des leeren Universums unter einem neuen Aspekt untersucht
werden.

\vspace{1cm}\centerline{{\large\bf 3.}}

Das wohl bekannteste Problem der Allgemeinen Relativit"atstheorie stellt ihre
Quantisierung dar, die man gerne durchgef"uhrt sehen m"ochte, bevor man eine
Vereinheitlichung mit den drei "ubrigen Grundkr"aften der Natur vornehmen kann. Vor
einer solchen "`gro\3en"' Vereinheitlichung sollten alle vier Einzeltheorien in einem
einheitlichen formalen System darstellbar sein, d.h. konkret: Jede Einzeltheorie
sollte als eine separate konsistente Eichtheorie klassisch und quantentheoretisch
formulierbar sein. Paradoxerweise macht hier aber gerade die Gravitation --- im
Unterschied zu den drei restlichen Kr"aften --- v"ollig unerwarteterweise gro\3e
Schwierigkeiten. Denn w"ahrend die Eichtheorien f"ur die elektromagnetische, schwache
und starke Kraft problemlos in einen einheitlichen {\em geometrischen} Rahmen
($\leadsto$ Faserb"undeltheorie) gebracht werden k"onnen, f"ugt sich hier
ausgerechnet die Allgemeine Relativit"atstheorie nicht so ohne weiteres ein, obwohl
doch mit ihr der Anspruch verbunden wird, eine {\em Geometrisierung} der
Gravitationskraft zu liefern. Die hierbei auftretenden Schwierigkeiten sollen kurz
skizziert werden.

Bei den "ublichen Eichtheorien vom Yang-Mills-Higgs Typ kann jedes physikalische Feld
eindeutig als geometrisches Objekt in einem Faserb"undel "uber der Raum-Zeit
identifiziert werden. So entspricht z.B. dem Eichpotential die Konnexion des
Faserb"undels, der Eichfeldst"arke entspricht die Kr"ummung und einem Materiefeld
entspricht ein Schnitt des entsprechenden Tensorb"undels. Durch diese Entsprechung
sind alle drei Kategorien von geometrischen Objekten bzw. physikalischen Feldern klar
voneinander abgegrenzt. Im Falle der Allgemeinen Relativit"atstheorie ist man
zun"achst versucht die Christoffel-Symbole in analoger Weise mit der Konnexion in
einem Gl(4,{\bf R})-B"undel zu identifizieren, da sich eine Koordinatentransformation
der Raum-Zeit wie eine Gl(4,{\bf R})-Umeichung ihres Tangentialb"undels auswirkt.
Diese Identifizierung h"atte nun aber zur Folge, da\3 die Kr"ummung der Konnexion
${\cal GL}(4,{\bf R})$-wertig ausf"allt, was im Widerspruch zum "Aquivalenzprinzip
steht, welches eine ${\cal SO}(1,3)$-wertige Kr"ummung verlangt (d.h. die Kr"ummung
bewirkt eine Lorentz-Transformation). Als Ausweg aus diesem Dilemma mu\3 eine zweites
geometrisches Objekt der Konnexion zur Seite gestellt werden, n"amlich das metrische
Feld, durch dessen kovariante Konstanz die Holonomiegruppe der Konnexion von der
urspr"unglichen Eichgruppe Gl(4,{\bf R}) zur Lorentzgruppe SO(1,3) reduziert wird.
Dadurch wird aber die Konnexion von der Metrik eindeutig bestimmt, soda\3 die Rolle
des physikalischen Potentials nun von der Metrik anstelle der Konnexion "ubernommen
wird. Hierdurch wird die Analogie zu den Eichtheorien vom Yang-Mills-Higgs Typ
zerst"ort und der Eichstatus der Allgemeinen Relativit"atstheorie wird wieder zu
einem offenen Problem.

Auch zu diesem Problem bietet die vorliegende Arbeit einen L"osungsvorschlag an: es
kann gezeigt werden, da\3 die Allgemeine Relativit"atstheorie in der vorliegenden
Neuformulierung als eine SO(3)-Eichtheorie aufgefa\3t werden kann. Dabei ist die
zugrundeliegende SO(3)-Dynamik in einem {\em flachen} Raum inkonsistent und ben"otigt
zu ihrer Realisierung einen {\em gekr"ummten} Raum. Die Kr"ummung der Raum-Zeit
erscheint also hier nicht wie bei Einstein als physikalisches Postulat sondern als
eine mathematische Notwendigkeit.

\section{ Die Allgemeine Relativit"atstheorie als SO(3)-Eichtheorie}
\indent

Die vorliegende Arbeit soll eine Antwort darstellen sowohl auf die Erfolge als auch
auf die Probleme der Allgemeinen Relativit"atstheorie. Da die erw"ahnten Erfolge
durchweg auf der (zumindest n"aherungsweisen) G"ultigkeit der Einsteinschen
Gleichungen beruhen, m"ussen diese Bewegungsgleichungen auch in der modifizierten
Theorie ihre G"ultigkeit behalten. Andererseits sollen aber auch die oben
erl"auterten Schwachpunkte der Einsteinschen Theorie ber"ucksichtigt werden, was
bedeutet, da\3 die Einstein-Gleichungen eine neue Interpretation im Rahmen der
modifizierten Theorie erfahren m"ussen.

Die drei Grundpfeiler der modifizierten Gravitationstheorie sollen kurz dargestell
werden:\\

\noindent\bm{\alpha} {\bf ) Allgemeine Struktur der Theorie}
\\

Zun"achst sei darauf hingewiesen, da\3 die "`Kinematik"' der Gravitation durch einen
neuen Variablensatz beschrieben wird. Die Metrik {\bf G} als Tensor zweiter Stufe
tritt ihre fundamentale Bedeutung an ihre "`mikroskopischen"' Bestandteile (die
"`"Atherfelder"' \bm{\calB}, {\bf p}) ab. Die "Atherfelder sind Vektorfelder "uber der
Raum-Zeit und bilden eine bez"uglich der Riemannschen Metrik {\bf G} orthonormierte
Tetrade \bm{\calE}, deren zeitartiges Mitglied {\bf p} ("`{\em charakteristischer
Vektor}"') die 3-dimensionale "`{\em charakteristische Distribution}"' $\Deltadach$
definiert. Diese Distribution induziert ein Ebenenb"undel $\taudach$ ("`{\em
charakteristisches B"undel}"') "uber der Raum-Zeit, dessen 3-dimensionale, innere
Geometrie stets vom konstanten Kr"ummungstyp ist und das zusammen mit dem
charakteristischen Vektorfeld {\bf p} die 4-dimensionale Geometrie der Raum-Zeit
aufbaut. Dabei wirken die raumartigen "Atherfelder \bm{\calB}$_i$ (i = 1,2,3) als
"`Verschwei\3ungsformen"' f"ur die Einbettung des zu $\taudach$ isomorphen SO(3)
B"undels $\tauquer$ ("`{\em repr"asentatives B"undel}"') in das Tangentialb"undel
$\tau_4$ der Raum-Zeit.

Die Tetradenfelder \bm{\calB}$_i$, {\bf p} gen"ugen einem SO(3)-kovarianten System
von Bewegungsgleichungen erster Ordnung, das einerseits die Konstanz der Riemannschen
Metrik {\bf G} gew"ahrleistet und andrerseits den Riemannschen Kr"ummungstensor {\bf
R} eindeutig festlegt. Die Kr"ummung der Raum-Zeit-Mannigfaltigkeit ergibt sich
hierbei als Integrabilit"atsbedingung f"ur die SO(3)-"Atherdynamik.

Die weiteren Variablen {\bf M} der Gravitation gehen in die ersten Ableitungen der
"Atherfelder ein und unterteilen sich in zwei Kategorien: die "au\3eren $({\bf\Mex})$
und die inneren Variablen $({\bf\Min})$. Der Riemannsche Kr"ummungstensor {\bf R} ist
aus den "Atherfeldern \mbox{\bm{\calB},} {\bf p}, der SO(3)-Feldst"arke {\bf F} und
den inneren Variablen ${\bf\Min}$ aufgebaut, enth"alt aber keine "au\3eren Variablen
${\bf\Mex}$. Daher mu\3 --- um ein abgeschlossenes System von Bewegungsgleichungen
f"ur die Gravitationsvariablen zu erhalten --- die "Atherdynamik nur um eine
Bewegungsgleichung f"ur das SO(3)-Eichpotential {\bf A} und f"ur die inneren
Variablen ${\bf\Min}$ erg"anzt werden. Hierbei stellt die Bewegungsgleichung f"ur
{\bf A} eine {\em Wahl} der inneren Geometrie des charakteristischen B"undels dar,
wobei als einschr"ankende Bedingung nur die SO(3)-Bianchi-Identit"at f"ur die
Eichfeldst"arke {\bf F} von {\bf A} zu ber"ucksichtigen ist. Diese Vorgehen f"uhrt zu
einem System von Differentialgleichungen erster Ordnung f"ur das Eichpotential {\bf
A}. Aber auch f"ur die inneren Variablen ${\bf\Min}$ erh"alt man Bewegungsgleichungen
erster Ordnung; und zwar dadurch, da\3 man f"ur diesen Zweck gerade die Einsteinschen
Gleichungen heranzieht, deren neue Bedeutung damit klargelegt ist. Der
Einstein-Tensor {\bf E} enth"alt n"amlich die inneren Variablen ${\bf\Min}$
h"ochstens bis zur ersten Differentiationsordnung, soda\3 also letzten Endes f"ur den
gesamten Variablensatz $\{\mbox{\bm{\calB}},{\bf p},{\bf A},{\bf\Min}\}$ der
Gravitation (au\3er ${\bf\Mex}$, siehe unten) durchweg Bewegungsgleichungen erster
Ordnung gelten. Dies scheint uns ein wichtiger Unterschied zur Einsteinschen Theorie
zu sein, die ja bekanntlich ein System von Bewegungsgleichungen zweiter Ordnung f"ur
die Metrik {\bf G} fordert.

In die gekoppelte Dynamik des "Atherfeldes und des SO(3)-Eichfeldes geht der
Energie/Impuls der Materie nicht direkt ein. Diese Felder bilden demnach ein von der
Materie teilweise abgekoppeltes Untersystem, dessen charakteristische Variabilit"at
den "`schnellen"' "Anderungen der Materiedichte (auf planetarem, stellarem oder sogar
galaktischem Ma\3stab) nicht unbedingt folgen mu\3; f"ur realistische L"osungen der
Bewegungsgleichungen wird man im Newton'schen Sinne f"ur das "Ather- und
SO(3)-Untersystem eine typische Variabilit"at in kosmischen Gr"o\3enordnungen
erwarten ($\leadsto$ Hubble-"`Konstante"', siehe Kapitel \ref{Reduktion}).
Demgegen"uber stellen die inneren Variablen ${\bf\Min}$ die "`schnellen"'
Ver"anderlichen der Theorie dar, die den lokalen Schwankungen der Energie-Dichte der
Materie unmittelbar folgen, da sie "uber die Einstein-Gleichungen ja direkt an die
Materie gekoppelt sind. Auch die "au\3eren Variablen ${\bf\Mex}$ sind zu den
schnellen Variablen zu rechnen. Diese Eigenschaft "ubertr"agt sich sowohl auf die
3-Geometrie der charakteristischen Fl"achen, die von den "au\3eren Variablen
${\bf\Mex}$ zusammen mit dem langsamen SO(3)-Eichfeld {\bf A} aufgebaut wird, als
auch auf den Energie-Impuls von ${\bf\Mex}$\\

\noindent\bm{\beta} {\bf ) Energie-Impuls der Gravitation}
\\

Nachdem an die Stelle des metrischen Feldes mehrere andere Variable getreten sind,
fallen auch die Schwierigkeiten bei der Definition der Energie-Impulsdichte der
Gravitations-Wechselwirkung mit Hilfe der Metrik weg. Es ist klar, da\3 eine (oder
mehrere) der Variablen als Tr"ager der Gravitations-Energie in Frage kommen k"onnen.
Insbesondere bieten sich hier die "au\3eren Variablen ${\bf\Mex}$ an, die zwar nicht
unmittelbar in die 4-dimensionale Riemannsche Geometrie der Raum-Zeit eingehen, deren
Energie-Impulsinhalt ${\bf\Te}$ aber auf der rechten Seite der Einstein-Gleichungen
trotzdem die Raum-Zeit-Struktur beeinflussen kann. In der Tat zeigt eine Untersuchung
der einfachsten Feldkonfiguration ("`Grundzustand"' des leeren Universums $\leadsto$
de Sitter-Raum), da\3 ein kanonischer Formalismus mit der zugeh"origen Definition des
Energie-Impuls-Tensors f"ur das gekoppelte Yang-Mills-Higgs-System aus "Ather- und
Eichfeldern m"oglich ist. Obwohl hierbei die gesuchte Energie-Impulsdichte formal
durch die "Ather- und Eichfelder ausgedr"uckt werden kann, zeigt eine genauere
Betrachtung, da\3 der zugeh"orige Tensor ${\bf\Te}$ genau dann verschwindet, wenn
auch die "au\3eren Variablen ${\bf\Mex}$ verschwinden, wobei die altbekannte Form der
Einstein-Gleichungen zur"uckgewonnen wird. Dies kann als eine Rechtfertigung
daf"ur angesehen werden, den Energie-Impulsinhalt ${\bf\Te}$
den "au\3eren Variablen ${\bf\Mex}$ zuzuschreiben, was auch durch die M"oglichkeit
unterst"utzt wird, ${\bf\Te}$ vollst"andig durch die "au\3eren Variablen ${\bf\Mex}$
auszudr"ucken. Dabei besteht der "au\3ere Feldfreiheitsgrad der Gravitation in einem
SO(3) Eichvektor ${\bf C} = \{C_{i\mu}\}$, den die vorliegende Gravitationstheorie
gegen"uber der konventionellen Einsteinschen Version {\em zus"atzlich} aufweist. Dies
bedeutet, da\3 der Einsteinschen Gravitation als Spezialfall innerhalb der
vorliegenden SO(3)-Eichtheorie kein Energie-Impulsinhalt zukommt\footnote{Auf die
zahllosen Versuche in der Literatur, f"ur die Gravitation innerhalb des Rahmens der
Einsteinschen Theorie eine Energie-Impulsdichte zu definieren, wird hier nicht
eingegangen.}\\

\noindent\bm{\gamma} {\bf ) Das Vakuum}
\\

Entfernt man alle Materie und Strahlung aus dem Universum, so bleibt der leere Raum
zur"uck, der allerdings eine nicht-triviale Riemannsche Geometrie tragen kann
("`klassisches Vakuum"'). Mit dieser Riemannschen Struktur kann nach den oben
gemachten Bemerkungen ein nicht-verschwindender Energie-Impuls verkn"upft sein
($\leadsto$ "`Vakuum-Energie"'). Abweichend von den urspr"unglichen Einsteinschen
Vorstellungen nimmt man heute an, da\3 der Einflu\3 der Vakuum-Energie auf die
Raum-Zeit-Geometrie durch den sogenannten "`kosmologischen Term"' beschrieben wird
\cite{Wein89}. Die Vakuumenergie selbst wird mit der Nullpunkts-Energie der
verschiedenen Quantenfelder identifiziert. Unter allen m"oglichen
Vakuum-Konfigurationen kann man diejenigen mit der gr"o\3tm"oglichen Symmetrie als
den "`Grundzustand"' des leeren Universums betrachten ($\leadsto$ de Sitter-Raum)

Mit einiger Befriedigung k"onnen wir zu dieser Problematik anmerken, da\3 sich in
unserer SO(3)-Eichtheorie genau dieser Grundzustand mit sehr wenigen Annahmen
ergibt. Die zugeh"orige Vakuum-Energie entspricht gerade dem Energieinhalt der
"au\3eren Variablen ${\bf\Mex}$ in ihrer Vakuum-Konfiguration. Jedoch unterscheiden
sich die angeregten Vakuum-Zust"ande in der vorliegenden Theorie ganz entscheidend
von denjenigen der (um den kosmologischen Term erweiterten) Einsteinschen Theorie, da
in der letzteren der Vakuum-Tensor stets proportional zur Metrik {\bf G} ist.
(Proportionalit"atsfaktor = "`Kosmologische Konstante"'). Diese eingeschr"ankte
Gestalt des Vakuum-Tensors ist aber aus der Sicht eines hierarchischen,
kosmologischen Modells nicht haltbar, da sich die verschiedenen Teilbereiche des
Universums in jeweils unterschiedlichen Entwicklungsphasen bez"uglich der Expansion
oder Kontraktion befinden k"onnten und somit die "`kosmologische Konstante"' nicht
"uberall im Universum denselben Wert annehmen kann. Andererseits ist eine
raumzeitlich variable kosmologische Konstante mit den Einsteinschen Gleichungen nicht
vertr"aglich, soda\3 aus unserer Sicht das Vakuum-Problem im Rahmen der Einsteinschen
Theorie nur eine unbefriedigende L"osung erf"ahrt.

F"ur die SO(3)-Eichtheorie der Gravitation wird zun"achst einmal der Grundzustand des
leeren Universums eingehend untersucht. Zur Disposition stehen hier zwei Kanditaten
mit unterschiedlich hoher Symmetrie, die aber beide dem "`perfekten kosmologischen
Prinzip"' gen"ugen: (i) der de Sitter-Raum als eine 4-dimensionale Pseudosph"are
$S^4_{(1)}$ im 5-dimensionalen flachen Raum ($\leadsto$ Unabh"angige
Symmetrie-Rotationen $\frac{1}{2}\cdot 5\cdot 4 = 10$) und (ii) die
pullback-Geometrie einer 3-Kugel $S^3_{(-1)}$ im 4-dimensionalen flachen Raum
($\leadsto$ Symmetrie-Operationen $\frac{1}{2}\cdot 4\cdot 3 = 6$). Obwohl der
$S^3_{(-1)}$-Kanditat eine sehr enge geometrische Verwandtschaft zum flachen
Minkowski-Raum aufweist (er entsteht aus diesem durch Foliation), geben wir doch dem
$S^4_{(1)}$-Kanditaten den Vorzug, weil der zugeh"orige Vakuum-Tensor aufgrund der
hohen Symmetrie proportional zur Metrik {\bf G} wird und somit wie ein
"`kosmologischer"' Term wirkt. Innerhalb dieser Konfiguration gibt es aber immer noch
zwei M"oglichkeiten, n"amlich die Expansion und die Kontraktion des Universums. Durch
eine Stabilit"atsanalyse kann gezeigt werden, da\3 nur die Expansion stabil ist,
w"ahrend die kontraktive Phase nach einigen Planck-Zeiten notwendigerweise in einer
Singularit"at endigen mu\3. Dieses Resultat kann auch als eine Erkl"arung f"ur die
Fixierung der (kosmischen) Zeitrichtung und f"ur die heute zu beobachtende Expansion
des Universums betrachtet werden.


\chapter{Die Reduktion des GL(4,{\bf R})-B"undels}
\indent \label{Reduktion}

Einsteins urspr"ungliche Version seiner "`Allgemeinen Relativit"atstheorie"'
basiert auf einer Verallgemeinerung des Relativit"atsprinzips der "`Speziellen
Relativit"atstheorie"'. Dort wird das physikalische Raum-Zeit Kontinuum als eine
4-Mannigfaltigkeit ${\bf M}_4$ beschrieben, deren triviales Tangentialb"undel $\taukr$
mit der {\em Minkowski Metrik} {\bf g} und der kanonischen flachen Konnexion
${\bf\Gammakr}$ ausgestattet ist. Diese besondere Form der Raum-Zeit Geometrie
erlaubt die Einf"uhrung eines speziellen Bezugssystem \bm{\calE} welches geometrisch
gesehen bez"uglich ${\bf\Gammakr}$ einen globalen kovariant konstanten, orthonormalen
Schnitt des zu $\taukr$ assozierten Prinzipalb"undels $\lambdakr$, darstellt. Dieses
Bezugssystem gew"ahrleistet nun die Existenz einer speziellen Klasse von
Koordinatensystemen ({\em Inertialsysteme}) von ${\bf M}_4$, deren Basisvektoren
$\left(\right.$\bm{\partial}$_\mu\left.\!\right)$ gerade mit den Tetradenvektoren
$\left(\right.$\bm{\calE}$_\mu\left.\!\right)$ von \bm{\calE}
"ubereinstimmen:
\alpheqn
\begin{eqnarray}
\label{calEpartial1}
\mbox{\bm{\calE}}_\mu\equiv\mbox{\boldmath$\partial$}_\mu &=&
{\calE^\alpha}_\mu\,\mbox{\boldmath$\partial$}_\alpha\\
\label{g2calEg}
g_{\alpha\beta}\,{\calE^\alpha}_\mu{\calE^\beta}_\nu &=& g_{\mu\nu}\ .
\end{eqnarray}
\reseteqn
Die Bedeutung dieser Raum-Zeit Geometrie f"ur die Physik in ${\bf M}_4$ wird durch
das {\em Relativit"atsprinzip} beschrieben, welches besagt, da\3 die physikalischen
Gesetze (Bewegungsgleichungen, Erhaltungsgesetze usw.) forminvariant bez"uglich
einer (zul"assigen) "Anderung \mbox{\bm{\calE}$\to$\bm{\calE}$^\prime$} des globalen Bezugssystem sein
m"ussen:
\begin{equation}
\mbox{\boldmath$\calE$}^\prime_\mu = \mbox{\boldmath$\calE$}_\nu\,{\Lambda^\nu}_\mu\ .
\label{ELambdaEstr}
\end{equation}
Hierbei ist $\Lambda$ ein Element der (eigentlichen) Lorentzgruppe SO(1,3), welche
aus diesem Grund die fundamentale Symmetriegruppe der speziellen Relativit"atstheorie
ist.

Trotz des gro\3en Erfolges dieser Theorie, beispielsweise in der Beschreibung der
elektromagnetischen Ph"anomene, war Einstein nie ganz zufrieden, da er keine
M"oglichkeit sah, wie die Gravitationswechselwirkung in konsistenter Weise in die
spezielle Relativit"atstheorie einzuf"ugen w"are. Aus diesem Grund schien es ihm
unvermeidbar zu sein, die Theorie durch eine Verallgemeinerung der Symmetriegruppe
SO(1,3) abzu"andern. Da alle wichtigen physikalischen Objekte (z.B. 4-Impuls,
4-Kraft, Energie-Impulsdichte usw.) im Tangentialb"undel $\tau_4$ der Raum-Zeit (oder
dessen assozierten B"undeln) liegen, wo ${\cal GL}^+(4,{\bf R})$ als die
korrespondierende Strukturgruppe wirkt, ist es naheliegend diese Gruppe als die
externe Symmetriegruppe f"ur eine allgemeinere Theorie zu ben"utzen, welche die
Gravitation mitber"ucksichtigt. Sobald man f"ur die physikalischen Gesetze die neue
Symmetriegruppe gew"ahlt hat, ist die Bestimmung der zugrundeliegenden geometrischen
Struktur nahezu eindeutig: Die Bedingung \rf{ELambdaEstr} f"ur "aquivalente
Bezugssysteme wird dahingehend abgeschw"acht, da\3 das Element $\Lambda$ nun aus der
neuen Strukturgruppe GL$^+$(4,{\bf R}) stammt, was zum verallgemeinerten
Relativit"atsprinzip f"uhrt. Man sieht ferner, da\3 solch eine Vergr"o\3erung der
Symmetriegruppe die Verwendung der oben ausgezeichneten Bezugssysteme verbietet und
die Metrik {\bf g} \rf{g2calEg} durch ein allgemeineres Objekt ${\bf G}({\bf x})$
ersetzt werden mu\3, welches einen globalen Schnitt des Quotientenb"undels
(Strukturgruppe GL$^+$(4,{\bf R})/SO(1,3)) von $\mbox{R}^1$-wertigen bilinearen,
symmetrischen Abbildungen des Tangentialraumes der Raum-Zeit Mannigfaltigkeit
darstellt:
\begin{equation}
G_{\mu\nu} = g_{\alpha\beta}\,{\calE^\alpha}_\mu{\calE^\beta}_\nu\ .
\label{Gg2calE}
\end{equation}
Die Tetradenkonstituenten \bm{\calE}$^\alpha$ k"onnen als ein (nicht notwendigerweise
globaler) Schnitt des Tangentialb"undels $\tau_4$ der Raum-Zeit betrachtet werden.
Der Schnitt {\bf G} \rf{Gg2calE} mu\3 nat"urlich von einer Pseudo-Riemannschen
Struktur herr"uhren, das bedeutet da\3 {\bf G} global durch einen geeigneten Wechsel
des Bezugssystems \bm{\calE} in die Minkowski Metrik {\bf g} transformiert werden
kann, so da\3 die Relation \rf{g2calEg}, aber nicht \rf{calEpartial1}, g"ultig
bleibt. Dies ist ein etwas heikler Punkt, da nicht jede 4-Mannigfaltigkeit solch eine
Pseudo-Riemannsche Struktur zul"asst. Eine notwendige und hinreichende Bedingung
hierf"ur ist, da\3 die Strukturgruppe GL$^+$(4,{\bf R}) des Tangentialb"undels auf
ihre Untergruppe SO(1,3) reduzierbar ist. Mit anderen Worten, das Tangentialb"undel
$\tau_4$ mu\3 die Errichtung eines B"undelatlases erlauben, dessen
"Ubergangsfunktionen ausschlie\3lich SO(1,3)-wertig sind.~\footnote{Es entstehen keine
Probleme, wenn man eine {\em Riemannsche} Struktur einf"uhren will, da GL$^+$(4,{\bf
R}) immer auf ihre maximale kompakte Untergruppe SO(4) reduzierbar ist.} Diese
geometrische Forderung, welche der Raum-Zeit Struktur auferlegt werden mu\3,
erm"oglicht physikalisch gesehen gerade das "Aquivalenzprinzip (siehe weiter unten).

Nachdem die Raum-Zeit Mannigfaltigkeit mit der Metrik {\bf G} ausgestattet ist,
taucht die Frage nach einer ${\cal GL}(4,{\bf R})$-Konnexion ${\bf\Gamma}$ auf,
welche die Verallgemeinerung der kanonisch flachen Konnexion ${\bf\Gammakr}$
darstellt und mit der man die physikalischen (und geometrischen) Objekte
differenzieren kann. Es liegt nun kein Grund vor, warum die Metrik {\bf G} kovariant
konstant bez"uglich dem gew"ahlten ${\bf\Gamma}$ sein sollte, vielmehr wird man
folgende Relation ansetzen:
\begin{equation}
\nabla_\lambda\,G_{\mu\nu} = -2Q_{\mu\nu\lambda}\ ,
\label{nablaGQ}
\end{equation}
wobei {\bf Q} den bis jetzt noch nicht n"aher spezifizierten {\em
Nicht-Metrizit"atstensor} bezeichnet. L"osen wir diese Gleichung nach der Konnexion
${\bf\Gamma}$ auf, so ergibt sich
\begin{equation}
{\Gamma^\sigma}_{\nu\lambda} = {\Gammaschl^\sigma}_{\nu\lambda} + {K^\sigma}_{\nu
\lambda} + {Q^\sigma}_{\nu\lambda}\ ,
\label{GamGamschlpKQ}
\end{equation}
wobei die Kontorsion {\bf K} ihre Werte in der Lie-Algebra ${\cal SO}(1,3)$ der
Lorentzgruppe SO(1,3) annimmt
\begin{equation}
{K^\sigma}_{\nu\lambda} = {Z^\sigma}_{\nu\lambda} - {{Z_\nu}^\sigma}_\lambda +
{Z_{\lambda\nu}}^\sigma + {{Q_\lambda}^\sigma}_\nu - {Q_{\lambda\nu}}^\sigma
\label{KZQ}
\end{equation}
und die Torsion {\bf Z} von ${\bf\Gamma}$ wie gew"ohnlich definiert ist als
\begin{equation}
{Z^\sigma}_{\nu\lambda} = {\Gamma^\sigma}_{[\nu\lambda]}\ .
\label{ZGamma}
\end{equation}
Die Kontorsion und die Nicht-Metrizit"at f"uhren somit zu einer schief-symmetrischen
Korrektur der Levi-Civita Konnexion ${\bf\Gammaschl}$ von {\bf G}
\begin{equation}
{\Gammaschl^\sigma}_{\nu\lambda} = \frac{1}{2}G^{\sigma\mu}\left(\partial_\lambda
G_{\mu\nu} + \partial_\nu G_{\mu\lambda} - \partial_\mu G_{\nu\lambda}\right)\ .
\label{GammaschlG}
\end{equation}
Auf diese Weise kommt man zu einer allgemeinen geometrischen Struktur der Raum-Zeit
Mannigfaltigkeit; und es war nun Einsteins geniale Vermutung, da\3 diese
Verallgemeinerungs-Prozedur gerade um einen Schritt zu weit getrieben wurde.

Die gesuchte einschr"ankende Bedingung f"ur die allgemeine Raum-Zeit Geometrie
erh"alt man durch ein physikalisches Argument, n"amlich das {\em "Aquivalenzprinzip}.
Laut diesem Prinzip forderte Einstein, da\3 die Gravitation eine fiktive Kraft ist,
welche nur aufgrund eines ungeeigneten Bezugssytems entsteht. Das bedeutet, da\3 man
die Gravitationskraft lokal immer zum Verschwinden bringen kann, vorausgesetzt es
wird ein lokales Inertialsystem zugrunde gelegt, in welchem die physikalischen
Gesetze ihre speziell-relativistische Form annehmen, wobei dann das spezielle
Relativit"atsprinzip {\em lokal} gilt. Aus diesem Grund kann man nun die Menge aller
lokalen Inertialsysteme ben"utzen, um ein bis auf eine Lorentztransformation
eindeutiges Bezugssystem $\hat{\mbox{\bm{\calE}}}$ zu konstruieren, welches
\rf{g2calEg}, aber nicht unbedingt \rf{calEpartial1}, gen"ugt. Man beachte, da\3
diese physikalische Forderung mit der geometrischen Bedingung f"ur die Existenz einer
Pseudo-Riemannschen Struktur "uber der Raum-Zeit Mannigfaltigkeit vertr"aglich ist
(vgl. die Diskussion nach Gleichung \rf{Gg2calE}).

Die Philosophie des "Aquivalenzprinzips mu\3 nun in eine mathematische Bedingung f"ur
die allgemeine Konnexion ${\bf\Gamma}$ \rf{GamGamschlpKQ} gebracht werden. Dazu
beachte man, da\3 die physikalischen Gr"o\3en bez"uglich des oben erw"ahnten
Lorentz'schen Bezugssystems $\hat{\mbox{\bm{\calE}}}$ als SO(1,3) Eichobjekte
erscheinen, welche mit ${\bf\Gamma}$ parallel transportiert werden k"onnen und zwar
auf eine Weise, welche die lokalen Lorentzinvarianten (z.B. Ruhemasse der Teilchen,
Spin usw.) unbeeinflu\3t l"a\3t und sie deshalb in der gesamten Raum-Zeit als globale
Invarianten darstellt. Daraus l"a\3t sich folgern, da\3 die Endposition eines solchen
Objektes, wenn man es entlang einer geschlossenen Kurve parallel transportiert, im
Lorentz-Orbit, welcher durch die Startposition geht, enthalten sein mu\3. Diese
Forderung bedeutet, da\3 die Kr"ummungs-2-Form {\bf R} von ${\bf\Gamma}$
\begin{equation}
{R^\sigma}_{\nu\lambda\mu} = \partial_\lambda{\Gamma^\sigma}_{\nu\mu} - \partial_\mu
{\Gamma^\sigma}_{\nu\lambda} + {\Gamma^\sigma}_{\rho\lambda}{\Gamma^\rho}_{\nu\mu} -
{\Gamma^\sigma}_{\rho\mu}{\Gamma^\rho}_{\nu\lambda}
\label{RGamma}
\end{equation}
bez"uglich eines holonomen Bezugssystems ihre Werte in der Lie-Algebra ${\cal SO}(1,3)$
der Lorentz\-gruppe annimmt. Das hei\3t, da\3 die Konnexion ${\bf\Gamma}$ die
Lorentzgruppe als Holonomie-Gruppe besitzt. Aus diesem Grund wird nach dem Theorem
von Ambrose-Singer \cite{KoNo69} die Gruppe SO(1,3) zur Strukturgruppe des
Holonomie-Unterb"undels des Tangentialb"undels $\tau_4$ der Raum-Zeit. Die notwendige
und hinreichende Bedingung daf"ur, da\3 die Lorentzgruppe die Holonomie-Gruppe der
Konnexion ${\bf\Gamma}$ ist, besteht im Verschwinden der Nicht-Metrizit"at {\bf Q}
\rf{nablaGQ}
\begin{equation}
Q_{\mu\nu\lambda} \equiv 0\ .
\label{Q0}
\end{equation}
Dies kann man leicht durch Betrachtung des symmetrischen Teils der ${\cal GL}(4,{\bf
R})$-wertigen Kr"ummungs 2-Form {\bf R} von ${\bf\Gamma}$ einsehen:
\begin{equation}
R_{(\mu\nu)\sigma\lambda} = -2{Z^\rho}_{\sigma\lambda}\,Q_{\mu\nu\rho} +
\left(\nabla_\sigma Q_{\mu\nu\lambda} - \nabla_\lambda Q_{\mu\nu\sigma}\right)\ .
\label{RZnablaQ}
\end{equation}

Das "Aquivalenzprinzip hat uns somit zu einer {\em metrischen Theorie der
Gravitation} gef"uhrt, welche jedoch nicht notwendigerweise zu einer Riemannschen
Raum-Zeit f"uhrt, wie Einstein urspr"unglich vermutete. Der Grund liegt darin, da\3
wir eine nicht verschwindende Torsion {\bf Z} in der Theorie haben. An diesem Punkt
gibt es nun zwei verschiedene M"oglichkeiten der weiteren Vorgehensweise:
\begin{enumerate}
\item[i)]
Die Verfechter der Torsionstheorien \cite{He76,He76_b} argumentieren, da\3 die
Torsion {\bf Z} von ${\bf\Gamma}$ in der Theorie bleiben mu\3, damit der Spineinflu\3
der Materie auf die Raum-Zeit, "ahnlich wie derjenige der Masse, ber"ucksichtigt
werden kann. Das bedeutet, da\3 die physikalische Raum-Zeit-Mannigfaltigkeit eine
Riemann-Cartan Struktur besitzt. Ein starkes Argument zugunsten dieses Gesichtpunktes
ist die Tatsache, da\3 die Einstein-Gleichungen f"ur eine Riemannsche Raum-Zeit
inkonsistent werden, wenn sie auf Materie mit Spin angewendet werden.
\cite{He76,He76_b}
\item[ii)]
Andererseits kann man auf Einsteins urspr"unglicher Version des "Aquivalenzprinzips
bestehen, welches besagt, da\3 die Raum-Zeit lokal wie ein Minkowski Raum
\{${\bf\Gammakr}$, {\bf g}\} aussehen soll. Diese Forderung f"uhrt jedoch dazu,
da\3 man die Torsion identisch Null setzen mu\3 wie man leicht aufgrund folgender
Argumentation sieht: F"uhren wir die Metrik {\bf G} durch ein geeignetes
{\em holonomes} Bezugssystem $\left\{\right.$\bm{\partial}$_\mu\left.\right\}$ lokal
in die Standardform des Minkowski Tensors {\bf g} $=\mbox{diag}\left(1,-1,-1,-1\right)$
"uber, dann verschwindet in einer Torsionstheorie die Riemann-Cartan Konnexion
${\bf\Gamma}$ im allgemeinen nicht! Deshalb wird die flache kanonische Konnexion
${\bf\Gammakr}$ eines Minkowski Raumes nicht einmal {\em lokal} durch ${\bf\Gamma}$
angen"ahert. Versucht man umgekehrt die Riemann-Cartan Konnexion ${\bf\Gamma}$ durch
"Ubergang zu einem anderen holonomen Bezugssystem $\hat{\mbox{\bm{\calE}}}$
\begin{equation}
\hE_\alpha = {\hE^\mu}_\alpha\,\mbox{\boldmath$\partial$}_\mu
\label{2calEpartial}
\end{equation}
zum Verschwinden zu bringen, was aufgrund des inhomogenen Transformationsgesetzes
f"ur die Konnexion immer m"oglich ist:
\begin{eqnarray}
{\omega^\beta}_{\alpha\nu} &=& {\hF^\beta}_\mu\,\partial_\nu{\hE^\mu}_\alpha +
{\hF^\beta}_\lambda\,{\Gamma^\lambda}_{\mu\nu}\,{\hE^\mu}_\alpha\\
{\hF^\beta}_\mu\,{\hE^\mu}_\alpha &=& {g^\beta}_\alpha\nonumber\ ,
\end{eqnarray}
so nimmt die Metrik {\bf G} im allgemeinen nicht die Form an, welche man von einer
flachen Minkowski Metrik {\bf g} erwartet!
\end{enumerate}

Obwohl die Torsionstheorien bis heute eingehend untersucht wurden und anscheinend
frei von Inkonsistenzen sind, werden wir jedoch die zweite, oben aufgef"uhrte
M"oglichkeit untersuchen. Es scheint, als ob das "Aquivalenzprinzip nur in einer
torsionsfreien Theorie in seinem urspr"unglichen Sinn realisiert werden kann. Da aber
davon ausgegangen wird, da\3 der Spin der Materie f"ur das Vorhandensein von Torsion
verantwortlich ist,
m"ussen wir einen Mechanismus finden, welcher den Einflu\3 desselben
auf die Raum-Zeit Geometrie verhindert ({\em Spinkompensation}). Ein solches
Kompensationsprinzip ergibt sich zwanglos, wenn man die reduzierte Strukturgruppe
SO(1,3) vollends zur gew"ohnlichen Rotationsgruppe SO(3) reduziert. Diese
Reduktionsprozedur kann man als Hinweis auf eine spontane Symmetriebrechung des
Raum-Zeit-Kontinuums verstehen. Einige Autoren hat die vorausgehende
Symmetriebrechung Gl(4,{\bf R}) $\to$ SO(1,3) dazu veranlasst das Gravitationsfeld
{\bf G}({\bf x}) als Higgs-Goldstone Feld aufzufassen \cite{IvSa83}. Allerdings ist
die reduzierte Strukturgruppe SO(1,3) nicht kompakt. Aus diesem Grund ist eine
weitere Reduktion des reduzierten Prinzipalb"undels $\hat{\lambda}_4$ immer m"oglich
\cite{Iwa49} und zwar so, da\3 die Strukturgruppe dieses kleinsten B"undels
$\lambdaquer$ gerade identisch der maximalen kompakten Untergruppe SO(3) der
Lorentzgruppe ist. Somit zeigt sich, da\3 die gew"ohnliche Rotationsgruppe in drei
Dimensionen die kleinstm"ogliche Strukturgruppe f"ur die Allgemeine
Relativit"atstheorie ist. F"ur $\tau_4$ existiert deshalb ein B"undelatlas dessen
"Ubergangsfunktionen SO(3)-wertig sind. Die Holonomie-Gruppe der Konnexion
${\bf\Gamma}$ bleibt jedoch die Lorentzgruppe SO(1,3) wie es das "Aquivalenzprinzip
vorschreibt.

\vspace{2mm}
{\bf Wir fassen nun die Reduktion zu SO(3) nicht nur als einen mathematischen Kunstgriff 
auf, sondern wir nehmen an, da\3 die   physikalische Raum-Zeit tats"achlich
einer spontanen Symmetrie\-brechung unterworfen ist, (unabh"angig von der Anwesenheit
von Materie)}. 
\vspace{2mm}

Dies bedeutet, da\3 wir als dynamische Variable einen globalen Schnitt
{\bf p} des Quotientenb"undels (Strukturgruppe SO(1,3)/SO(3)) verwenden werden,
welcher zusammen mit der $\tau_4$ Fasermetrik {\bf G} die Symmetriebrechung zu SO(3)
induziert und dabei einer Bewegungsgleichung gen"ugt, welche die lokale Foliation der
Raum-Zeit auf dynamische Weise beschreibt. Die Fasermetrik {\bf G} aus $\tau_4$,
deren Existenz wir hier voraussetzen (siehe unten), stellt sicher, da\3 der
zeitartige Schnitt {\bf p} ({\em charakteristisches Vektorfeld}) aus $\tau_4$ auf
Eins normiert werden kann:
\begin{equation}
G_{\mu\nu}\,p^\mu p^\nu \equiv p_\mu p^\mu = +1\ .
\label{Gpp1}
\end{equation}
Das charakteristische Vektorfeld {\bf p} definiert also eine bestimmte
3-Distribution in der Raum-Zeit: die sogenannte {\em charakteristische Distribution}
$\Deltadach$. {\bf p} und $\Deltadach$ definieren nun
eine lokale Zerlegung der Raum-Zeit in einen
3-Raum und eine Zeit\-richtung, wobei diese (1+3)-Zerlegung aufgrund der
Bewegungsgleichung f"ur den charakteristischen Vektor {\bf p} als eine dynamische
Gr"o\3e betrachtet werden mu\3. Um die weiteren dynamischen Variablen zu definieren
beachten wir, da\3 sich die charakteristische Distribution $\Deltadach$ durch drei
orthonormale Schnitte \bm{\calB_i} (i,j,k = 1\ldots 3) von $\tau_4$ aufspannen
l"a\3t:
\alpheqn
\begin{eqnarray}
\label{2calBg}
\calB_{i\mu}\,{\calB_j}^\mu &=& g_{ij}\\
\label{calBp0_1}
\calB_{i\mu}p^\mu &=& 0\ ,
\end{eqnarray}
\reseteqn
so da\3 ihr Quadrat \bm{\calB}$^2$ senkrecht zu {\bf p} projiziert
\alpheqn
\begin{eqnarray}
\label{Proj1}
\calB_{i\mu}\,{\calB^i}_\nu &=& \calB_{\mu\nu}\\
\label{Proj2}
\calB_{\mu\nu}\,{\calB^\nu}_\lambda &=& \calB_{\mu\lambda}\\
\calB_{\mu\nu}\,p^\nu &=& 0\ .
\label{calBp0_2}
\end{eqnarray}
\reseteqn
Hierbei k"onnen wir nun umgekehrt wir die Riemannsche Metrik {\bf G} folgenderma\3en
aus den SO(3) Objekten {\bf p}, \bm{\calB_i} aufbauen, die wir dann als die
eigentlichen dynamischen Variablen der Gravitation ansehen wollen:
\begin{equation}
G_{\mu\nu} = \calB_{\mu\nu} + p_\mu p_\nu\ .
\label{GcalBpp}
\end{equation}

Nach dieser Philosophie verliert also die Metrik {\bf G} in der neuen
Gravitationstheorie ihre fundamentale Bedeutung und erscheint nur noch als ein
sekund"ares makroskopisches Objekt, welches sich aus den eigentlichen fundamentalen
Gr"o\3en \bm{\calB_i} und {\bf p} zusammensetzt, die wir im folgenden als {\em
"Atherfelder} bezeichnen wollen. Die dynamischen Gleichungen f"ur das
Gravitationsfeld {\bf G} beziehen sich deshalb auf dessen mikroskopische
Konstituenten \bm{\calB_i} und {\bf p}, wobei die Gleichungen \rf{Gpp1} bis
\rf{calBp0_2} Zwangsbedingungen darstellen, welche mit den Bewegungsgleichungen
vertr"aglich sein m"ussen. Eine weitere Einschr"ankung ist die Metrizit"atsbedingung
\begin{equation}
\nabla_\lambda G_{\mu\nu}\equiv 0\ ,
\label{nablaG0}
\end{equation}
welche allein jedoch nicht ausreicht um die rein Riemannsche Natur der Raum-Zeit zu
garantieren, wie man aus den obigen Argumenten erkennen kann. Aus diesem Grund mu\3
die mikroskopische Dynamik so aufgebaut sein, da\3 der Spinkompensationseffekt
auftritt. Ein einfaches Beispiel daf"ur werden wir in Kapitel \ref{Spinkomp}
kennenlernen.

Abgesehen von dieser letzteren Bedingung kann die allgemeine Form der mikroskopischen
Gleichungen leicht angegeben werden. Dazu definieren wir f"ur das raumartige
"Atherfeld \bm{\calB_i} eine (eich- und koordinaten-) kovariante Ableitung \bm{\calD}
durch
\begin{eqnarray}
\label{DefcalD}
\calD_\mu\calB_{i\nu} &=& \nabla_\mu\calB_{i\nu} + {\epsilon_i}^{jk}A_{j\mu}
\calB_{k\nu}\\
&=& D_\mu\calB_{i\nu} - {\Gamma^\lambda}_{\nu\mu}\calB_{i\lambda}\nonumber\ ,
\end{eqnarray}
wobei \bm{\nabla} ({\bf D}) die koordinaten- (eich-) kovariante Ableitung bezeichnet.
Die dynamischen Gleichungen f"ur die "Atherfelder haben nun die folgende Form:
\alpheqn
\begin{eqnarray}
\label{Dgl1}
\calD_\mu\calB_{i\nu} &=& \calH_{i\nu\mu}\left(\mbox{\boldmath$\calB$},{\bf p},
{\bf M}\right)\\
\label{Dgl2}
\nabla_\mu\,p_\nu&=&H_{\nu\mu}\left(\mbox{\boldmath$\calB$},{\bf p},{\bf M}\right)\ .
\end{eqnarray}
\reseteqn
Die Objekte \bm{\calH_i} und {\bf H} auf der rechten Seite m"ussen als Funktionen der
"Atherfelder und zus"atzlicher Feldvariablen {\bf M} betrachtet werden, welche mit
den "Atherfeldern "uber eine Gleichung erster Ordnung gekoppelt sind
\begin{equation}
\calD_\mu{\bf M} = \calM_\mu\left(\mbox{\boldmath$\calB$},{\bf p},{\bf M}\right)\ ,
\label{Dgl3}
\end{equation}
die selbst wiederum aus den Einsteinschen Gleichungen abzuleiten ist. Dadurch
behalten die Einsteinschen Gleichungen nach wie vor ihre G"ultigkeit (siehe unten).
Den Gleichungen \rf{Dgl1}, \rf{Dgl2} und \rf{Dgl3} m"ussen wir nun noch einen
Ansatz f"ur die SO(3) Feldst"arke {\bf F$_i$} hinzuf"ugen
\begin{equation}
F_{i\mu\nu} = {\cal F}_{i\mu\nu}\left(\mbox{\bm{\calB}},{\bf p},{\bf M}\right)\ ,
\label{Dgl4}
\end{equation}
der konsistent ist mit der Bianchi Identit"at
\begin{equation}
\calD_\lambda F_{i\mu\nu}+\calD_\mu F_{i\nu\lambda}+\calD_\nu F_{i\lambda\mu}=0\ .
\label{3calDF}
\end{equation}

Die Gleichungen \rf{Dgl1} bis \rf{Dgl4} stellen nun ein geschlossenes System erster
Ordnung dar, welches die mikroskopische Dynamik von gravitierender Materie bestimmt.
Im Falle der Abwesenheit von Materie liegt ein Universum vor, dessen Gravitation nur
auf sich selbst wirkt. Diesen Fall werden wir in Kapitel \ref{SelbstWW} anhand eines
Beispiels untersuchen. Die Objekte \bm{\calH_i} und {\bf H} m"ussen nun so gew"ahlt
werden, da\3 die Bedingungen \rf{Gpp1} bis \rf{calBp0_2} und \rf{nablaG0} f"ur die
"Atherfelder durch die mikroskopische Dynamik \rf{Dgl1} und \rf{Dgl2} erf"ullt
werden, wobei sich zus"atzlich eine verschwindende Torsion $({\bf Z} = 0)$ ergeben
mu\3. Dies bedeutet aber keineswegs, da\3 das Ph"anomen der Torsion in unserer
Theorie "uberhaupt keine Rolle spielen wird. Ganz im Gegenteil: Wir werden zeigen,
da\3 die in \rf{DefcalD} auftretende ${\cal SO}$(3)-Konnexion {\bf A} zu einer
Torsion f"uhrt, die in engem Zusammenhang mit den Spindichten der "Atherfelder
\bm{\calB}, {\bf p} steht. Allerdings handelt es sich hierbei um eine 3-dimensionale
Oberfl"achentorsion, die den streng Riemannschen Charakter der 4-dimensionalen
Raum-Zeit unber"uhrt l"a\3t.

Durch eine einfache Rechnung erhalten wir aus der oben erw"ahnten Forderung folgende
drei Bedingungen
\alpheqn
\begin{eqnarray}
\label{Bed1}
H_{\mu\nu}\,p^\mu &=& 0\\
\label{Bed2}
\calH_{i\mu\lambda}\,p^\mu + \calB_{i\mu}\,{H^\mu}_\lambda &=& 0\\
\label{Bed3}
\calH_{i\nu\lambda}\,{\calB_j}^\nu + \calH_{j\nu\lambda}\,{\calB_i}^\nu &=& 0\ .
\end{eqnarray}
\reseteqn
Aus den ersten zwei Bedingungen \rf{Bed1} und \rf{Bed2} erkennt man, da\3 {\bf H}
vollst"andig durch \bm{\calH_i} und die "Atherfelder bestimmt wird:
\begin{equation}
H_{\nu\lambda} = -{\calB^i}_\nu\,\calH_{i\mu\lambda}\,p^\mu\ .
\label{HcalBcalHp}
\end{equation}

Der Tensor ${\bf H} = \{H_{\nu\lambda}\}$, den wir im folgenden als {\em
Hubble-Tensor} bezeichnen werden, bestimmt die "au\3ere Geometrie der
charakteristischen Distribution $\Deltadach$. Im Falle eines symmetrischen Tensors
$(H_{\mu\nu}=H_{\nu\mu})$ ist der charakteristische Vektor ein Gradient $(p_\mu =
\partial_\mu\theta)$ eines skalaren Feldes $\theta$, welches man als "`{\em
Universalzeit}"' betrachten kann. Die charakteristische Distribution $\Deltadach$
wird in diesem Fall integrabel und die Integralfl"achen k"onnen als der "`{\em
absolute Raum}"' angesehen werden. Dabei erhalten die {\em charakteristischen
Linien} (= Integrallinien des charakteristischen Vektorfeldes {\bf p}) die Bedeutung
von Weltlinien der Punkte des absoluten Raumes. Verfolgen wir ein 3-Volumenelement
des absoluten Raumes, das stets aus denselben absoluten Punkten besteht, entlang der
charakteristischen Str"omung, so erhalten wir aus Gl. \rf{Dgl2} f"ur seine relative
Volumendehnung pro Zeiteinheit $(\theta)$ die {\em Hubble-"`Konstante"'} H zu
\begin{equation}
H := \frac{1}{3}\,{H^\mu}_\mu\ ,
\label{HspurH}
\end{equation}
die im allgemeinen aber nicht konstant sein wird. Sie stellt ein erstes, einfaches
Beispiel dar, wie die Geometrie der Raum-Zeit aus den Bewegungsgleichungen \rf{Dgl1}
- \rf{Dgl4} zu konstruieren ist. Insbesonders kann aus diesem Gleichungssystem der
{\em Riemannsche} Kr"ummungstensor {\bf R} eindeutig konstruiert werden. Dabei zeigt
sich, da\3 {\bf R} nur von der inneren Kr"ummung $({\bf\Fquer})$ der
charakteristischen Fl"achen \mbox{$(\theta = \mbox{konst.})$} und dem Hubble-Tensor
{\bf H} abh"angt.

Hier ist nun die Tatsache wichtig, da\3 {\bf H} durch eine Projektion l"angs der
charakteristischen Richtung {\bf p} aus dem Objekt \bm{\calH} gewonnen wird (vgl.
\rf{HcalBcalHp}). Dies hat zur Folge, da\3 im allgemeinen ein Teil der
Feldfreiheitsgrade {\bf M} von \bm{\calH} herausprojiziert wird; diese nennen wir die
{\em "au\3eren Feldvariablen} ${\bf\Mex}$, da sie offensichtlich nicht direkt in die
innere Geometrie der Raum-Zeit eingehen. Dadurch wird aber der Riemannsche
Kr"ummungstensor {\bf R} nur noch eine Funktion der "Atherfelder \bm{\calB}, {\bf p}
und der {\em inneren Feldvariablen} ${\bf\Min}$: ${\bf R} = {\bf R}$(\bm{\calB},{\bf
p},${\bf\Min}$). Dieser funktionale Zusammenhang "ubertr"agt sich nat"urlich auch auf
den Einstein-Tensor {\bf E}: ${\bf E} = {\bf E}$(\bm{\calB},{\bf p},${\bf\Min}$).
Damit entsteht nun die Frage, welche Bedeutung die "au\3eren Feldvariablen
${\bf\Mex}$ f"ur die vorliegende Theorie haben.

Die Antwort auf diese Frage ergibt sich aus der Untersuchung der geometrischen
Struktur des Vakuums. Wenn man sich n"amlich auf den Standpunkt stellt, da\3 nach
Entfernung aller Materie aus dem Universum das sogenannte "`Quantenvakuum"' ( =
Nullpunkt-Fluktuationen aller Quantenfelder) als nicht zu unterdr"uckender Rest
zur"uckbleibt, so 
wird man auch zugestehen m"ussen, da\3 dessen Energie-Impuls-Inhalt ${\bf\Tnull}$ die
geometrische Struktur des leeren Universums gem"a\3 den Einsteinschen Gleichungen
beeinflussen sollte:\footnote{ Die Felder werden in geometrischen Einheiten gemessen
\cite{MiWh57}, so da\3 ihr physikalischer Energie-Impulstensor ${\bf\tilde{T}}$ die
Dimension ${({\rm L"ange})}^{-4}$ annimmt und dann mit {\bf T} bezeichnet wird. ({\bf
T}$={(\hbar c)}^{-1}{\bf\tilde{T}})$. Aus diesem Grund enthalten die Einsteinschen
Feldgleichungen das Quadrat der Planck L"ange $L_p = {(\hbar kc^{-3})}^{\frac{1}{2}}$
anstelle der Newtonschen Gravitationskonstanten k.}
\begin{equation}
E_{\mu\nu} = 8\pi\,L^2_P\,\Tnull_{\mu\nu}\ .
\label{ELpTnull}
\end{equation}
Mi\3t man also dem Quantenvakuum einen nicht-verschwindenden Energie-Impuls-Inhalt
${\bf\Tnull}$ bei, so kann die Vakuum-Struktur nicht mehr in der flachen
Minkowski-Geometrie bestehen! Andererseits sollte man erwarten, da\3 das
Quantenvakuum die Raum-Zeit auf eine etwas andere Art und Weise kr"ummt als dies die
normale Materie bewirkt. Man w"urde dem Quantenvakuum aufgrund seiner
Unvermeidbarkeit gerne eine Zwischenstellung zwischen einer rein geometrischen
Gr"o\3e (wie z.B. der Metrik {\bf G}) und einer reinen Materievariable (wie z.B. der
Massedichte ${\scr\calM}$ oder dem Druck ${\scr\calP}$) zuschreiben. Gerade f"ur
einen solchen Zweck bieten sich aber nun die "au\3eren Feldvariablen ${\bf\Mex}$
unserer Theorie an! Denn einerseits handelt es sich hierbei um geo\-metrische
Gr"o\3en, welche die kovariante Ableitung des Vierbein-Feldes \bm{\calB}, {\bf p}
bestimmen (vgl. \rf{Dgl1}, \rf{Dgl2}), andererseits gehen sie aber definitionsgem"a\3
nicht in die Riemannsche Geo\-metrie der Raum-Zeit ein. Daher liegt es nahe, diese
zus"atzlichen Freiheitsgrade ${\bf\Mex}$ als Variable mit nicht-verschwindendem
Vakuum-Erwartungswert ("`Vakuumvariable"') zu betrachten, welche die geometrische
Struktur des Vakuums beschreiben k"onnen. Bezeichnet man also mit ${\bf\Te}$ den
Energie-Impuls-Inhalt der "au\3eren Variablen ${\bf\Mex}$, so sollte diese Gr"o\3e
ebenso wie ${\bf\TM}$ f"ur normale Materie das Raumzeit-Kontinuum nach der
Einsteinschen Vorstellung kr"ummen:
\begin{eqnarray}
\label{ETeTm}
E_{\mu\nu} &=& 8\pi\,L^2_P\left(\Te_{\mu\nu} + \TM_{\mu\nu}\right)\\
\label{allgFormTe}
\bigg[\,\Te_{\mu\nu} &=& \Te_{\mu\nu}\left(\mbox{\bm{\calB}},{\bf p},{\bf\Mex}\right)
\,\bigg]
\nonumber\ .
\end{eqnarray}
Selbst wenn man alle normale Materie aus dem Universum entfernt $({\bf\TM}\to 0)$, so
bleibt dennoch ein essentiell dynamisches Gebilde zur"uck $({\bf\Te}\ne 0)$, das wir
das "`{\em klassische Vakuum}"' nennen. Falls man ${\bf\Mex}$ noch zu den
geometrischen Variablen rechnen will, k"onnte man dem Tensor ${\bf\Te}$ die Bedeutung
eines Energie-Impuls-Inhaltes der Gravitationskraft zuerkennen. Auf eine "ahnliche
Weise kann man die Spindichte ${\bf\Sigma}$ des Gravitationsfeldes definieren. Eine
wichtige Konsequenz der modifizierten Einstein-Gleichungen \rf{ETeTm} ist die
Tatsache, da\3 der Materietensor ${\bf\TM}$ nicht notwendigerweise divergenzfrei sein
mu\3. Eine eventuell auftretende Matthisson-Kraft der Materie \mbox{${\bf\tM}\,(:=
\nabla \cdot{\bf\TM}\ne 0)$} kann nun durch eine Verzerrung des Vakuums
("`Vakuum-Polarisation"') wieder kompensiert werden! Vermutlich wird aber die
"`{\em Vakuumpolarisation}"' nur in exotischen Situationen bedeutsam werden (z.B. in
der Kosmologie, bei schwarzen L"ochern, Gravitation der Elementarteilchen), soda\3
die klassischen Tests der Allgemeinen Relativit"atstheorie durch die vorliegende
Modifikation der Einsteinschen Theo\-rie nicht ber"uhrt werden. Im "ubrigen k"onnen
die gew"ohnlichen Einsteinschen Gleichungen aus den unsrigen \rf{ETeTm},
zur"uckgewonnen werden, indem man die "au\3eren Variablen verschwinden l"a\3t
\mbox{$({\bf\Mex}\to 0)$}, wobei auch der Spin ${\bf\Sigma}$ und der Energie-Impuls
${\bf\Te}$ der Gravitation wieder verschwinden.

Die Untersuchung der Eigenschaften eines solchen nicht-trivialen Vakuums bildet den
Gegenstand der nachfolgenden Ausf"uhrungen \mbox{$({\bf\TM} = 0)$}. Angesichts des
zur Verf"ugung stehenden zeitlichen Rahmens konnten die Effekte der
Vakuumpolarisation nicht mehr untersucht werden. Einige interessante Ph"anomene sind
von dem Umstand zu erwarten, da\3 die Gravitation "uber die mikroskopische
Kraftdichte ${\bf\te} = -{\bf\tM}$ Energie und Impuls auf die Materie "ubertragen
kann, was in der konventionellen Einsteinschen Theorie nicht m"oglich ist, da dort
wegen des verschwindenden Energie-Impulses der Gravitation $({\bf\Te}\equiv~0)$ auch
die auf die Materie ausge"ubte Kraftdichte ${\bf\tM}$ identisch verschwinden mu\3!


\chapter{Die Geometrie der (1+3) Zerlegung}
\indent
\label{RiemStr}

Wie wir gesehen haben, ist die im letzten Abschnitt beschriebene Aufspaltung der
Raum-Zeit in eine Zeitrichtung und eine 3-Untergeometrie das eigentliche dynamische
Objekt. Da die 3-Untergeometrie auf diese Weise durch die 4-Geometrie der Raum-Zeit
induziert wird, k"onnen wir nat"urlich umgekehrt die 4-Geometrie aus der (1+3)
Aufspaltung der Raum-Zeit und der entsprechenden 3-Untergeometrie aufbauen.

Wir werden dies in vier Schritten erreichen. Zuerst wird die 3-dimensionale
Oberfl"achenkonnexion hergeleitet, welche durch die 4-dimensionale Riemannsche
Geometrie induziert wird. Sodann werden wir die korrespondierende 3-Kr"ummung mit
Hilfe eines B"undelisomorphismus untersuchen, wobei ein zus"atzlicher Freiheitsgrad
auftaucht, der auch in den mikroskopischen Feldgleichungen vorhanden ist und einer
Oberfl"achentorsion entspricht. Ferner werden wir die Euklidische Foliation n"aher
untersuchen, welche einer speziellen Raum-Zeit Aufspaltung entspricht.

Der n"achste Schritt besteht in einer ausf"uhrlichen Diskussion der Rolle, welche die
Torsion in der vorliegenden Theorie spielt. Obwohl die vierdimensionale Raum-Zeit
einen streng Riemannschen Charakter besitzt, ist die Torsion der dreidimensionalen
Untergeometrie eine wichtige Gr"o\3e und steht in enger Beziehung zur Spindichte der
"Atherfelder. Schlie\3lich zeigen wir, da\3 die mikroskopische Dynamik \rf{Dgl1} -
\rf{Dgl3} nicht nur die Geometrie der Raum-Zeit eindeutig festlegt, sondern auch
gegen"uber der konventionellen Einsteinschen Theorie noch einen zus"atzlichen
Feldfreiheitsgrad $({\bf\Mex})$ aufweist, der dem Gravitationsfeld als der
geometrischen Manifestation des Quantenvakuums eine nicht-triviale Spin- und
Energie-Impuls-Dichte verleiht.

\section{Die charakteristische Konnexion}
\indent

Die Riemannsche Konnexion ${\bf\Gamma}$ des Tangentialb"undels $\tau_4$ der Raum-Zeit
induziert eine Konnexion ${\bf\Gammadach}$ im reduzierten Unterb"undel $\taudach$,
die sog. {\em charakteristische Konnexion}, durch nat"urliche Projektion
($\stackrel{\cap}{{\bf P}}$) auf die charakteristische Distribution $\Deltadach$.
Aufgrund der verschwindenden Torsion von ${\bf\Gamma}$ mu\3 die Einschr"ankung der
Oberfl"achentorsion ${\bf\Zdach}$ ({\em charakteristische Torsion}) auf $\Deltadach$
ebenso verschwinden \cite{Che73}:
\alpheqn
\begin{equation}
\Zdach\Big|_{\Deltadach}\equiv 0\ ,
\end{equation}
in Komponentenschreibweise
\begin{eqnarray}
\label{ZdachGammadach}
{\Zdach^\lambda}_{\mu\nu} &:=& {\Gammadach^\lambda}_{[\mu\nu]}\\
\label{calBZdach2calB}
{\calB^\rho}_\lambda\,{\Zdach^\lambda}_{\mu\nu}\,{\calB^\mu}_\sigma{\calB^\nu}
_\kappa &=& 0\ .
\end{eqnarray}
\reseteqn
Die entsprechende koordinaten-kovariante Ableitung \bm{\nabladach}, die {\em
charakteristische Ableitung}, erh"alt man aus der Riemannschen Ableitung \bm{\nabla}
durch folgende Zerlegung
\begin{equation}
\mbox{\bm{\nabladach}} = \mbox{\bm{\ndp}} + \mbox{\bm{\nds}}\ .
\label{nabladachnstpnsts}
\end{equation}
Der parallele Anteil ${\bf\ndp}$ wird hierbei mit Hilfe der Riemannschen
Ableitung \bm{\nabla} und einer Projektion ${\bf\Pcap}$ auf $\Deltadach$
definiert
\begin{equation}
{\bf\ndp} = \stackrel{\cap}{\bf P}\circ\mbox{\bm{\nabla}}\circ\stackrel{\cap}
{\bf P}\ .
\label{ndpPdachnablaPdach}
\end{equation}
Beispielsweise lautet die parallele Ableitung eines beliebigen $\tau_4$ Schnittes
{\bf U}
\begin{equation}
\ndp_\mu\,U_\nu = {\calB^\lambda}_\nu\,\nabla_\mu\left({\calB_\lambda}^\sigma\,
U_\sigma \right)\ .
\label{nstpU}
\end{equation}
Den orthogonalen Anteil ${\bf\nds}$ in \rf{nabladachnstpnsts} erhalten wir,
indem wir uns vergegenw"artigen, da\3 \bm{\nabladach} aufgrund der durch den
charakteristischen Vektor {\bf p} induzierten (1+3)-Aufspaltung der Raum-Zeit nur auf
SO(3)-Skalare wirkt. Es ergibt sich somit folgende Definition
\begin{equation}
\nds_\mu\,U_\nu = p_\nu\,\partial_\mu\left(p^\lambda U_\lambda\right)\ .
\label{nstsU}
\end{equation}
Damit k"onnen wir f"ur einen beliebigen Tangentenvektor {\bf U} dessen SO(3)
Vektoranteil {\bf u} und SO(3) Skalaranteil u definieren durch
\begin{eqnarray}
\label{Defsigma}
u^i &=& {\calB^i}_\mu U^\mu\\
\label{psiUp}
u &=& U^\mu p_\mu\nonumber\ ,
\end{eqnarray}
das hei\3t {\bf U} l"a\3t sich nun folgenderma\3en zerlegen
\begin{equation}
U_\mu = u^i\calB_{i\mu} + u\,p_\mu\ .
\label{ZerlU}
\end{equation}
Die charakteristische Ableitung von {\bf U} ergibt sich somit zu
\begin{eqnarray}
\label{nabladachU}
\nabladach_\lambda U_\mu &=& \left(\bar{D}_\lambda u^i\right)\calB_{i\mu} +
\left(\partial_\lambda u\right)p_\mu\\
&\equiv& \partial_\lambda\,U_\mu - {\Gammadach^\nu}_{\mu\lambda}\,U_\nu\ .\nonumber
\end{eqnarray}
Man beachte hierbei, da\3 die SO(3) kovariante Ableitung ${\bf\Dquer}$ nicht mit der
in den mikroskopischen  Feldgleichungen \rf{Dgl1} bis \rf{Dgl3} auftauchenden
Ableitung {\bf D} \rf{DefcalD} "ubereinzustimmen braucht. Wir werden diesen
interessanten Punkt sp"ater noch eingehender untersuchen.

Aus der Definition \rf{nabladachU} der charakteristischen Ableitung
\bm{\nabladach} l"a\3t sich die kovariante Konstanz der "Atherfelder bez"uglich
${\bf\Gammadach}$ folgern:
\alpheqn
\begin{eqnarray}
\label{nabladachp0}
\nabladach_\lambda\,p_\mu &=& 0\\
\label{calDhalbcalB0}
\calDhalb_\lambda\calB_{i\mu} &=& 0\ .
\end{eqnarray}
\reseteqn
Die Ableitung \bm{\calDhalb} ist nun genauso wie \bm{\calD} in Gleichung \rf{DefcalD}
definiert, jedoch mit dem Unterschied, da\3 die zwei Konnexionen ${\bf\Gamma}$, {\bf
A} durch ${\bf\Gammadach}$ und ${\bf\Aquer}$ ersetzt werden m"ussen. Die Bedingungen
der kovarianten Konstanz \rf{nabladachp0}, \rf{calDhalbcalB0} sind konsistent mit den
Bedingungen \rf{2calBg} und \rf{calBp0_1}. Wir werden sp"ater sehen, da\3 die
Bedingungen \rf{nabladachp0} und \rf{calDhalbcalB0} einen affinen
B"undelisomorphismus beschreiben, welcher in der (1+3) Aufspaltung eine wichtige
Rolle spielt. Bis dahin bemerken wir nur, da\3 der Projektor \bm{\calB}$\,^2$
aufgrund von \rf{calDhalbcalB0} kovariant konstant ist
\begin{equation}
\nabladach_\lambda\calB_{\mu\nu} = 0
\label{nabladachcalB0}
\end{equation}
und deshalb als Fasermetrik ${\bf\Gdach}$ von $\taudach$ verwendet werden kann.
Da nun aber beide Anteile \bm{\calB} und {\bf p} der Riemannschen Metrik {\bf G}
\rf{GcalBpp} konstant sind, gilt ebenfalls
\begin{equation}
\nabladach_\lambda G_{\mu\nu} = 0\ .
\label{nabladachG0}
\end{equation}
Das hei\3t die Metrik {\bf G} ist kovariant konstant bez"uglich beider Konnexionen
${\bf\Gamma}$ und ${\bf\Gammadach}$. Nach Gleichung \rf{GamGamschlpKQ} unterscheiden
sich die Konnexionen daher durch einen Kontorsionsterm ${\bf\Kdach}$
\begin{equation}
{\Gammadach^\lambda}_{\mu\nu} = {\Gamma^\lambda}_{\mu\nu} +
{\Kdach^\lambda}_{\mu\nu} \ .
\label{GammadachGammaKdach}
\end{equation}
Die Kontorsion ${\bf\Kdach}$ bestimmt nun die Torsion ${\bf\Zdach}$ durch
\begin{equation}
{\Zdach^\lambda}_{\mu\nu} = {\Kdach^\lambda}_{[\mu\nu]}\ .
\label{ZdachKdach}
\end{equation}
F"ur ${\bf\Kdach}$ kann man eine physikalische Interpretation angeben. Dazu zerlegen
wir ${\bf\Kdach}$ in einen parallelen $({\bf\Kdp})$ und einen orthogonalen Teil
$({\bf\Kds})$:
\begin{equation}
{\Kdach^\lambda}_{\mu\nu} = {\Kdp^\lambda}_{\mu\nu} + {\Kds^\lambda}_{\mu\nu}\ .
\label{KdachKdpKds}
\end{equation}
Der orthogonale Teil l"a\3t sich nun mit Hilfe von \rf{Dgl2}, \rf{nabladachp0}
und der Ableitung des charakteristischen Vektors folgenderma\3en ausdr"ucken:
\begin{eqnarray}
\label{Kdppnablap}
{\Kds^\lambda}_{\mu\nu}&=& p^\lambda\,\nabla_\nu\,p_\mu-p_\mu\nabla_\nu\,
p^\lambda\\
&=& p^\lambda H_{\mu\nu} - p_\mu {H^\lambda}_\nu\nonumber\ .
\end{eqnarray}
Definieren wir nun die Spindichten der "Atherfelder nach \cite{He76}
\alpheqn
\begin{eqnarray}
\label{DefSigmap}
\Sigmap^{\mu\nu\lambda} &=& \frac{1}{4\pi}\left[p^\mu\left(\nabla^\lambda p^\nu\right)
-p^\nu\left(\nabla^\lambda p^\mu\right)\right]\\
\label{DefSigmacalB}
\SigmacalB^{\mu\nu\lambda} &=& -\frac{1}{4\pi}\left[{\calB_i}^\mu\left(\calD^\lambda
\calB^{i\nu}\right) - {\calB_i}^\nu\left(\calD^\lambda\calB^{i\mu}\right)\right]\ ,
\end{eqnarray}
\reseteqn
dann ist der orthogonale Teil ${\bf\Kds}$ identisch mit der Spindichte des
charakteristischen Vektorfeldes
\begin{equation}
{\Kds^\lambda}_{\mu\nu} = 4\pi\,\,{\Sigmap^\lambda}_{\mu\nu}\ .
\label{KdsSigmap}
\end{equation}
Man kann nat"urlich annehmen, da\3 eine "ahnliche Beziehung f"ur die Spindichte
${\bf\SigmacalB}$ des \bm{\calB} Feldes und den parallelen Teil ${\bf\Kdp}$ der
Kontorsion ${\bf\Kdach}$ existiert; dies wird im folgenden n"aher untersucht.

Aufgrund der Definition \rf{nabladachU} ist die Anwendung der Ableitung
\bm{\nabladach} nicht nur auf $\taudach$ Objekte beschr"ankt. Das hei\3t, da\3 die
charakteristische Konnexions 1-Form ${\bf\Gammadach}$ auch Komponenten entlang
der charakteristischen Richtung besitzt und demzufolge eine entsprechende Zerlegung
zul"a\3t:
\begin{equation}
{\Gammadach^\lambda}_{\mu\nu}={\gammadach^\lambda}_{\mu\nu}+p^\lambda\,z_{\mu\nu}
\ .
\label{Gammadachgammadachpz}
\end{equation}
Hierbei bezeichnet \bm{\gammadach} den {\em wesentlichen Teil} von
${\bf\Gammadach}$
\begin{equation}
{\gammadach^\lambda}_{\mu\nu}={\calB^\lambda}_\sigma{\Gammadach^\sigma}_{\mu\nu}\ .
\label{gammadachcalBGammadach}
\end{equation}
\bm{\gammadach} ist im allgemeinen nicht symmetrisch, sondern wird vielmehr
den parallelen Teil der charakteristischen Torsion ${\bf\Zdach}$ festlegen:
\alpheqn
\begin{eqnarray}
\label{ZdachZdpZds}
{\Zdach^\lambda}_{\mu\nu} &=& {\Zdp^\lambda}_{\mu\nu} + {\Zds^\lambda}_{\mu\nu}\\
\label{Zdpgammadach}
{\Zdp^\lambda}_{\mu\nu} &=& {\gammadach^\lambda}_{[\mu\nu]}\ .
\end{eqnarray}
\reseteqn
Der orthogonale Teil ${\bf\Zds}$
\begin{equation}
{\Zds^\lambda}_{\mu\nu} = p^\lambda z_{[\mu\nu]}
\label{Zdspz}
\end{equation}
l"a\3t sich mit der Integrabilit"atsbedingung f"ur $\Deltadach$ in Beziehung setzen.
Und zwar gilt
\begin{equation}
\nabladach_{[\lambda}p_{\mu]} \equiv p_{[\lambda}f_{\mu]} + {\Zdach^\rho}_{\lambda
\mu}\,p_\rho\ .
\label{IntegrBed}
\end{equation}
Den hier auftauchenden {\em Frobeniusvektor} {\bf f} k"onnen wir als einen
${\bf\taudach}$ Schnitt betrachten und ohne Beschr"ankung der Allgemeinheit annehmen
\begin{equation}
p^\mu\,f_\mu = 0\ .
\label{pf0}
\end{equation}
Aus \rf{IntegrBed} schlie\3t man im Falle der Integrabilit"at von $\Deltadach$
\begin{equation}
z_{[\mu\nu]} = f_{[\mu} p_{\nu]}
\label{zpf}
\end{equation}
und somit
\begin{equation}
{\Zds^\lambda}_{\mu\nu} = p^\lambda\,f_{[\mu}p_{\nu]}\ .
\label{Zdspfp}
\end{equation}
Man beachte, da\3 die Einschr"ankung des orthogonalen Anteils ${\bf\Zds}$ auf
$\Deltadach$ verschwindet, was dann aufgrund von \rf{calBZdach2calB} auch f"ur den
parallelen Anteil ${\bf\Zdp}$ gelten mu\3.

\newpage
{\large\bf III.2\ Die innere Kr"ummung des charakteristischen B"undels}
\addtocounter{section}{1}
\addcontentsline{toc}{section}{Die innere Kr"ummung des charakteristischen B"undels}
\indent
\label{IntrKrue}

Aufgrund der Einbettung in das Tangentialb"undel der Raum-Zeit besitzt das
charakteristische B"undel $\taudach$ die Gruppe GL(3,{\bf R}) als nat"urliche
Strukturgruppe. Aus diesem Grund stimmt dessen Eichgruppe, welche als die kleinst
m"oglichste Untergruppe der Strukturgruppe definiert ist, mit der gew"ohnlichen
Rotationsgruppe SO(3), der maximalen kompakten Untergruppe von GL(3,{\bf R}),
"uberein. Demnach ist es m"oglich, die intrinsische $\taudach$ Geometrie in der Form
einer SO(3) Eichtheorie zu beschreiben!

Die grundlegenden SO(3) Objekte sind die "Atherfelder \bm{\calB}$_i$, welche sich
homogen transformieren:
\begin{eqnarray}
\label{HomTrafo}
\calB^{\,\prime}_{i\mu} &=& {S^j}_i\calB_{j\mu}\\
{\bf S} &=& \{ {S^j}_i \} \in SO(3)\nonumber
\end{eqnarray}
Wir k"onnen nun mit Hilfe der "Atherfelder die Einbettung eines SO(3)-B"undels
$\tauquer$ in das Tangentialb"undel $\tau_4$ definieren. Da $\tauquer$, das sog. {\em
repr"asentative B"undel}, aufgrund seiner 
Konstruktion isomorph zu $\taudach$ ist, mu\3 eine 1-1 Abbildung $[B] =
\left\{[\Bquer], [\Bdach]\right\}$ existieren, welche die folgenden Relationen
erf"ullt
\alpheqn
\begin{eqnarray}
\label{DefBquer}
[\Bquer] &:& \taudach \to \tauquer\\
\label{DefBdach}
[\Bdach] &:& \tauquer \to \taudach
\end{eqnarray}
\reseteqn
soda\3 gilt
\alpheqn
\begin{eqnarray}
\label{idtauquer}
[\Bquer]\circ[\Bdach] &=& {\mbox{id}}_{\tauquer}\\
\label{idtaudach}
[\Bdach]\circ[\Bquer] &=& {\mbox{id}}_\taudach\ .
\end{eqnarray}
\reseteqn
In Komponenten geschrieben erkennt man, da\3 diese Abbildungen die Schnitte
\bm{\nu}(x) $\in \tauquer$ und {\bf V}(x) $\in \taudach$ zueinander in Beziehung
setzen:
\alpheqn
\begin{eqnarray}
\label{nucalBV}
\nu_i &=& \calB_{i\mu}V^\mu\\
\label{VcalBnu}
V_\mu &=& \calB_{i\mu}\nu^i\ .
\end{eqnarray}
\reseteqn
Dabei ist die G"ultigkeit von \rf{idtauquer} und \rf{idtaudach} aufgrund der
Gleichungen \rf{2calBg} bis \rf{calBp0_2} gew"ahrleistet. Ferner mu\3 die Aufspaltung
der B"undeltangentialr"aume in einen horizontalen und vertikalen Unterraum mit der
Abbildung [B] vertr"aglich sein, d.h. es mu\3 gelten:
\alpheqn
\begin{eqnarray}
\label{DBBnabla}
{\bf\Dquer}\circ[\Bquer] &=& [\Bquer]\,\circ\mbox{\bm{\nabladach}}\\
\label{nablaBBD}
\mbox{\bm{\nabladach}}\circ[\Bdach] &=& [\Bdach]\circ{\bf\Dquer}\ .
\end{eqnarray}
\reseteqn
Hierbei geh"ort zur SO(3) kovarianten Ableitung ${\bf\Dquer}$ aus $\tauquer$ eine
${\cal SO}(3)$ wertige Konnexion ${\bf\Aquer}$, {\em welche im allgemeinen nicht mit
der Konnexion {\bf A} "ubereinstimmt}, die in den mikroskopischen Feldgleichungen
\rf{Dgl1} bis \rf{Dgl3} ben"utzt wird. Die Relationen \rf{DBBnabla} und \rf{nablaBBD}
lauten in Komponenten
\alpheqn
\begin{eqnarray}
\label{DBBnablaKomp}
\Dquer_\mu\nu_i &=& \calB_{i\lambda}\left(\nabladach_\mu V^\lambda\right)\\
\label{nablaBBDKomp}
\nabladach_\mu V_\lambda &=& \calB_{i\lambda}\left(\Dquer_\mu\nu^i\right)
\end{eqnarray}
\reseteqn
und sind aufgrund von Gleichung \rf{calDhalbcalB0} erf"ullt.

Im letzten Abschnitt haben wir die Beziehung zwischen den zwei Konnexionen
${\bf\Gamma}$ und ${\bf\Gammadach}$ verdeutlicht. Nun wollen wir untersuchen, welche
Relation zwischen {\bf A} und ${\bf\Aquer}$ besteht. Dazu beachten wir, da\3 die
Konstruktion von \bm{\calDhalb} aus $\calDquerF$ durch einen "ahnlichen Proze\3
wie bei der Bildung von \bm{\nabladach} aus \bm{\nabla} erm"oglicht wird.
Das bedeutet, wir k"onnen schreiben
\begin{equation}
\label{calDhalbaufsp}
\mbox{\bm{\calDhalb}}\,\, = \mbox{\bm{\cDhp}} + \mbox{\bm{\cDhs}}\ ,
\end{equation}
wobei symbolisch wieder gilt
\alpheqn
\begin{eqnarray}
\label{cDhpPcapDquerPcap}
\mbox{\bm{\cDhp}} &=& {\bf\Pcap}\circ\calDquerF\circ{\bf\Pcap}\\
\label{cDhspDquerp}
\mbox{\bm{\cDhs}} &=& {\bf p}\otimes{\bf\Dquer}\circ {\bf p}\bullet\ .
\end{eqnarray}
\reseteqn
Die Bedingung der kovarianten Konstanz \rf{calDhalbcalB0} l"a\3t sich leicht
erf"ullen, wenn wir $\calDquerF$ folgenderma\3en definieren
\begin{equation}
\calDquerF := {\bf p}\otimes {\bf p}\bullet\mbox{\bm{\calD}}
\label{DefcalDquer}
\end{equation}
bzw. in Komponenten
\begin{equation}
\calDquer_\mu\calB_{i\nu} = p_\nu p^\lambda\left(\calD_\mu\calB_{i\lambda}\right)\ .
\label{calDquercalB1}
\end{equation}
Die Konnexion ${\bf\Aquer}$, welche Gleichung \rf{calDquercalB1} gen"ugt, erhalten
wir durch den Ansatz
\begin{equation}
\Aquer_{i\mu} = A_{i\mu} + C_{i\mu}
\label{AquerAC}
\end{equation}
wobei wir den Eichvektor {\bf C} noch bestimmen m"ussen. Mit diesem Ansatz ergibt
sich f"ur die linke Seite von \rf{calDquercalB1}
\begin{equation}
\calDquer_\mu\calB_{i\nu} = \calD_\mu\calB_{i\nu} + {\epsilon_i}^{jk} C_{j\mu}
\calB_{k\nu}
\label{calDquercalB2}
\end{equation}
und somit
\begin{equation}
C_{j\mu} = \frac{1}{2}\,{\epsilon_j}^{mn}\,\calB_{m\sigma}\left(\calD_\mu{\calB_n}
^\sigma\right)\ .
\label{C}
\end{equation}

Dem repr"asentativen B"undel $\tauquer$ mit der Konnexion ${\bf\Aquer}$ kann man also
ein neues B"undel $\tauschl$ mit der Konnexion {\bf A} zuordnen, wobei ${\bf\Aquer}$
das ${\cal SO}$(3) Analogon der charakteristischen Konnexion ${\bf\Gammadach}$
darstellt; somit kann man {\bf A} als {\em geometrischen Teil} der Konnexion
${\bf\Aquer}$ betrachten. Der {\em dynamische Teil} {\bf C} von ${\bf\Aquer}$ ist
unabh"angig von der Riemannschen Konnexion ${\bf\Gamma}$ und beeinflu\3t deshalb
nicht direkt die 4-Geometrie der Raum-Zeit! Dagegen wird die Kinematik der Triade
\bm{\calB}$_i$ wesentlich durch den dynamischen Teil {\bf C} "uber die
mikroskopischen Feldgleichungen \rf{Dgl1} bis \rf{Dgl4} mitbe\-stimmt. Aufgrund
dieses Sachverhaltes k"onnen die "au\3eren Variablen ${\bf\Mex}$ mit dem Eichvektor
{\bf C} identifiziert werden! Aus der Zerlegung \rf{AquerAC} der Konnexion {\bf A} in
einen geometrischen und einen dynamischen Teil erhalten wir eine analoge Aufspaltung
der $\tauquer$ Kr"ummung ${\bf\Fquer}$:
\begin{equation}
\Fquer_{i\mu\nu} = F_{i\mu\nu} + \calD_\mu C_{i\nu} - \calD_\nu C_{i\mu} +
{\epsilon_i}^{jk}C_{j\mu}C_{k\nu}\ .
\label{FquerFC}
\end{equation} Diese Darstellung der Kr"ummung ist nun geeignet um die Funktion
${\cal F}_i$\, (\bm{\calB},{\bf p},{\bf M}) \rf{Dgl4}, welche das System der
mikroskopischen Feldgleichungen schlie\3t, zu untersuchen. Der dynamische Teil von
{\bf F}, welcher aus dem Eichvektor {\bf C} besteht, ist nun schon durch Gleichung
\rf{C} festgelegt. Das bedeutet wir d"urfen nur noch "uber den geometrischen Teil
{\bf F} von ${\bf\Fquer}$ verf"ugen! Offensichtlich charakterisiert ${\bf\Fquer}$
die 3-Geometrie der charakteristischen Fl"achen. Die einfachste Wahl f"ur
${\bf\Fquer}$ ist
\begin{equation}
\Fquer_{i\mu\nu} \equiv 0\ .
\label{Fquer0}
\end{equation}
Das hei\3t die charakteristischen Fl"achen besitzen eine Euklidische 3-Geometrie
({\em Euklidische Foliation}). Nebenbei bemerkt existiert kein Flachheitsproblem f"ur
solch ein Universum \cite{FlachhPr}. Die Kr"ummung {\bf F} \rf{FquerFC} besteht jetzt
nur aus dem dynamischen Teil und erf"ullt damit automatisch die Bianchi-Identit"at
\rf{3calDF}. Die so festgelegten charakteristischen Fl"achen k"onnen als {\em
absoluter Raum} aufgefasst werden, welcher durch eine spontane lokale Aufspaltung der
Raum-Zeit in einen 3-Raum $\Deltadach_x$ und eine Zeitrichtung ${\bf p}_x$ entsteht.
Dieser absolute Raum wird dabei mit einer Euklidischen Untergeometrie ausgestattet.
Verbinden wir diese (1+3) Aufspaltung mit dem "Aquivalenzprinzip, so ist die Existenz
eines ausgezeichneten Lorentz-Bezugssystems garantiert und kommt Newton's absoluter
Vorstellung von Raum und Zeit recht nahe.
\section{Die Torsion}
\indent

Wie die Untersuchungen "uber die Spindichte der "Atherfelder und die Integrabilit"at
von $\Deltadach$ im vorletzten Abschnitt gezeigt haben, spielt die Torsion trotz der
Tatsache, da\3 die 4-Geometrie der Raum-Zeit in unserer Theorie torsionsfrei ist,
eine gewisse Rolle. Aus diesem Grund wollen wir die Effekte der Torsion f"ur den
allgemeineren Fall der Nichtintegrabilit"at der charakteristischen Distribution
$\Deltadach$ berechnen.

Der parallele Anteil ${\bf\Kdp}$ der charakteristischen Kontorsion ${\bf\Kdach}$
\rf{KdachKdpKds}  kann aus der Konstanz von \bm{\calB}$_i$ \rf{calDhalbcalB0}
bestimmt werden. Wir erhalten
\begin{equation}
\calDquer_\mu\calB_{i\nu} = {\Kdach^\lambda}_{\nu\mu}\,\calB_{i\lambda}
\label{calDquercalBKdachcalB}
\end{equation}
und daraus mit Hilfe des orthogonalen Teils ${\bf\Kds}$ \rf{Kdppnablap}
\begin{equation}
\Kdp_{\sigma\nu\mu} = 0\ .
\label{KdpZerlegt}
\end{equation}
Das bedeutet wir k"onnen Gleichung \rf{KdsSigmap} verallgemeinern zu
\begin{equation}
{\Kdach^\lambda}_{\mu\nu} = 4\pi\;{\Sigmap^\lambda}_{\mu\nu}\ .
\label{KdachSigmaphi}
\end{equation}
Es scheint nun etwas seltsam zu sein, da\3 sich die charakteristische Kontorsion
${\bf\Kdach}$ ausschlie\3lich aus der Spindichte ${\bf\Sigmap}$ des
charakteristischen Vektorfeldes zusammensetzt, w"ahrend doch die Gleichung
\rf{KdsSigmap} zu der Vermutung berechtigt, da\3 sich der parallele Teil ${\bf\Kdp}$
aus der Spindichte ${\bf\SigmacalB}$ \rf{DefSigmacalB} des \bm{\calB} Feldes
zusammensetzt und demnach die totale Kontorsion ${\bf\Kdach}$ von der totalen
\mbox{Spindichte ${\bf\Sigma}$} abh"angt:
\begin{equation}
\Sigma^{\mu\nu\lambda} = \Sigmap^{\mu\nu\lambda} + \SigmacalB^{\mu\nu\lambda} =
\frac{1}{2\pi}\epsilon^{ijk}{\calB_i}^\mu {C_j}^\lambda {\calB_k}^\nu\ .
\label{SigmaSigmapSigmaB}
\end{equation}
Man beachte jedoch, da\3 die zwei m"oglichen Definitionen von ${\bf\SigmacalB}$
davon abh"angen, ob man die totale Konnexion {\bf A} wie in Gleichung
\rf{DefSigmacalB}, oder nur den geometrischen Teil ${\bf\Aquer}$ verwendet. Es gilt
die folgende Beziehung zwischen den zwei Ausdr"ucken f"ur ${\bf\SigmacalB}$:
\begin{eqnarray}
\SigmacalBquer^{\rho\mu\nu} &=& 2\;\SigmacalB^{\rho\mu\nu}+{\calB^\rho}_\kappa\;
\SigmacalB^{\mu\kappa\nu} - {\calB^\mu}_\kappa\;\SigmacalB^{\rho\kappa\nu}\\
&=& p^\mu p_\kappa\;\SigmacalB^{\rho\kappa\nu}-p^\rho p_\kappa\;\SigmacalB^{\mu\kappa
\nu} \nonumber\ .
\end{eqnarray}
Da die charakteristische Kontorsion ${\bf\Kdach}$ ein Objekt aus dem B"undel
$\taudach$ ist, sollte sie sich auf die Spindichte $\bar{\bf\Sigma}\left(:={\bf
\SigmacalBquer}+{\bf\Sigmap}\right)$ beziehen und nicht auf die totale Dichte
${\bf\Sigma}$ \rf{SigmaSigmapSigmaB}. Eine einfache Rechnung zeigt jedoch, da\3 der
geometrische Teil $\bar{\bf\Sigma}$ der totalen Dichte ${\bf\Sigma}$ verschwindet
$\left(\bar{\bf\Sigma}\equiv 0\right)$; im Gegensatz zu ${\bf\Kdach}$! Deshalb ist es
nun klar, warum der parallele Teil ${\bf\Kdp}$ und die Spindichte
${\bf\SigmacalBquer}$ keiner so einfachen Beziehung gehorchen k"onnen wie es f"ur den
orthogonalen Teil ${\bf\Kds}$ und ${\bf\Sigmap}$ in Gleichung \rf{KdsSigmap} der Fall
ist.

Die charakteristische Kontorsion ${\bf\Kdach}$ l"a\3t sich mit Hilfe des
Hubble-Tensors ausdr"ucken
\begin{equation}
{\Kdach^\lambda}_{\mu\nu} = p^\lambda H_{\mu\nu} - p_\mu{H^\lambda}_\nu\ ,
\label{KdachpH}
\end{equation}
womit die charakteristische Torsion die folgende Gestalt annimmt
\begin{equation}
{\Zdach^\sigma}_{\nu\lambda} \equiv {\Kdach^\sigma}_{[\nu\lambda]}
= p^\sigma H_{[\nu\lambda]} + {H^\sigma}_{[\nu}p_{\lambda]}\nonumber\ .
\label{ZdachKdach2}
\end{equation}
In dieser Form treten die Torsionseigenschaft \rf{ZdachGammadach},
\rf{calBZdach2calB} und die Zerlegung \rf{ZdachZdpZds}, \rf{Zdpgammadach} deutlich
hervor. Die allgemeine Beziehung \rf{KZQ} zwischen Torsion und Kontorsion ist im
Falle eines verschwindenden Nichtmetrizit"atstensors {\bf Q} f"ur ${\bf\Zdach}$ und
${\bf\Kdach}$ identisch erf"ullt.

Im Hinblick auf konkrete Anwendungen ist es vorteilhaft, den Hubble-Tensor in einen
parallelen und einen senkrechten Anteil zu zerlegen
\alpheqn
\begin{eqnarray}
\label{Hhfp}
H_{\nu\lambda} = h_{\nu\lambda} + f_\nu\,p_\lambda\\
\label{hp}
h_{\nu\lambda}\,p^\nu = h_{\nu\lambda}\,p^\lambda = 0\ .
\end{eqnarray}
\reseteqn
Den rein parallelen Teil {\bf h} wollen wir {\em reduzierten Hubble-Tensor} nennen.
Ferner zerlegen wir {\bf h} in einen symmetrischen und antisymmetrischen Teil:
\alpheqn
\begin{eqnarray}
\label{hhplushminus}
h_{\nu\lambda} &=& \hplus_{\nu\lambda}\, + \hminus_{\nu\lambda}\\
\label{hplushplus}
\hplus_{\nu\lambda} &=& \hplus_{\lambda\nu}\\
\label{hminushminus}
\hminus_{\nu\lambda} &=& -\hminus_{\lambda\nu}
\end{eqnarray}
\reseteqn
Die Integrabilit"atsbedingung f"ur die charakteristische Distribution $\Deltadach$
besteht offensichtlich im Verschwinden des antisymmetrischen Anteils $({\bf\hminus}
\equiv 0)$. Damit ergibt sich die charakteristische Torsion ${\bf\Zdach}$ zu
\begin{equation}
{\Zdach^\sigma}_{\nu\lambda} = p^\sigma\hminus_{\nu\lambda} + p^\sigma f_{[\nu}\,
p_{\lambda]} + {h^\sigma}_{[\nu}\,p_{\lambda]}\ ,
\label{ZdachNACHh}
\end{equation}
woraus wir sofort ihren parallelen und orthogonalen Teil (vgl. \rf{ZdachZdpZds})
entnehmen k"onnen:
\alpheqn
\begin{eqnarray}
\label{ZdsNACHh}
{\Zds^\sigma}_{\nu\lambda} &=& p^\sigma\hminus_{\nu\lambda} + p^\sigma f_{[\nu}\,
p_{\lambda]}\\
\label{ZdpNACHh}
{\Zdp^\sigma}_{\nu\lambda} &=& {h^\sigma}_{[\nu}\,p_{\lambda]}\ .
\end{eqnarray}
\reseteqn
Im integrablen Fall $({\bf\hminus}\equiv 0)$ erhalten wir wieder \rf{Zdspfp}.

Es scheint nun etwas sonderbar zu sein, da\3 auch die urspr"ungliche Spindichte
${\bf\SigmacalB}$ \rf{DefSigmacalB} des \bm{\calB}\ Feldes nicht in der
charakteristischen Kontorsion ${\bf\Kdach}$ \rf{KdachSigmaphi} auftaucht. Der Grund
besteht darin, da\3 die charakteristische Konnexion auf den charakteristischen
Fl"achen torsionsfrei sein mu\3 \mbox{(vgl. \rf{ZdachGammadach})}. Dies wird im
allgemeinen nur durch die Spindichte ${\bf\Sigmap}$ \rf{DefSigmap} des {\bf p} Feldes
aber nicht durch ${\bf\SigmacalB}$ \rf{DefSigmacalB} gew"ahrleistet.

Trotzdem bewirkt die Spindichte des \bm{\calB} Feldes eine gewisse Kontorsion. Um
diese Tatsache zu untersuchen, wollen wir die Konnexion ${\bf\Gammaasymp}$
einf"uhren, welche "ahnlich wie ${\bf\Gammadach}$ im charakteristischen B"undel
$\taudach$ nun in einem B"undel $\tauasymp_4$ wirkt, jedoch {\bf A} als ${\cal
SO}$(3) Gegenst"uck in einem B"undel $\tauschl$ besitzt. Das bedeutet wir betrachten
einen zweiten B"undelisomorphismus zwischen $\tauasymp_4$ und $\tauschl$ wobei die
Gleichungen \rf{DefBquer} bis \rf{nablaBBDKomp} weiterhin gelten, allerdings mit den
Ersetzungen $\left\{\tauquer, [\Bquer], [\Bdach], \mbox{\bm{\nabladach}}, \calDquerF,
\mbox{\bm{\calDhalb}}\right\}\to \left\{\tauschl, [\Bschl], [\Basymp],
\mbox{\bm{\nablaasymp}}, \mbox{\bm{\calD}}, \calDasympF \right\}$. Die Gleichungen
\rf{DBBnabla} und \rf{nablaBBD} lauten jetzt beispielsweise
\alpheqn
\begin{eqnarray}
\label{DBBnabla2}
{\bf D}\circ[\Bschl] &=& [\Bschl]\circ\mbox{\bm{\nablaasymp}}\\
\label{nablaBBD2}
\mbox{\bm{\nablaasymp}}\circ[\Basymp] &=& [\Basymp]\circ{\bf D}\ .
\end{eqnarray}
\reseteqn
Der affine B"undelisomorphismus bezieht sich jetzt also auf die Konnexionen ${\bf
A}$, ${\bf\Gammaasymp}$ anstatt auf ${\bf\Aquer}$, ${\bf\Gammadach}$ wie im ersten
Fall. ${\bf\Gammaasymp}$ soll jedoch weiterhin eine Oberfl"achenkonnexion sein, d.h
es gilt:
\begin{eqnarray}
\label{nablaasymp0}
\nablaasymp_\mu p_\nu &=& 0\\
\nablaasymp_\mu\calB_{\nu\lambda} &=& 0\ ,
\end{eqnarray}
sowie
\begin{equation}
\calDasymp_\mu\calB_{i\nu} = 0\ .
\label{calDasympcalB0}
\end{equation}
Damit erhalten wir f"ur die {\em relative Kontorsion} ${\bf\Kcup}$ von
${\bf\Gammaasymp}$ bez"uglich ${\bf\Gammadach}$
\begin{equation}
{\Kcup^\lambda}_{\mu\nu}:={\Gammadach^\lambda}_{\mu\nu} -
{\Gammaasymp^\lambda}_{\mu\nu} = \epsilon^{ijk}{\calB_i}^\lambda\,C_{j\nu}\,
\calB_{k\mu}\ .
\label{KcupcalBC}
\end{equation}

Diese Gleichung zeigt die Bedeutung der Kontorsion ${\bf\Kcup}$. Sie beeinhaltet
gerade den in der Konnexion {\bf A} \rf{AquerAC} eingef"uhrten Freiheitsgrad in Form
des Eichvektors {\bf C}. Setzen wir nun {\bf C} aus Gleichung \rf{C} in
\rf{KcupcalBC} ein, so erhalten wir
\begin{equation}
{\Kcup^\lambda}_{\mu\nu} = \frac{1}{2}\left[{\calB^\lambda}_\sigma{\calB^i}_\mu
\left(\calD_\nu{\calB_i}^\sigma\right) - \calB_{\mu\sigma}\calB^{i\lambda}
\left(\calD_\nu{\calB_i}^\sigma\right)\right]\ .
\label{KcupExpl}
\end{equation}
Die {\em relative Torsion} ${\bf\Zcup}$ definieren wir nun als
\begin{equation}
{\Zcup^\lambda}_{\mu\nu} := {\Kcup^\lambda}_{[\mu\nu]} =
\frac{1}{2}\,\epsilon^{ijk}\calB_{i\lambda} \left(C_{j\nu}\calB_{k\mu} - C_{j\mu}
\calB_{k\nu}\right)
\label{DefZcup}
\end{equation}
so da\3 die Beziehung \rf{KZQ} im Fall verschwindender Nicht-Metrizit"at {\bf Q} f"ur
${\bf\Zcup}$ und ${\bf\Kcup}$ gerade erf"ullt wird. Die Gleichung \rf{KcupExpl} k"onnen
wir nat"urlich auch mit Hilfe der Spindichten ${\bf\Sigma}$ der "Atherfelder
ausdr"ucken:
\begin{equation}
{\Kcup^\lambda}_{\mu\nu} = -2\pi\left[{\SigmacalB^\lambda}_{\mu\nu} +
{\calB^\lambda}_\sigma\;{{\SigmacalB_\mu}^\sigma}{}_\nu - \calB_{\mu\sigma}\;
{\SigmacalB^{\lambda\sigma}}_\nu\right]\ .
\label{KcupSigmaBcalB}
\end{equation}
Damit ist die zur Spindichte ${\bf\SigmacalB}$ geh"orige Kontorsion aufgefunden.
Verbinden wir nun die Gleichungen \rf{GammaasympGammaKasymp} und \rf{KcupcalBC}, so
ergibt sich die {\em totale Kontorsion} ${\bf\Kasymp}$ zu
\begin{equation}
{\Kasymp^\lambda}_{\mu\nu} := {\Kdach^\lambda}_{\mu\nu} - {\Kcup^\lambda}_{\mu\nu}
\label{Kasympkdachkcup}
\end{equation}
welche die Abweichung der torsionsfreien Konnexion ${\bf\Gamma}$ von der
modifizierten Oberfl"achenkonnexion ${\bf\Gammaasymp}$ beschreibt:
\begin{equation}
{\Gammaasymp^\lambda}_{\mu\nu} = {\Gamma^\lambda}_{\mu\nu} +
{\Kasymp^\lambda}_{\mu\nu}\ .
\label{GammaasympGammaKasymp}
\end{equation}
Die {\em totale Torsion} ${\bf\Zasymp}$ lautet somit
\begin{equation}
{\Zasymp^\lambda}_{\mu\nu} = {\Kasymp^\lambda}_{[\mu\nu]} = {\Zdach^\lambda}_{\mu\nu}
- {\Zcup^\lambda}_{\mu\nu}\ .
\label{ZasympExpl}
\end{equation}
Im allgemeinen wird die totale Torsion ${\bf\Zasymp}$ die Oberfl"achenbedingung
\rf{calBZdach2calB} nicht erf"ullen. Aus diesem Grund bestimmen die mikroskopischen
Feldgleichungen \rf{Dgl1} bis \rf{Dgl2} nicht nur die Raum-Zeit Geometrie, sondern
f"uhren auch einen neuen Freiheitsgrad ein, dessen geometrische Bedeutung in der
zus"atzlichen Oberfl"achen-Kontorsion ${\bf\Kcup}$ liegt. Die mikroskopischen
Feldgleichungen bestimmen also nicht nur die Verkn"upfung der gravitierenden Materie
mit der Raum-Zeit Geometrie wie es in der Einsteinschen Theorie der Fall ist,
sondern enthalten auch einen zus"atzlichen Eichvektor {\bf C} nicht-Riemannschen
Ursprungs, dessen physikalische Bedeutung zu unserer Verf"ugung steht. {\em
Entsprechend unserer oben dargelegten Philosophie \rf{KdsSigmap} messen wir dem neuen
Feldfreiheitsgrad {\bf C} die Bedeutung einer Spin-Variablen zu!}

\section{Die Kr"ummung}
\indent

Die Zerlegung der Raum-Zeit in die charakteristischen Fl"achen erm"oglicht es, die
4-Kr"ummung {\bf R} aus der inneren und "au\3eren Kr"ummung dieser 3-Fl"achen
aufzubauen. Dadurch werden wir, insbesondere im Falle der Euklidischen Foliation, den
Einflu\3 der 3-Kr"ummung auf die 4-Geometrie untersuchen k"onnen. Au\3erdem werden
wir f"ur den Hubble-Tensor Bedingungen finden, mit denen wir eine streng Riemannsche
Raum-Zeit erhalten, deren Kr"ummung {\bf R} folgende zwei Bianchi-Identit"aten
erf"ullen mu\3
\alpheqn
\begin{eqnarray}
\label{Bianchi1}
R_{\mu[\sigma\nu\lambda]} &\equiv& 0\\
\label{Bianchi2}
R_{\mu\sigma[\nu\lambda;\rho]} &\equiv& 0\ .
\end{eqnarray}
\reseteqn

Die Zerlegung der Riemannschen Konnexion ${\bf\Gamma}$ \rf{GammadachGammaKdach} in
die charakteristische Konnexion ${\bf\Gammadach}$ und Kontorsion ${\bf\Kdach}$
induziert nun eine analoge Aufspaltung der Kr"ummung {\bf R} \rf{RGamma}:
\begin{equation}
{R^\lambda}_{\mu\sigma\nu} = {\Rdach^\lambda}_{\mu\sigma\nu}
+ \nabladach_\nu{\Kdach^\lambda}_{\mu\sigma}
- \nabladach_\sigma{\Kdach^\lambda}_{\mu\nu}
+ {\Kdach^\lambda}_{\rho\sigma}{\Kdach^\rho}_{\mu\nu}
- {\Kdach^\lambda}_{\rho\nu}{\Kdach^\rho}_{\mu\sigma}
-2{\Zdach^\rho}_{\sigma\nu}{\Kdach^\lambda}_{\mu\rho}\ .
\label{R_nach_dach}
\end{equation}
Die 4-Kr"ummung l"a\3t sich somit vollst"andig durch Gr"o\3en der charakteristischen
3-Geometrie darstellen. Um die rechte Seite von Gleichung \rf{R_nach_dach}
in Abh"angigkeit von ${\bf\Fquer}$ und {\bf H} zu berechnen, ben"utzen wir die
Identit"at
\begin{equation}
\left(\calDhalb_\nu\calDhalb_\lambda - \calDhalb_\lambda\calDhalb_\nu\right)
\calB_{i\mu} = {\epsilon_i}^{jk}\Fquer_{j\nu\lambda}\calB_{k\mu} -
{\Rdach^\rho}_{\mu\nu\lambda}\calB_{i\rho} + 2{\Zdach^\rho}_{\nu\lambda}\left(
\calDhalb_\rho\calB_{i\mu}\right)\ ,
\label{2calDhalbcalB}
\end{equation}
welche sich wegen der kovarianten Konstanz von ${\calB}_{i\mu}$ \rf{calDhalbcalB0}
vereinfacht zu
\begin{equation}
\calB_{i\rho}{\Rdach^\rho}_{\mu\nu\lambda} = {\epsilon_i}^{jk}\Fquer_{j\nu\lambda}
\calB_{k\mu}\ .
\label{calBRdachFquercalB}
\end{equation}

Um nun diese Gleichung weiter nach ${\bf\Rdach}$ aufzul"osen, ben"utzen wir die
Tatsache, da\3 der charakteristische Vektor ebenfalls kovariant konstant ist
\rf{nabladachp0}. Die Identit"at
\begin{equation}
\left(\nabladach_\nu\nabladach_\lambda - \nabladach_\lambda\nabladach_\nu\right)p_\mu
= - {\Rdach^\rho}_{\mu\nu\lambda}p_\rho
\label{2nabladachp}
\end{equation}
vereinfacht sich somit zu
\begin{equation}
{\Rdach^\rho}_{\mu\nu\lambda}\,p_\rho = 0\ .
\label{Rdachp0}
\end{equation}
Dieses Ergebnis besagt nun zusammen mit der Metrizit"atsbedingung \rf{nabladachG0},
da\3 die Holonomie-Gruppe der Oberfl"achenkonnexion ${\bf\Gammadach}$ wie die
gew"ohnliche Rotationsgruppe SO(3) auf die charakteristische Distribution
$\Deltadach$ wirkt. Aus Gleichung \rf{calBRdachFquercalB} erhalten wir daher f"ur
${\bf\Rdach}$
\begin{equation}
\Rdach_{\sigma\mu\nu\lambda} = \epsilon^{ijk}\calB_{i\sigma}\Fquer_{j\nu\lambda}
\calB_{k\mu}\ .
\label{Rdach2calBFquer}
\end{equation}

Setzen wir nun noch unsere Gleichungen f"ur die Torsion \rf{ZdachKdach} und
Kontorsion \rf{KdachpH} in \rf{R_nach_dach} ein, so l"a\3t sich {\bf R} durch die
innere Kr"ummung ${\bf\Fquer}$ und den Hubble-Tensor {\bf H} ausdr"ucken
\begin{eqnarray}
\label{R_expl}
R_{\mu\sigma\nu\lambda} &=& \epsilon^{ijk}\calB_{i\mu}\Fquer_{j\nu\lambda}
\calB_{k\sigma} + H_{\mu\lambda}H_{\sigma\nu} - H_{\mu\nu}H_{\sigma\lambda}\\
&+& p_\mu\left(\nabla_\lambda H_{\sigma\nu} - \nabla_\nu H_{\sigma\lambda}\right) -
p_\sigma\left(\nabla_\lambda H_{\mu\nu} - \nabla_\nu H_{\mu\lambda}\right)\nonumber
\ .
\end{eqnarray}
Im Falle einer Euklidischen Foliation $({\bf\Fquer} \equiv 0)$ wird die 4-Kr"ummung
vollst"andig durch den Hubble-Tensor {\bf H} festgelegt. Der entgegengesetzte Fall
liegt vor, wenn der charakteristische Vektor {\bf p} bez"uglich ${\bf\Gamma}$
konstant ist (dann ist ${\bf H}\equiv 0$, s. \rf{Dgl2}), so da\3 die 4-Kr"ummung nur
durch die 3-Kr"ummung ${\bf\Fquer}$ der charakteristischen Fl"achen bestimmt wird.
Wir werden weiter unten zwei Beispiele f"ur diese Grenzf"alle betrachten.

Gleichung \rf{R_expl} erlaubt nun die Zerlegung der Kr"ummung {\bf R} in einen
parallelen $({\bf\Rpara})$ und einen senkrechten Teil $({\bf\Rperp})$:
\begin{equation}
R_{\mu\sigma\nu\lambda} = \Rpara_{\mu\sigma\nu\lambda} +
\Rperp_{\mu\sigma\nu\lambda}
\label{RRparaPsenkr}
\end{equation}
mit
\begin{equation}
\Rpara_{\mu\sigma\nu\lambda} = \epsilon^{ijk}\calB_{i\mu}\Fquer_{j\nu\lambda}
\calB_{k\sigma} + H_{\mu\lambda}H_{\sigma\nu} - H_{\mu\nu}H_{\sigma\lambda}
\label{RparaExpl}
\end{equation}
und
\begin{equation}
\Rperp_{\mu\sigma\nu\lambda} =
p_\mu\left(\nabla_\lambda H_{\sigma\nu} - \nabla_\nu H_{\sigma\lambda}\right) -
p_\sigma\left(\nabla_\lambda H_{\mu\nu} - \nabla_\nu H_{\mu\lambda}\right)\ .
\label{RperpExpl}
\end{equation}
Diese Zerlegung wird f"ur die folgenden Betrachtungen sehr n"utzlich sein. Man
beachte jedoch, da\3 die parallele Kr"ummung ${\bf\Rpara}$, obwohl sie in $\Deltadach$
wirkt, nicht mit der charakteristischen Kr"ummung ${\bf\Rdach}$ \rf{Rdach2calBFquer}
identisch ist.

Nachdem die allgemeine Gestalt der Riemannschen Kr"ummung {\bf R} bekannt ist,
k"onnen wir uns nun der Frage zuwenden, wodurch der Riemannsche Charakter der
Raum-Zeit gew"ahrleistet wird. Das bedeutet, da\3 die beiden Bianchi-Identit"aten
\rf{Bianchi1} und \rf{Bianchi2} einschr"ankende Bedingungen f"ur den Hubble-Tensor
und die 3-Geometrie liefern m"ussen. Durch Einsetzen k"onnen wir uns leicht davon
"uberzeugen, da\3 die zweite Bianchi-Identit"at \rf{Bianchi2} automatisch erf"ullt
ist, vorausgesetzt die analoge Bianchi-Identit"at f"ur die 3-Kr"ummung ${\bf\Fquer}$
gilt:
\begin{equation}
\calDquer_\rho\Fquer_{j\nu\lambda} + \calDquer_\nu\Fquer_{j\lambda\rho} +
\calDquer_\lambda\Fquer_{j\rho\nu} \equiv 0\ .
\label{Bianchi3}
\end{equation}
Gleichung \rf{Bianchi2} liefert also keine weitere Einschr"ankung f"ur die
mikroskopischen Feldgleichungen.

Dagegen erhalten wir aus der ersten Bianchi-Identit"at \rf{Bianchi1} f"ur den
orthogonalen Teil ${\bf\Rperp}$ eine Bedingung an den Hubble-Tensor {\bf H}:
\begin{equation}
\nabla_{[\lambda}H_{\sigma\nu]} \equiv 0\ .
\label{nablaH0}
\end{equation}
F"ur die Herleitung der aus dem parallelen Teil ${\bf\Rpara}$ von {\bf R} stammenden
Bedingung nehmen wir zur Vereinfachung den speziellen Fall einer Euklidischen
Foliation der Raum-Zeit an. Damit erhalten wir f"ur den reduzierten Hubble-Tensor
(s. \rf{hhplushminus} bis \rf{hminushminus})
\begin{equation}
h_{\mu[\lambda}\hminus_{\sigma\nu]} \equiv 0\ .
\label{hhminus0}
\end{equation}
Da der schiefsymmetrische Teil ${\bf\hminus}$ von {\bf h} einem $\taudach$ Schnitt
{\bf l} "aquivalent ist, gilt
\begin{equation}
\hminus_{\mu\nu} = \epsilon_{\mu\nu\lambda\sigma}\,l^\lambda\,p^\sigma\ .
\label{hminusepsilonlp}
\end{equation}
Die Bedingung \rf{hhminus0} kann daher auch in folgender Weise formuliert werden
\begin{equation}
\hplus_{\mu\nu}\,l^\nu = 0\ .
\label{hplusl0}
\end{equation}
Das bedeutet nichts anderes als da\3 der symmetrische Teil ${\bf\hplus}$ von {\bf h}
im Falle der Nichtintegrabilit"at der charakteristischen Distribution $\Deltadach$
und bei Euklidischer Foliation der Raum-Zeit einen 2-dimensionalen Tensor darstellen
mu\3.

Ferner l"a\3t sich aus der ersten Bianchi-Identit"at f"ur ${\bf\Rpara}$ eine weitere
Bedingung herleiten:
\begin{equation}
h_{\mu[\lambda}f_{\sigma]} + f_\mu\hminus_{\lambda\sigma} + H_{\mu[\sigma;\lambda]} =
k_{\mu[\sigma}p_{\lambda]}\ .
\label{hfhminusHkp}
\end{equation}
Hier ist {\bf k} ein reines $\taudach$ Objekt, d.h.
\begin{equation}
k_{\mu\nu}p^\mu = k_{\mu\nu}p^\nu = 0\ .
\label{2kp0}
\end{equation}
Um {\bf k} n"aher zu bestimmen, permutieren wir die Indizes in Gleichung
\rf{hfhminusHkp} zyklisch, wobei sich {\bf k} als eine symmetrische Gr"o\3e
herausstellt:
\begin{equation}
k_{\mu\nu} = k_{\nu\mu}\ .
\label{2k}
\end{equation}
Der Frobeniusvektor {\bf f} koppelt dabei folgenderma\3en an den schiefsymmetrischen
Teil ${\bf\hminus}$ von {\bf h}
\begin{equation}
f_{[\mu}\hminus_{\lambda\sigma]} = 0\ .
\label{fhminus0}
\end{equation}
Diese Bedingung kann aber nur erf"ullt werden, wenn ${\bf\hminus}$ die Gestalt
\begin{equation}
\hminus_{\lambda\sigma} = f_{[\lambda}w_{\sigma]}
\label{hminusfw}
\end{equation}
annimmt, wobei {\bf w} ein $\taudach$ Schnitt ist
\begin{equation}
w_\sigma p^\sigma = 0\ ,
\label{wp0}
\end{equation}
welcher wie der Frobeniusvektor {\bf f} orthogonal zu {\bf l} sein mu\3
\begin{equation}
w_\mu l^\mu = f_\mu l^\mu = 0\ .
\label{wlfl0}
\end{equation}
Die Bi-Vektoren ${\bf p}\wedge{\bf l}$ und ${\bf f}\wedge{\bf w}$ sind also,
abgesehen von einem skalaren Faktor, dual zueinander.

Fassen wir nun diese Ergebnisse zusammen, so l"a\3t sich der Riemannsche
Kr"ummungstensor {\bf R} in einer Form ausdr"ucken, welche unmittelbar die Symmetrien
erkennen l"a\3t, die zur Erf"ullung der ersten Bianchi-Identit"at \rf{Bianchi1}
notwendig sind:
\begin{eqnarray}
\label{R_eukl_foliation}
\frac{1}{2}R_{\mu\sigma\nu\lambda} &=&
\hplus_{\mu[\lambda}\hplus_{\nu]\sigma} - \hminus_{\mu[\lambda}
\hminus_{\nu]\sigma}\\
&+&
2p_{[\mu}\hplus_{\sigma][\nu}f_{\lambda]} + 2p_{[\nu}\hplus_{\lambda][\mu}
f_{\sigma]}\nonumber\\ &+&
\hminus_{\nu\lambda}p_{[\mu}f_{\sigma]} + \hminus_{\mu\sigma}p_{[\nu}
f_{\lambda]}\nonumber\\ &+&
p_\mu p_{[\lambda}k_{\nu]\sigma} - p_\sigma p_{[\lambda}k_{\nu]\mu}\nonumber\ .
\end{eqnarray}
Man beachte jedoch, da\3 {\em dieses Ergebnis nur im speziellen Fall einer
Euklidischen Foliation gilt.}

Als n"achstes wollen wir den Einsteinschen Tensor {\bf E}
\begin{equation}
E_{\mu\nu} = R_{\mu\nu} - \frac{1}{2}R\,G_{\mu\nu}
\label{Etensor}
\end{equation}
untersuchen, da wir diese Gr"o\3e "uber die Einsteinschen Feldgleichungen mit dem
Energie-Impuls Tensor {\bf T} der Materie verkn"upfen k"onnen.
\begin{equation}
E_{\mu\nu} = 8\pi L^2_P T_{\mu\nu}\ .
\label{ELpT}
\end{equation}
Wir f"uhren folgende Abk"urzungen ein
\alpheqn
\begin{eqnarray}
\label{Defwp2}
\wp_{\sigma\lambda} &:=& {h^\mu}_\mu\hplus_{\sigma\lambda} -
{\hplus^\mu}_\lambda\hplus_{\sigma\mu} - {\hminus^\mu}_\lambda\hminus_{\sigma\mu}\\
\label{Defwp1}
\wp_\lambda &:=& {h^\mu}_\mu\,f_\lambda - h_{\lambda\mu}\,f^\mu\ .
\end{eqnarray}
\reseteqn
Mit diesen Gr"o\3en schreibt sich der Einstein-Tensor
\begin{equation}
E_{\sigma\lambda}=-\wp_{\sigma\lambda}+\frac{1}{2}{\wp^\mu}_\mu\,
\calB_{\sigma\lambda} + \frac{1}{2}{\wp^\mu}_\mu\,p_\sigma p_\lambda
+ {k^\mu}_\mu\calB_{\sigma\lambda} - k_{\sigma\lambda} - p_\sigma\wp_\lambda -
p_\lambda\wp_\sigma\ .
\label{Ewp}
\end{equation}
Da unsere Theorie eine universale Zeitrichtung (entlang {\bf p}) beeinhaltet,
definieren wir die Energie $\calE$ als
\begin{equation}
\calE = T_{\mu\nu}\,p^\mu p^\nu = \frac{1}{8\pi L^2_P}\,p^\mu p^\nu E_{\mu\nu} =
\frac{1}{16\pi L^2_P}\,{\wp^\mu}_\mu\ .
\label{calETpp}
\end{equation}
Eine einfache Rechnung ergibt f"ur die Spur von ${\bf\wp}$
\begin{eqnarray}
\label{trwp}
\mbox{Sp}\,{\bf\wp} = {\wp^\mu}_\mu &=& {h^\mu}_\mu\,{h^\sigma}_\sigma -
h^{\mu\sigma} h_{\mu\sigma}\\
&=& \left(\mbox{Sp}\,{\bf h}\right)^2 - \left(\mbox{Sp}\,{\bf h}^2\right)\nonumber
\end{eqnarray}
Wir fordern hier, da\3 die Energie positiv sein soll, d.h.
\begin{equation}
\left(\mbox{Sp}\,{\bf h}\right)^2 \ge \left(\mbox{Sp}\,{\bf h}^2\right)\ ,
\label{posEnBed}
\end{equation} was eine physikalische Erg"anzung der vorausgegangenen geometrischen
Bedingungen f"ur den Hubble-Tensor im Falle einer Euklidischen Foliation darstellt.


\chapter{Das klassische Vakuum}
\label{Unterkonnexion}
\label{SelbstWW}
\indent

Der Vakuumzustand des Universums l"a\3t sich in der herk"ommlichen Beschreibung der
Gravitation ohne Probleme definieren. Man geht davon aus, da\3 die Raum-Zeit
Mannigfaltigkeit die normale Minkowskische (flache) Geometrie annimmt, wenn alle
Materie aus dem Universum entfernt wird. Diese flache Struktur wird nun h"ochstens
etwas durch Gravitationswellen gekr"ummt, wobei nach deren Aussterben wieder ein
Minkowskischer Raum vorliegt.

Diese Vorstellung scheint nun jedoch etwas zu einfach zu sein: Betrachten wir
n"amlich die Nullpunktsenergie der verschiedenen elementaren Teilchenfelder, so
sollten diese aufgrund der Einsteinschen Feldgleichungen ebenfalls eine Kr"ummung des
Raumes hervorrufen. Es w"are in der Tat nur schwer verst"andlich, wenn die
Nullpunktsenergie den Raum nicht ebenso kr"ummen w"urde, wie dies alle anderen Formen
von Masse und Energie tun. Aus diesem Grund liegt es nahe, das klassische Vakuum
nicht mit einer flachen Minkowski-Struktur zu beschreiben, sondern mit Hilfe eines
nicht-trivialen, pseudo-Riemannschen Raumes.

Eine in der Literatur weitverbreitete Vorgehensweise besteht darin, da\3 man den
Einsteinschen Feldgleichungen einen kosmologischen Term hinzuf"ugt
\begin{equation}
E_{\mu\nu} = 8\pi L^2_P\left(\TM_{\mu\nu} + \Lambda_c G_{\mu\nu}\right).
\label{ETMLambdaG}
\end{equation}
$\Lambda_c$ ist die sogenannte "`kosmologische Konstante"'. Diese Gr"o\3e beschreibt
die Energie-Impulsdichte ${\bf\Tnull}$ der Nullpunkt-Feldfluktuationen in der
klassischen N"aherung:
\begin{equation} \Tnull_{\mu\nu} = \Lambda_c\,G_{\mu\nu}\ .
\label{T0LambdaG}
\end{equation}
Hier ist ${\bf\Tnull}$ proportional zur Metrik {\bf G}, da die Vakuumstruktur eine
maximale (d.h. lokale Lorentz-) Symmetrie besitzen mu\3 ($\leadsto$ de-Sitter-Raum,
s.u.).

An diesem Punkt wollen wir nun versuchen, ein leistungsf"ahigeres Modell zu
entwickeln. Wenn das Universum eine nicht-triviale Vakuumstruktur besitzt, mu\3 diese
als ein wichtiger dynamischer Bestandteil der Gravitationstheorie angesehen werden.
Dies f"uhrt uns sofort zu zwei wichtigen Fragen:
\begin{itemize}
\item[i)] Welche dynamischen Gleichungen bestimmen die Vakuumstruktur?
\item[ii)] Wie ver"andert sich das Vakuum, wenn Materie in das leere Universum
hineingebracht wird ("`Vakuum-Polarisation"')?
\end{itemize}
In diesem Kapitel wollen wir uns mit der ersten Frage besch"aftigen, w"ahrend die
zweite Frage Gegenstand einer zuk"unftigen Untersuchung sein wird. Es scheint nun
etwas schwierig zu sein, auf der Grundlage der Einsteinschen Gleichungen eine
dynamische Struktur des Vakuums zu entwickeln, da zu den Feldgleichungen f"ur den
materiefreien Fall $({\bf\TM\equiv 0})$ \rf{ETMLambdaG} viele L"osungen existieren,
wogegen das Vakuum wegen seiner hohen Symmetrie als eindeutig angesehen werden mu\3.
Dies hat zur Folge, da\3 man aufgrund einer Maximalforderung bez"uglich der Symmetrie
den Vakuum-Grundzustand aus der gesamten L"osungsmannigfaltigkeit erst heraussuchen
mu\3. Wir werden jedoch sehen, da\3 mit Hilfe der neuen Feldgleichungen \rf{Dgl1},
\rf{Dgl2} dieses Problem direkter gel"ost werden kann. Dazu mu\3 nur die Form der
Eichtensoren \bm{\calH}$_i$ und \bm{\calF}$_i$, sowie des Hubble Tensors {\bf H} als
Funktion der "Atherfelder \bm{\calB}$_i$ und {\bf p} festgelegt werden, was aber
aufgrund formaler Einschr"ankungen zu einem nahezu eindeutigen Ergebnis f"uhrt. Als
Folge ergibt sich die Vakuumdynamik des "Atherfeldsystems, wobei der
Energie-Impulstensor des Vakuums ${\bf\Tnull}$ mit demjenigen der "Atherfelder
${\bf\Te}$ in ihrer Vakuumkonfiguration identifiziert werden kann. D.h es gilt
${\bf\Tnull} \equiv {\bf\Te}\left|_{Vac}\right.$. Diese Annahmen f"uhren zu einem
Raum konstanter Kr"ummung, welcher einer Euklidischen Foliation unterworfen ist. Die
urspr"ungliche Vorstellung, da\3 das leere Universum nicht gekr"ummt ist bewahrheitet
sich also teilweise, da der Vakuum-gekr"ummte 4-Raum eine flache 3-Foliation besitzt.

Auf diese Weise k"onnen wir mit relativ wenigen und einfachen Annahmen --- und vor
allem ohne irgendwelche Symmetrieforderungen --- einen hochsymmetrischen Grundzustand
des klassischen Vakuums definieren (de-Sitter-Universum). Gleichzeitig bietet sich
dabei eine Beschreibungsm"oglichkeit der angeregten Vakuum-Zust"ande von niedriger
Symmetrie an, indem man den inneren $({\bf\Min})$ und "au\3eren $({\bf\Mex})$
Feldvariablen gewisse Abweichungen von ihren (konstanten) Grundzustands-Werten
erlaubt.

\section{Teleparallelismus und Vakuum}
\label{Vakuumkandidat1}
\indent

Die einfachste Struktur f"ur ein leeres Universum erhalten wir im Falle eines
verschwindenden Hubble-Tensors {\bf H} und Eichtensors \bm{\calH}$_i$. Dies wird uns
zu einem Raum mit Teleparallelismus f"uhren, das hei\3t es wird m"oglich sein im
Tangentialb"undel der Raum-Zeit Mannigfaltigkeit eine flache Konnexion einzuf"uhren.
Wir werden sehen, da\3 die Spindichten ${\bf\SigmacalB}$ und ${\bf\Sigmap}$
\rf{DefSigmap}, \rf{DefSigmacalB}, sowie die Expansionsrate H \rf{HspurH} unter diesen
Vorraussetzungen identisch verschwinden, so da\3 sich eine sehr spezielle Struktur
der Raum-Zeit ergibt, welche sich als geeignetes Modell f"ur ein leeres Universum
anbietet.

Aus den mikroskopischen Gleichungen ersehen wir, da\3 die "Atherfelder unter den
obigen Annahmen kovariant konstant sind:
\alpheqn
\begin{eqnarray}
\label{nablap0}
\nabla_\mu p_\nu &=& 0\\
\label{calDcalB0}
\calD_\mu \calB_{i\nu} &=& 0\ .
\end{eqnarray}
\reseteqn
Setzen wir die Funktion \bm{\calF} \rf{Dgl4} identisch Null, so ist der Minkowski
Raum eine spezielle L"osung dieses Gleichungssystems, wobei sich die Riemannsche
Metrik {\bf G} auf die Minkowski Metrik ${\bf g} = \mbox{diag}(1,-1,-1,-1)$
reduziert und die Konnexion ${\bf\Gamma}$ in die (flache) kanonische Konnexion
${\bf\Gammakr}$ "ubergeht. Die Komponenten der "Atherfelder k"onnen wir ansetzen zu
$p_\nu\equiv {g^0}_\nu$, $\calB_{i\nu} \equiv g_{i\nu}$. Der Teleparallelismus wird
somit trivialerweise realisiert.

Nun existiert jedoch noch eine zweite nichttriviale L"osung von \rf{nablap0},
\rf{calDcalB0} Aus dem Verschwinden des Hubble-Tensors l"a\3t sich n"amlich nicht
folgern, da\3 der Kr"ummungstensor ebenfalls verschwindet. Betrachten wir Gleichung
\rf{R_expl}, so reduziert sich die 4-Kr"ummung {\bf R} auf die 3-Kr"ummung
${\bf\Rdach}$: ${\bf R} \equiv{\bf\Rdach}=:{\bf\Rstern}$. Das bedeutet die
4-Geometrie der Raum-Zeit wird nun zu einer reinen Oberfl"achengeometrie
$({\bf\Gamma} \equiv {\bf\Gammadach} =: {\Gammastern})$ deren Kr"ummung sich aus
Gleichung \rf{R_expl} ergibt:
\begin{equation}
\Rstern_{\sigma\mu\nu\lambda} = \epsilon^{ijk}\calB_{i\sigma}\Fquer_{j\nu\lambda}
\calB_{k\mu}\ .
\label{Rstern2calBFquer}
\end{equation}
Die charakteristische Kontorsion ${\bf\Kdach}$ \rf{KdachpH} und der Eichvektor {\bf
C} \rf{C} verschwinden dagegen, d.h der Spinfreiheitsgrad der Gravitation ist in
diesem Fall nicht angeregt. Die erste Bianchi Identit"at \rf{Bianchi1} mu\3
nat"urlich f"ur die charakteristische Kr"ummung erf"ullt sein. Wir erhalten somit
aufgrund von Gleichung \rf{Rstern2calBFquer} die Bedingung
\begin{equation}
\epsilon^{ijk}\calB_{i\sigma}\Fquer_{j[\nu\lambda}\calB_{k\mu]}\equiv 0\ .
\label{eps2calBFquer0}
\end{equation}

F"ur die weiteren "Uberlegungen wollen wir uns auf die folgende L"osung der Gleichung
\rf{eps2calBFquer0} beschr"anken:
\begin{equation}
\Fquer_{j\nu\lambda} = \zeta{\epsilon_j}^{kl}\calB_{k\nu}\calB_{l\lambda}\ ,
\label{Fquerzeta2calB}
\end{equation}
wobei $\zeta$ eine skalare Gr"o\3e ist, welche aufgrund der Bianchi Identit"at
\rf{Bianchi3} konstant sein mu\3. Wir setzen also
\begin{eqnarray}
\label{zetasigmascrc}
\zeta &=& \frac{\sigma}{\scrc^2}\\
(\sigma &=& \plmi 1)\nonumber\ .
\end{eqnarray}
Der konstante L"angenparameter $\scrc$ renormiert die "Atherfelder gem"a\3
\begin{equation} B_{i\mu} := \frac{1}{\scrc}\calB_{i\mu}\ ,
\label{BcalB}
\end{equation}
so da\3 die Unterkr"ummung ${\bf\Fquer}$ \rf{Fquerzeta2calB} und die
charakteristische Kr"ummung ${\bf\Rstern}$ \rf{Rstern2calBFquer} die folgende Gestalt
annehmen
\alpheqn
\begin{eqnarray}
\label{Fquer2calB}
\Fquer_{j\nu\lambda} &=& \frac{\sigma}{\scrc^2}{\epsilon_j}^{kl}\calB_{k\nu}
\calB_{l\lambda}\\
\label{Rstern2calB}
\Rstern_{\sigma\mu\nu\lambda} &=& \frac{\sigma}{\scrc^2}\left(\calB_{\sigma\nu}
\calB_{\lambda\mu} - \calB_{\sigma\lambda}\calB_{\nu\mu}\right)\ .
\end{eqnarray}
\reseteqn
Aus der kovarianten Konstanz der Gr"o\3e \bm{\calB} \rf{calDcalB0} ergibt sich
\begin{equation}
D_\mu B_{i\nu} = D_\nu B_{i\mu}\ ,
\label{2DB}
\end{equation}
was wiederum die G"ultigkeit der Bianchi Identit"at f"ur ${\bf\Fquer}$ zu Folge hat.

Wir haben nun eine ganz spezielle Geometrie erhalten, welche wir in Kapitel VI
n"aher untersuchen werden. An dieser Stelle wollen wir nur deren wichtigste
Eigenschaften kurz aufz"ahlen: Aufgrund der kovarianten Konstanz der Gr"o\3e
\bm{\calB} \rf{calDcalB0} ergibt sich die Konstanz der $\taudach$ Fasermetrik
\bm{\calB}$^2$ (vgl.\rf{nabladachcalB0})
\begin{equation}
\nablastern_\lambda\,\calB_{\mu\nu} = 0\ .
\label{nablasterncalB0}
\end{equation}
Damit ist die Konstanz der Riemannschen Kr"ummung ${\bf\Rstern}$ gew"ahrleistet
\begin{equation}
\nablastern_\rho\,\Rstern_{\sigma\mu\nu\lambda} = 0\ .
\label{nablasternRstern0}
\end{equation}
Diese Gleichung besagt das wir einen (lokal) symmetrischen Raum vorliegen haben.
Ferner sehen wir, da\3 der Weylsche Tensor {\bf W} von ${\bf\Rstern}$ verschwindet:
\begin{equation}
{W^{\nu\mu}}_{\kappa\lambda} = {\Rstern^{\nu\mu}}_{\kappa\lambda} -\left(
{\Rstern^\nu}_{[\kappa}{G^\mu}_{\lambda]} - {\Rstern^\mu}_{[\kappa}{G^\nu}_{\lambda]}
\right) + \frac{1}{3}\,\Rstern\,{G^\nu}_{[\kappa}{G^\mu}_{\lambda]} = 0\ .
\label{WRsternG}
\end{equation}
Diese Eigenschaft des Weylschen Tensors ist nun gerade die notwendige und
hinreichende Bedingung f"ur die Existenz einer konformal flachen Geometrie. Berechnen
wir den Ricci-Tensor nach Gleichung \rf{Rstern2calB}, so ergibt sich
\begin{equation}
\Rstern_{\mu\lambda} = \sigma\frac{2}{\scrc^2}\calB_{\mu\lambda} =
\frac{1}{3}\Rstern\,\calB_{\mu\lambda}\ .
\label{RsterncalB}
\end{equation}

Diese Beziehung zwischen dem Ricci Tensor und der Fasermetrik \bm{\calB}$^2$ besagt,
da\3 die $\taudach$ B"undelgeometrie von einem dreidimensionalen Einsteinschen Raum
herr"uhrt. Nach einem Theorem der Riemannschen Geometrie ist jeder Einsteinsche,
konformal flache Raum ein Raum konstanter Kr"ummung. In unserem Fall erkennt man mit
Hilfe der Gleichung \rf{Rstern2calB} sofort die G"ultigkeit diese Theorems, da dort
${\bf\Rstern}$, ausgedr"uckt mit Hilfe der Fasermetrik \bm{\calB}$^2$, die f"ur einen
Raum mit konstanter Kr"ummung notwendige Gestalt hat. Aus diesem Grund ist die
gesamte $\taudach$ B"undelgeometrie, welche hier mit der totalen 4-Geometrie der
Raum-Zeit "ubereinstimmt, die Pullback-Geometrie eines Modellraumes mit konstanter
Kr"ummung. Ein Beispiel f"ur solch einen Modellraum ist die dreidimensionale
(pseudo-) Sph"are $S^3_{(\sigma)}$. Die Berechnung der Pullback-Geometrie l"a\3t sich
leicht bewerkstelligen und wird au\3erdem die Entstehung einer flachen Konnexion
begr"unden, welche den Teleparallelismus erzeugt.

Diese flache Konnexion, welche wir $\omegakr$ nennen wollen, ist folgenderma\3en
definiert
\begin{equation}
\omkrsig_\mu = \Aquer_{i\mu}\,L^i + B_{i\mu}\,l^i_{(\sigma)}\ ,
\label{omkrsigAquerLBl}
\end{equation}
wobei $\{L^i,l^i_{(\sigma)}\}$ die Rotations- und Boostgeneratoren der
vierdimensionalen Rotationsgruppen SO(4) bzw. SO(1,3) sind, je nachdem ob $\sigma$
positiv oder negativ ist:
\alpheqn
\begin{eqnarray}
\left[L^i,L^j\right] &=& {\epsilon^{ij}}_k\,L^k\\
\left[l^i_{(\sigma)},L^j\right] &=& {\epsilon^{ij}}_k\,l^k\\
\left[l^i_{(\sigma)},l^j_{(\sigma)}\right] &=& \sigma{\epsilon^{ij}}_k\,L^k\ .
\end{eqnarray}
\reseteqn
Die Liealgebra-wertige 1-Form \bm{\omkrsig} \rf{omkrsigAquerLBl} wirkt als eine
Konnexion im Prinzipalb"undel $\lambdasigma$ der $\gsigma$-orthonormalen Tetraden
$\{\hE\}$ "uber der Raum-Zeit mit der Metrik $\gsigma = \mbox{diag}
(\sigma,-1,-1,-1)$. F"ur $\sigma = +1$ ist $\hat{\lambda}_4$ identisch mit dem
Prinzipalb"undels $\lambdakr$, assoziert zum Tangentialb"undel $\taukr$ der
Minkowskischen Raum-Zeit. $\omkrsig$ ist eine flache Konnexion, wie durch Berechnen
der zugeh"origen Kr"ummung \bm{\Omkrsig} mit Hilfe der Gleichungen \rf{Fquer2calB}
and \rf{2DB} leicht festgestellt werden kann. Aufgrund dieser Tatsache erhalten wir
eine eindeutige globale L"osung $\hat{\mbox{\bm{\calE}}}$(x) der Cartanschen
Strukturgleichungen
\begin{equation}
\partial_\mu\hat{\mbox{\bm{\calE}}}_\alpha = \hat{\mbox{\bm{\calE}}}_\beta\;
{\omkrsig^\beta}_{\alpha\mu}\ ,
\label{Cartan}
\end{equation}
so da\3 das Prinzipalb"undel $\lambdasigma$ tats"achlich trivial ist und deshalb
einen eindeutigen Teleparallelismus definiert. Die Strukturgleichungen \rf{Cartan}
k"onnen nun nach den Konnexionskoeffizienten {\bf A} und {\bf B} aufgel"ost werden:
\alpheqn
\begin{eqnarray}
\label{AhEdhE}
A_{i\mu}&=&\frac{1}{2}\,{\epsilon_i}^{jk}\left(\hE_j\cdot\partial_\mu\hE_k\right)\\
\label{BhEdhE}
B_{i\mu} &=& -\left(\hE_i\cdot\partial_\mu\hE_0\right)\ .
\end{eqnarray}
\reseteqn
Aus diesen Gleichungen l"a\3t sich die G"ultigkeit der Trivialisierungsbedingungen
\rf{Fquer2calB} und \rf{2DB} direkt verifizieren.

Da wir einen globalen Schnitt $\hat{\mbox{\bm{\calE}}}$ von $\hat{\lambda}_4$ jetzt
kennen, l"a\3t sich die erste Komponente \mbox{$\hat{\mbox{\bm{\calE}}}_0 :=
{\bf\ndach}$} als ein Schnitt im korrespondierenden (pseudo-) Sph"arenb"undel
$(\gsigma(\ndach,\ndach) = \sigma)$ auffassen, welcher die {\em Gau\3-Abbildung}
$[\ndach]$ der Raum-Zeit ${\bf E}_{1,3}$ auf eine  Einheits (pseudo-) Sph"are
$S^3_{(\sigma)}$ definiert \cite{BrSo86}
\begin{equation}
[\ndach] : E_{1,3} \to S^3_{(\sigma)}\ .
\label{EtoS}
\end{equation}
Dieser Schnitt definiert auch die {\em repr"asentative Distribution} $\Deltaquer$,
welche das repr"asentative B"undel $\tauquer$ festlegt (vgl. die Untersuchung in
[\ref{IntrKrue}]). $\tauquer$ wird durch die Pullback-Abbildung des Tangentialb"undels
von $S^3_{(\sigma)}$ erzeugt. Es zeigt sich also, da\3 die Raum-Zeit als
Pullback-Geometrie bez"uglich der Gau\3 Abbildung $[\ndach]$ von $S^3_{(\sigma)}$
aufgefasst werden kann. Die Fasermetrik \bm{\calB}$^2$ entspricht der ersten
Fundamentalform von $S^3_{(\sigma)}$ und hat daher die folgende Gestalt
\begin{equation}
\calBstern_{\mu\nu} = \scrc^2\left(\partial_\mu\ndach\right)\left(\partial_\nu\ndach
\right)\ .
\label{calBstern2dn}
\end{equation}
Das Linienelement der Raum-Zeit wird somit
\begin{eqnarray}
\label{dsstern}
d\stern{s}{}^2 &=& \Gstern_{\mu\nu}dx^\mu dx^\nu\\
&=& \left(p_\mu dx^\mu\right)^2 + \scrc^2\left(d{\bf\ndach}\right)^2\nonumber\\
&=& \left(d\theta\right)^2 - \scrc^2
\left\{\begin{array}{l}
(d\kappa)^2+\sinh^2\kappa\,(d{\bf\rdach})^2,\hspace{5mm} \sigma = +1\nonumber\\
(d\kappa)^2+\sin^2\kappa\,(d{\bf\rdach})^2,\hspace{5mm} \sigma = -1\nonumber\\
\end{array}\right.\ .
\end{eqnarray}
Dabei erzeugt die "`Universalzeit"' $\theta$ den charakteristischen Vektor {\bf p}
gem"a\3
\begin{equation}
p_\mu = \partial_\mu\theta\ .
\label{pdtheta}
\end{equation}
Der erzeugende Schnitt ${\bf\ndach}$ wurde folgenderma\3en parametrisiert
\begin{equation}
{\bf\ndach} = \left\{\begin{array}{l}
\cosh\kappa\,{\bf\tdach} + \sinh\kappa\,{\bf\rdach},\hspace{5mm} \sigma = +1\\
\cos\kappa\,{\bf\tdach} + \sin\kappa\,{\bf\rdach},\hspace{5mm} \sigma = -1\ ,\\
\end{array}\right.
\end{equation}
wobei das Linienelement der 2-Sph"are wie gew"ohnlich mit Hilfe der sph"arischen
Polarkoordinaten $\vartheta,\varphi$ ausgedr"uckt werden kann
\begin{equation}
(d{\bf\rdach})^2 = d\vartheta^2 + \sin^2\vartheta\,d\varphi^2\ .
\label{spherlinel}
\end{equation}

In dieser Form verwendet man das Linienelement \rf{dsstern} im "`Standardmodell"' der
Kosmologie \cite{MiThWh73}. Es beschreibt ein geschlossenes $(\sigma = -1)$ bzw.
offenes $(\sigma = +1)$ Universum, das mit der Topologie $R^1\otimes S^3_{(\sigma)}$
ausgestattet ist. Es gibt jedoch einen wichtigen Unterschied: in unserem Fall ist der
Radius $\scrc$ der (pseudo-) Sph"are konstant ,wogegen im Standardmodell der Radius
einer bestimmten Bewegungsgleichung unterworfen ist, welche aus den Einsteinschen
Gleichungen abgeleitet werden kann und die Entwicklung des Universums beschreibt. Das
unrealistische Verhalten unseres Modells beruht auf der Wahl eines verschwindenden
Hubble-Tensors {\bf H} (vgl. \rf{nablap0}). Es ist nun trotzdem interessant, den
Energie-Impulstensor {\bf T} dieses Modelluniversums mit Hilfe der Einsteinschen
Gleichungen \rf{ELpT} zu untersuchen. Der Einstein Tensor ${\bf\stern{E}}$
\rf{Etensor} l"a\3t sich mit Hilfe des Ricci Tensors \rf{RsterncalB} folgenderma\3en
ausdr"ucken
\begin{equation}
\stern{E}_{\mu\nu}=-\frac{\sigma}{\scrc^2}\left(\calB_{\mu\nu}+3p_\mu p_\nu\right)\ .
\label{EsterncalBp}
\end{equation}
Dieser Einstein Tensor erzeugt einen Energie-Impuls von der Form wie ihn eine ideale
Fl"ussigkeit besitzt
\begin{equation}
\Tid_{\mu\nu} = {\scr\calM}p_\mu p_\nu - {\scr\calP}\calB_{\mu\nu}\ .
\label{Tid}
\end{equation}
Dabei sind der Druck ${\scr\calP}$ und die Massen-Energiedichte wie folgt verkn"upft
\begin{equation}
\stern{\scr\calP} = - \frac{1}{3}\stern{\scr\calM}\ .
\label{Zustgl}
\end{equation}

Abgesehen von dem unphysikalischen Minuszeichen ist dies die Zustandsgleichung f"ur
ein ideales Gas von masselosen Teilchen. Trotz des etwas unrealistischen Verhaltens
des Modelluniversums erkennt man den Einflu\3 der reinen 3-Geometrie auf den
Energie-Impulsinhalt der Raum-Zeit, welcher in der Erzeugung von Druck
$\stern{\scr\calP}$ und Energiedichte $\stern{\scr\calM}$ besteht:
\alpheqn
\begin{eqnarray}
\label{pstern}
\stern{\scr\calP} &=& \frac{\sigma}{8\pi L_P^2\scrc^2}\\
\label{mstern}
\stern{\scr\calM} &=& -\frac{3\sigma}{8\pi L_P^2\scrc^2}\ .
\end{eqnarray}
\reseteqn

Unser Modell ist aber aus einem weiteren Grund physikalisch nicht akzeptabel.
Aufgrund des Auftretens des charakteristischen Vektors {\bf p} im Vakuumtensor
\rf{Tid} wird ein spezielles Lorentzbezugssytem ausgezeichnet! Zwar ist gegen das
Auftauchen eines solchen Bezugssystems im Falle des Vorhandenseins von Materie nichts
einzuwenden, denn wir k"onnen mit Hilfe der Materieverteilung ein solches
Bezugssystem sicher errichten, jedoch mu\3 der Vakuumzustand des Universums lokal die
volle Lorentzsymmetrie besitzen und somit kann der Vakuum Energie-Impulstensor nur
proportional zur Metrik {\bf G} sein. Mit anderen Worten: die Foliation darf nicht
explizit im Energie-Impulstensor auftauchen. Wir werden sehen, da\3 sich dies in der
Tat erreichen l"a\3t, wenn die 4-Kr"ummung {\bf R} \rf{R_expl} ausschlie\3lich durch
die {\em extrinische} Kr"ummung der charakteristischen Fl"achen erzeugt wird. D.h wir
werden den Fall einer Euklidischen Foliation untersuchen. Da der Vakuumzustand
hochsymmetrisch sein mu\3, werden wir ihn, wie im obigen Beispiel, mit der
Pullback-Geometrie einer 3-(pseudo) Sph"are $S^3_{(\sigma)}$ austatten.
$S^3_{(\sigma)}$ ist parallelisierbar und deshalb wollen wir versuchen eine
Verallgemeinerung innerhalb dieser speziellen Klasse von R"aumen zu finden.

\section{Parallelisierbare R"aume}
\indent

Um eine hinreichend allgemeine Klasse von parallelisierbaren Raum-Zeiten zu finden,
kombinieren wir die Gleichungen \rf{calDquercalBKdachcalB} \rf{KdachpH}:
\begin{equation}
\calDquer_\mu\calB_{i\nu} = -p_\nu\,{H^\lambda}_\mu\,\calB_{i\lambda}
\label{calDquercalBpHcalB}
\end{equation}
und w"ahlen dann den reduzierten Hubble-Tensor {\bf h} \rf{Hhfp} zu
\begin{equation}
h_{\mu\nu} = -\frac{1}{l}\calB_{\mu\nu}\ .
\label{hlcalB}
\end{equation}
Damit erhalten wir f"ur \rf{calDquercalBpHcalB} einen einfacheren Ausdruck
\begin{equation}
\calDquer_\mu\calB_{i\nu} = \frac{1}{l}\,\calB_{i\mu}p_\nu - \varphi_i p_\mu p_\nu\ .
\label{calDquer2calBpvarphi}
\end{equation}
Die L"ange $l$ ist eine Raum-Zeit Funktion und der Higgsskalar \bm{\varphi} das
$\tauquer$ Bild des $\taudach$ Schnittes {\bf f} (Frobenius Vektor) bez"uglich des
B"undelisomorphismus $[B]$ \rf{DefBquer}, \rf{DefBdach}.
\begin{equation}
\varphi_i = \calB_{i\mu}f^\mu\ .
\label{varphicalBf}
\end{equation}
Aufgrund der Symmetrie des reduzierten Hubble-Tensors {\bf h} \rf{hlcalB} k"onnen wir
einen integrierenden Faktor $\lambda$ einf"uhren, so da\3 eine Universalzeit $\theta$
existiert:
\begin{equation}
\partial_\mu\theta = \lambda\,p_\mu\ .
\label{dthetalambdap}
\end{equation}

Wir definieren nun den {\em Radius} $\calR\ (=\calR(\theta))$ des Universums als
Funktion der \mbox{Universalzeit $\theta$} durch
\begin{equation}
\frac{\dot{\calR}}{\calR} = -\frac{1}{l\lambda} \equiv \frac{H}{\lambda}\ .
\label{calR}
\end{equation}
Das erm"oglicht es uns, die "Atherfelder {\bf B} umzuskalieren
\begin{equation}
B_{i\mu} := \calR^{-1}\calB_{i\mu}\ ,
\label{BcalRcalB}
\end{equation}
wobei deren Ableitung symmetrisch wird
\begin{equation}
\calDquer_\mu B_{i\nu}=\frac{1}{l\lambda\calR}\left[\calB_{i\mu}(\partial_\nu\theta)
+\calB_{i\nu}(\partial_\mu\theta)\right] - \frac{1}{\calR}\,\varphi_i p_\mu p_\nu\ ,
\label{cqlDquerB}
\end{equation}
d.h.
\begin{equation}
\Dquer_\mu B_{i\nu} - \Dquer_\nu B_{i\mu} = 0\ .
\label{2DquerB0}
\end{equation}
Im vorangehenden Beispiel \rf{nablap0}, \rf{calDcalB0} wurde schon erw"ahnt, da\3 die
Gleichung \rf{2DquerB0} die eine H"alfte der notwendigen und hinreichenden
Bedingungen f"ur die Existenz einer trivialen Konnexion $\omkrsig$ f"ur das
Prinzipalb"undel $\lambdasigma$ darstellt. Wir k"onnen deshalb diese Bedingung durch
Hinzunahme der entsprechenden anderen H"alfte vervollst"andigen
\begin{equation}
\Fquer_{i\mu\nu} = \sigma{\epsilon_i}^{jk}B_{j\mu}B_{k\nu}\ .
\label{Fquersigma2B}
\end{equation}
Damit ist automatisch die Bianchi Identit"at erf"ullt
\begin{equation}
\calDquer_\lambda\Fquer_{i\mu\nu} + \calDquer_\mu\Fquer_{i\nu\lambda} +
\calDquer_\nu\Fquer_{i\lambda\mu} = 0\ .
\end{equation}
Die Argumentation, welche von \rf{omkrsigAquerLBl} nach \rf{dsstern} gef"uhrt hat,
l"a\3t sich nat"urlich auch auf unser vorliegendes Beispiel anwenden. Der Radius des
Universums ist jetzt aber eine Funktion der Universalzeit $\theta$. Das Linienelement
\rf{dsstern} verallgemeinert sich deshalb zu
\begin{equation}
ds^2 = \left(\frac{d\theta}{\lambda}\right)^2 - \calR^2 (\theta)
\left\{\begin{array}{l}
(d\kappa)^2+\sinh^2\kappa\,(d{\bf\rdach})^2,\hspace{5mm} \sigma = +1\nonumber\\
(d\kappa)^2+\sin^2\kappa\,(d{\bf\rdach})^2,\hspace{5mm} \sigma = -1\nonumber\ .\\
\end{array}\right.
\label{ds}
\end{equation}
Um zu einer Form des Linienelementes zu gelangen, wie man sie in der Standardkosmologie
findet, ben"utzen wir einen verschwindenden Frobeniusvektor ({\bf f}$\equiv 0$), so
da\3 sich $\lambda$ auf 1 reduziert.

Wir kombinieren nun die $S^3_{(\sigma)}$ Pullback-Geometrie mit der Bedingung f"ur
eine Euklidische Foliation, um eine physikalisch akzeptable Vakuumstruktur f"ur das
leere Universum zu erhalten.

\section{Euklidische Foliation}
\label{EuklFol}
\indent

Die Euklidischen Foliation ist nicht nur rein formal interessant, sondern besitzt
auch einen wichtigen philosophischen Aspekt. Akzeptiert man die Tatsache, da\3 die
Gravitationsselbstwechselwirkungen im Vakuumzustand, ebenso wie alle anderen
Wechselwirkungen, einen bestimmten Energie-Impulsbetrag ${\bf\Te}$ besitzen, so
l"a\3t sich ${\bf\Te}$ in die rechte Seite der Einsteinschen Feldgleichungen
einsetzen. Wir erhalten dann ein inhomogenes Gleichungssystem, welches das leere
Universum (den klassischen Vakuumzustand) bestimmt
\begin{equation}
E_{\mu\nu} = 8\pi L^2_P \Te_{\mu\nu}\ .
\label{ELpTG}
\end{equation}
In diesem Fall kann der Vakuumzustand aber nicht mit dem flachen Minkowski Raum
$(E_{\mu\nu}\equiv 0)$ der Einsteinschen Theorie "ubereinstimmen. Der Vakuumzustand
mu\3 deshalb mit einer nicht-trivialen Riemannschen Geometrie ausgestattet sein!
Jedoch sollte aufgrund der besonderen Rolle der Gravitationswechselwirkungen ein
gewisser Unterschied existieren in der Art und Weise, wie die Vakuumenergie den
Raum im Vergleich zu normaler Materie kr"ummt. Eine L"osung dieses Problems besteht
darin, das Vakuum der Raum-Zeit durch die Gravitationskr"afte so kr"ummen zu lassen,
da\3 die charakteristischen 3-Fl"achen flach bleiben. Mit anderen Worten, die
Gleichung \rf{Fquer0} mu\3 gelten. Ist nun Materie vorhanden, so m"ussen die
Einsteinschen Gleichungen eine L"osung ergeben, welche die charakteristische
Kr"ummung ${\bf\Rdach}$ \rf{Rdach2calBFquer} auschlie\3lich auf den Einflu\3 realer
Materie zur"uckf"uhrt. Wir werden sp"ater ein Beispiel hierf"ur angeben, aber zuerst
wollen wir den physikalisch richtigen Vakuumzustand f"ur den Fall finden, da\3 nur
die "Atherfelder im Universum vorhanden sind.

F"ur die Vakuumform \bm{\calHnull}$_i$ der Funktion \bm{\calH}$_i$ machen wir den
Ansatz
\begin{eqnarray}
\label{calHnull}
\calHnull_{i\mu\lambda} &=& \frac{1}{l}\,\calB_{i\lambda}p_\mu + \frac{1}{L}\,
{\epsilon_i}^{jk}\calB_{j\lambda}\calB_{k\mu}\\
(l,L &=& \mbox{const.})\nonumber\ .
\end{eqnarray}
Der Hubble-Tensor {\bf H} wird sodann nach Gleichung \rf{HcalBcalHp}
\begin{equation}
\Hnull_{\mu\nu} = - \frac{1}{l}\calB_{\mu\lambda}\ .
\label{Hnull}
\end{equation}
Durch Vergleich von \rf{calHnull} und \rf{Hnull} erkennen wir den Unterschied in der
Bedeutung der L"angenparameter $L$ und $l$: Der Hubble-Tensor {\bf H} \rf{Hnull}
enth"alt die L"ange $L$ nicht. Aus diesem Grund ist $L$ ein "au\3erer Parameter des
Gravitationsfeldes, wogegen $l$ die Bedeutung eines inneren Parameters hat. Wir
sehen, da\3 die Symmetrie von {\bf H,} genauso wie im zuvor untersuchten Modell, die
Existenz einer Universalzeit $\theta$ \rf{dthetalambdap} garantiert. Wir k"onnen
deshalb in Gleichung \rf{dthetalambdap} $\lambda\equiv 1$ setzen. Um die
vollst"andige Vakuumdynamik zu erhalten, machen wir f"ur die Funktion \bm{\calF}$_i$
\rf{Dgl4} den folgenden Ansatz
\begin{eqnarray}
\label{calFcalBp}
\calF_{i\mu\nu} &=& f_{||}{\epsilon_i}^{jk}\calB_{j\mu}\calB_{k\nu} + 2f_{\perp}
\calB_{i[\mu}p_{\nu]}\\
(f_{||},f_{\perp} &=& \mbox{const.})\nonumber\ .
\end{eqnarray}
Die Bianchi Identit"at \rf{3calDF} lautet somit
\begin{eqnarray}
\label{3calDcalFcalBp}
0 &\stackrel{!}{=}& \calD_\lambda\calF_{i\mu\nu} + \calD_\mu\calF_{i\nu\lambda} +
\calD_\nu\calF_{i\lambda\mu}\\
&=& 2\left(\frac{f_{\perp}}{L}-\frac{f_{||}}{l}\right){\epsilon_i}^{jk}
\left( p_\lambda\calB_{j\mu}\calB_{k\nu} + p_\mu\calB_{j\nu}\calB_{k\lambda} +
p_\nu\calB_{j\lambda}\calB_{k\mu}\right)\nonumber\ .
\end{eqnarray}
Die L"angenparameter $\{f_{||},f_{\perp}\}$ m"ussen daher folgenderma\3en mit den
Gr"o\3en $\{l,L\}$ verkn"upft sein
\begin{equation}
\frac{f_{||}}{l} = \frac{f_{\perp}}{L}\ .
\label{fparalfsenkL}
\end{equation}

Der Riemannsche Kr"ummungstensor {\bf R} \rf{R_eukl_foliation} unseres
Modelluniversums (\rf{calHnull} - \rf{calFcalBp}) l"a\3t sich leicht berechnen, wenn
wir ber"ucksichtigen, da\3 der Frobeniusvektor {\bf f} \rf{Hhfp} und der
schief-symmetrische Teil ${\bf\hminus}$ \rf{hminushminus} des Hubble-Tensors {\bf H}
identisch verschwinden. F"ur den symmetrische Tensor {\bf k} \rf{hfhminusHkp}
erhalten wir dagegen
\begin{equation}
\knull_{\mu\nu} = \frac{1}{l^2}\calB_{\mu\nu}\ .
\label{knullcalB}
\end{equation}
Daraus bestimmt sich {\bf R} zu
\begin{eqnarray}
\label{Rnull}
\frac{1}{2}\,\Rnull_{\mu\sigma\nu\lambda} &=& \hplus_{\mu[\lambda}\hplus_{\nu]\sigma} +
p_\mu p_{[\lambda}k_{\nu]\sigma} - p_\sigma p_{[\lambda}k_{\nu]\mu}\\
&=& \frac{1}{l^2}\,G_{\mu[\lambda}G_{\nu]\sigma}\nonumber\ .
\end{eqnarray}
Die Vakuumgeometrie wird also durch einen Raum mit konstanter Kr"ummung realisiert,
welcher nach den mikroskopischen Feldgleichungen
\begin{eqnarray}
\label{MikroGl1}
\calD_\lambda\calB_{i\mu} &=& \frac{1}{l}\,\calB_{i\lambda}p_\mu + \frac{1}{L}\,
{\epsilon_i}^{jk}\calB_{j\lambda}\calB_{k\mu}\\
\label{MikroGl2}
\nabla_\lambda p_\mu &=& -\frac{1}{l}\calB_{\mu\lambda}
\end{eqnarray}
in die charakteristischen 3-Fl"achen zerlegt ist.

Als n"achstes wollen wir uns davon "uberzeugen, da\3 dies eine Euklidische Foliation
ist. Dazu berechnen wir den Eichvektor {\bf C} \rf{C} und erhalten
\begin{equation}
\Cnull_{i\mu} = \frac{1}{2}{\epsilon_i}^{jk}\calB_{j\sigma}\,{{\calHnull_k}{}^\sigma}
{}_\mu = - \frac{1}{L}\calB_{i\mu}\ .
\label{Cnull}
\end{equation}
Mit Hilfe von Gleichung \rf{FquerFC} ergibt sich f"ur die intrinsische Kr"ummung
${\bf\Fquer}$ des repr"asentativen B"undels $\tauquer$ der folgenden Ausdruck
\begin{equation}
\Fquer_{i\mu\nu} = \left(f_{||}-\frac{1}{L^2}\right){\epsilon_i}^{jk}\calB_{j\mu}
\calB_{k\nu} + \left(f_{\perp}-\frac{1}{lL}\right)\left(\calB_{i\mu}p_\nu -
\calB_{i\nu}p_\mu\right)\ .
\label{FquercalBp}
\end{equation}
Die charakteristischen 3-Fl"achen besitzen also genau dann eine Euklidische Geometrie
$({\bf\Fquer} \equiv 0)$, wenn wir verlangen
\alpheqn
\begin{eqnarray}
\label{fparaL}
f_{||} &= \frac{1}{L^2}\\
\label{fsenklL}
f_{\perp} &= \frac{1}{lL}\ .
\end{eqnarray}
\reseteqn
Die Funktion \bm{\calF} \rf{calFcalBp} wird damit
\begin{equation}
\calFnull_{i\mu\nu} = \frac{1}{L^2}{\epsilon_i}^{jk}\calB_{j\mu}\calB_{k\nu} +
\frac{1}{lL}\left(\calB_{i\mu}p_\nu - \calB_{i\nu}p_\mu\right)\ .
\label{calFnull}
\end{equation}

Fassen wir zusammen: Der klassische Vakuumzustand des Universums ist ein Raum
konstanter Kr"ummung \rf{Rnull}, welcher nach den mikroskopischen Gleichungen
\rf{MikroGl1} und \rf{MikroGl2} sowie nach \rf{calFnull} in flache 3-Hyperfl"achen
zerlegt wird ({\em Euklidische Foliation}). Der korrespondierende Einstein-Tensor
${\bf\Enull}$ ergibt sich nach Gleichung \rf{Ewp} zu
\begin{equation}
\Enull_{\mu\nu} = \frac{3}{l^2}\,G_{\mu\nu}
\label{EnulllG}
\end{equation}
woraus wir mit Hilfe der Vakuum-Einsteingleichungen
\begin{equation}
\Enull_{\mu\nu} = 8\pi L^2_P \Tnull_{\mu\nu}
\label{EnullTnull}
\end{equation}
einen nichtverschwindenden Vakuum Energie-Impulstensor ${\bf\Tnull}$ erhalten
\begin{equation}
\Tnull_{\mu\nu} = \frac{3}{8\pi L^2_P l^2}G_{\mu\nu}\ .
\label{TnullG}
\end{equation}
Dieser Energie-Impulstensor ist physikalisch akzeptabel, da in ihm der
charakteristische Vektor {\bf p}, welcher den "Atherflu\3 beschreibt, nicht
auftaucht. Mit anderen Worten, der "Atherwind ist nicht me\3bar! Trotzdem taucht nun
die Frage auf woher der nichtverschwindende Energie-Impulstensor herr"uhrt. Da die
"Atherfelder die einzigen Objekte sind, welche im materiefreien Universum existieren,
sind sie es, die den Energie-Impulsinhalt des Vakuums erzeugen. Wir werden diese
Annahme sogleich beweisen, vorher wollen wir jedoch die "Atherspindichte untersuchen.

Die charakteristische Kontorsion ${\bf\Kdach}$ l"a\3t sich leicht aus Gleichung
\rf{KdachpH} bestimmen:
\begin{equation}
{\Kdach^\lambda}_{\mu\nu} = - \frac{1}{l}\left(p^\lambda\calB_{\mu\nu} - p_\mu
{\calB^\lambda}_\nu\right)\ .
\label{KdachlpcalB}
\end{equation}
Daraus erhalten wir unmittelbar f"ur die charakteristische Torsion ${\bf\Zdach}$
\rf{ZdachKdach2} den Ausdruck
\begin{equation}
{\Zdach^\sigma}_{\nu\lambda} = -\frac{1}{l}{\calB^\sigma}_{[\nu}p_{\lambda]}\ ,
\label{ZdachlcalB}
\end{equation}
welcher vertr"aglich ist mit dem Oberfl"achentheorem \rf{ZdachGammadach},
\rf{calBZdach2calB}. Die relative Kontorsion ${\bf\Kcup}$ dagegen reduziert sich zu
einer reinen Oberfl"achengr"o\3e, welche mit der relativen Torsion ${\bf\Zcup}$
"ubereinstimmt:
\begin{eqnarray}
\label{Zcup3calB}
{\Zcup^\lambda}_{\mu\nu} = {\Kcup^\lambda}_{\mu\nu} &=& -\frac{1}{L}\,\epsilon^{ijk}
{\calB_i}^\lambda\calB_{j\nu}\calB_{k\mu}\\
\label{Zcuppdual}
&\equiv& -\frac{1}{L}\,{\pdual^\lambda}_{\nu\mu}\nonumber\ .
\end{eqnarray}
Die gesamte Spindichte ${\bf\Sigma} = {\bf\Sigmap} + {\bf\SigmacalB}$ stimmt mit dem
Poincar\'edual ${\bf\pdual}$ des charakteristischen Vektorfeldes {\bf p} "uberein
\begin{equation}
\Sigma_{\mu\nu\lambda} = \frac{1}{2\pi L}\,\pdual_{\mu\nu\lambda}\ .
\label{SigmapdualL}
\end{equation}
Damit erhalten wir eine sehr einfache Beziehung zwischen der relativen Torsion
${\bf\Zcup}$ und dem Spin ${\bf\Sigma}$
\begin{equation}
{\Zcup^\lambda}_{\mu\nu} = 2\pi{\Sigma^\lambda}_{\mu\nu}\ .
\label{ZcupSigma}
\end{equation}
Dieses bemerkenswerte Ergebnis besagt, da\3 die flachen 3-Hyperfl"achen mit einer
nicht-trivialen Oberfl"achentorsion ausgestattet sind, welche f"ur $L\to\infty$
verschwindet, wobei wir ebenfalls $l\to\infty$ gesetzt haben (vgl. weiter unten). Die
Vakuumdynamik \rf{MikroGl1}, \rf{MikroGl2} wird in diesem Grenzfall trivial und
die Riemannsche Vakuumstruktur reduziert sich zum flachen Minkowski Raum. Wir sehen
also, da\3 der Spinparameter $L$ f"ur die kompliziertere Vakuumstruktur
verantwortlich ist.

\section{Die Vakuumenergie}
\indent

Die Vakuumdynamik \rf{MikroGl1}, \rf{MikroGl2} und \rf{calFnull} ist der
Ausgangspunkt f"ur die Definition der entsprechenden Vakuum Energie-Impulsdichte
${\bf\Tnull}$. Wir werden nat"urlich die Vakuumdichte ${\bf\Tnull}$ mit dem
Energie-Impulsinhalt ${\bf\Te}$ der "Atherfelder in der Vakuumkonfiguration
identifizieren. Um den Energie-Impulstensor ${\bf\Te}$ in einer beliebigen
"Atherkonfiguration zu erhalten, liegt es nahe den kanonischen Formalismus
anzuwenden. Wenn das "Atherfeldsystem im flachen Raum bekannt w"are, k"onnte man es
dann mit Hilfe des "Aquivalenzprinzips, welches f"ur Gleichungen erster Ordnung gilt,
in den gekr"ummten Raum "ubertragen. Leider funktioniert dies in unserem Fall nicht,
da das Gleichungssystem erster Ordnung \rf{MikroGl1}, \rf{MikroGl2} und \rf{calFnull}
im flachen Raum inkonsistent ist. Der Grund liegt darin, da\3 es die Riemannsche
Struktur der Raum-Zeit eindeutig bestimmt (vgl. auch \ref{RiemStr}). Wir m"ussen
deshalb zuerst ein System von Gleichungen zweiter Ordnung aus der Vakuumdynamik
konstruieren. Auf dieses Gleichungssystem wenden wir dann den kanonischen Formalismus
im flachen Raum an und erhalten so den Energie-Impulsinhalt ${\bf\Te}$ der
"Atherfelder.

Durch eine weitere Differentiation des Systems \rf{MikroGl1}, \rf{MikroGl2} erreichen
wir eine Entkopplung der Feldvariablen:
\alpheqn
\begin{eqnarray}
\label{2nablap}
\nabla_\lambda\nabla^\lambda p_\mu &=& -\frac{3}{l^2}\,p_\mu\\
\label{2calDcalB}
\calD_\lambda\calD^\lambda\calB_{i\mu} &=& -\frac{1}{\calL^2}\calB_{i\mu}\\
\Big(\frac{1}{\calL^2} &:=& \frac{1}{l^2} - \frac{2}{L^2}\Big)\nonumber\ .
\end{eqnarray}
\reseteqn
Die Divergenz des Feldst"arketensors \bm{\calFnull} \rf{calFnull} ergibt
\begin{eqnarray}
\label{calDFnull}
\calD^\mu\,\calFnull_{i\mu\nu} &=& -{\epsilon_i}^{jk}\,\Bnull_{j\mu}\left(\calD_\nu\,
{\Bnull_k}^\mu\right)\\ \Big(\Bnull_{i\mu} &:=&
\frac{1}{L}\,\left(1+\frac{L^2}{l^2}\right)^{\frac{1}{2}} \calB_{i\mu}\equiv
\scrc^{-1}\calB_{i\mu} \Big)\nonumber
\end{eqnarray}
und schlie\3t das Gleichungssystem zweiter Ordnung. Diese Gleichungen vom
Klein-Gordon bzw. Yang-Mills-Higgs Typ sind nun formal entkoppelt und w"urden in
einer quantisierten Version dieser Theorie die Existenz von Spin-1 Teilchen nach sich
ziehen, welche die Bestandteile des "Athers sind. Ein Teilchen dieser Art wollen wir
"`{\em Gravon}"' nennen. Die typische Gravonmasse l"a\3t sich aus Gleichung
\rf{2nablap} herleiten:
\begin{equation}
m_G = \sqrt{3}\frac{\hbar}{|l|c}\ .
\label{mG}
\end{equation}
Interessant hierbei ist, da\3 diese Massenformel die Tr"agheits-Eigenschaft eines
Elementar-Teilchens mit einer globalen Eigenschaft der Raum-Zeit verkn"upft, n"amlich
mit der Hubble-Konstanten $H$ (vgl. \rf{calR}). Eine Beziehung dieser Art ist unter
dem Begriff {\em Mach'sches Prinzip} bekannt! Ben"utzen wir den heutigen Wert f"ur
die Hubble-Konstante
\begin{equation}
H \equiv \frac{\dot{\calR}}{\calR} = -\frac{1}{l} \approx \frac{1}{1.7\cdot 10^{28}
[\mbox{cm}]}\ ,
\label{Hnumerisch}
\end{equation}
so ergibt sich f"ur das Gravon eine Masse von
\begin{equation}
m_G \approx 3\cdot 10^{-66}[\mbox{g}] \approx 10^{-39}m_{el}\ .
\label{mgmel}
\end{equation}
D.h das Gravon ist praktisch masselos. Wie man der Klein-Gordongleichung
\rf{2calDcalB} entnehmen kann, mu\3 der L"angenparameter $l$ immer kleiner als der
Spinparameter $L$ sein damit wir einen reellen Wert f"ur die Gravonmasse erhalten.
Aus den Gleichungen zweiter Ordnung f"ur die Vakuumdynamik \rf{2nablap},
\rf{2calDcalB} und \rf{calDFnull} k"onnen wir nun mit Hilfe des kanonischen
Formalismus den korrespondierenden Energie-Impulstensor ${\bf\Te}$ f"ur das gesamte
"Atherfeldsystem konstruieren: \footnote{Eine ausf"uhrliche Ableitung findet sich in
Kapitel \ref{Spinkomp}}
\alpheqn
\begin{eqnarray}
\label{SummeT}
\Te_{\mu\nu} &=& \Tp_{\mu\nu} + \TcalB_{\mu\nu} + \TF_{\mu\nu}\\
\label{Tp}
\Tp_{\mu\nu} &=& \frac{1}{4\pi}\left(\nabla_\mu p_\lambda\right)\left(\nabla_\nu
p^\lambda\right) - G_{\mu\nu}\Lambda_p\\
\label{TcalB}
\TcalB_{\mu\nu} &=& -\frac{1}{4\pi}\left(\calD_\mu\calB_{i\lambda}\right)
\left(\calD_\nu\calB^{i\lambda}\right) - G_{\mu\nu}\Lambda_B\\
\label{TcalF}
\TcalF_{\mu\nu} &=& \frac{1}{4\pi}{\calF^i}_{\mu\lambda}{\calF_{i\nu}}^\lambda -
G_{\mu\nu}\Lambda_\calF\ ,
\end{eqnarray}
\reseteqn
mit den folgenden Ausdr"ucken f"ur die Lagrangedichten $\Lambda$
\alpheqn
\begin{eqnarray}
\label{Lambdap}
\Lambda_p &=& \frac{1}{8\pi}\left[\left(\nabla^\mu p^\nu\right)\left(\nabla_\mu p_\nu
\right) - \frac{3}{l^2}\,p^\mu p_\mu\right]\\
\label{LambdacalB}
\Lambda_\calB &=& -\frac{1}{8\pi}\left[\left(\calD^\mu\calB^{i\nu}\right)
\left(\calD_\mu\calB_{i\nu}\right) - \frac{1}{\calL^2}\,\calB^{i\mu}\calB_{i\mu}
\right]\\
\label{LambdacalF}
\Lambda_\calF &=& \frac{1}{16\pi}\,\calF_{i\mu\nu}\calF^{i\mu\nu}\ .
\end{eqnarray}
\reseteqn
Ersetzen wir hier die Ableitungen erster Ordnung der "Atherfelder durch die aus der
Vakuumdynamik \rf{MikroGl1} und \rf{MikroGl2} folgenden Ausdr"ucke und den Eichtensor
\bm{\calF} durch Gleichung \rf{calFnull}, dann wird die "Atherdichte ${\bf\Te}$
\begin{equation} \Te_{\mu\nu} =
\frac{3}{8\pi}\frac{1+{\dis\frac{L^2}{l^2}}}{L^4}G_{\mu\nu}\ . \label{TeLlG}
\end{equation}

F"ur das Vakuum sind nun die "Atherdichte ${\bf\Te}$ und die Vakuumdichte
${\bf\Tnull}$ \rf{TnullG} identisch. Aufgrund dieser Identifikation erhalten wir die
folgende Bedingung f"ur die L"angenparameter $\{l,L\}$:
\begin{equation}
L^2_P = L^2 \frac{L^2}{L^2+l^2}\ .
\label{LpLl}
\end{equation}
Dieser Ausdruck ist eine Bestimmungsgleichung f"ur den Spinparameter $L$ als Funktion
des inneren Parameters $l$.

Werden beide L"angenparameter $\{l,L\}$ unendlich, so verschwinden die "Atherdichte
${\bf\Te}$ \rf{TeLlG} genauso wie der Vakuum Einstein-Tensor ${\bf\Enull}$
\rf{EnulllG} und die Einsteingleichungen \rf{ELpTG} sind trivialerweise erf"ullt. Es
liegt dann ein triviales Vakuum (Minkowskische Raum-Zeit) vor. Ein endlicher Wert von
$l$ erzwingt jedoch aufgrund der Einsteingleichungen \rf{ELpTG} ebenfalls einen
endlichen Wert f"ur $L$. Wir sehen also, da\3 die Existenz eines nichttrivialen
Vakuums von der klassischen Anregung eines Spinfreiheitsgrades des "Atherfeldes
herr"uhrt!

Physikalisch gesehen wird das nichttriviale Vakuum durch das folgende Linienelement
charakterisiert $(\theta\equiv t)$:
\begin{equation}
ds^2=dt^2-\left(\frac{\calR(t)}{\calR_*}\right)^2\left(dx^2 + dy^2 +dz^2\right)\ .
\label{dscalRdxdydz}
\end{equation}
Den Radius $\calR$ des Universums erhalten wir durch Integration der Hubble-Konstanten
\rf{calR} $(\lambda = 1)$
\begin{equation}
\calR(t) = \calR(t_*)\exp\left(-\frac{t-t_*}{l}\right)\ .
\label{calRexp}
\end{equation}
Das Universum ist also offen und expandiert $(l<0)$ bzw. kontrahiert $(l>0)$
exponentiell bez"uglich eines Referenzzeitpunktes $t_*$. Da die Energiedichte
$\calUnull$ des Vakuums konstant ist (vgl. \rf{TeLlG}), gilt
\begin{equation}
\calUnull := \Tnull_{\mu\nu}p^\mu p^\nu = \frac{3}{8\pi L^2\scrc^2}\ .
\label{calUnull}
\end{equation}
Die Energie, welche sich in einem 3-Volumen $V=\calR^3$ befindet, nimmt exponentiell
ab bzw. zu
\begin{equation}
\Unull = \calUnull\cdot V = \frac{3\calR^3_*}{8\pi L^2\scrc^2}\exp\left(-3\,
\frac{t-t_*}{l}\right)\ .
\label{Unull}
\end{equation}
Die Arbeit $\delta{\cal A}$, welche an dem 3-Volumen $V$ durch den Vakuumdruck
$\calPnull$
\begin{equation}
\calPnull = - \calUnull = -\frac{3}{8\pi L^2\scrc^2}
\label{calPnull}
\end{equation}
verrichtet wird, betr"agt
\begin{equation}
\delta{\cal A} = -\calPnull\delta V = + \calUnull\delta V\ .
\label{deltacalA}
\end{equation}
Demnach ist das Energieerhaltungsgesetz erf"ullt
\begin{equation}
\frac{d\Unull}{dt} - \frac{d{\cal A}}{dt} = 0\ .
\label{dUnulldt}
\end{equation}
D.h die Expansion bzw. Kontraktion des Vakuums ist adiabatisch.

In der Literatur wird die vorliegende Vakuum-Konfiguration als
{\em "`deSitter-Universum"'} bezeichnet. In der herk"ommlichen Gravitationstheorie
\cite{MiThWh73} beschreibt diese Konfiguration eine L"osung der modifizierten
Einstein-Gleichungen \rf{ETMLambdaG} f"ur ein leeres Universum $({\bf\TM}\equiv 0)$,
welches eine nicht-verschwindende kosmologische Konstante besitzt $(\Lambda_c \ne 0)$.
Dabei werden folgende Annahmen gemacht:
\begin{itemize}
\item[i)] Das Universum expandiert bzw. kontrahiert stetig und
\item[ii)] sieht f"ur alle Beobachter, die sich mit der kosmologischen
Fl"ussigkeit mitbewegen, immer gleich aus. Dabei spielt der Standort und der
Zeitpunkt der Beobachtung keine Rolle ("`{\em perfektes kosmologisches Prinzip}"')
\end{itemize}

Innerhalb der vorliegenden mikroskopischen Gravitationstheorie ist diese
konventionelle Betrachtungsweise jedoch h"ochst fragw"urdig: Das deSitter-Universum
wird, wie oben erw"ahnt, als ein leeres Universum betrachtet. Es existiert keine
kosmologische Fl"ussigkeit, welche als Bezugssystem f"ur einen hypothetischen
Beobachter ben"utzt werden kann. Die einzige Behauptung die man aufstellen kann, ist
die, da\3 das deSitter-Universum spezielle Klassen von Beobachtern {\em zul"a\3t},
f"ur welche das Universum, wie in i) und ii) aufgef"uhrt, erscheint. ($\leadsto$
Euklidische Foliation des deSitter-Universums). Die verschiedenen Klassen lassen
sich durch einen SO(1,4)-Boost, welcher auf den einbettenden 5-dimensionalen
pseudo-Euklidischen Raum ${\bf E}_{1,4}$ wirkt, ineinander "uberf"uhren. Das
deSitter-Universum wird dabei als 4-dimensionale pseudo-Sph"are in ${\bf E}_{1,4}$
eingebettet.

Die mathematische Auswahl eines speziellen Beobachters aufgrund von
Symmetrie"uberlegungen scheint vom physikalischen Standpunkt doch etwas unmotiviert
zu sein. In unserer mikroskopischen Theorie ist dieser Schwachpunkt jedenfalls nicht
vorhanden, da diese auf dem Postulat aufgebaut ist, da\3 die Raum-Zeit spontan in
Zeitrichtung $({\bf p} = {\bf d}\theta)$ und einen 3-dimensionalen Raum
$(\Deltadach)$ aufspaltet, wobei die mikroskopische Dynamik \rf{MikroGl1} und
\rf{MikroGl2} der Raum-Zeit {\em automatisch} eine deSitter-Struktur mit
Euklidischer Foliation auferlegt. Aus diesem Grund ben"otigen wir die zwei
Forderungen i) und ii) nicht, um in unserer Theorie den Vakuumzustand zu definieren.
Der Vakuumzustand wird vielmehr durch Wahl der Objekte \bm{\calH}$_i$ und {\bf H}
\rf{calHnull}, \rf{Hnull} eindeutig festlegt. Diese einfachste Wahl bietet sich von
selbst an, wenn wir uns ins Ged"achtnis rufen, da\3 au\3er den "Atherfeldern
\mbox{\bm{\calB}$_i$}, {\bf p} keine anderen Gr"o\3en zur Konstruktion der Objekte
\bm{\calH}$_i$, {\bf H} im leeren Universum vorhanden sind.

Fassen wir zusammen. Wir haben gezeigt, da\3 die mikroskopische Dynamik \rf{MikroGl1},
\rf{MikroGl2} und \rf{calFnull} eine Riemannsche Struktur der Raum-Zeit liefert, welche
alle Eigenschaften eines klassischen Vakuums besitzt:
\begin{itemize}
\item[i)] Die Raum-Zeit Geometrie ist hochsymmetrisch (Symmetriegruppe: SO(1,4)) und
wird durch einen parallelisierbaren Riemannschen Raum konstanter Kr"ummung erzeugt.
Anzahl der Symmetrie-Operationen: $\frac{1}{2}\cdot 5\cdot 4 = 10$
\item[ii)] Der Vakuumgrundzustand ist eindeutig und seine Energie-Impulsdichte
${\bf\Tnull}$ ist proportional zur Metrik {\bf G}.
\item[iii)] Die Vakuumdichte ${\bf\Tnull}$ l"a\3t sich mit der kanonischen
Energie-Impulsdichte ${\bf\Te}$ der "Atherfelder identifizieren.
\item[iv)] Da die "Atherfelder die Bestandteile der Riemannschen Metrik {\bf G} sind,
k"onnen wir die "Atherdichte ${\bf\Te}$ ebenfalls mit dem Energie-Impuls der
Gravitationskr"afte identifizieren. Letztere erzeugen dann nach den Einsteinschen
Gleichungen \rf{ELpTG} die Raumkr"ummung in {\em fast} der gleichen Art und Weise wie
die "ubrige Materie. Der Unterschied besteht darin, da\3 der Energie-Impulstensor der
Gravitation die charakteristischen Fl"achen nicht {\em intrinsisch} sondern nur {\em
extrinsisch} kr"ummt!
\end{itemize}

\section{Vakuum-Anregungen}
\indent

Im letzten Abschnitt haben wir festgestellt, da\3 der Grundzustand des klassischen
Vakuums, innerhalb der mikroskopischen Gravitationstheorie, eine dynamische Rolle
spielt. Die n"achste Frage, die sich nun unmittelbar stellt, mu\3 sich daher auf
m"ogliche Abweichungen vom Vakuum-Grundzustand beziehen. (Quantenfeldtheoretisch
gesehen w"urden solche angeregte Zust"ande einer nichtverschwindenden Anzahl von
Gravonen entsprechen). Diese Abweichungen vom Grundzustand besitzen nicht mehr die
hohe Symmetrie des Grundzustandes, d.h der Energie-Impulstensor wird aufgrund der
Vakuumanregung nicht mehr proportional zum Metriktensor {\bf G} sein (vgl.
\rf{TnullG}). Diese Vakuumanregungen k"onnen also nicht mehr durch einen
kosmologischen Term beschrieben werden. Wir wollen dies im folgenden anhand eines
speziellen Typs einer klassischen Vakuumanregung untersuchen. Dieser Vakuumzustand
ergibt sich durch eine leichte Verallgemeinerung der Grundzustands-Tensoren
\bm{\calHnull} und ${\bf\Hnull}$ \rf{calHnull}, \rf{Hnull}:
\alpheqn
\begin{eqnarray}
\label{calHvarphiphicalB}
\calH_{i\mu\lambda} &=& \varphi\calB_{i\lambda}p_\mu + \phi{\epsilon_i}^{jk}
\calB_{j\lambda}\calB_{k\mu}\\
\label{HvarphicalB}
H_{\mu\lambda} &=& - \varphi\calB_{\mu\lambda}\ .
\end{eqnarray}
\reseteqn
Die skalaren Felder $\varphi$ und $\phi$ sind willk"urliche Raum-Zeit Funktionen.
Der Einfachheit halber beschr"anken wir uns auf den speziellen Fall, da\3 beide
Felder nur von der Universalzeit $\theta$ abh"angen, d.h. wir setzen
\alpheqn
\begin{eqnarray}
\label{dvarphidotvarphip}
\partial_\mu\varphi &=& \dot{\varphi}\,p_\mu\\
\label{dphidotphip}
\partial_\mu\phi &=& \dot{\phi}\,p_\mu\ .
\end{eqnarray}
\reseteqn
Damit das Gleichungssystem \rf{calHvarphiphicalB}, \rf{HvarphicalB} geschlossen ist,
ben"otigen wir noch einen Ansatz f"ur die SO(3) Feldst"arke \bm{\calF}. Ein
naheliegender Ansatz ist die Gleichung \rf{calFcalBp}, wobei die Gr"o\3en $f_\perp$
und $f_{||}$ jetzt aber Funktionen "uber der Raum-Zeit sind. Die verallgemeinerten
Gleichungen f"ur die Vakuum-Dynamik lauten also
\alpheqn
\begin{eqnarray}
\label{VerallgMikroGl1}
\calD_\lambda\calB_{i\mu} &=& \varphi\calB_{i\lambda}p_\mu + \phi
{\epsilon_i}^{jk}\calB_{j\lambda}\calB_{k\mu}\\
\label{VerallgMikroGl2}
\nabla_\lambda p_\mu &=& -\varphi\calB_{\mu\lambda}\\
\label{VerallgFeldst}
\calF_{i\mu\lambda} &=& f_{||}{\epsilon_i}^{jk}\calB_{j\mu}\calB_{k\lambda} +
2f_{\perp}\calB_{i[\mu}p_{\lambda]}\ .
\end{eqnarray}
\reseteqn

Das Problem besteht nun darin, nichttriviale L"osungen \bm{\calB}, {\bf p}, $\varphi$
und $\phi$ zu finden, wobei die geometrische Bedeutung der skalaren Gr"o\3e $\varphi$
durch die Hubble-Konstante gegeben ist \mbox{(vgl. \rf{calR}, $\lambda = 1$)}
\begin{equation}
H\equiv \frac{1}{3}\,{H^\mu}_\mu = -\varphi = \frac{\dot{\calR}}{\calR}\ .
\label{Hvarphi}
\end{equation}
Die nichttrivialen L"osungen von \rf{VerallgMikroGl1} - \rf{VerallgFeldst}
beschreiben dann ein dynamisches Universum, dessen zeitliche Entwicklung durch den
Energie-Impulsinhalt der Vakuumanregungen bestimmt wird. Geometrisch gesehen bestehen
diese Anregungen aus etwas komplizierteren, spontanen Aufspaltungen der
4-dimensionalen Raum-Zeit in einen 3-Raum und eine Zeitrichtung. Die Dynamik dieser
(1+3) Zerlegung wird aber nur teilweise durch die mikroskopischen Gleichungen
\rf{VerallgMikroGl1} - \rf{VerallgFeldst} bestimmt. Das Gleichungssystem mu\3 durch
Bewegungsgleichungen f"ur $\varphi$ und $\phi$ vervollst"andigt werden, welche wir
aus den Einsteinschen Gleichungen \rf{ELpTG} f"ur die Vakuumdynamik herleiten
m"ussen. D.h. wir ben"otigen den Einsteintensor {\bf E} und den Energie-Impulstensor
${\bf\Te}$ der angeregten "Ather-Konfiguration. Diese zwei Gr"o\3en sind dann in die
Einsteingleichungen einzusetzen, um ein zus"atzliches Gleichungssystem f"ur die
inneren und "au\3eren Variablen $\varphi$, $\phi$ zu erhalten.

Wir ben"otigen dazu eine allgemeine Definition der Energie-Impulsdichte ${\bf\Te}$
f"ur eine beliebige Konfiguration des "Athersystems. Um zu einer solchen
Verallgemeinerung zu gelangen, betrachten wir nochmals den Vakuum-Grundzustand des
vorausgegangenen Abschnittes. Man sieht leicht, da\3 die Lagrangedichten
$\Lambda_\calB$ \rf{LambdacalB} und $\Lambda_p$ \rf{Lambdap} f"ur den Grundzustand
\rf{MikroGl1}, \rf{MikroGl2} verschwinden. Aus diesem Grund ist es naheliegend die
folgende, allgemeinere Definition f"ur die Energie-Impulsdichte des "Athersystems zu
verwenden:
\begin{eqnarray}
\label{VerallgTe}
\Te_{\mu\nu} = \frac{1}{4\pi\scrc^2}\left(H_{\lambda\mu}{H^\lambda}_\nu -
\calH_{i\lambda\mu}{\calH^{i\lambda}}_\nu\right)+ \frac{1}{4\pi}\left(f_{i\mu\lambda}
{{f^i}_\nu}{}^\lambda - \frac{1}{4}G_{\mu\nu}f_{i\rho\sigma}f^{i\rho\sigma}\right)\\
(\scrc = \mbox{konst.})\nonumber\ .
\end{eqnarray}
Hierbei haben wir von der Tatsache Gebrauch gemacht, da\3 die ${\cal SO}$(3)
Kr"ummung ${\bf\Fquer}$ der charakteristischen Fl"achen verschwindet und der
Eichtensor \bm{\calFnull} \rf{calFnull} deshalb allein durch den Eichvektor {\bf C}
\rf{C}, wie in Gleichung \rf{FquerFC}, ausgedr"uckt werden kann. Wir schlagen
nun vor, da\3 der Eichvektor {\bf C} in dieser Form zur obigen verallgemeinerten
Definition f"ur ${\bf\Te}$ beitr"agt, d.h. wir identifizieren
\begin{equation}
f_{i\mu\nu} := -\left(\calD_\mu C_{i\nu} - \calD_\nu C_{i\mu} + {\epsilon_i}^{jk}
C_{j\mu}C_{k\nu}\right)\ .
\label{fcalDC}
\end{equation}

Der interessante Punkt hierbei ist, da\3 durch die Definition \rf{VerallgTe} der
Energie-Impulsinhalt des "Athers ausschlie\3lich vom zus"atzlichen Freiheitsgrad {\bf
C} herr"uhrt. Dies ist ganz analog zum "Atherspin (vgl. ${\bf\Sigma}$,
\rf{SigmaSigmapSigmaB} Um diesen Effekt deutlich zu machen, brauchen wir nur den
ersten Teil ${\bf\Tp} + {\bf\TcalB}$ von ${\bf\Te}$ \rf{VerallgTe}, welcher die
Tensoren {\bf H} und \bm{\calH} enth"alt, mit Hilfe von {\bf C} auszudr"ucken, da in
der zweiten H"alfte von ${\bf\Te}$ nur {\bf f} \rf{fcalDC} enthalten ist. Dr"ucken
wir also den Hubble-Tensor {\bf H} mit Hilfe des Eichtensors \bm{\calH} \rf{HcalBcalHp}
aus, so ergibt sich
\begin{equation}
H_{\lambda\mu}{H^\lambda}_\nu = \calH_{i\sigma\mu}{\calH^i}_{\rho\nu}\,p^\sigma
p^\rho\ .
\label{2H2calH2p}
\end{equation}
Damit erhalten wir f"ur den ersten Teil von ${\bf\Te}$ \rf{VerallgTe}
\begin{equation}
H_{\lambda\mu}{H^\lambda}_\nu - \calH_{i\lambda\mu}{\calH^{i\lambda}}_\nu = -
\calH_{i\sigma\mu}{\calH^i}_{\rho\nu}\,\calB^{\sigma\rho}\ .
\label{2H4calHcalB}
\end{equation}
Als n"achstes spalten wir den Eichtensor \bm{\calH} in den charakteristischen, von
{\bf C} unabh"angigen Teil $\bar{\mbox{\bm{\calH}}}$ und in einen {\bf C} abh"angigen
Teil auf:
\begin{equation}
\calDquer_\mu\calB_{i\nu} \equiv \bar{\calH}_{i\nu\mu} = \calH_{i\nu\mu} +
{\epsilon_i}^{jk}C_{j\mu}\calB_{k\nu}\ .
\label{ZerlegcalH}
\end{equation}
Man beachte, da\3 der {\bf C} abh"angige Teil herausf"allt, wenn wir den
Hubble-Tensor {\bf H} \rf{HcalBcalHp} berechnen. Aus diesem Grund taucht {\bf C} auch
"uberhaupt nicht in der Riemannschen Kr"ummung {\bf R} auf, welche ausschlie\3lich
eine Funktion des Hubble-Tensors {\bf H} und der 3-Kr"ummung ${\bf\Fquer}$ ist (vgl.
\rf{R_expl}). Vergleichen wir die allgemeing"ultige Zerlegung
des Eichtensors \bm{\calH}
\begin{equation}
\calH_{i\nu\mu} = p_\nu p^\lambda\,\calH_{i\lambda\mu} + {\calB^\lambda}_\nu
\calH_{i\lambda\mu}
\label{calHpcalB}
\end{equation}
mit der vorausgehenden Relation \rf{ZerlegcalH}, so ergibt sich unmittelbar
\begin{equation}
{\calB^\lambda}_\nu\calH_{i\lambda\mu} = - {\epsilon_i}^{jk}C_{j\mu}\calB_{k\nu}\ .
\label{calBcalHCcalB}
\end{equation}
Somit erhalten wir aus \rf{2H4calHcalB}, f"ur den ersten Teil der
Energie-Impulsdichte ${\bf\Te}$ \rf{VerallgTe}
\begin{equation}
H_{\lambda\mu}{H^\lambda}_\nu - \calH_{i\lambda\mu}{\calH^{i\lambda}}_\nu =
2{C^i}_\mu C_{i\nu}\ .
\label{2H2calH2C}
\end{equation}
Damit findet man f"ur ${\bf\Te}$ in "Ubereinstimmung mit der obigen Behauptung:
\begin{equation}
\Te_{\mu\nu} = \frac{1}{2\pi\scrc^2}\,{C^i}_\mu C_{i\nu} + \frac{1}{4\pi}\left(
f_{i\mu\lambda}{{f^i}_\nu}{}^\lambda - \frac{1}{4}\,G_{\mu\nu} f_{i\rho\sigma}
f^{i\rho\sigma}\right)\ .
\label{Te2C4fG}
\end{equation}
Wir sehen, da\3 das "Athersystem i.a. nur dann einen nicht-verschwindenden
Energie-Impuls ${\bf\Te}$ (und Spin \bm{\Sigma}) besitzt, wenn der zus"atzliche
Freiheitsgrad {\bf C} angeregt ist.

Dieses etwas unerwartete Ergebnis legt nun eine neue Auffassung "uber den
Energie-Impulsinhalt der Gravitationskraft nahe. W"ahrend die fr"uheren Beziehungen
\rf{SummeT} - \rf{TcalF} dem "Ather-System (\bm{\calB}, {\bf p}) einen
Energie-Impulsinhalt zugesprochen haben, n"amlich ${\bf\Te} + {\bf\TcalB}$, so wird
jetzt anhand der Beziehung \rf{2H2calH2C} deutlich, da\3 dieser Energie-Anteil
eigentlich dem Eichfeld {\bf C} zuzuordnen ist! Daher wird man sich jetzt aufgrund
der Beziehung \rf{Te2C4fG} auf den Standpunkt stellen, da\3 der Energie-Impulsinhalt
der Gravitationskraft einzig und allein vom {\bf C}-Feld, also den "au\3eren
Variablen ${\bf\Mex}$, getragen wird. Rein formal l"a\3t sich aber ${\bf\Te}$ wegen
der Beziehung \rf{C} auch durch das "Atherfeld \bm{\calB} ausdr"ucken!

Als Beispiel betrachten wir die Vakuum-Dynamik \rf{VerallgMikroGl1},
\rf{VerallgMikroGl2}. Der Ausdruck f"ur den Eichvektor ${\bf\Cnull}$ \rf{Cnull}
verallgemeinert sich zu
\begin{equation}
C_{i\mu} = -\phi\calB_{i\mu}
\label{CphicalB}
\end{equation}
und gibt der skalaren Gr"o\3e $\phi$ die Bedeutung einer extrinsischen Variablen,
welche in der Energie-Impulsdichte ${\bf\Te}$, aber nicht in der Riemannschen
Kr"ummung {\bf R} auftaucht. F"ur das Feld {\bf f} \rf{fcalDC} erhalten wir daher
\begin{equation}
f_{i\mu\nu} = \phi^2 {\epsilon_i}^{jk}\calB_{j\mu}\calB_{k\nu} -
\left(\dot{\phi}-\varphi\phi\right)\left(\calB_{i\mu}p_\nu - \calB_{i\nu} p_\mu
\right)\ .
\label{fexpl}
\end{equation}
Mit diesem Ergebnis k"onnen wir nun den entsprechenden Energie-Impulstensor
${\bf\Tf}$ berechnen und erhalten
\begin{eqnarray}
\label{Tf}
\Tf_{\mu\nu} &:=& \frac{1}{4\pi}\left(f_{i\mu\lambda}{{f^i}_\nu}{}^\lambda -
\frac{1}{4}\,G_{\mu\nu}f_{i\rho\sigma}f^{i\rho\sigma}\right)\\
&=&\frac{1}{8\pi} \left[\left(\dot{\phi}-\varphi\phi\right)^2 + \phi^4 \right]
\left(3p_\mu p_\nu - \calB_{\mu\nu} \right)\nonumber\ .
\end{eqnarray}
Diese Form des Energie-Impulstensors entspricht dem einer idealen Fl"ussigkeit
\begin{equation}
\Tid_{\mu\nu} = {\scr{\cal M}}p_\nu p_\mu - {\scr{\cal P}}\calB_{\mu\nu}\ ,
\label{Tid3}
\end{equation}
wobei die Energiedichte ${\scr\calMf}$ und der Druck ${\scr\calPf}$ durch folgende
Ausdr"ucke gegeben sind
\alpheqn
\begin{eqnarray}
\label{calMf}
\calMf &=& \frac{3}{8\pi}\left[\left(\dot{\phi}-\varphi\phi\right)^2+\phi^4\right]\\
\label{calPf}
\calPf &=& \frac{1}{8\pi}\left[\left(\dot{\phi}-\varphi\phi\right)^2+\phi^4\right]\ .
\end{eqnarray}
\reseteqn
Demnach gehorchen ${\scr\calMf}$ und ${\scr\calPf}$ einer thermodynamischen
Zustandsgleichung f"ur ein ideales Gas von masselosen Teilchen (n"amlich den SO(3)
Eichbosonen des {\bf C} Feldes):
\begin{equation}
\calPf = \frac{1}{3}\,\calMf\ .
\label{calPfcalMf}
\end{equation}
Der Energie-Impulsinhalt des \bm{\calB} und {\bf p} Feldes stellt sich als ein reiner
Drucktensor heraus:
\begin{eqnarray}
\label{TcalBTp}
\TcalB_{\mu\nu} + \Tp_{\mu\nu} &:=& \frac{1}{2\pi\scrc^2}{C^i}_\mu C_{i\nu}\\
&=& \frac{1}{2\pi\scrc^2}\,\phi^2\calB_{\mu\nu}\nonumber\ .
\end{eqnarray}
Damit ergibt sich der totale Energie-Impulsinhalt des gesamten "Athersystems zu
\begin{eqnarray}
\label{TotEnImp}
\Te_{\mu\nu} &=& \TcalB_{\mu\nu} + \Tp_{\mu\nu} + \Tf_{\mu\nu}\\
&=& \frac{3}{8\pi}\left[\left(\dot{\phi}-\varphi\phi\right)^2+\phi^4\right]
p_\mu p_\nu - \frac{1}{8\pi}\left[\left(\dot{\phi}-\varphi\phi\right)^2+\phi^4 -
\frac{4}{\scrc^2}\phi^2\right]\calB_{\mu\nu}\nonumber\ .
\end{eqnarray}
Der im letzten Abschnitt diskutierte Grundzustand \rf{TnullG} kann hieraus wieder
zur"uckgewonnen werden durch die Substitution:
\alpheqn
\begin{eqnarray}
\label{phitoL}
\phi &\to& \frac{1}{L} = \mbox{konst.}\\
\label{varphitol}
\varphi &\to& \frac{1}{l} = \mbox{konst.}\ .
\end{eqnarray}
\reseteqn

\section{Die Konstruktion der Riemannschen Kr"ummung}
\indent

Um zu einer L"osung des Gleichungssystems f"ur die verallgemeinerte Vakuumdynamik
\rf{VerallgMikroGl1} - \rf{VerallgFeldst} zu gelangen, m"ussen wir zuerst den
Riemannschen Kr"ummungstensor {\bf R} aus diesen Gleichungen bestimmen. Zun"achst
"uberzeugen wir uns von deren Widerspruchsfreiheit, das hei\3t, da\3 die
Bianchi-Identit"at \rf{3calDF} erf"ullt sein mu\3. Eine direkte Berechnung ergibt
f"ur den Ansatz \rf{VerallgFeldst}
\begin{equation}
\label{BianchicalF}
\calD_\lambda \calF_{i\mu\nu} + \calD_\mu \calF_{i\nu\lambda} + \calD_\nu
\calF_{i\lambda\mu} = \left[\dot{f}_{||} - 2\left(\varphi f_{||} - \phi f_{\perp}
\right) \right] {\epsilon_i}^{jk}\left(p_\lambda\calB_{j\mu}\calB_{k\nu} +
p_\mu\calB_{j\nu} \calB_{k\lambda} + p_\nu\calB_{j\lambda}\calB_{k\mu} \right)\ .
\end{equation}
Wir haben dabei angenommen, da\3 die skalaren Felder $f_{||}$, $f_{\perp}$, "ahnlich
wie die skalaren Gr"o\3en $\varphi$, $\phi$ \rf{dvarphidotvarphip}, \rf{dphidotphip},
auf jeder charakteristischen Fl"ache konstant sind
\alpheqn
\begin{eqnarray}
\label{dfpara}
\partial_\mu f_{||} &=& \dot{f}_{||}\,p_\mu\\
\label{dfsenkr}
\partial_\mu f_{\perp} &=& \dot{f}_{\perp}\,p_\mu\ .
\end{eqnarray}
\reseteqn
Anstelle von Bedingung \rf{fparalfsenkL} gilt jetzt der allgemeinere Ausdruck
\begin{equation}
\dot{f}_{||} = 2\left(\varphi\,f_{||} - \phi\,f_{\perp}\right)\ .
\label{fpkt}
\end{equation}

Als n"achstes berechnen wir die charakteristische ${\cal SO}$(3) Kr"ummung
${\bf\Fquer}$ \rf{FquerFC}, welche eine Verallgemeinerung von \rf{FquercalBp} ist:
\begin{equation}
\Fquer_{i\mu\nu} = \bar{f}_{||}\,{\epsilon_i}^{jk}\calB_{j\mu}\calB_{k\nu} +
\bar{f}_{\perp}\left(\calB_{i\mu}p_\nu - \calB_{i\nu}p_\mu\right)\ ,
\label{VerallgFquer}
\end{equation}
wobei folgende Abk"urzungen benutzt wurden:
\alpheqn
\begin{eqnarray}
\label{fparaquer}
\bar{f}_{||} &:=& f_{||} - \phi^2\\
\label{fsenkrquer}
\bar{f}_{\perp} &:=& f_{\perp} - \left(\varphi\phi - \dot{\phi}\right)\ .
\end{eqnarray}
\reseteqn
Im Gegensatz zum Grundzustand verschwindet die charakteristische Kr"ummung
${\bf\Rdach}$ \rf{Rdach2calBFquer} nicht:
\begin{equation}
\Rdach_{\mu\sigma\nu\lambda} = \bar{f}_{||}\left(\calB_{\mu\nu}\calB_{\sigma\lambda}
- \calB_{\mu\lambda}\calB_{\sigma\nu}\right) + \bar{f}_{\perp}
\left(\pdual_{\mu\sigma\lambda} p_\nu - \pdual_{\mu\sigma\nu} p_\lambda\right)\ .
\label{VerallgRdach}
\end{equation}
F"ur den parallelen Teil ${\bf\Rpara}$ \rf{RparaExpl} von {\bf R} erhalten wir deshalb
den folgenden, allgemeineren Ausdruck
\begin{equation}
\Rpara_{\mu\sigma\nu\lambda} =
\left(\bar{f}_{||}-\varphi^2\right)\left(\calB_{\mu\nu}\calB_{\sigma\lambda} -
\calB_{\mu\lambda}\calB_{\sigma\nu}\right) + \bar{f}_{\perp}
\left(\pdual_{\mu\sigma\lambda} p_\nu - \pdual_{\mu\sigma\nu} p_\lambda\right)\ .
\label{VerallgRpara}
\end{equation}
Der orthogonale Teil \rf{RperpExpl} ergibt sich zu
\begin{equation}
\Rperp_{\mu\sigma\nu\lambda} = \left(\varphi^2 - \dot{\varphi}\right)
\left( \calB_{\mu\lambda}p_\sigma p_\nu + \calB_{\sigma\nu}p_\mu p_\lambda -
\calB_{\sigma\lambda}p_\mu p_\nu - \calB_{\mu\nu}p_\sigma p_\lambda\right)\ .
\label{VerallgRperp}
\end{equation}
Man beachte, da\3 hier wie erwartet nur der intrinsische Skalar $\varphi$, aber nicht
die extrinsische Gr"o\3e $\phi$ im Riemannschen Kr"ummungstensor \mbox{${\bf R} =
{\bf\Rperp} + {\bf\Rpara}$} auftaucht.

Der totale Riemannsche Kr"ummungstensor {\bf R} mu\3 nun die zwei
Bianchi-Identit"aten \rf{Bianchi1} und \rf{Bianchi2} erf"ullen. Aufrund der ersten
Bianchi-Identit"at mu\3 der zweite Term in der charakteristischen Kr"ummung
${\bf\Rdach}$ \rf{VerallgRdach} verschwinden, d.h
\begin{equation}
\bar{f}_{\perp} \equiv 0\ .
\label{fsenkrquer0}
\end{equation}
Die ${\cal SO}$(3) Kr"ummung ${\bf\Fquer}$ \rf{VerallgFquer} vereinfacht sich deshalb
zu
\begin{equation}
\Fquer_{i\mu\nu} = \bar{f}_{||}\,{\epsilon_i}^{jk}\calB_{j\mu}\calB_{k\nu}\ ,
\label{Fquerfquer2calB}
\end{equation}
d.h. die charakteristischen Fl"achen werden mit der Geometrie einer 3-(pseudo-)
Sph"are $S^3_{(\sigma)}$ ausgestattet. Mit Hilfe der Bianchi-Identit"at f"ur
${\bf\Fquer}$ k"onnen wir eine Beziehung zwischen dem Skalar $\bar{f}_{||}$ und dem
Radius von $S^3_{(\sigma)}$ herleiten. Zu diesem Zweck ist es vorteilhaft, von der
kovarianten Konstanz von \bm{\calB} Gebrauch zu machen. Dazu gehen wir von der
charakteristischen Ableitung $\calDhalbF$ \rf{calDhalbcalB0} aus, welche sich jedoch
auf die charakteristische Konnexion ${\bf\Gammadach}$ mit nicht-verschwindender
Torsion ${\bf\Zdach}$ \rf{ZdachKdach2} bezieht. Deshalb mu\3 die homogene
Bianchi-Identit"at durch ihre torsive Verallgemeinerung ersetzt werden:
\begin{equation}
\calDhalb_\lambda\Fquer_{i\mu\nu} + \calDhalb_\mu\Fquer_{i\nu\lambda} +
\calDhalb_\nu\Fquer_{i\lambda\mu} = 2\left(
\Fquer_{i\nu\sigma}{\Zdach^\sigma}_{\mu\lambda} +
\Fquer_{i\mu\sigma}{\Zdach^\sigma}_{\lambda\nu} +
\Fquer_{i\lambda\sigma}{\Zdach^\sigma}_{\nu\mu}\right)\ ,
\label{VerallgBianchi}
\end{equation}
mit der charakteristischen Torsion ${\bf\Zdach}$
\begin{equation}
{\Zdach^\sigma}_{\nu\lambda} = - \frac{1}{2}\,\varphi\left({\calB^\sigma}_\nu
p_\lambda - {\calB^\sigma}_\lambda p_\nu\right)\ .
\label{ZdachphicalBp}
\end{equation}
Auch hier h"angt wiederum die charakteristische Torsion ${\bf\Zdach}$ ausschlie\3lich
von der intrinsischen Gr"o\3e $\varphi$ ab. Durch Einsetzen "uberzeugt man sich, da\3
${\bf\Zdach}$ dem Oberfl"achentheorem \rf{calBZdach2calB} gen"ugt. Die relative
Torsion \rf{DefZcup} dagegen ist eine Funktion der extrinsischen Variablen $\phi$
\begin{equation}
{\Zcup^\lambda}_{\mu\nu} = \phi\,{\pdual^\lambda}_{\mu\nu}
\label{Zcupphipdual}
\end{equation}
und gehorcht erwartungsgem"a\3 dem Theorem nicht. Setzen wir nun die ${\cal SO}$(3)
Kr"ummung ${\bf\Fquer}$ \rf{Fquerfquer2calB} und die charakteristische Torsion
${\bf\Zdach}$ \rf{ZdachphicalBp} in die inhomogene Bianchi-Identit"at
\rf{VerallgBianchi} ein, so ergibt sich f"ur die Gr"o\3e $\bar{f}_{||}$ die folgende
Bedingung
\begin{equation}
\dot{\bar{f}}_{||} = 2\varphi\bar{f}_{||}\ .
\label{fquerpktvarphifquer}
\end{equation}
Um diese Gleichung zu integrieren, ben"utzen wir die Tatsache, da\3 die
verallgemeinerte Vakuum-Dynamik \rf{VerallgMikroGl1} - \rf{VerallgFeldst} die
Bedingung der Parallelisierbarkeit \rf{calDquer2calBpvarphi} (mit $\lambda = 1$)
erf"ullt. Damit ergibt sich die folgende Relation zwischen $\varphi$ und der
relativen Expansionsrate des Universums:
\begin{equation}
\varphi = - \frac{\dot{\calR}}{\calR}\ ,
\label{varphicalRpktcalR}
\end{equation}
mit der L"osung
\begin{equation}
\bar{f}_{||} = \frac{\sigma}{\calR^2}\ .
\label{fparasigmacalR}
\end{equation}
F"ur $\sigma = 0$ erh"alt man eine Euklidische Foliation der Raum-Zeit, wogegen wir
f"ur \mbox{$\sigma = \plmi 1$} die (pseudo-) sph"arische Geometrie
erhalten. Mit Hilfe dieser Ergebnisse k"onnen wir die endg"ultige Form der
charakteristischen Kr"ummung finden
\alpheqn
\begin{eqnarray}
\label{FquersigmacalR2calB}
\Fquer_{i\mu\nu} &=& \frac{\sigma}{\calR^2}\,{\epsilon_i}^{jk}\calB_{j\mu}
\calB_{k\nu}\\
\label{RdachsigmacalR2calB}
\Rdach_{\mu\sigma\nu\lambda} &=& \frac{\sigma}{\calR^2}
\left(\calB_{\mu\nu}\calB_{\sigma\lambda}-\calB_{\mu\lambda}\calB_{\sigma\nu}\right)\ .
\end{eqnarray}
\reseteqn
Diese Ausdr"ucke unterscheiden sich von den Gleichungen des fr"uheren
Vakuum-Kanditaten \rf{Fquer2calB}, \rf{Rstern2calB} nur durch die Variabilit"at des
Radius $\calR$!

Nachdem jetzt alle Bestandteile \rf{VerallgRdach} - \rf{VerallgRperp} des
Kr"ummungstensors  bekannt sind, k"onnen wir diesen angeben:
\begin{eqnarray}
\label{VerallgR}
R_{\mu\sigma\nu\lambda} &=& \left(\frac{\sigma}{\calR^2}-\varphi^2\right)
\left(G_{\mu\nu}G_{\sigma\lambda}-G_{\mu\lambda}G_{\sigma\nu}\right)\\
&+&\left(\frac{\sigma-\dot{\calR}^2}{\calR^2}+\frac{\ddot{\calR}}{\calR}\right)
\left(G_{\mu\lambda}p_\sigma p_\nu + G_{\sigma\nu}p_\mu p_\lambda -
G_{\mu\nu}p_\lambda p_\sigma - G_{\sigma\lambda}p_\mu p_\nu\right)\nonumber\ .
\end{eqnarray}
Man "uberzeugt sich leicht, da\3 die zwei Bianchi-Identit"aten \rf{Bianchi1} und
\rf{Bianchi2} erf"ullt sind. Eine zweite, indirekte Methode daf"ur besteht darin,
da\3 wir den Einsteintensor {\bf E} berechnen
\begin{equation}
E_{\mu\nu}=\left(\frac{\dot{\calR}^2-\sigma}{\calR^2}+2\,\frac{\ddot{\calR}}
{\calR}\right)\calB_{\mu\nu} + 3\,\frac{\dot{\calR^2}-\sigma}{\calR^2}\,p_\mu p_\nu\ ,
\label{VerallgE}
\end{equation}
welcher eine verschwindende Divergenz besitzen mu\3 \mbox{(\bm{\nabla}$\cdot{\bf E}
\equiv$} 0). Mit dem Energie-Impuls-Tensor ${\bf\Te}$ \rf{TotEnImp} und dem
Einstein-Tensor \rf{VerallgE} k"onnen jetzt die Einstein-Gleichungen f"ur das leere
Universum hingeschrieben werden und dessen Verhalten untersucht werden (siehe
Kapitel VII).


\chapter{Vakuum-Dynamik und Spin-Kompensation}
\indent
\label{Spinkomp}

Nach der Definition des Vakuums im vorhergehenden Kapitel soll nun im Detail
untersucht werden, wie der Riemannsche Charakter des Vakuums mit der Existenz einer
nicht-verschwindenden Vakuum-Spindichte $({\bf\Sigma\ne 0})$ vertr"aglich ist. Das
Problem besteht dabei darin, da\3 spintragende Felder -- wie z.B. unser "Atherfeld
\bm{\calB}, {\bf p} -- eine nicht-verschwindende Mathisson'sche Kraftdichte im
gekr"ummten Raum erfahren. Das hat aber zur Folge, da\3 die Einsteinschen
Feldgleichungen inkonsistent werden. Der "ubliche Ausweg besteht darin, den
Riemannschen Charakter der Raum-Zeit aufzugeben und einen Raum mit Torsion und
Kr"ummung zugrunde zu legen (Riemann-Cartan-Raum), in dem die Divergenz des
Einstein-Tensors nicht notwendigerweise verschwinden mu\3. In Gegensatz zu diesem
Ausweg, wollen wir aber an der Idee des Riemannschen Raumes festhalten und
innerhalb diese Rahmens eine andere L"osung des Problems aufzeigen.

Unsere L"osung besteht darin, statt der Einstein-Gleichungen die mikroskopischen
Feld\-gleichungen \rf{Dgl1} bis \rf{Dgl4} als die fundamentale Dynamik anzunehmen. Es
wird f"ur den Vakuumzustand explizit nachgewiesen, da\3 die L"osungen dieses
Gleichungssystems erster Ordnung genau diejenige Teilmenge der L"osungen des
zugeh"origen 2.Ordnungs-Systems (Klein-Gordon-Gleichungen) identifizieren, auf der
die Mathisson'sche Kraftdichte verschwindet. Dadurch erweist sich der klassische
Vakuumzustand als ein einfaches Demonstrationsbeispiel f"ur den Effekt der
Spin-Kompensation im Riemannschen Raum. Dieser Effekt erm"oglicht also durch\-aus
die Existenz spin-tragender Felder in einem torsionfreien Raum. Zwar produziert der
"Atherspin auch in unserer Theorie eine Torsion, jedoch handelt es sich hierbei um
die Torsion einer Unterkonnexion (n"amlich der charakteristischen Konnexion
${\bf\Gammadach}$, siehe Kapitel \ref{RiemStr}). Die Konnexion ${\bf\Gamma}$
des Tangentialb"undels $\tau_4$ der Raum-Zeit bleibt nach wie vor streng Riemann'sch.

Durch diese Ergebnis scheint uns das oft ge"au\3erte Argument der Anh"anger der
Torsionstheorien, da\3 n"amlich das Spinph"anomen notwendigerweise zu einer
Riemann-Cartan-Struktur der Raum-Zeit f"uhren m"usse, widerlegt zu sein. Wegen der
Bedeutung dieses Resultates wurde das folgende Kapitel soweit wie m"oglich
unabh"angig aufgebaut, so da\3 es auch ohne Kenntnis der "ubrigen Teile der Arbeit
gelesen werden kann.

\section{Das Spin-Problem in der Einsteinschen Theorie}
\indent
\label{VakummKanditat1}

Der Ausgangspunkt in der Allgemeinen Relativit"atstheorie ist die Annahme, da\3 der
metrische Tensor {\bf G} das fundamentale Objekt ist, mit dem sich die anziehende
Materie beschreiben l"a\3t. In diesem Sinne sind die Einsteinschen Feldgleichungen
\begin{eqnarray}
\label{ET}
E_{\mu\nu} = 8\pi\frac{k}{c^4}{\tilde{T}}_{\mu\nu}\equiv 8\pi L^2_p\,T_{\mu\nu}\\
\left(E_{\mu\nu} := R_{\mu\nu} - \frac{1}{2}R\,G_{\mu\nu}\right)\nonumber
\end{eqnarray}
Differentialgleichungen zweiter Ordnung zur Bestimmung des metrischen Feldes {\bf G}
bei gegebenem Energie-Impulstensor {\bf T}. Im Makroskopischen, d.h. f"ur die
Beschreibung von Sternen, Galaxien usw. ist dies richtig, doch versagt die Theorie
schon im klassischen Bereich, wenn man versucht, das Gravitationsfeld an elementare
Teilchen zu koppeln. Im folgenden wollen wir dieses Ph"anomen genauer untersuchen. Im
Riemannschen Raum verschwindet die Divergenz des Einstein Tensors {\bf E} identisch
\begin{equation}
\nabla^\mu E_{\mu\nu}\equiv 0\ .
\label{nablaE}
\end{equation}
Deshalb mu\3 der symmetrische Energie-Impulstensor eine verschwindende Divergenz
besitzen
\begin{equation}
\nabla^\mu T_{\mu\nu} = 0\ .
\end{equation}
F"ur makroskopische Materie ist dies ein plausibles Resultat, im mikroskopischen
Bereich taucht jedoch das Problem auf, wie man sich einen symmetrischen
Energie-Impuls Tensor beschafft, dessen Divergenz im gekr"ummten Raum verschwindet.
In der Literatur wird dieses Problem mehr oder weniger "ubergangen und nur auf
einige Beispiele verwiesen, bei denen das Problem nicht auftaucht.\footnote{Ein
Ausnahme bildet das Buch von V.de Sabbata und M.Gasperini: "`Introduction to
Gravitation"', World Scientific (1985), in welchem die Autoren die Einf"uhrung der
Torsion bevorzugen.}

Bevor wir diese Schwierigkeiten im einzelnen untersuchen, wollen wir jedoch kurz
einen unproblematischen Fall betrachten. Ein solcher liegt bei einem Skalarfeld und
einem reinen Eichfeld vor. Betrachten wir beispielsweise den symmetrischen
Energie-Impulstensor ${\bf\TF}$ f"ur ein SO(3) Eichfeld {\bf F} im flachen Raum
\footnote{Wir ben"utzen die Minkowski Metrik $g_{\mu\nu} = $diag\,(1,-1,-1,-1).}
\begin{equation}
\TF_{\rho\kappa} = \frac{1}{4\pi}\left(g^{\lambda\sigma}{F^i}_{\rho\lambda}
F_{i\kappa\sigma} - \frac{1}{4}g_{\rho\kappa}g^{\lambda\sigma}g^{\mu\nu}
F_{i\lambda\mu}{F^i}_{\sigma\nu}\right)\ ,
\label{TF}
\end{equation}
so erhalten wir dessen Ausdruck im gekr"ummten Raum mit Hilfe des
"Aquivalenzprinzips \mbox{(\bm{\partial}$\to$\bm{\nabla}}, {\bf g}$\to${\bf G}) zu
\begin{equation}
\calTF_{\rho\kappa} = \frac{1}{4\pi}G^{\lambda\sigma}{F^i}_{\rho\lambda}
F_{i\kappa\sigma} - G_{\rho\kappa}\Lambda_F\ ,
\label{calTF}
\end{equation}
wobei die Lagrange-Dichte $\Lambda_F$ durch
\begin{equation}
\Lambda_F=\frac{1}{16\pi}G^{\lambda\sigma}G^{\mu\nu}F_{i\lambda\mu}{F^i}_{\sigma\nu}
\label{LambdaF}
\end{equation}
gegeben ist. Die kovariante Divergenz dieses Tensors ist
\begin{equation}
\nabla_\rho\,\calTF^{\rho\kappa}=\frac{1}{4\pi}\left({\cal D}_\rho{F_i}^{\rho\lambda}
\right){F^{i\kappa}}_\lambda\ .
\label{nablacalTF}
\end{equation}
\bm{\cal D} ist die allgemeine (d.h. eich- und koordinaten-) kovariante Ableitung:
\begin{eqnarray}
\label{calD}
{\cal D}_\sigma F^{i\rho\lambda} &=& \nabla_\sigma F^{i\rho\lambda} + \epsilon^{ijk}
A_{j\sigma}{F_k}^{\rho\lambda}\\
&=& D_\sigma F^{i\rho\lambda} + {\Gamma^\rho}_{\kappa\sigma}F^{i\kappa\lambda} +
{\Gamma^\lambda}_{\kappa\sigma}F^{i\rho\kappa}\nonumber\ .
\end{eqnarray}
Die damit berechnete Divergenz \rf{nablacalTF} verschwindet wie erwartet
\begin{equation}
\nabla_\rho\,\calTF^{\rho\kappa} = 0\ ,
\label{nablacalTF0}
\end{equation}
vorausgesetzt die homogenen Yang-Mills Gleichungen sind erf"ullt
\begin{equation}
{\cal D}_\rho {F_i}^{\rho\lambda} = 0\ .
\label{YM}
\end{equation}
Es entstehen somit keine Schwierigkeiten f"ur die Divergenzfreiheit \rf{nablaE}, wenn wir
die Yang-Mills Felder an das Gravitationsfeld koppeln. Der hier betrachtete Fall des
reinen Eichfeldes ist jedoch etwas zu einfach, da der Energie-Impulstensor
${\bf\calTF}$ \rf{calTF} die Konnexionskoeffizienten ${\bf\Gamma}$ nicht enth"alt.
Der Grund hierf"ur ist, da\3 sich letztere herausheben, wenn man das Feld {\bf F}
mittels des "Aquivalenzprinzips vom flachen- in den gekr"ummten Raum "ubertr"agt:
\begin{eqnarray}
\label{F}
F_{i\mu\nu} &=& \nabla_\mu A_{i\nu} - \nabla_\nu A_{i\mu} +
{\epsilon_i}^{jk}A_{j\mu}A_{k\nu}\\
&\equiv&
\partial_\mu A_{i\nu} - \partial_\nu A_{i\mu} + {\epsilon_i}^{jk}A_{j\mu}A_{k\nu}
\nonumber
\end{eqnarray}
Taucht nun die Konnexion ${\bf\Gamma}$ nun im Energie-Impulstensor {\bf T} auf, dann
produziert die Divergenz von {\bf T} Ableitungen von ${\bf\Gamma}$, welche die Form
eines zus"atzlichen Kr"ummungsterms haben und daf"ur verantwortlich sind, da\3 die
Divergenz nicht mehr verschwindet.

Um dies an einem Beispiel zu verdeutlichen, betrachten wir ein SO(3) Materiefeld
{\bf B}, welches wir durch die folgende Verallgemeinerung der Yang-Mills Gleichung an
ein Eichfeld {\bf F} koppeln wollen:
\begin{equation}
{\cal D}^\mu F_{i\mu\nu} = -{\epsilon_i}^{jk} B_{j\mu}\left({\cal D}_\nu{B_k}^\mu
\right)\ .
\label{calDFBcalDB}
\end{equation}
Weiter nehmen wir an, da\3 das neue Feld {\bf B} der Klein-Gordon Gleichung im
gekr"ummten Raum gehorcht
\begin{equation}
{\cal D}^\lambda{\cal D}_\lambda B_{i\mu} = -\frac{1}{{\cal L}^2}B_{i\mu}\ .
\label{calDcalDB}
\end{equation}
Der L"angenparameter ${\cal L}$ hat die Bedeutung einer Comptonwellenl"ange
\begin{equation}
{\cal L} = \frac{\hbar}{Mc}\ ,
\label{calL}
\end{equation}
wobei $M$ die Masse des zugeh"origen Eichbosons bezeichnet. Wir k"onnen nun das
"Aquivalenzprinzip anwenden, um den {\em kanonischen} Energie-Impuls\-tensor
${\bf\TB}$ im gekr"ummten Raum zu berechnen
\begin{equation}
\TB_{\rho\kappa} = -\frac{1}{4\pi}\left({\cal D}^\rho B^{i\sigma}\right)
\left({\cal D}^\kappa B_{i\sigma}\right) - G^{\rho\kappa}\Lambda_B\ ,
\label{TB}
\end{equation}
mit der Lagrange-Dichte
\begin{equation}
\Lambda_B = -\frac{1}{8\pi}\left(\left({\cal D}_\mu B_{i\nu}\right)
\left({\cal D}^\mu B^{i\nu}\right) - \frac{1}{{\cal L}^2}B_{i\mu}B^{i\mu}\right)\ .
\label{LambdaB}
\end{equation}
Der Tensor ${\bf\TB}$ enth"alt nun explizit die Konnexion ${\bf\Gamma}$. Aus diesem
Grund produziert dessen Divergenz zus"atzliche, aus der Kr"ummung {\bf R} und der
Spindichte ${\bf\SigmaB}$, aufgebaute Terme:
\begin{equation}
\nabla_\rho\,\TB^{\rho\kappa} = -\frac{1}{4\pi}{\epsilon_i}^{jk}{F_{j\nu}}^\kappa
B_{k\mu}\left({\cal D}^\nu B^{i\mu}\right) - \frac{1}{2}{R^\kappa}_{\nu\mu\rho}
\,\SigmaB^{\rho\mu\nu}\ ,
\label{nablaTB}
\end{equation}
wobei der Spin (Drehimpuls) ${\bf\SigmaB}$ wie folgt definiert ist
\alpheqn
\begin{equation}
\SigmaB^{\lambda\nu\mu} = -\frac{\partial\Lambda_B}{\partial\left({\cal D}_\mu
B_{i\sigma}\right)}{(L^{\nu\lambda})^\rho}_\sigma\,B_{i\rho}
\label{defSigmaB}
\end{equation}
\begin{equation}
{(L^{\nu\lambda})^\rho}_\sigma = g^{\nu\rho}{g^\lambda}_\sigma - g^{\lambda\rho}
{g^\nu}_\sigma\ .
\end{equation}
\reseteqn
F"ur das {\bf B} Feld finden wir mit Gleichung \rf{LambdaB}
\begin{equation}
\SigmaB^{\lambda\nu\mu} = -\frac{1}{4\pi}\left(\left({\cal D}^\mu B^{i\nu}\right)
{B_i}^\lambda - \left({\cal D}^\mu B^{i\lambda}\right){B_i}^\nu\right)\ .
\label{SigmaBB}
\end{equation}
Betrachtet man den gesamten Energie-Impulstensor des Eich- und Materiefeldes, so
bereitet der erste Term auf der rechten Seite von \rf{nablaTB} keine Schwierigkeiten,
da er sich mit der rechten Seite von \rf{nablacalTF} aufgrund der inhomogenen
Yang-Mills Gleichung \rf{calDFBcalDB} hinweghebt:
\begin{equation}
\nabla_\rho\left(\calTF^{\rho\kappa} + \TB^{\rho\kappa}\right) =
-\frac{1}{2}{R^\kappa}_{\nu\mu\rho}\,\SigmaB^{\rho\mu\nu}\ .
\label{nablacalTFTB}
\end{equation}
Der verbleibende Kr"ummungsterm, die {\em Mathisson Kraftdichte}, verhindert nun ganz
offensichtlich das Verschwinden der Divergenz in \rf{nablacalTFTB}. Das hei\3t, da\3
der kanonische Energie-Impulstensor des gekoppelten Systems nicht mehr f"ur die
Einstein-Gleichungen \rf{ET} verwendet werden kann. Daher sind diese Gleichungen mit
der kanonischen Feldtheorie im flachen Raum inkonsistent, oder aber das
"Aquivalenzprinzip versagt hier.

Der gew"ohnliche Weg dieses Problem zu beseitigen besteht darin, da\3 man das
"Aquivalenzprinzip scheinbar aufgibt und einen neuen Energie-Impulstensor
\bm{\cal T} (den sog. {\em metrischen} Energie-Impulstensor) einf"uhrt, wobei dessen
Divergenz aufgrund seiner Konstruktion verschwindet. Diesen neuen Tensor \bm{\cal T}
erh"alt man, indem man das Wirkungsintegral $W$ bez"uglich der Metrik {\bf G}
variiert:
\begin{equation}
\delta_G\, W = -\frac{1}{2}\int{\cal T}^{\rho\kappa}\left(\delta G_{\rho\kappa}
\right)\sqrt{-G}\;d^4 x\ .
\label{Wirkung0}
\end{equation}
Paradoxerweise wird hier das "Aquivalenzprinzip, obwohl zur Konstruktion des
geeigneten Energie-Impulstensors gerade erst aufgegeben, trotzdem f"ur das
Wirkungsintegral benutzt:
\begin{equation}
W = \int\Lambda\sqrt{-G}\;d^4 x\ .
\label{Wirkung}
\end{equation}
F"ur das vorliegende System der gekoppelten Felder setzt sich die Lagrange-Dichte
$\Lambda$ aus dem Eichteil $\Lambda_F$ und dem Materieteil $\Lambda_B$ zusammen
\begin{equation}
\Lambda = \Lambda_F + \Lambda_B\ .
\label{Lambda}
\end{equation}
Der modifizierte Tensor ${\bf\calTF}$ f"ur den Eichteil ist, aufgrund seiner
Eichinvarianz, identisch mit dem fr"uheren Tensor \bm{\calTF} \rf{calTF}. Im Falle
eines Materiefeldes unterscheidet sich der neue Tensor \bm{\calTB} jedoch vom
kanonischen ${\bf\TB}$ \rf{TB} durch einen nicht-trivialen Term ${\bf\tB}$
\begin{equation}
\calTB^{\rho\kappa} = \TB^{\rho\kappa} + \tB^{\rho\kappa}\ ,
\label{calTBTBtB}
\end{equation}
welcher folgende Gestalt besitzt
\begin{equation}
\tB^{\rho\kappa} = -\frac{1}{2}\nabla_\nu\left(\SigmaB^{\rho\nu\kappa} +
\SigmaB^{\kappa\nu\rho} - \SigmaB^{\kappa\rho\nu}\right)\ .
\label{tB}
\end{equation}
Aufgrund dieses Korrekturterms verschwindet jetzt die Divergenz des gesamten,
modifizierten Tensors, d.h.
\begin{equation}
\nabla_\rho\left(\calTF^{\rho\kappa} + \calTB^{\rho\kappa}\right) =
\nabla_\rho\left(\calTF^{\rho\kappa} + \TB^{\rho\kappa}\right) +
\nabla_\rho\tB^{\rho\kappa} = 0\ .
\label{nablacalTFcalTB}
\end{equation}
Vergleichen wir diesen Ausdruck mit \rf{nablacalTFTB} so ergibt sich
\begin{equation}
\nabla_\rho\,\tB^{\rho\kappa}=+\frac{1}{2}{R^\kappa}_{\nu\mu\rho}\,
\SigmaB^{\rho\mu\nu}\ .
\label{nablatBRSigmaB}
\end{equation}
F"ur die folgenden Betrachtungen ist es n"utzlich, die Herleitung von Gleichung
\rf{nablatBRSigmaB} im einzelnen zu zeigen. Dazu beachte man, da\3 die Spindichte
${\bf\SigmaB}$ \rf{SigmaBB} schiefsymmetrisch ist
\begin{equation}
\SigmaB^{\rho\nu\kappa} = - \SigmaB^{\nu\rho\kappa}\ ,
\label{SigmaBSchiefsymm}
\end{equation}
und ihre Divergenz aufgrund der Klein-Gordon Gleichung \rf{calDcalDB} verschwindet:
\begin{equation}
\nabla_\kappa\,\SigmaB^{\rho\nu\kappa} = 0\ .
\label{nablaSigmaB}
\end{equation}
Aus diesen Eigenschaften k"onnen wir folgende Relationen herleiten
\alpheqn
\begin{eqnarray}
\label{Rel1}
\nabla_\rho\nabla_\nu\,\SigmaB^{\rho\nu\kappa} &=&
-\frac{1}{2}{R^\kappa}_{\sigma\nu\rho}\,\SigmaB^{\rho\nu\sigma}\\
\nabla_\rho\nabla_\nu\,\SigmaB^{\kappa\nu\rho} &=&
{R^\kappa}_{\sigma\rho\nu}\,\SigmaB^{\sigma\nu\rho}\ ,
\end{eqnarray}
\reseteqn
aus welchen wir, unter Ber"ucksichtigung der ersten Bianchi Identit"at
\begin{equation}
{R^\kappa}_{[\rho\nu\sigma]} = 0,
\end{equation}
die gesuchte Gleichung \rf{nablatBRSigmaB} erhalten. Der {\em metrische}
Energie-Impulstensor f"ur das gekoppelte System besitzt also, wie erwartet, eine
verschwindende Divergenz \rf{nablacalTFcalTB}.

Obwohl dieses Ergebnis in der Literatur weit verbreitet ist, liegt doch eine
unbefriedigende Situation vor:
\begin{enumerate}
\item Der kanonische Tensor ${\bf\TB}$ l"a\3t sich mit Hilfe des "Aquivalenzprinzips
nicht vom flachen in den gekr"ummten Raum "ubertragen, obwohl ${\bf\TB}$ eine in der
Literatur allgemein akzeptierte Gr"o\3e innerhalb der kanonischen Feldtheorie des
flachen Raumes ist.
\item Man mu\3 vielmehr durch Variation des Wirkungsintegrals einen neuen
Materietensor ${\bf\calTB}$ konstruieren, wobei aber der Integrand des Wirkungsintegrals
durch das "Aquivalenzprinzip bestimmt wird.
\item Dieser neue Tensor ${\bf\calTB}$ unterscheidet sich vom kanonischen ${\bf\TB}$
durch einen Spinterm ${\bf\tB}$, welcher auf den ersten Blick f"ur eine direkte
Wechselwirkung zwischen Spin und Raumkr"ummung verantwortlich zu sein scheint
(Mathisson-Kraft). Da der Energie-Impulsinhalt von ${\bf\tB}$ nun im modifizierten
Tensor ${\bf\calTB}$, welcher eine verschwindende Divergenz besitzt, absorbiert ist,
ist man versucht, dieses Verfahren allzu voreilig als eine L"osung des Problems zu akzeptieren.
\end{enumerate}

Diese Interpretation des Korrekturterms ${\bf\tB}$ ist jedoch h"ochst
widerspr"uchlich, da
\begin{enumerate}
\item der Korrekturterm im Grenzfall des flachen Raumes nicht verschwindet, und daher
in diesem Grenzfall zu einer anderen Energie-Impuls-Lokalisierung f"uhrt.
\item Ferner gehorcht das spinnende Teilchen einer anderen Bewegungsgleichung, da
die Mathisson Kraft nicht mehr existent ist.
\end{enumerate}

Aufgrund dieser Problematik in der Einsteinschen Relativit"atstheorie scheint es
unvermeidbar zu sein einige Verbesserungen anzubringen. In diesem Zusammenhang ist
die Idee, die Torsion mit einzubeziehen, eine elegante M"oglichkeit, das Problem des
spinnenden Teilchens zu l"osen \cite{He76}. Bevor jedoch solch eine grundlegende
"Anderung der Einsteinschen Theorie unternommen wird, sollte man versuchen einen
Mechanismus zu finden, der den Einflu\3 des Spins auf die Raum-Zeit Geometrie
innerhalb der Einsteinschen Theorie zu unterdr"ucken vermag.

\section{Das Kompensationsfeld}
\indent

Diese Schwierigkeit weist einige "Ahnlichkeit mit dem Problem auf, in einer
{\em global} eichinvarianten Eichtheorie {\em lokale} Eichinvarianz zu erhalten.
In solch einer Theorie produziert eine lokale Eichtransformation des Materiefeldes
st"orende Terme, welche die Ableitung des ben"utzten Elementes der Eichgruppe
enthalten. Um diese zus"atzlichen Terme zu beseitigen, wird ein Kompensationsfeld
(das Eichpotential {\bf A}) in die Theorie eingef"uhrt, wobei dessen inhomogene Terme
sich mit den St"ortermen hinwegheben. Als Ergebnis erhalten wir ein gekoppeltes
System von Eich- und Materiefelder, welche zusammen die gew"unschte lokale
Eichinvarianz besitzen.

Es ist nun naheliegend, diese Vorgehensweise auch auf unsere Problematik anzuwenden.
Das bedeutet, da\3 wir dem Materiefeld {\bf B} ein neues Feld {\bf W} zuordnen
m"ussen, so da\3 der Spinterm ${\bf\tB}$, welcher das "Aquivalenzprinzip
zerst"ort, durch geeignete, von {\bf W} stammende Terme, kompensiert wird. Daraus
resultiert ein System von Feldern {\bf B} und {\bf W}, wobei deren kanonische
Feldtheorie mit Hilfe des "Aquivalenzprinzips konsistent vom flachen in den
gekr"ummten Raum "ubertragen werden kann.

Dieser Kompensationsmechanismus basiert auf einer engen Beziehung zwischen den
Feldern {\bf B}, {\bf W} und der Metrik {\bf G}. Das Eichanalogon dieser Beziehung
besteht in der gemeinsamen Eichinvarianz des Materie- und Kompensationsfeldes.
Unser Ansatz ruht nun auf der Annahme, da\3 sich der Metriktensor {\bf G} aus den
Feldern {\bf B} und {\bf W} zusammensetzt:\footnote{Der konstante L"angenparameter
$\scrc$ wird eingef"uhrt um die Feldvariablen dimensionlos zu halten:
\bm{\calB} := $\scrc${\bf B}, {\bf p} = $\scrc${\bf W}}
\begin{eqnarray}
\label{GBBWW}
G_{\mu\nu} &=& \scrc^2\left(B_{i\mu}{B^i}_\nu + W_\mu W_\nu\right)\\
&\equiv& \calB_{i\mu}{\calB^i}_\nu + p_\mu p_\nu\nonumber\ .
\end{eqnarray}
Dieser Ausdruck enth"alt nun eine neue, die Interpretation des metrischen Feldes {\bf
G} in der Allgemeinen Relativit"atstheorie betreffende Philosophie! Bisher wurde die
Metrik, abgesehen von der Konnexion ${\bf\Gamma}$, als {\em Potential} betrachtet.
Dieser Standpunkt scheint f"ur den unklaren Eichstatus der Einsteinschen Theorie
verantwortlich zu sein\cite{IvSa83}. Der Ansatz \rf{GBBWW} erm"oglicht es nun,
dem metrischen Feld {\bf G} die Bedeutung eines {\em makroskopischen} Materiefeldes
zu geben, welches durch die {\em mikroskopischen} Felder {\bf B} und {\bf W}
erzeugt wird. Diese "`doppelte"' Betrachtungsweise (makroskopisch und mikroskopisch)
der Allgemeinen Relativit"atstheorie f"uhrt nun zu einigen interessanten Fragen:
\begin{enumerate}
\item Welche Gestalt mu\3 die mikroskopische Dynamik besitzen, damit die folgende
makroskopische Bedingung f"ur {\bf G} erf"ullt ist:
\begin{equation} \nabla_\lambda G_{\mu\nu} = 0\ .
\label{item1}
\end{equation}
\item Besitzen die {\em mikroskopischen} Felder \bm{\calB} und {\bf W} einen {\em
makroskopischen} Energie-Impulstensor {\bf T}, welcher in die rechte Seite der
{\em makroskopischen} Feldgleichungen \rf{ET} eingesetzt werden kann? Ist dies
dann konsistent mit dem "Aquivalenzprinzip?
\end{enumerate}

\section{Die mikroskopische Dynamik}
\indent

Um die erste Frage zu beantworten, nehmen wir an, da\3 die Felder {\bf B} und
{\bf W} auf mikroskopischem Niveau an weitere Felder koppeln, die jedoch nicht
explizit in der Metrik {\bf G} auftauchen. F"ur die gegenw"artigen Betrachtungen
beschr"anken wir uns auf den Fall einer reinen Selbstkopplung von {\bf B} und
{\bf W}. Wir werden sp"ater sehen, da\3 wir mit dieser Einschr"ankung einen sehr
speziellen Riemannschen Raum erhalten, n"amlich einen Raum mit konstanter Kr"ummung.
Unser Ansatz f"ur die mikroskopischen Feldgleichungen lautet also:
\alpheqn
\begin{equation}
{\cal D}_\lambda\calB_{i\mu} = \frac{1}{l}\calB_{i\lambda}\,p_\mu + \frac{1}{L}
{\epsilon_i}^{jk}\calB_{j\lambda}\calB_{k\mu}
\label{MikrGl1}
\end{equation}
\begin{equation}
\nabla_\lambda\,p_\mu = -\frac{1}{l}\calB_{i\mu}{\calB^i}_\lambda
\label{MikrGl2}
\end{equation}
\reseteqn
Die konstanten L"angenparameter $l, L$ sind frei w"ahlbar. F"ur $l, L \to \infty$
erhalten wir die Gleichungen f"ur trivialisierbare Eichfelder (s. Kapitel III).

Wir wollen nun die Folgerungen, die sich aus den Gleichungen \rf{MikrGl1} und
\rf{MikrGl2} ergeben, genau untersuchen. Man sieht leicht, da\3 die Bedingung
\rf{item1} in der Tat erf"ullt ist. Ferner erlauben die mikroskopischen Feldgleichungen
die Auferlegung einer weiteren Bedingung:
\begin{equation}
\calB_{i\mu}p^\mu = 0\ .
\label{calBp}
\end{equation}
Wir k"onnen also annehmen, da\3 der charakteristische Vektor {\bf p} immer zeitartig
ist, w"ahren die drei raumartigen "`Materiefelder"' \bm{\calB_i} , $(i = 1, 2, 3)$
die sogenannte {\em charakteristische Distribution} $\Deltadach$ aufspannen. Weiter
ersehen wir aus \rf{MikrGl1} und \rf{MikrGl2}, da\3 der charakteristische Vektor der
Gradient einer skalaren Funktion $\Theta(x)$ ist
\begin{equation}
p_\mu = \partial_\mu\Theta\ .
\label{pdTheta}
\end{equation}
Dieser Skalar kann als eine Art von "`Universalzeit"' betrachtet werden, wobei die
Integralfl"achen ($\Theta(x) = $const.) von $\Deltadach$ als "`absoluter Raum"'
interpretiert werden k"onnen. Da die L"ange von {\bf p} wegen \rf{calBp} und
\rf{MikrGl2} konstant bleibt, k"onnen wir sie auf Eins normieren
\begin{equation}
G_{\mu\nu}p^\mu p^\nu \equiv p^\mu p_\mu = 1\ .
\label{pp1}
\end{equation}
Der Einheitsvektor {\bf p} l"a\3t sich nun als Vierergeschwindigkeit eines sich
bewegenden Mediums, dem sog. {\em "Ather}, betrachten, dessen Energie-Impulstensor
wir sogleich untersuchen wollen. Aus den Gleichungen \rf{MikrGl1}, \rf{MikrGl2}
und \rf{calBp} entnehmen wir, da\3 die Stromlinien des "Athers, die sog. {\em
charakteristischen Linien}, Geod"aten sind:
\begin{equation}
p^\mu\nabla_\mu p_\nu = 0\ .
\label{Geodaeten}
\end{equation}
Ein Beobachter, der sich entlang der charakteristischen Linien frei bewegt, kann
demzufolge bez"uglich des "Athers als "`lokal in Ruhe"' betrachtet werden. Die
charakteristischen Linien sind aus diesem Grund die Weltlinien der Punkte des
absoluten Raumes.

F"ur die nachfolgenden Betrachtungen ist es n"utzlich, wenn wir die mikroskopischen
Feldgleichungen \rf{MikrGl1} und \rf{MikrGl2} mit Hilfe des (Poincar\'e)
Duals ${\bf\pdual}$ von {\bf p} ausdr"ucken
\begin{equation}
\pdual_{\mu\nu\lambda} := \epsilon_{\mu\nu\lambda\sigma}\,p^\sigma\ .
 \label{defpdual}
\end{equation}
Der \bm{\epsilon} Tensor wird hierbei wie gew"ohnlich definiert:
\begin{eqnarray}
\epsilon_{\mu\nu\lambda\sigma} = \sqrt{-G}\,[\mu,\nu,\lambda,\sigma]\\
\left(\nabla_\kappa\,\epsilon_{\mu\nu\lambda\sigma} = 0\right)\nonumber\ ,
\label{defepsilon}
\end{eqnarray}
wobei die Klammer $[\dots]$ das "ubliche Permutationssymbol bezeichnet. Alternativ
l"a\3t sich ${\bf\pdual}$ mit Hilfe der \bm{\calB} Felder schreiben
\begin{equation}
\pdual_{\mu\nu\lambda} = \epsilon^{ijk}\calB_{i\mu}\calB_{j\nu}\calB_{k\lambda}\ ,
\label{pdualcalB}
\end{equation}
woraus wir den topologischen Charakter innerhalb der SO(3) Eichtheorie erkennen
k"onnen ("`Gau\3 Strom"' \cite{BrSo87}). Wir erhalten somit aus \rf{MikrGl1} und
\rf{MikrGl2}
\begin{equation}
\nabla_\kappa\,\pdual_{\mu\nu\lambda}=\frac{3}{l}\pdual_{\kappa[\mu\nu}p_{\lambda]}.
\label{nablapdual}
\end{equation}
Hieraus folgert man leicht, da\3 die Divergenz von ${\bf\pdual}$ verschwindet:
\begin{equation}
\nabla^\mu\,\pdual_{\mu\nu\lambda} = 0\ .
\label{nablapdual0}
\end{equation}
Wir werden sp"ater sehen, da\3 das Objekt ${\bf\pdual}$ im Kompensationsmechanismus
eine wichtige Rolle spielt.

\section{Der Energie-Impuls des Vakuums}
\indent

Wenden wir uns nun der zweiten Frage zu, so m"ussen wir untersuchen, ob die
mikroskopischen Konstituenten \bm{\calB} und {\bf p} der Metrik {\bf G} einen
Energie-Impulstensor {\bf T} erzeugen. Wenn dies der Fall ist, so wird der
Energie-Impuls, nach den Einsteinschen Gleichungen \rf{ET}, eine Kr"ummung {\bf R}
der Raum-Zeit erzeugen. Diese Vermutung l"a\3t sich nachpr"ufen, indem wir beide Seiten
der Einsteinschen Feldgleichung berechnen und uns "uberzeugen, da\3 sie
"ubereinstimmen. In diesem Abschnitt wollen wir nun die rechte Seite der
Feldgleichungen berechnen. Auf diese Weise werden wir sehen, wie der
Kompensationsmechanismus wirkt.

Es liegt nahe den Energie-Impulstensor {\bf T} der mikroskopischen Felder {\bf B}
und {\bf W} aus den korrespondierenden dynamischen Gleichungen \rf{MikrGl1} und
\rf{MikrGl2} herzuleiten, indem wir {\bf T} im flachen Raum konstruieren und dann
mittels des "Aquivalenzprinzips in den gekr"ummten Raum "ubertragen. Leider ist das
jetzt nicht mehr m"oglich! Der Grund liegt darin, da\3 die Gleichungen \rf{MikrGl1}
und \rf{MikrGl2} im flachen Raum inkonsistent sind. Um dies zu sehen, berechnen wir
die alternierende Ableitung von {\bf p}
\begin{equation}
\nabla_{[\sigma}\nabla_{\lambda]}p_\mu = -\frac{1}{l^2}\calB_{i\mu}
{\calB^i}_{[\sigma}p_{\lambda]}\ .
\label{altAblp}
\end{equation}
woraus offensichtlich im flachen Raum (\bm{\nabla} $\to$ \bm{\partial}) ein
Widerspruch entsteht!

Erinnern wir uns jedoch an die Klein-Gordon Gleichung \rf{calDcalDB}, welche zweiter
Ordnung ist, und f"ur die wir ohne Schwierigkeiten einen assozierten
Energie-Impulstensor konstruieren k"onnen (vgl.\rf{TB} und \rf{LambdaB}). Wir werden
deshalb versuchen, Gleichungen zweiter Ordnung aus den Gleichungen erster
Ordnung herzuleiten. In der Tat erhalten wir so die Klein-Gordon Gleichung f"ur
{\bf B}, vorausgesetzt die Comptonwellenl"ange ${\cal L}$ \rf{calL} gehorcht
der Bedingung
\begin{equation}
\frac{1}{{\cal L}^2} = \frac{1}{l^2} - \frac{2}{L^2}\ .
\label{calLlL}
\end{equation}
Ebenso gen"ugt auch das charakteristische Vektorfeld {\bf p} einer Feldgleichung vom
Klein-Gordon Typ:
\begin{equation}
\nabla_\lambda\nabla^\lambda p_\mu = -\frac{3}{l^2}p_\mu\ .
\label{nablanablap}
\end{equation}
Es ist nun naheliegend, f"ur das korrespondierende Boson eine Masse $m$ einzuf"uhren:
\begin{equation}
m = \sqrt{3}\frac{\hbar}{lc}\ .
\label{m}
\end{equation}
Obwohl die mikroskopischen Felder in den Gleichungen zweiter Ordnung entkoppelt sind,
besteht doch eine schwache Wechselwirkung "uber die Raum-Zeit Geometrie, zu welcher
beide Felder "uber die Metrik {\bf G} \rf{GBBWW} beitragen.

Die Energie-Impulstensoren f"ur die Dynamik zweiter Ordnung lassen sich nun leicht
angeben: Au\3er dem Ergebnis \rf{TB} f"ur ${\bf\TB}$ finden wir f"ur den kanonischen
Energie-Impulstensor ${\bf\TW}$ des {\bf W} Feldes
\begin{eqnarray}
\label{TW}
\TW^{\rho\kappa} &=& \frac{1}{4\pi}\left(\nabla^\rho W_\nu\right)
\left(\nabla^\kappa W^\nu\right) - G^{\rho\kappa}\Lambda_ W\\
\label{LambdaW}
\Lambda_ W &=& \frac{1}{8\pi}\left(\left(\nabla_\mu W_\nu\right)
\left(\nabla^\mu W^\nu\right) - \frac{3}{l^2} W^\mu W_\mu\right)\ .
\end{eqnarray}
Wie erwartet verschwindet die Divergenz von ${\bf\TW}$ nicht, sondern ergibt analog
zu \rf{nablaTB}
\alpheqn
\begin{eqnarray}
\label{nablaTW}
\nabla_\rho\,\TW^{\rho\kappa} &=& -\frac{1}{2}{R^\kappa}_{\nu\mu\rho}\,
\SigmaW^{\nu\mu\rho}\\
\label{SigmaW}
\SigmaW^{\rho\mu\nu} &=& \frac{1}{4\pi}\left(\left(\nabla^\nu W^\mu\right)
 W^\rho - \left(\nabla^\nu  W^\rho\right) W^\mu\right)\ .
\end{eqnarray}
\reseteqn
Da nun die einzelnen Energie-Impulstensoren bekannt sind, k"onnen wir den gesamten
Energie-Impulstensor {\bf T} des "Athers aufstellen:
\begin{equation}
T^{\rho\kappa} = \calTF^{\rho\kappa} + \TB^{\rho\kappa} + \TW^{\rho\kappa}\ .
\label{T}
\end{equation}
Wir k"onnen {\bf T} nun f"ur einen ersten Test der Kompensation ben"utzen, welche,
wenn sie funktioniert, folgendes Resultat ergeben mu\3
\begin{equation}
\nabla_\rho T^{\rho\kappa} = 0\ .
\label{nablaT}
\end{equation}
Aus \rf{nablaTW} und \rf{nablacalTFTB} erhalten wir jedoch
\begin{eqnarray}
\label{nablaTRSigma}
\nabla_\rho T^{\rho\kappa} &=& -\frac{1}{2}{R^\kappa}_{\nu\mu\rho}\Sigma^{\rho\mu\nu}\\
(\Sigma^{\rho\mu\nu} &:=& \SigmaB^{\rho\mu\nu} + \SigmaW^{\rho\mu\nu})\ ,
\nonumber
\end{eqnarray}
wobei die rechte Seite nicht von vornherein verschwindet. Verwenden wir aber die
Gleichungen erster Ordnung f"ur die gesamte Spindichte ${\bf\Sigma}$ (${\bf\Sigma} =
{\bf\SigmaB} + {\bf\SigmaW}$), so ergibt sich hierf"ur
\begin{equation}
\Sigma^{\rho\mu\nu} = -\frac{1}{2\pi\scrc^2 L}\,\pdual^{\nu\mu\rho}\ .
\label{Sigmapdualscrc}
\end{equation}
Ben"utzen wir weiter die erste Bianchi Identit"at \rf{Bianchi1}, so erhalten wir
doch noch das gew"unschte Ergebnis \rf{nablaT}. Dies ist der erste Erfolg des
Kompensationsmechanismus in Verbindung mit den Gleichungen erster Ordnung
\rf{MikrGl1}, \rf{MikrGl2}. Man beachte, da\3 sich die Spindichten ${\bf\Sigma}$ der
{\bf B} und {\bf W} Felder nicht zu Null aufzusummieren brauchen; {\em es gen"ugt
wenn der gesamte Spin proportional einem axialen Vektor ist.}

Wir k"onnen nun, ohne auf einen Widerspruch zu sto\3en, den kanonischen Tensor {\bf
T} des "Atherfeldsystems f"ur die Einstein Gleichungen verwenden. Dieses Ergebnis
besagt nun zun"achst noch nicht da\3 {\bf T} mit seinem modifizierten Analogon
\bm{\cal T} aus Gleichung \rf{Wirkung0} "ubereinstimmt: Betrachten wir dazu den
modifizierten Tensor \bm{\calTW} f"ur das {\bf W} Feld, so unterscheidet sich dieser
von seinem kanonischen Analogon ${\bf\TW}$ \rf{TW} durch einen Korrekturterm
${\bf\tW}$
\begin{equation}
\calTW_{\rho\kappa} = \TW_{\rho\kappa} + \tW_{\rho\kappa}\ ,
\label{calTW}
\end{equation}
welcher analog zu ${\bf\tB}$ aufgebaut ist:
\begin{equation}
\tW^{\rho\kappa} = -\frac{1}{2}\nabla_\nu\left(\SigmaW^{\rho\nu\kappa} +
\SigmaW^{\kappa\nu\rho} - \SigmaW^{\kappa\rho\nu}\right)\ .
\label{tW}
\end{equation}
Da auch dieser Korrekturterm eingef"uhrt wurde um die nicht verschwindende Divergenz
von {\bf T} zu beseitigen, mu\3 er im allgemeinen Fall ungleich Null sein. Diese
Tatsache k"onnen wir nun ben"utzen, um den Kompensationsmechanismus auf den gesamten
modifizierten Tensor \bm{\cal T} anzuwenden:
\begin{equation}
{\cal T}_{\rho\kappa}=\calTF_{\rho\kappa}+\calTB_{\rho\kappa}+\calTW_{\rho\kappa}
\ .
\label{calT}
\end{equation}
Dieser Tensor unterscheidet sich von seinem kanonischen Analogon {\bf T} durch
einen Korrekturterm {\bf t}
\begin{equation}
{\cal T}_{\rho\kappa} = T_{\rho\kappa} + t_{\rho\kappa}\ ,
\label{calTTt}
\end{equation}
welcher sich aus der Summe der einzelnen Korrekturterme ${\bf\tW}$ und ${\bf\tB}$
zusammensetzt
\begin{equation}
t^{\rho\kappa} = -\frac{1}{2}\nabla_\nu \left(\Sigma^{\rho\nu\kappa} +
\Sigma^{\kappa\nu\rho} - \Sigma^{\kappa\rho\nu}\right)\ .
\label{tSigma}
\end{equation}
Ben"utzen wir nun das Ergebnis \rf{Sigmapdualscrc} f"ur ${\bf\Sigma}$, so erhalten wir
aufgrund von \rf{nablapdual0} unmittelbar die Identit"at der beiden
Energie-Impulstensoren:
\begin{equation}
{\cal T}_{\rho\kappa} \equiv T_{\rho\kappa}
\label{calTT}
\end{equation}
Wir sehen, da\3 das Kompensationsprinzip in Verbindung mit den Feldgleichungen erster
Ordnung all die erw"ahnten Inkonsistenzen beseitigt!

Bevor wir nun den Energie-Impulstensor {\bf T} in die Einsteinschen Feldgleichungen
einsetzen, wollen wir uns noch mit dessen speziellen Eigenschaften besch"aftigen.
Es lassen sich n"amlich alle Ableitungen erster Ordnung der Felder durch die Felder
selber ausdr"ucken. Da die Lagrangeterme $\Lambda_B$ \rf{LambdaB} und $\Lambda_ W$
\rf{LambdaW} identisch verschwinden, erhalten wir f"ur die kanonischen
Energie-Impulstensoren
\alpheqn
\begin{equation}
\TB_{\rho\kappa} = -\frac{1}{4\pi\scrc^2{\cal L}^2}\calB_{\rho\kappa}
\label{TBcalB}
\end{equation}
\begin{equation}
\TW_{\rho\kappa} = \frac{1}{4\pi\scrc^2 l^2}\calB_{\rho\kappa}\ .
\label{TWcalB}
\end{equation}
\reseteqn
Hierbei haben wir den Projektor \bm{\cal B} der charakteristischen Distribution
$\Deltadach$ benutzt
\begin{eqnarray}
\label{calBBB}
\calB_{\rho\kappa} = \calB_{i\rho}{\calB^i}_\kappa\\
\left(\calB_{\rho\kappa}{\calB^\kappa}_\lambda = \calB_{\rho\lambda}\right)\ .
\nonumber
\end{eqnarray}
Um die physikalische Bedeutung dieser Tensoren zu zeigen, nehmen wir an, da\3 der
"Ather eine ideale Fl"ussigkeit ist und folgenden Energie-Impulstensor besitzt
\begin{eqnarray}
\label{Tid2}
\Tid_{\mu\nu} &=& \left({\scr{\cal M}} + {\scr{\cal P}}\right)p_\nu p_\mu -
{\scr{\cal P}}\,G_{\mu\nu}\\
&\equiv& {\scr{\cal M}}\,p_\mu p_\nu - {\scr{\cal P}}\,\calB_{\mu\nu}\ .\nonumber
\end{eqnarray}
Vergleichen wir dies mit den Tensoren aus \rf{TBcalB} und \rf{TWcalB}, so sehen
wir da\3 die zeitartigen Konstituenten des "Athers einen negativen (konstanten)
Druck produzieren
\begin{equation}
{^{(W)}}{\scr{\cal P}} = -\frac{1}{4\pi\scrc^2 l^2}\ ,
\label{pW}
\end{equation}
w"ahrend der Druck der raumartigen Konstituenten positiv ist
\begin{equation}
{^{(B)}}{\scr{\cal P}} = +\frac{1}{4\pi\scrc^2 {\cal L}^2}\ .
\label{pB}
\end{equation}
Es scheint nun etwas sonderbar zu sein, da\3 keine der beiden Konstituenten zur
Energiedichte ${\scr{\cal M}}$ des "Athers beitr"agt. Wir werden jedoch sehen, da\3 nur der
gesamte Energie-Impulstensor {\bf T}, welcher den Eichteil \bm{\calTF} \rf{calTF}
enth"alt, eine physikalische Bedeutung besitzt. Bevor wir den Einflu\3 von
\bm{\calTF} auf {\bf T} untersuchen, m"ussen wir uns mit dem Yang-Mills Feld {\bf F}
besch"aftigen.

\section{Das Eichfeld}
\indent

\label{Intbed1}
Das in \rf{calD} ben"utzte Eichpotential {\bf A} beeinflu\3t nun die Metrik {\bf G}
nicht direkt. Trotzdem ist es m"oglich, da\3 das Eichfeld {\bf A} "uber den
Energie-Impulstensor \bm{\calTF} \rf{calTF} die Kr"ummung {\bf R} beeinflu\3t. Da
wir \bm{\cal T} f"ur die Einstein-Gleichungen \rf{ET} ben"otigen, m"ussen wir ihn
analog zu ${\bf\TB}$ und ${\bf\TW}$ mit Hilfe des charakteristischen Vektors
{\bf p} und dem Projektor \bm{\calB} \rf{calBBB} ausdr"ucken.

Wir w"ahlen f"ur die Feldst"arke {\bf F} \rf{F} folgenden Ansatz
\begin{equation}
F_{i\mu\nu} = f_{||}\,{\epsilon_i}^{jk}\calB_{j\mu}\calB_{k\nu} + f_{\perp}
\left(\calB_{i\mu}p_\nu - \calB_{i\nu}p_\mu\right)\ .
\label{Fff}
\end{equation}
Die Konstanten $f_{||}$ und $f_{\perp}$ m"ussen wir nun bestimmen. Die erste
Forderung, die der Ansatz \rf{Fff} erf"ullen mu\3, ist die Bianchi Identit"at
\begin{equation}
{\cal D}_{[\lambda}F_{i\mu\nu]} = 0\ .
\label{Bianchi}
\end{equation}
Da die kovariante Ableitung aller Objekte in {\bf F} aufgrund der mikroskopischen
Gleichungen bekannt ist, kann die kovariante Ableitung von {\bf F} direkt mit Hilfe
der "Atherfelder ausgedr"uckt werden:
\begin{equation}
{\cal D}_\lambda F_{i\mu\nu} = 2\left(\frac{f_{||}}{L} + \frac{f_{\perp}}{l}\right)
\calB_{i[\nu}\calB_{\mu]\lambda} + 2\left(\frac{f_{||}}{l} -
\frac{f_{\perp}}{L}\right) {\epsilon_i}^{jk}\calB_{j\lambda}\calB_{k[\nu}p_{\mu]}\ .
\label{calDF1}
\end{equation}
Die Bianchi-Identit"at \rf{Bianchi} kann jedoch nur dann erf"ullt werden, wenn der
zweite Term auf der rechten Seite von \rf{calDF1} verschwindet, d.h.
\begin{equation}
f_{\perp} = \frac{L}{l}f_{||}\ .
\label{fLlf}
\end{equation}
Die Feldst"arke nimmt daher folgende Gestalt an (wir setzen $f_{||}\equiv f$)
\begin{equation}
F_{i\mu\nu} = f\,{\epsilon_i}^{jk}\calB_{j\mu}\calB_{k\nu} + 2f\frac{L}{l}
\calB_{i[\mu}p_{\nu]}\ .
\label{Ff}
\end{equation}
Die n"achste Bedingung, welcher {\bf F} gehorchen mu\3, ist die inhomogene Yang-Mills
Gleichung \rf{calDFBcalDB}. Ben"utzen wir das Ergebnis \rf{Ff}, so erhalten wir
f"ur {\bf F}
\begin{equation}
{\cal D}^\mu F_{i\mu\nu} = -f\left(1+\frac{L^2}{l^2}\right){\epsilon_i}^{jk}
\calB_{i\mu}\left({\cal D}_\nu{\calB_k}^\mu\right)\ ;
\label{calDF2}
\end{equation}
dadurch wird die Gr"o\3e $f$ fixiert:
\begin{equation}
f = \scrc^{-2}{\left(1 + \frac{L^2}{l^2}\right)}^{-1}\ .
\label{fcLl}
\end{equation}
Nachdem nun die Feldst"arke {\bf F} bekannt ist, k"onnen wir den von {\bf F}
produzierten Energie-Impulsinhalt ${\bf\calTF}$ untersuchen. Wir erhalten mit
Hilfe von \rf{fLlf} und \rf{fcLl} aus \rf{calTF}:
\begin{equation}
\calTF_{\rho\kappa} = \frac{1}{8\pi\left({\dis 1+\frac{L^2}{l^2}}\right)\scrc^4}
\left(3p_\rho p_\kappa - \calB_{\rho\kappa}\right)\ .
\label{calTFLlpcalB}
\end{equation}
Vergleichen wir dieses Ergebnis mit dem Tensor ${\bf\Tid}$ \rf{Tid}, so erhalten
wir f"ur den Druck ${}^{(F)}{\scr{\cal P}}$ und die Energiedichte ${{}^{(F)}\scr{\cal
M}}$ des Eichfeldes
\begin{equation}
^{(F)}{\scr{\cal P}} = \frac{1}{3}\;{}^{(F)}{\scr{\cal M}} = {\dis\frac{1}{8\pi\left(
1+{\dis\frac{L^2}{l^2}}\right)\scrc^4}}\ .
\label{FpFM}
\end{equation}
Dies ist nun gerade die Zustandsgleichung f"ur ein Gas aus masselosen Teilchen
(n"amlich den masselosen Eichbosonen {\bf A}, welche die Wechselwirkung der {\bf B}
Bosonen in der quantisierten Theorie vermitteln). Ein weiterer interessanter Aspekt
zeigt sich, wenn wir uns den gesamten Energie-Impulstensor {\bf T} \rf{T} anschauen.
Dieser ergibt sich mit \rf{TBcalB},\rf{TWcalB} und \rf{calTFLlpcalB} zu
\begin{equation}
T_{\rho\kappa}=\frac{3}{8\pi\scrc^4\left(1+{\dis\frac{L^2}{l^2}}\right)}p_\rho p_\kappa
+\frac{1}{8\pi\scrc^2}\left(\frac{4}{L^2} - \frac{1}{\scrc^2
\left(1+{\dis\frac{L^2}{l^2}}\right)}\right)\calB_{\rho\kappa}\ .
\label{T2}
\end{equation}
Wenn der L"angenparameter $\scrc$, den wir von Hand in die Theorie eingef"ugt
hatten, den folgenden Wert annimmt:
\begin{equation}
\scrc = L{\left(1+\frac{L^2}{l^2}\right)}^{-\frac{1}{2}}\ ,
\label{cLl}
\end{equation}
so wird der gesamte Energie-Impulstensor proportional dem Metriktensor {\bf G}
\begin{equation}
T_{\rho\kappa} = \frac{3}{8\pi}\frac{1+{\dis\frac{L^2}{l^2}}}{L^4}\,G_{\rho\kappa}\ .
\label{TG}
\end{equation}
Ein Tensor dieser Art wird "ublicherweise f"ur das Quantenvakuum angesetzt!
\cite{MiThWh73}. Wir k"onnen nun den "Ather als Quantenvakuum identifizieren, wobei
die mikroskopischen Felder {\bf B} und {\bf W} die Vakuumserwartungswerte der
entsprechenden Quantenfelder sind. Der interessante Punkt dabei ist, da\3 der
L"angenparameter $\scrc$ den speziellen Wert \rf{cLl} annehmen {\em mu\3}. Diese
Tatsache wird wichtig, wenn wir die linke Seite der Einstein-Gleichung \rf{ET}
untersuchen.

\pagebreak
\section{Die geometrische Struktur des Vakuums}
\indent

\label{Intbed2}
Die mikroskopische Dynamik bestimmt nun nicht nur den Energie-Impulsinhalt der
mikroskopischen Felder, sondern auch die Geometrie des Raumes in dem sich diese
Felder befinden. Wie wir gesehen haben sind die Gleichungen erster Ordnung im flachen
Raum inkonsistent. Aus diesem Grund ben"otigen wir einen gekr"ummten Raum in dem sie
konsistent wirken k"onnen. Um diesen Raum zu finden, m"ussen wir dessen Kr"ummung
{\bf R} konstruieren, so da\3 die allgemeine Identit"at
\begin{equation}
\nabla_{[\sigma}\nabla_{\lambda]}p_\mu=-\frac{1}{2}{R^\rho}_{\mu\sigma\lambda}
\,p_\rho\label{altAblp2}
\end{equation}
mit der Gleichung \rf{altAblp}, welche eine notwendige Integrabilit"atsbedingung
f"ur die mikroskopischen Felder darstellt, vertr"aglich ist.

Obige Gleichung legt, wenn sie auf \rf{altAblp} angewandt wird, die Riemannsche
Kr"ummung {\bf R} nicht eindeutig fest. Deshalb zerlegen wir {\bf R} in zwei Teile
\begin{equation}
{R^\rho}_{\mu\sigma\lambda} = {\Rpara^\rho}_{\mu\sigma\lambda} +
{\Rperp^\rho}_{\mu\sigma\lambda}\ ,
\label{RRparaRperp}
\end{equation}
wobei der erste Term parallel zur charakteristischen Distribution ist
\begin{equation}
{\Rpara^\rho}_{\mu\sigma\lambda}\,p_\rho = 0\ .
\label{Rparap}
\end{equation}
Der "ubrigbleibende Term ${\bf\Rperp}$ wird dann durch \rf{altAblp},\ \rf{altAblp2}
eindeutig festgelegt und hat folgende Gestalt:
\begin{equation}
{\Rperp^{\rho\mu}}_{\sigma\lambda} = {\left(\frac{2}{l}\right)}^2 p^{[\rho}
{\calB^{\mu]}}_{[\sigma}p_{\lambda]}\ .
\label{Rperp}
\end{equation}
Auf "ahnliche Weise l"a\3t sich ${\bf\Rpara}$ finden. Wir m"ussen dazu nur die
alternierenden Ableitungen der raumartigen Felder \bm{\calB} anstelle der zeitartigen
{\bf p} verwenden und erhalten
\begin{equation}
D_{[\sigma}D_{\lambda]}\calB_{i\mu} = \frac{1}{2}{R^\rho}_{\mu\sigma\lambda} +
\left(\frac{1}{l^2} + \frac{1}{L^2}\right)\calB_{i[\sigma}\calB_{\lambda]\mu} +
\frac{1}{l^2}p_\mu\calB_{i[\sigma}p_{\lambda]} + \frac{1}{lL}{\epsilon_i}^{jk}
\calB_{k\mu}\calB_{j[\sigma}p_{\lambda]}\ .
\label{DDcalB}
\end{equation}
Ben"utzen wir die allgemeing"ultige Gleichung
\begin{equation}
D_{[\sigma}D_{\lambda]}\calB_{i\mu} = \frac{1}{2}{\epsilon_i}^{jk}F_{j\sigma\lambda}
\calB_{k\mu}
\label{DDcalBFcalB}
\end{equation}
f"ur die linke Seite von \rf{DDcalB}, so l"a\3t sich \rf{DDcalBFcalB} mit Hilfe von
\rf{Ff} auch folgenderma\3en formulieren
\begin{equation}
D_{[\sigma}D_{\lambda]}\calB_{i\mu} = f\,\calB_{i[\sigma}\calB_{\lambda]\mu} +
\frac{fL}{l}{\epsilon_i}^{jk}\calB_{k\mu}\calB_{j[\sigma}p_{\lambda]}\ .
\label{DDcalBcalB}
\end{equation}
Vergleichen wir nun die Gleichungen \rf{DDcalB} und \rf{DDcalBcalB} miteinander,
so k"onnen wir einige wichtige Schlu\3folgerungen ziehen: Da die rechten Seiten
beider Gleichungen identisch sein m"ussen, mu\3 der Parameter $f$ den Wert
\begin{equation}
f = \frac{1}{L^2}
\label{fL}
\end{equation}
annehmen. Kombiniert man diese Forderung mit \rf{fcLl}, so ergibt sich gerade
der spezielle Wert \rf{cLl} f"ur den L"angenparameter $\scrc$\,! Ferner hebt
sich der orthogonale Anteil ${\bf\Rperp}$ \rf{Rperp} in \rf{DDcalB} heraus, und nur
der parallele Teil ${\bf\Rpara}$ bleibt "ubrig
\begin{equation}
{\Rpara^\rho}_{\mu\sigma\lambda} = \frac{2}{l^2}{\calB^\rho}_{[\lambda}
\calB_{\sigma]\mu}\ .
\label{RparaBB}
\end{equation}
Es sind nun beide Teile ${\bf\Rpara}$ und ${\bf\Rperp}$ bekannt, und wir k"onnen
{\bf R} angeben:
\begin{equation}
{R^\rho}_{\mu\sigma\lambda} = \frac{2}{l^2}{G^\rho}_{[\lambda}G_{\sigma]\mu}\ .
\end{equation}
Dieser Kr"ummungstensor geh"ort zu einem Raum konstanter Kr"ummung.

Nachdem wir nun die Geometrie des Riemannschen Raumes bestimmt haben, l"a\3t sich die
linke Seite der Einstein-Gleichung \rf{ET} angeben. Berechnen wir zuerst den Ricci
Tensor
\begin{equation}
R_{\mu\lambda}\equiv{R^\rho}_{\mu\rho\lambda} = -\frac{3}{l^2}G_{\mu\lambda}\ ,
\label{RlG}
\end{equation}
und den Kr"ummungsskalar
\begin{equation}
R\equiv{R^\lambda}_\lambda = -\frac{12}{l^2}\ ,
\label{Rl}
\end{equation}
so erhalten wir f"ur den Einstein-Tensor {\bf E}
\begin{equation}
E_{\mu\nu}\equiv R_{\mu\nu} - \frac{1}{2}R\,G_{\mu\nu} = \frac{3}{l^2}G_{\mu\nu}\ .
\label{ElG}
\end{equation}
Offensichtlich ist hier {\bf E} proportional zum gesamten Energie-Impuls-Tensor
{\bf T} \rf{TG}, soda\3 die Einsteinschen Feldgleichungen \rf{ET} automatisch
erf"ullt sind, vorausgesetzt zwischen den L"angenparametern $L,l$ und der
Planckl"ange $L_p$ gilt die folgende Beziehung
\begin{equation}
L^2_P = L^2\frac{L^2}{L^2+l^2}\ .
\label{PlanckLl}
\end{equation}
Dieses Ergebnis scheint den Gleichungen erster Ordnung \rf{MikrGl1} und \rf{MikrGl2}
in der Allgemeinen Relativit"atstheorie eine tiefere Bedeutung zu geben. Denn die
mikroskopischen Gleichungen verm"ogen sowohl die Feldgleichungen zweiter Ordnung
als auch die Raum-Zeit Geometrie so zu bestimmen, da\3 die herk"ommliche und die
in diesem Kapitel dargelegte Vorgehensweise innerhalb der Einsteinschen Theorie
"ubereinstimmen. Wir k"onnen uns sogar auf den Standpunkt stellen, da\3 die
Einsteinschen Gleichungen aus den Gleichungen erster Ordnung {\em ableitbar} sind.
Dabei werden zus"atzlich die Schwierigkeiten in der Wahl des richtigen
Energie-Impulstensors elegant beseitigt.

Es ist wichtig sich vor Augen zu halten, da\3 die Argumentation, welche von der
mikroskopischen Dynamik zu den makroskopischen Einsteinschen Gleichungen f"uhrt,
nicht umkehrbar ist. Denn geht man von den Einsteinschen Gleichungen \rf{ET} mit dem
Energie-Impulstensor \bm{\cal T} \rf{calT} aus, so erh"alt man nicht notwendigerweise
die mikroskopischen Gleichungen \rf{MikrGl1} und \rf{MikrGl2}, die daher als eine
sehr spezielle L"osung der gekoppelten Einstein-Klein-Gordon Gleichung anzusehen
sind. Au\3erdem fordert man in der Einsteinschen Theorie die Bedingung, da\3 der Raum
flach wird ($G_{\mu\nu}\to g_{\mu\nu}$) wenn der gesamte Energie-Impuls-Inhalt des
Universums verschwindet. Dies l"a\3t sich nun aber mit unserer Betrachtungsweise
nicht vereinbaren, denn l"a\3t man alle "Atherfelder {\bf F}, {\bf B} und {\bf W}
gegen Null gehen, so strebt die Metrik {\bf G} \rf{GBBWW} ebenfalls gegen Null und
zerst"ort den metrischen Charakter der Raum-Zeit Mannigfaltigkeit. Das hei\3t es ist
in der vorliegenden Theorie kein flaches Universum denkbar, welches v"ollig frei von
jeglicher physikalischer Dynamik existieren k"onnte.

Die Gleichung \rf{PlanckLl} zeigt nochmals auf einfache Weise den Zusammenhang der
drei verschiedenen L"angenskalen unserer Gravitationstheorie auf. Der
L"angenparameter $L$ definiert aufgrund von \rf{Cnull} eine typische L"angenskala
f"ur das SO(3) Vektorfeld {\bf C}, die nach unserer allgemeinen Philosophie in
Kapitel \ref{Reduktion} zwischen der kosmischen Skala $(l\approx 10^{28} cm)$ und der
Planck-L"ange $L_p$ $(\approx 10^{-33} cm)$ liegen mu\3. In der Tat findet man aus
\rf{PlanckLl}
\begin{equation}
L\approx \sqrt{L_p\,l}\approx 10^{-2} cm
\label{WertL}
\end{equation}
Dieser Wert ist in Einklang mit unserer Vorstellung, da\3 das {\bf C}-Feld die
Gravitationsph"anomene im Zwischenbereich zwischen dem kosmischen und dem
Quantenma\3stab beschreiben sollte.


\chapter{Riemannsche Struktur von trivialisierbaren Eichfeldern}
\indent

Der erste Vakuum-Kanditat in Abschnitt \ref{Vakuumkandidat1} hat sich dadurch
ausgezeichnet, da\3 die Riemannsche Konnexion ${\bf\Gamma}$ der 4-dimensionalen
Raum-Zeit mit der Oberfl"achen-Konnexion ${\bf\Gammadach}$ "ubereinstimmte. Diese
spezielle Eigenschaft der Riemannschen Geometrie war hinreichend f"ur die
Parallelisierbarkeit der Raum-Zeit. Anschaulich gesprochen bedeutet diese
Parallelisierbarkeit, da\3 man sich die Riemannschen Objekte (wie z.B. die Metrik
{\bf G}) als ein Feld von Gr"o\3en "uber dem (pseudo-) Euklidischen Raum vorstellen
kann, d.h. die entsprechende Riemannsche Geometrie darf man sich als eine Feldtheorie
im "ublichen Sinne "uber dem flachen Raum vorstellen. In diesem Fall --- {\em aber
auch nur in diesem} --- kann man also den physikalischen Inhalt der Allgemeinen
Relativit"atstheorie dahingehend modifizieren, da\3 man die Riemannsche Struktur als
ein reines Bewegungsgesetz f"ur gravitierende Materie im {\em flachen} Raum versteht,
ganz "ahnlich wie die Maxwellsche Elektrodynamik die Ausbreitung des
elektromagnetischen Feldes und die Bewegung der elektrischen Ladungen im flachen Raum
beschreibt. Bei dieser Interpretation der Einsteinschen Gravitationstheorie verliert
die Gravitationskraft anscheinend ihre Sonderstellung unter den "ubrigen
Grundkr"aften der Natur, weil die von Einstein so erfolgreich durchgef"uhrte
Geometrisierung der Gravitationskraft wieder verloren geht! Es gibt sogar Autoren
\cite{Wein72}, die diesen "`{\em entgeometrisierenden}"' Standpunkt gegen"uber der
{\em gesamten} Relativit"atstheorie Einsteins in ihrem vollen Umfang einnehmen und
damit bezweifeln, da\3 Masse und Energie das Raum-Zeit-Kontinuum wirklich kr"ummen.
Diese Auffassung ist aber von kompetenter Seite zur"uckgewiesen worden \cite{Mack81}
und hat sich in der Literatur nicht durchgesetzt.

Aber selbst in dem hier vorliegenden Spezialfall des ersten Vakuum-Kanditaten in
Abschnitt \ref{Vakuumkandidat1}, bei dem die entgeometrisierende Sichtweise
berechtigt erscheint, k"onnen wir zeigen, da\3 die Riemannsche Struktur "uber dem
flachen Raum dennoch eine geometrische Bedeutung hat! Diese mu\3 sich naturgem"a\3
auf eine gewisse Eigenschaft des {\em flachen Raumes} beziehen, was angesichts dessen
trivialer Geometrie verwunderlich erscheinen mag: wie kann ein "`{\em flacher}"' Raum
eine "`{\em Kr"ummung}"' hervorbringen? Die Antwort ist einfach: durch das in Kapitel
\ref{Reduktion} beschriebene {\em lokale Aufspaltungs-Prinzip}. Das hei\3t, in jedem
Punkt {\bf x} der Mannigfaltigkeit existiert eine ausgezeichnete Zeitrichtung
${\bf\ndach}_x$, die einen entsprechenden 3-dimensionalen Unterraum $\Deltaquer_x$
erzeugt. (Man beachte, da\3 es sich hier nicht um die charakteristische Aufspaltung
$({\bf p}_x, \Deltadach_x)$ handelt.)

Im folgenden werden wir Schritt f"ur Schritt die von dieser (1+3)-Aufspaltung des
flachen Raumes erzeugte Riemannsche Geometrie aufbauen.

\section{Das inverse Problem und trivialisierbare SO(3) Eichfelder}
\indent

Das inverse Problem bez"uglich der {\em Euklidischen Foliation eines Riemannschen
Raumes} besteht darin, den Euklidischen Raum so aufzubl"attern, da\3 eine Riemannsche
Struktur entsteht. Man k"onnte diesen Proze\3 die {\em Riemannsche Foliation eines
Euklidischen Raumes} nennen. Eine Foliation dieser letzten Art wird durch ein ${\cal
SO}(3)$ Eichpotential {\bf A} und einen zugeh"origen Eichvektor {\bf B} beschrieben,
wobei beide eine trivialisierbare Eichfeld-Konfiguration bestimmen. Es wird sich
jedoch zeigen, da\3 wir nur bei einer integrablen charakteristischen Distribution
$\Deltadach$ zu einer rein Riemannschen Struktur gelangen. Im nicht-integrablen Fall
liegt eine Riemann-Cartan Struktur vor, das hei\3t der entsprechende Raum besitzt
eine nicht-verschwindende Torsion. Zuerst wollen wir aber einige geometrische
Begriffe f"ur trivialisierbare Eichfelder n"aher erl"autern.

Wir k"onnen stets "uber ${\bf E_4}$ bzw ${\bf M_4}$ als Basisraum ein triviales
Prinzipalb"undel $\lambdakr$ bilden, welches die verschwindende Kr"ummung
${\bf\Omegakr}$ und die flache Konnexion \bm{\omegakr} besitzt. \bm{\omegakr}
l"a\3t sich folgenderma\3en nach den {\bf A} und {\bf B} zerlegen
\begin{equation}
\mbox{\bm{\omegakr}} = {\bf A}_i L^i + {\bf B}_i l^i ,\ \ i = 1,2,3
\label{omegakrALBl}
\end{equation}
wobei $L^i$ und $l^i$ mit Hilfe der ${\cal SO}$(4) bzw. ${\cal SO}$(1,3)
Standardgeneratoren zu bilden sind \cite{Heck89}. Nach Voraussetzung verschwindet die
$\lambdakr$ B"undelkr"ummung
\begin{equation}
{\Omegakr}_{\mu\nu} =
\partial_\mu\omegakr_\nu - \partial_\nu\omegakr_\mu + \left[\omegakr_\mu,
\omegakr_\nu\right] = 0\ .
\end{equation}
Setzen wir hier nun die Zerlegung \rf{omegakrALBl} ein, und f"uhren eine "`innere"'
SO(3) Kr"ummung {\bf F} ein
\begin{equation}
F_{i\mu\nu} = \partial_\mu A_{i\nu}-\partial_\nu A_{i\mu}+{\epsilon_i}^{jk}
A_{j\mu}A_{k\nu}\ ,
\end{equation}
so ergibt sich durch Koeffizientenvergleich f"ur {\bf A} und {\bf B}
\alpheqn
\begin{equation}
F_{i\mu\nu} = g_{00}\,{\epsilon_i}^{jk} B_{j\mu}B_{k\nu}
\label{triva}
\end{equation}
\begin{equation}
D_\mu B_{i\nu} = D_\nu B_{i\mu}
\label{trivb}
\end{equation}
\reseteqn
Diese Gleichungen sind nun gerade die notwendigen und hinreichenden Bedingungen
f"ur die Trivialisierbarkeit der SO(3) Eichfelder {\bf A}. Durch die Mitnahme von
$g_{00}$ in \rf{triva} ber"ucksichtigen wir, da\3 die Trivialisierbarkeit sowohl
"uber den Basisr"aumen ${\bf E_4}$ als auch ${\bf M_4}$ erkl"art werden kann.

Errichtet man nun an einem beliebigen Punkt in diesem Basisraum ein Vierbein
\mbox{$({\hat{e}_0}^\mu \equiv \ndach_\mu, {e_i}^\mu)$\ ,} welches folgende
Orthonormalit"atsbedingungen erf"ullt:
\alpheqn
\begin{eqnarray}
\ndach_\mu {e^\mu}_i &=& 0\\
g_{\mu\nu}{e^\mu}_i{e^\nu}_j &=& g_{ij}\ ,
\end{eqnarray}
\reseteqn
so definiert $\ndach_\mu$ als Normalenvektor die sog. repr"asentative Distribution
$\Deltaquer$, welche dann von den drei ${e^\mu}_i$ aufgespannt wird.
{\bf A} und {\bf B} lassen sich nun durch diese Vektorfelder ausdr"ucken. Um dies zu
zeigen bemerken wir, da\3 die Konnexion \bm{\omegakr} von $\lambdakr$ die "Anderung
dieses Vierbeins bei Parallelverschiebung beschreibt:
\begin{equation}
\partial_\lambda{\bf e}_\mu = {\omegakr^\gamma}_{\mu\lambda}{\bf e}_\gamma\ .
\end{equation}
Mit Hilfe der Zerlegung \rf{omegakrALBl} erhalten wir daraus\footnote{Wir beschreiben
hier gleichzeitig die Aufspaltung der (pseudo-) Euklidischen R"aume ${\bf E}_4$,
${\bf M}_4$ (Metrik: $g = diag (\plmi 1,-1,-1,-1)$)}
\alpheqn
\begin{equation}
A_{i\mu} = \frac{1}{2}g_{\lambda\nu}\,{\epsilon^{jk}}_i {e^\lambda}_j \partial_\mu
{e^\nu}_k
\end{equation}
\begin{equation}
B_{i\mu} = g_{00}\,g_{\lambda\nu} {e^\lambda}_i \partial_\mu \hat{n}^\nu\ ,
\label{ABe}
\end{equation}
\reseteqn
was sich durch direktes Einsetzen in die Trivialisierungsbedingungen \rf{triva} und
\rf{trivb} verifizieren l"a\3t. Wir sehen nun, da\3 {\bf A} die Verschiebung des
Dreibeins ${\bf e}_\mu$ innerhalb $\Deltaquer$ regelt, w"ahrend {\bf B} den Verlauf des
Normalenvektorfeldes ${\bf\ndach}$ kennzeichnet. ${\bf\ndach}$ und damit $\Deltaquer$
sind SO(3) eichinvariant, da SO(3) nur auf die ${e^\mu}_i$ wirkt und $\Deltaquer$ in
sich "uberf"uhrt.\cite{Heck89}

Weiter definieren wir die charakteristische Distribution $\Deltadach$, die von den
$B_{i\mu}$ aufgespannt werden soll. Der charakteristische Vektor ${\bf\pdach}$
fungiert hierbei als Normalenvektor, das hei\3t es gilt
\begin{equation}
B_{i\mu}\pdach^\mu = 0\ .
\label{Bp}
\end{equation}
$\Deltadach$ ist ebenfalls SO(3) invariant. Denn aufgrund des tensoriellen Verhaltens
von {\bf B} unter SO(3) Eichtransformationen sind die umgeeichten {\bf B}' nur
Linearkombinationen der alten {\bf B}. Wir wollen nun den Zusammenhang zwischen den
beiden Distributionen $\Deltadach$ und $\Deltaquer$ untersuchen. Aus der
Orthogonalit"atsbedingung \rf{Bp} und Gleichung \rf{ABe} erhalten wir
\begin{equation}
\left(\pdach^\mu\partial_\mu{\ndach}^\nu\right) {e_i}_\nu = 0\ .
\end{equation}
Da ${\bf\ndach}$ normiert ist, gilt au\3erdem
\begin{equation}
\left(\partial_\mu{\ndach}^\nu\right)\ndach_\nu = 0\ ,
\end{equation}
also insgesamt
\begin{equation}
\pdach^\mu\left(\partial_\mu{\ndach}^\nu\right) = 0\ .
\end{equation}
D.h. der Normalenvektor ${\bf\ndach}$ der repr"asentativen Distribution $\Deltaquer$
wird entlang der Integrallinien von ${\bf\pdach}$, den sogenannten charakteristischen
Linien, bez"uglich der Konnexion \bm{\omegakr} parallel verschoben.

\section{Konstruktion des charakteristischen B"undels}
\label{Eig1}
\subsection{Trivialisierbarkeit und Torsion}
\label{TrivUndTors}
\indent

Wir wollen uns nun wieder dem Problem, der Bestimmumg der Riemannschen-, bzw.
Riemann-Cartanschen Struktur, zuwenden. Nach Kapitel \ref{Reduktion} l"a\3t sich die
Metrik des Raumes folgenderma\3en aus {\bf B} und ${\bf\pdach}$ aufbauen:
\footnote{Um die Metrik dimensionslos zu halten, f"uhren wir einen L"angenparameter
$\scrc$ auf der rechten Seite der Gleichung ein}
\begin{equation}
G_{\mu\nu} = \scrc^2 B_{i\mu}{B^i}_\nu + g_{00} \pdach_\mu\pdach_\nu
\label{GBBpp}
\end{equation}
Aus der kovarianten Ableitung von {\bf G} erhalten wir mit Hilfe der Definition der
allgemein kovarianten Ableitung
\begin{equation}
\nabla_\lambda G_{\mu\nu} =
\scrc^2\left({\cal D}_\lambda B_{i\mu}\right){B^i}_\nu+\scrc^2\left({\cal D}_\lambda
{B^i}_\nu\right)B_{i\mu} +
\left(\nabla_\lambda\pdach_\mu\right)\pdach_\nu + \pdach_\mu\left(\nabla_\lambda
\pdach_\nu\right)\ .
\label{nablaG}
\end{equation}
Die Bedingung der kovarianten Konstanz von {\bf G} ist nun erf"ullt wenn gilt
\begin{equation}
{\cal D}_\lambda B_{i\mu} = 0
\label{calDB}
\end{equation}
bzw ausgeschrieben
\begin{eqnarray}
D_\lambda B_{i\mu} &=& {\Gamma^\sigma}_{\mu\lambda} B_{i\sigma}
\label{DBGamB}
\end{eqnarray}
sowie
\begin{equation}
\nabla_\lambda\pdach_\mu = 0\ .
\label{nablaP}
\end{equation}
Es existieren nun entscheidende Einschr"ankungen f"ur die bis jetzt nicht n"aher
festgelegte ${\cal GL}(4,{\bf R})$ wertige Konnexion ${\bf\Gamma}$ \footnote{In
diesem Kapitel ist ${\bf\Gamma}$ die Riemann-Cartan Verallgemeinerung der
Riemannschen Konnexion ${\bf\Gamma}$ aus Kapitel \ref{Reduktion}}

Die erste Einschr"ankung f"ur ${\bf\Gamma}$ r"uhrt von der Tatsache her, da\3 wir
aufgrund der Gleichungen \rf{triva} und \rf{trivb} von trivialisierbaren
Eichfeldern ausgehen.

Definiert man nun die Torsion {\bf Z} wie "ublich \footnote{Die Indexklammer
$[\cdots]$ bedeutet die totale Antisymetrisierung der eingeschlossenen Indizes.}
\begin{equation}
{Z^\lambda}_{\mu\nu} = \frac{1}{2}\left({\Gamma^\lambda}_{\mu\nu} -
{\Gamma^\lambda}_{\nu\mu}\right) \equiv {\Gamma^\lambda}_{[\mu\nu]}\ ,
\label{Zgamma}
\glossary{${\bf Z} Torsion}
\end{equation}
so gilt f"ur einen beliebigen Gradientenvektor {\bf C}
\begin{equation}
C_\mu = \partial_\mu C
\label{CdC}
\end{equation}
folgende Identit"at \cite{KoNo69}
\begin{equation}
\nabla_{[\mu}C_{\nu]} = {Z^\lambda}_{\mu\nu} C_\lambda\ .
\label{DCZC}
\end{equation}
Benutzt man nun Gleichung \rf{DBGamB} f"ur die zweite Trivialisierbarkeitsbedingung
\rf{trivb}, so erh"alt man folgende einschr"ankende Bedingungsgleichung f"ur die
Torsion {\bf Z} von ${\bf\Gamma}$:
\begin{equation}
{Z^\lambda}_{\mu\nu} B_{i\lambda} = 0\ .
\label{ZB}
\end{equation}
Da der charakteristische Vektor ${\bf\pdach}$ definitionsgem"a\3 durch die
extrinsische Kr"ummung annihiliert wird, vgl. \rf{Bp}, erh"alt man f"ur den
Torsionstensor {\bf Z} folgenden Ausdruck
\begin{equation}
{Z^\lambda}_{\mu\nu} = g_{00}\;\pdach^\lambda\pdach_\sigma {Z^\sigma}_{\mu\nu}\ .
\label{ZppZ}
\end{equation}
Diese spezielle Form der Torsion, l"a\3t sich nun nicht aus Gleichung \rf{DBGamB}
herleiten, vielmehr wird durch diese Gleichung nur der sogenannte {\em wesentlichen
Teil} \bm{\gamma} von ${\bf\Gamma}$ festgelegt:
\begin{equation}
{\gamma^\lambda}_{\mu\nu} = {\Pdach^\lambda}_\sigma\;{\Gamma^\sigma}_{\mu\nu}
\label{gamPGam}
\end{equation}
wobei wir den Projektor ${\bf\Pdach}$ auf die charakteristische 3-Ebene wie
folgt definiert haben:
\alpheqn
\begin{eqnarray}
\label{P}
{\Pdach^\lambda}_\sigma={g^\lambda}_\sigma - g_{00}\;\pdach^\lambda \pdach_\sigma\\
\label{pp}
\pdach^\lambda \pdach_\lambda = g_{\mu\nu}\;\pdach^\mu\pdach^\nu = g_{00}\ .
\end{eqnarray}
\reseteqn
Somit k"onnen wir nun die zweite Trivialisierbarkeitsbedingung \rf{trivb} mit Hilfe
des wesentlichen Teils \bm{\gamma} ausdr"ucken
\begin{equation}
D_\mu B_{i\nu} = {\gamma^\lambda}_{\nu\mu} B_{i\lambda}\ ,
\label{DBgamB}
\end{equation}
wobei \bm{\gamma} nun symmetrisch sein mu\3
\begin{equation}
{\gamma^\lambda}_{\mu\nu} = {\gamma^\lambda}_{\nu\mu}\ .
\end{equation}
Die Konnexion ${\bf\Gamma}$ wird also nicht vollst"andig durch das Eichpotential
{\bf A} und derem assozierten Kr"ummungsfeld {\bf B} \footnote{Gleichbedeutend
hierzu verwenden wir den Begriff "`extrinsische Kr"ummung"'} ($\equiv{\bf B_i},
\ i = 1,2,3 $) festgelegt. Der verbleibende Freiheitsgrad ist die Komponente parallel
zu ${\bf\pdach}$, wobei sich die Torsion {\bf Z} \rf{ZppZ} gerade als
schiefsymmetrischer Teil eben dieses Freiheitsgrades erweist. Dies erm"oglicht es uns
die unbestimmte Komponente von ${\bf\Gamma}$ so festzulegen, da\3 eine konsistente
Riemann-Cartan-Struktur ensteht; wie im folgenden gezeigt wird. Die so vollst"andig
bestimmte {\em Standard-Konnexion} ${\bf\Gamma}$ besitzt eine nichtverschwindende
Torsion {\bf Z}, so da\3 wir im allgemeinen keine streng Riemannsche Raum-Zeit
Struktur zugrunde legen k"onnen.

\subsection{Die B"undel-Konnexion}
\indent

In diesem Abschnitt wollen wir die geometrische Bedeutung der ${\cal GL}(4,{\bf R})$
Konnexion ${\bf\Gamma}$ mit Hilfe des charakteristischen Vektorfeldes ${\bf\pdach}$
untersuchen. Dazu verlassen wir das charakteristische B"undel $\taudach$ und
untersuchen die Trivialisierbarkeitsbedingungen \rf{triva} und \rf{trivb} mit Hilfe
der aus dem repr"asentative B"undel stammenden Gr"o\3en. Die kovarianten Ableitungen
der $\tauquer$ B"undelschnitte ${e^\nu}_i$ k"onnen mit Hilfe der extrinsischen
Kr"ummung aus $\tauquer$ folgenderma\3en ausgedr"uckt werden:
\begin{equation}
D_\mu {e^\nu}_i = - B_{i\mu}\ndach^\nu\ .
\label{De}
\end{equation}
Betrachten wir uns nun die zweite Bedingung \rf{trivb}. Da der
Normalenvektor ${\bf\ndach}$ eine eichinvariante Gr"o\3e ist erh"alt man aus
\rf{ABe} mit Hilfe von \rf{De}
\begin{equation}
D_\mu B_{i\nu} = g_{00}\;g_{\lambda\sigma} {e^\lambda}_i \partial_\mu \partial_\nu
\ndach^\sigma\ ,
\label{DBen}
\end{equation}
woraus man unmittelbar die Identit"at \rf{trivb} erkennt.

Wir k"onnen nun eine wichtige Schlu\3folgerung aus Gleichung \rf{DBen} ziehen.
Kombiniert man sie n"amlich mit \rf{DBgamB}, so ergibt die Kontraktion
mit ${\bf\pdach}$ eine Beziehung zwischen der Normalen ${\bf\ndach}$ von
$\Deltaquer$ und dem wesentlichen Teil \bm{\gamma} von ${\bf\Gamma}$:
\begin{equation}
\pdach^\nu\partial_\nu\partial_\mu\ndach^\lambda = {\gamma^\sigma}_{\mu\nu}\,
\pdach^\nu\partial_\sigma\ndach^\lambda\ .
\label{pngampn}
\end{equation}
Andererseits impliziert die Definition des charakteristische Vektorfeldes
\begin{equation}
{\Pdach^\mu}_\nu\partial_\mu\ndach^\lambda = \partial_\nu\ndach^\lambda \ .
\label{Pdn}
\end{equation}
Differenziert man diese Gleichung und kontrahiert sie mit ${\bf\pdach}$, so ergibt
sich ein der Gleichung \rf{pngampn} "ahnlicher Ausdruck:
\begin{equation}
\pdach^\nu\partial_\nu\partial_\mu\ndach^\lambda = \left(\partial_\mu
{\Pdach^\sigma}_\nu\right)\pdach^\nu\left(\partial_\sigma\ndach^\lambda\right) \ .
\end{equation}
Ein Vergleich mit \rf{pngampn} ergibt als Resultat
\begin{equation}
\partial_\lambda\pdach^\sigma + {\gamma^\sigma}_{\nu\lambda}\,\pdach^\nu = 0.
\label{dgam}
\end{equation}
Dies ist ein interessantes Ergebnis: Denn k"onnte der wesentliche Teil
\bm{\gamma} durch die Konnexion ${\bf\Gamma}$ ersetzt werden, so w"are der
Einheitsvektor kovariant konstant bez"uglich der ${\cal GL}(4,{\bf R})$
Konnexion ${\bf\Gamma}$, welche durch Gleichung \rf{DBGamB} unabh"angig vom
charakteristischen Vektorfeld eingef"uhrt wurde. Da\3 dies in der Tat m"oglich ist,
l"a\3t sich folgenderma\3en zeigen. Wir spalten die Konnexion ${\bf\Gamma}$ in ihren
wesentlichen Teil \bm{\gamma} und einen "`Rest"' {\bf z} auf
\begin{equation}
{\Gamma^\lambda}_{\mu\nu}={\gamma^\lambda}_{\mu\nu}+\pdach^\lambda\,z_{\mu\nu}.
\label{Gamgamz}
\end{equation}
Letzterer gehorche der Bedingung
\begin{equation}
z_{\mu\nu}\;\pdach^\mu = 0 .
\label{zp}
\end{equation}
Der charakteristische Vektor ist nun in der Tat kovariant konstant
\begin{equation}
\nabla_\lambda\pdach^\sigma = \partial_\lambda\pdach^\sigma +
{\Gamma^\sigma}_{\nu\lambda}\pdach^\nu = 0\ ,
\label{nablap1}
\end{equation}
w"ahrend die Torsion {\bf Z} folgender Gleichung gen"ugt
\begin{equation}
{Z^\lambda}_{\mu\nu} = \pdach^\lambda z_{[\mu\nu]}  .
\label{Zpz}
\end{equation}
Die kovariante Konstanz von ${\bf\pdach}$ ist au\3erdem konsistent mit der
Normalisierungsbedingung \rf{pp} und der Orthogonalit"atsbedingung \rf{Bp} aufgrund
folgender Identit"at
\begin{equation}
D_\mu\left(B_{i\lambda}\pdach^\lambda\right) \equiv \left({\cal D}_\mu
B_{i\lambda}\right)\pdach^\lambda+B_{i\lambda}\left(\nabla_\mu\pdach^\lambda\right) .
\label{DBp}
\end{equation}
Die geometrische Bedeutung der Standard-Konnexion ${\bf\Gamma}$ wird nun deutlich.
${\bf\Gamma}$ ist die (lokale) Konnexionsform im charakteristischen B"undel
$\taudach$. Diese Konnexion l"a\3t sich ferner aus "`extrinsischer"' Sicht
mit Hilfe eines karte"-sischen Koordinatensystems im einbettenden Raum beschreiben.
Betrachtet man zum Beispiel einen Schnitt $V_\mu({\bf x})$ aus dem B"undel $\taudach$,
so gilt
\begin{equation}
\pdach^\mu V_\mu = 0\ ,
\label{pV}
\end{equation}
Die Bedingung der kovarianten Konstanz von ${\bf\pdach}$, Gleichung \rf{nablap1},
garantiert da\3 die Ableitung von {\bf V} ebenfalls ein Objekt aus $\taudach$ ist:
\begin{equation}
\pdach^\mu\nabla_\lambda V_\mu = 0 \ .
\label{pnablaV}
\end{equation}
Im allgemeinen ist die so gefundene Standard-Konnexion nicht identisch der Konnexion,
die man durch Reduktion der kanonischen Konnexion \bm{\omegakr} auf eine im
${\bf E_4}$ oder ${\bf M_4}$ eingebettete Distribution, erh"alt. Man erkennt dies
leicht an folgendem Beispiel: Die t'Hooft-Polyakov Monopol L"osung \cite{BrSo87} f"ur
ein gekoppeltes SO(3) Yang-Mills-Higgs System enth"alt ein trivialisierbares Eichfeld,
welches gleichzeitig statisch in dem Sinne ist, da\3 die charakteristischen Linien
bez"uglich \bm{\omegakr} gerade Linien sind. Das hei\3t, die charakteristischen
Fl"achen sind flache 3-er Hyperebenen (z.b. orthogonal zur $x^0$ Achse). Die
Oberfl"achenreduktion von \bm{\omegakr} auf diese Hyperebenen ergibt eine Konnexion
mit verschwindender Kr"ummung im Widerspruch zur nichttrivialen t'Hooft-Polyakov
L"osung f"ur das Eichfeld, dessen Feldst"arketensor ungleich Null ist.
Andererseits gibt es Konfigurationen, bei denen sowohl ${\bf\Gamma}$ als auch {\bf A}
"ubereinstimmen, und aus diesem Grunde SO(3) Reduktionen von \bm{\omegakr} sind. Ein
Beispiel ist die Euklidische Ein-Meron L"osung \cite{BrSo86}, bei welcher die
Distributionen $\Deltaquer$ und $\Deltadach$ in eine sph"arische Distribution
"ubergehen, d.h. die Integralfl"achen sind in beiden F"allen konzentrische 3-Sph"aren
um den Ort des Merons.

\subsection{Die B"undelkr"ummung}
\indent

Nachdem nun die Bedeutung von ${\bf\Gamma}$ als eine B"undelkonnexion in $\taudach$
erkannt wurde, besteht der n"achste Schritt in der Auffindung der korrespondierenden
B"undelkr"ummung {\bf R}, die wie "ublich \cite{KoNo69} definiert ist:
\begin{equation}
{R^\sigma}_{\nu\mu\lambda} = \partial_\mu {\Gamma^\sigma}_{\nu\lambda} -
\partial_\lambda {\Gamma^\sigma}_{\nu\mu} +
{\Gamma^\sigma}_{\kappa\mu}{\Gamma^\kappa}_{\nu\lambda} -
{\Gamma^\sigma}_{\kappa\lambda}{\Gamma^\kappa}_{\nu\mu} \ .
\label{defR}
\end{equation}
Obwohl wir die genaue Gestalt der Konnexion ${\bf\Gamma}$ bis jetzt noch nicht
kennen, ist es doch m"oglich einige Aussagen "uber die ihre Kr"ummung {\bf R} zu
machen.

Dazu kombinieren wir die Identit"at
\begin{equation}
\nabla_{[\lambda}\nabla_{\mu]}\,\pdach^\sigma = \frac{1}{2}{R^\sigma}_{\nu\lambda\mu}
\,\pdach^\nu+{Z^\nu}_{\lambda\mu}\nabla_\nu\pdach^\sigma
\end{equation}
mit der Bedingung der kovarianten Konstanz von ${\bf\pdach}$ \rf{nablap1}, und
erhalten
\begin{equation}
{R^\sigma}_{\nu\lambda\mu}\,\pdach^\nu = 0\ .
\label{Rp}
\end{equation}
Diese Bedingung ist also automatisch erf"ullt. Damit {\bf R}, in "Ubereinstimmung
mit seiner Bedeutung als B"undelkr"ummung in $\taudach$, vollst"andig innerhalb
der charakteristischen Distribution $\Deltadach$ wirkt, m"ussen wir zus"atzlich
fordern, da\3 {\bf R} durch den charakteristischen Vektor ${\bf\pdach}$ bez"uglich
des gesamten ersten Indexpaares annihiliert wird. Das hei\3t, es mu\3 zus"atzlich
gelten:
\begin{equation}
\pdach_\sigma {R^\sigma}_{\nu\mu\lambda} = 0\ .
\label{pR}
\end{equation}
Diese Bedingung wird nun im allgemeinen nicht erf"ullt sein. Sie repr"asentiert
vielmehr eine weitere Einschr"ankung f"ur die Standard-Konnexion ${\bf\Gamma}$.
Das ist nicht weiter schlimm, da wir in ${\bf\Gamma}$ ja einen bis jetzt unbestimmten
Teil {\bf z} haben, welchen wir dazu verwenden wollen, Gleichung \rf{pR} zu
erf"ullen.

Betrachten wir uns dazu die Aufspaltung von ${\bf\Gamma}$ in den wesentlichen Teil
\bm{\gamma} und den Rest {\bf z}, wobei letzterer der Bedingung \rf{zp} unterworfen
ist, so l"a\3t sich \rf{pR} in eine Gleichung f"ur {\bf z} umschreiben.
Setzt man dazu Gleichung \rf{Gamgamz} in den definierenden Ausdruck \rf{defR} f"ur
{\bf R} ein, so ergibt sich eine komplizierte Differentialgleichung f"ur {\bf z}
\begin{equation}
\partial_\mu z_{\nu\lambda} - \partial_\lambda z_{\nu\mu} =
g_{00}\,{\gamma^\sigma}_{\nu\lambda}
\left(\partial_\mu\pdach_\sigma - g_{00}\,z_{\sigma\mu}\right) -
g_{00}\,{\gamma^\sigma}_{\nu\mu}
\left(\partial_\lambda\pdach_\sigma - g_{00}\,z_{\sigma\lambda}\right)
\label{dglz}
\end{equation}
Es l"a\3t sich jedoch eine einfache L"osung angeben, welche \rf{zp} gehorcht:
\begin{equation}
z_{\sigma\mu} = g_{00}\, \partial_\mu\pdach_\sigma\ .
\end{equation}
Diese Gleichung kann man auch mit Hilfe von \rf{dgam} umschreiben
\begin{equation}
z_{\sigma\mu} = -g_{00}\,\gamma_{\sigma\nu\mu}\pdach^\nu\ .
\end{equation}
Somit ergibt sich f"ur die Standard-Konnexion ${\bf\Gamma}$
\begin{equation}
{\Gamma^\lambda}_{\mu\nu} = {\gamma^\lambda}_{\mu\nu} - g_{00}\,
\pdach^\lambda\pdach^\sigma \gamma_{\mu\sigma\nu}\ ,
\label{Gamgamppgam}
\end{equation}
und f"ur die dazugeh"orige Torsion \rf{Zpz}
\begin{equation}
{Z^\lambda}_{\mu\nu} = -g_{00}\,\pdach^\lambda\partial_{[\mu}\pdach_{\nu]}\ .
\label{Zpdp}
\end{equation}
Die L"osung des Problems \rf{pR} in der Gestalt der Standard-Konnexion
${\bf\Gamma}$ \rf{Gamgamppgam} ist immer m"oglich. Da ${\bf\Gamma}$ auf einer
speziellen L"osung der Differentialgleichung f"ur {\bf z} \rf{dglz} basiert, ist es
denkbar, da\3 weitere Konnexionen existieren, deren Torsion {\bf Z} verschwindet.
Zu diesem Punkt werden wir sp"ater zur"uckkehren.

Der charakteristische Vektor ${\bf\pdach}$ ist nun in der Tat unter der oben
gefundenen Standard-Konnexion kovariant konstant; d.h. es gilt nicht nur
\rf{nablap1} sondern zus"atzlich auch noch
\begin{equation}
\nabla_\lambda\pdach_\sigma \equiv \partial_\lambda\pdach_\sigma -
{\Gamma^\rho}_{\sigma\lambda}\pdach_\rho = 0\ ,
\label{nablap2}
\end{equation}
wobei wir die Indizes momentan noch mit der Minkowskischen bzw. Euklidischen Metrik
$g_{\mu\nu}$ bewegt haben. Es gilt also
\begin{equation}
\pdach_\sigma = g_{\sigma\lambda}\pdach^\lambda,\
\gamma_{\mu\sigma\nu} = g_{\mu\lambda}{\gamma^\lambda}_{\sigma\nu} \ {\rm usw.}
\end{equation}

Die Metrik {\bf g} ist nat"urlich nicht kovariant konstant bez"uglich ${\bf\Gamma}$,
jedoch gelten auf jeden Fall die zwei Gleichungen \rf{nablap1} und \rf{nablap2}.
Aus diesem Grund kann man nun vermuten, da\3 eine Metrik {\bf G} existiert, welche
bez"uglich ${\bf\Gamma}$ kovariant konstant ist, und die wie {\bf g} agiert, wenn
man sie auf Vektoren anwendet, welche in die charakteristische Richtung zeigen.
Das hei\3t
\begin{equation}
G_{\mu\nu}\pdach^\mu\pdach^\nu  = g_{\mu\nu}\pdach^\mu\pdach^\nu = g_{00}\ .
\label{Gpp}
\end{equation}
Der Beweis dieser Annahme ist Gegenstand des folgenden Abschnittes.

\subsection{Die Fasermetrik}
\indent

Um die geometrischen Gr"o\3en des charakteristischen B"undels zu vervollst"andigen
ben"otigen wir eine Fasermetrik f"ur $\taudach$. Die Fasermetrik f"ur das
repr"asentative B"undel $\tauquer$ ist die 3-dimensionale Metrik {\bf g}, welche
bez"uglich der trivialisierbaren Konnexion {\bf A} kovariant konstant ist:
\begin{equation}
D_\lambda g_{ij} = 0\ .
\label{Dg}
\end{equation}
Ferner versuchen wir eine neue Fasermetrik {\bf G} f"ur das Tangentenb"undel
$\taukr$ zum Basisraum ${\bf E_4}$ bzw ${\bf M_4}$ zu finden, welche einerseits
\begin{equation}
\nabla_\lambda G_{\mu\nu} = 0
\label{nablaG2}
\end{equation}
gehorcht, und andererseits der Bedingung \rf{Gpp} gen"ugt. Die Einschr"ankung
${\bf\Gdach}$ von {\bf G} auf $\Deltadach$ kann man dann als Fasermetrik f"ur das
charakteristische B"undel $\taudach$ verwenden:
\begin{equation}
\Gdach_{\mu\nu} = G_{\mu\nu} - g_{00}\,\pdach_\mu\pdach_\nu
\label{GdachGpp}
\end{equation}
Die Gleichungen \rf{nablap2} und \rf{nablaG2} garantieren die kovariante Konstanz
von ${\bf\Gdach}$:
\begin{equation}
\nabla_\lambda \Gdach_{\mu\nu} = 0\ .
\label{nablaGdach}
\end{equation}
Es l"a\3t sich nun leicht ein Ausdruck f"ur ${\bf\Gdach}$ finden:
\begin{equation}
\Gdach_{\mu\nu} = \scrc^2 B_{i\mu} {B^i}_\nu \equiv \scrc^2 B_{\mu\nu}\ .
\label{defGdachBB}
\end{equation}
Die kovariante Konstanz \rf{nablaGdach} ist sofort gezeigt, wenn man ber"ucksichtigt,
da\3 gilt:
\begin{equation}
\nabla_\lambda B_{\mu\nu} = 0\ ;
\end{equation}
was wir mit Hilfe der Identit"at
\begin{equation}
\nabla_\lambda B_{\mu\nu} = \left({\cal D}_\lambda\,B_{i\mu}\right){B^i}_\nu +
B_{i\mu}\left({\cal D}_\lambda{B^i}_\nu\right) \ ,
\end{equation}
sowie der Bedingung \rf{calDB} f"ur die allgemein kovariante Konstanz der
extrinsischen Kr"ummung {\bf B} leicht zeigen k"onnen. Wir haben nun au\3er der
Metrik {\bf g} noch eine zweite Metrik {\bf G} im Tangentenb"undel $\taukr$, und
m"ussen aus diesem Grunde eine gewisse Vorsicht walten lassen, wenn wir Indizes
bewegen. Folgende Konvention gelte: Indizes, die mit {\bf G} bewegt werden,
kennzeichnen wir mit einem Punkt. Die folgenden Beispiele m"ogen dies verdeutlichen.
\alpheqn
\begin{equation}
\pdach_{\dot{\mu}} = G_{\mu\nu}\pdach^\nu = \pdach_\mu
\end{equation}
\begin{equation}
G^{\mu\nu} = \scrc^2 B^{\mu\nu} + g_{00}\,\pdach^\mu\pdach^\nu
\end{equation}
\begin{equation}
G^{\dot{\mu}\dot{\nu}} = \scrc^{-2}{\left(B^{-1}\right)}^{\mu\nu} +
g_{00}\,\pdach^\mu\pdach^\nu
\end{equation}
\begin{equation}
{\left(B^{-1}\right)}^{\mu\nu} B_{\nu\lambda} = {\Pdach^\mu}_\lambda
\end{equation}
\begin{equation}
G^{\dot{\mu}\dot{\nu}} G_{\nu\lambda} = {g^\mu}_\lambda\ .
\end{equation}
\reseteqn
Die Objekte $G^{\mu\nu}, B^{\mu\nu}, \Pdach_{\mu\nu}$ und $\Pdach^{\mu\nu}$ sind im
allgemeinen nicht konstant, es gilt vielmehr
\alpheqn
\begin{equation}
\nabla_\lambda G^{\dot{\mu}\dot{\nu}} = 0
\end{equation}
\begin{equation}
\nabla_\lambda {\left(B^{-1}\right)}^{\mu\nu} = 0
\end{equation}
\begin{equation}
\nabla_\lambda {\Pdach^\mu}_\nu = 0\ .
\end{equation}
\reseteqn
Wir wollen nun die aus der Existenz der Metrik {\bf G} folgenden Eigenschaften der
Kr"ummung {\bf R} untersuchen. Betrachten wir dazu die allgemeing"ultige Identit"at
\begin{equation}
\left(\nabla_\lambda\nabla_\sigma-\nabla_\sigma\nabla_\lambda\right)G_{\mu\nu}
\equiv -
{R^\rho}_{\mu\lambda\sigma}G_{\rho\nu} - {R^\rho}_{\nu\lambda\sigma}G_{\mu\rho} +
2 {Z^\rho}_{\lambda\sigma}\nabla_\rho G_{\mu\nu} \ ,
\end{equation}
und ber"ucksichtigen die kovariante Konstanz von {\bf G} \rf{nablaG}, so erhalten
wir die Schief-Symmetrie des Kr"ummungsoperators {\bf R}
\begin{equation}
R_{\dot{\nu}\mu\lambda\sigma} = - R_{\dot{\mu}\nu\lambda\sigma}\ .
\end{equation}
Wir sehen nun, da\3 die Geometrie des charakteristischen B"undels eine
{\em Riemann-Cartan} Struktur besitzt. Die Konnexion ${\bf\Gamma}$ l"a\3t sich nun in
diesem Riemann-Cartan Raum in einen Riemannschen Teil ${\bf\Gammaschl}$ und in die
sog. Kontorsion {\bf K} zerlegen \cite{He76}
\begin{equation}
{\Gamma^\lambda}_{\mu\nu} = {\Gammaschl^\lambda}_{\mu\nu} + {K^\lambda}_{\mu\nu}
\label{GamGamschlK}
\end{equation}
Der Riemannsche Teil ${\bf\Gammaschl}$ ist hierbei durch die Christoffel Symbole
gegeben
\begin{equation}
{\Gammaschl^\lambda}_{\mu\nu} = \frac{1}{2} G^{\dot{\lambda}\dot{\sigma}}
\left(\partial_\mu G_{\sigma\nu} + \partial_\nu G_{\sigma\mu} -
\partial_\sigma G_{\mu\nu}\right)\ .
\end{equation}
wogegen sich die Kontorsion {\bf K} folgenderma\3en aus der Torsion {\bf Z}
zusammensetzt:
\begin{equation}
{K^\lambda}_{\mu\nu} = {Z^\lambda}_{\mu\nu}+ {Z_{\dot{\mu}\nu}}^{\dot{\lambda}} +
{Z_{\dot{\nu}\mu}}^{\dot{\lambda}}\ .
\label{KZZZ}
\end{equation}

\section{Der B"undelisomorphismus}
\label{Eig2}
\indent

Die Beziehung der beiden B"undel $\tauquer$ und $\taudach$ zueinander ist enger als
es die vorausgehenden Betrachtungen zun"achst erscheinen lassen. Es wird sich in der
Tat zeigen, da\3 die beiden B"undel isomorph sind, und damit als identisch betrachtet
werden k"onnen.

Wir wollen in diesem Kapitel den B"undelisomorphismus untersuchen. Die Ergebnisse
benutzen wir dann um die Kr"ummung {\bf R} aus $\taudach$ mit Hilfe der extrinsischen
Kr"ummung {\bf B} aus $\tauquer$ auszudr"ucken. Dies wird uns zu einer
Verallgemeinerung der Dimeronkonfiguration f"uhren. Dabei zeigt sich, da\3 die
meisten Eigenschaften des Dimeron Falles diese Verallgemeinerungsprozedur
unver"andert "uberstehen. So vertr"agt sich zum Beispiel die Riemann-Cartan Struktur
formal mit den Bedingungen f"ur die Existenz eines konformal flachen, lokal
symmetrischen Einstein Raumes. Wir k"onnen daraus folgern, da\3 die
charakteristischen Fl"achen eine konstante Kr"ummung besitzen.

\subsection{Die B"undelabbildung}
\indent

Die beiden B"undel $\tauquer$ und $\taudach$ werden miteinander durch eine Abbildung
verkn"upft, die durch die extrinsische Kr"ummung {\bf B} von $\tauquer$ induziert
wird. Betrachten wir zun"achst die B"undelabbildung $[\Bdach]$:
\begin{equation}
[\Bdach]:  \tauquer \to \taudach \ .
\label{tauquerTotaudach}
\end{equation}
Durch diese Abbildung wird ein Schnitt ${\bf v} \in \tauquer$ in sein Bild ${\bf V}
\in \taudach$ "ubergef"uhrt, so da\3 gilt
\begin{equation}
V_\mu = \scrc B_{i\mu} v^i\ ,
\label{VBv}
\end{equation}
wobei der Parameter $\scrc$ aus Dimensionsgr"unden eingef"uhrt wurde. Diese
B"undelabbildung kann man nat"urlich auf alle Tensorprodukte erweitern. So l"a\3t
sich z.B. die Fasermetrik ${\bf\Gdach}$ \rf{defGdachBB} aus $\taudach$ als das Bild
der Fasermetrik {\bf g} \rf{Dg} darstellen:
\begin{equation}
\Gdach_{\mu\nu} = \scrc^2 B_{i\mu}B_{j\nu}g^{ij}\ .
\end{equation}
Aus diesem Grund ist das Skalarprodukt zweier Schnitte $({\bf u},{\bf v})\in\tauquer$
und ihrer Bilder $({\bf U},{\bf V})\in\taudach$ invariant bez"uglich $[\Bdach]$
\begin{equation}
{\bf G}(\bf{U},\bf{V}) = {\bf g}(\bf{u},\bf{v}) ,
\end{equation}
bzw. in Komponenten
\begin{equation}
G^{\dot{\mu}\dot{\nu}}U_\mu V_\nu = g_{ij} v^i u^j\ .
\end{equation}
Die inverse Abbildung $[\Bquer]$ bildet jeden Schnitt ${\bf V}\in\taudach$ nach
$\tauquer$ ab:
\begin{equation}
v_i = \scrc B_{i\mu} V^{\dot{\mu}}\ .
\label{vBV}
\end{equation}

Kombiniert man beide Abbildungen, so zeigt sich da\3 \rf{vBV} und \rf{VBv} die
Identit"atsabbildung f"ur die korrespondierenden Bildr"aume ergeben:
\alpheqn
\begin{eqnarray}
\label{BquerBdach}
[\Bquer]\circ[\Bdach] &=& \mbox{id}_{\tauquer}\\
\label{BdachBquer}
[\Bdach]\circ[\Bquer] &=& \mbox{id}_{\taudach}\ .
\end{eqnarray}
\reseteqn
Somit haben wir die 1-1 Beziehung der beiden B"undel gezeigt!

\subsection{Extrinsische und intrinsische Kr"ummung}
\indent

F"ur einen affinen B"undelisomorphismus \cite{KoNo69} gen"ugt es nicht, da\3 die
Fasern "uber jedem Punkt {\bf x} im Basisraum in 1-1 Beziehung stehen.
Zus"atzlich mu\3 die Aufspaltung in einen horizontalen und einen vertikalen
Unterraum mit der B"undelabbildung konsistent sein. Mit anderen Worten, der
Proze\3 der Parallelverschiebung mu\3 mit der B"undelabbildung vertauschen.
Dies ist in der Tat der Fall, falls die kovarianten Ableitungen mit der
B"undelabbildung kommutieren, das hei\3t es mu\3 gelten:
\alpheqn
\begin{eqnarray}
\label{BuendelAbb}
\mbox{\bm{\nabla}}\circ[\Bdach] &=& [\Bdach]\circ {\bf D}\\
{\bf D}\circ[\Bquer] &=& [\Bquer]\circ\mbox{\bm{\nabla}}\ .
\end{eqnarray}
\reseteqn
In Komponenten ausgedr"uckt lauten diese Bedingungen
\alpheqn
\begin{eqnarray}
\label{BuendelAbbKomp2}
\nabla_\lambda V_\mu = \scrc B_{i\mu}\left(D_\lambda v^i\right)\\
\label{BuendelAbbKomp1}
D_\lambda v_i = \scrc B_{i\mu}\nabla_\lambda V^{\dot{\mu}}
\end{eqnarray}
\reseteqn
Differenziert man die korrespondierenden Gleichungen \rf{VBv} und \rf{vBV}, so
erhalten wir
\alpheqn
\begin{equation}
\nabla_\lambda V_\mu \equiv \scrc \left({\cal D}_\lambda B_{i\mu}\right)v^i
+ \scrc B_{i\mu}\left( D_\lambda v^i\right)
\end{equation}
\begin{equation}
D_\lambda v_i \equiv \scrc\left({\cal D}_\lambda B_{i\mu}\right) V^{\dot{\mu}} +
\scrc B_{i\mu}\left(\nabla_\lambda V^{\dot{\mu}}\right) \ .
\end{equation}
\reseteqn
Benutzen wir weiter die Konstanzbedingung \rf{calDB} f"ur die extrinsische Kr"ummung
{\bf B} so zeigt sich, da\3 \rf{BuendelAbbKomp1} und \rf{BuendelAbbKomp2} erf"ullt
sind.

Aufgrund dieser Tatsache zeigt sich, da\3 die extrinsischen Kr"ummungskoeffizienten
sowohl in $\tauquer$ als auch in $\taudach$ wirken. Diese Gr"o\3en k"onnen deshalb
auch nur durch gleichzeitige Anwendung der trivialisierbaren Konnexion {\bf A}
{\em und} der Standard-Konnexion parallel verschoben werden. Die allgemein kovariante
Ableitung \bm{{\cal D}} \rf{calDB} ber"ucksichtigt dies und l"a\3t sich deshalb auf
alle gemischten Objekte mit beliebigem Rang anwenden. Die oben aufgef"uhrten
Transportgesetze mit \bm{\nabla} und {\bf D} sind also nur spezielle F"alle des
allgemeineren Transportes durch \bm{{\cal D}.}

Mit Hilfe dieses B"undelisomorphismus wollen wir nun die Beziehung der beiden
B"undel\-kr"ummungen untersuchen. Da sich die intrinsische Kr"ummung {\bf F} in
$\tauquer$ mit Hilfe der ersten Trivialisierungsbedingung \rf{triva} vollst"andig
durch die extrinsischen Kr"ummungsfelder {\bf B} ausdr"ucken l"a\3t, k"onnen wir die
Kr"ummung {\bf R} in $\taudach$ ebenfalls vollst"andig mit Hilfe von {\bf B}
formulieren. Um zu einer Beziehung zwischen {\bf R} und {\bf B} zu gelangen,
differenzieren wir die Gleichungen \rf{BuendelAbb} einmal. Wir erhalten symbolisch:
\alpheqn
\begin{eqnarray}
\mbox{\bm{\nabla}}\circ\mbox{\bm{\nabla}}\circ[\Bdach] &=& [\Bdach]\circ
\mbox{\bm{\calD}}\circ {\bf D}\\
\mbox{\bm{\calD}}\circ {\bf D}\circ[\Bquer] &=& [\Bquer]\circ\mbox{\bm{\nabla}}\circ
\mbox{\bm{\nabla}}\ .
\end{eqnarray}
\reseteqn
In Komponenten ausgedr"uckt, ergibt sich nach Schief-Symmetrisierung
\alpheqn
\begin{eqnarray}
\label{diffAbb}
\nabla_{[\sigma}\nabla_{\lambda]}V_\mu &=& \scrc B_{i\mu}{\cal D}_{[\sigma}
D_{\lambda]}v^i\\
{\cal D}_{[\sigma}D_{\lambda]}v_i &=& \scrc B_{i\mu}\nabla_{[\sigma}\nabla_{\lambda]}
V^{\dot{\mu}} \ .
\end{eqnarray}
\reseteqn
Ber"ucksichtigen wir noch die Identit"aten
\alpheqn
\begin{eqnarray}
\label{Id1}
{\cal D}_{[\sigma}D_{\lambda]}v^i &\equiv& \frac{1}{2}{\epsilon^{ij}}_k\,
F_{j\sigma\lambda} v^k + {Z^\rho}_{\sigma\lambda}\left(D_\rho v^i\right)\\
\nabla_{[\sigma}\nabla_{\lambda]}V_\mu &\equiv& - \frac{1}{2}{R^\rho}_{\mu\sigma
\lambda} V_\rho + {Z^\rho}_{\sigma\lambda}\left(\nabla_\rho V_\mu\right)\ ,
\end{eqnarray}
\reseteqn
sowie die B"undelabbildungen \rf{VBv} und \rf{vBV}, so erhalten wir die Beziehung
\begin{equation}
{R^\rho}_{\mu\sigma\lambda} = - \left({\Pdach^\rho}_\sigma B_{\mu\lambda} -
{\Pdach^\rho}_\lambda B_{\mu\sigma}\right)\ .
\label{RPB}
\end{equation}
Diese spezielle Gestalt der Kr"ummung {\bf R} hat wiederum die Form f"ur die
Euklidische Dimeron Konfiguration \cite{Br88}\,! Wir k"onnen {\bf R} als total
kovariantes Objekt schreiben, indem wir den ersten Index mit Hilfe der Fasermetrik
${\bf\Gdach}$ \rf{defGdachBB} senken:
\begin{equation}
R_{\dot{\mu}\nu\sigma\lambda} = -\scrc^2\left(B_{\mu\sigma}B_{\nu\lambda} -
B_{\mu\lambda}B_{\nu\sigma}\right)\ .
\label{RBB}
\end{equation}
In dieser Gestalt ist die Kr"ummung {\bf R} aus $\taudach$ vollst"andig mit Hilfe der
extrinsischen Kr"ummung {\bf B} aus $\tauquer$ ausgedr"uckt. Stehen die ersten beiden
Indizes oben, so l"a\3t sich {\bf R} durch den Euklidischen Projektor ${\bf\Pdach}$
allein ausdr"ucken:
\begin{equation}
{R^{\mu\dot{\nu}}}_{\sigma\lambda} = -\scrc^{-2}\left({\Pdach^\mu}_\sigma
{\Pdach^\nu}_\lambda - {\Pdach^\mu}_\lambda {\Pdach^\nu}_\sigma
\right)\ .
\label{RPP}
\end{equation}
{\bf R} gen"ugt den beiden Bedingungen \rf{Rp}, \rf{pR}, und l"a\3t au\3er dem
charakteristischen  Vektor ${\bf\pdach}$ keinen weiteren kovariant konstanten Vektor
zu! Insbesondere gibt es keinen $\taudach$ Schnitt $U$ dessen kovariante
Ableitung wie folgt lautet:
\begin{equation}
\nabla_\lambda U_\mu = N_\mu\pdach_\lambda .
\label{nablaUNp}
\end{equation}
Um diese Annahme zu beweisen, benutzen wir die Identit"aten \rf{Id1} und das
Ergebnis \rf{RPB} f"ur {\bf R} und erhalten
\begin{equation}
\nabla_{[\sigma}\nabla_{\lambda]}U_\mu = \pdach_{[\lambda}\nabla_{\sigma]}N_\mu =
B_{\mu[\lambda}U_{\sigma]}
\end{equation}
Da der letzte Term ${\bf\pdach}$ nicht enth"alt, mu\3 er identisch Null sein, was nur
m"oglich ist f"ur verschwindendes {\bf U}. Das hei\3t die Gleichung \rf{nablaUNp}
besitzt keine nicht-triviale L"osung f"ur {\bf U} und demzufolge gibt es au\3er
${\bf\pdach}$ keinen kovariant konstanten Schnitt. Die Konsequenz davon ist, da\3 die
kovariante Ableitung eines beliebigen Schnittes aus $\taudach$ in Wirklichkeit folgende
Gestalt haben mu\3:
\begin{equation}
\nabla_\lambda U_\mu = N_\mu\pdach_\lambda + M_{\mu\lambda}\ .
\label{nablaUNM}
\end{equation}
Aufgrund von \rf{pnablaV} k"onnen wir ohne Beschr"ankung der Allgemeinheit setzen:
\begin{equation}
\pdach^\mu N_\mu = \pdach^\mu M_{\mu\lambda} = \pdach^\lambda M_{\mu\lambda} = 0\ .
\end{equation}
Der Tensor {\bf M}  kann nun im Gegensatz zum Vektor {\bf N} niemals verschwinden.
Um daf"ur ein Beispiel zu geben, l"osen wir die Konstanzbedingung \rf{calDB} nach
der koordinaten kovarianten Ableitung ${\bf\nabla B}$ auf, und erhalten mit Hilfe von
\rf{nablaUNM}
\begin{equation}
\nabla_\lambda B_{i\mu} = N_{i\mu}\pdach_\lambda + M_{i\mu\lambda}\ ,
\label{nablaBNM}
\end{equation}
mit
\alpheqn
\begin{eqnarray}
N_{i\mu} &=& {\epsilon_i}^{jk}B_{k\mu}\left(\pdach^\sigma A_{j\sigma}\right)\\
M_{i\mu\lambda}&=&g_{00}{\epsilon_i}^{jk}B_{k\mu}A_{j\sigma}{\Pdach^\sigma}
_\lambda\ .
\end{eqnarray}
\reseteqn
Ben"utzen wir nun diejenige Eichung in der ${\bf A}\equiv{\bf B}$ ist (positive
Eichung siehe \cite{Ma85}), so reduziert sich \rf{nablaBNM} zu
\begin{equation}
\nabla_\lambda B_{i\mu} = F_{i\lambda\mu}\ .
\end{equation}
Wir sehen, da\3 der Vektor {\bf N}, aber nicht der Tensor {\bf M} in \rf{nablaBNM}
weggeeicht werden kann. Mit einer "ahnlichen Argumentation l"a\3t sich zeigen, da\3 es
au\3er ${\bf\Gdach}$ keinen weiteren kovariant konstanten Tensor zweiten Ranges in
$\taudach$ gibt.

\subsection{Verallgemeinerte Konformalit"at}
\label{VerallgKonf}
\indent

Die vorausgehenden Untersuchungen des B"undelisomorphismus k"onnen dazu ben"utzt
werden, um zu untersuchen, wie die konformalen Eigenschaften der Dimeron
Konfiguration zu verallgemeinern sind. Dazu erinnern wir uns, da\3 die notwendige und
hinreichende Bedingung f"ur die konformale Flachheit eines Riemannschen Raumes in
mindestens vier Dimensionen das Verschwinden des Weylschen Tensors {\bf W} ist
\cite{Cho82}.
\begin{equation}
{W^{\nu\dot{\mu}}}_{\kappa\lambda}={R^{\nu\dot{\mu}}}_{\kappa\lambda}-\frac{2}{n-2}
\left({{\cal R}^{\dot{\nu}}}_{[\kappa}{G^{\dot{\mu}}}_{\lambda]}-
{{\cal R}^{\dot{\mu}}}_{[\kappa}{G^{\dot{\nu}}}_{\lambda]}\right) +
\frac{2}{(n-1)(n-2)}\,{\cal S}\,{G^{\dot{\nu}}}_{[\kappa}{G^{\dot{\mu}}}_{\lambda]}\ .
\label{W}
\end{equation}
Streng genommen gilt dieses Kriterium nur f"ur einen Riemannschen Raum, doch liegt
die M"oglichkeit nahe, es formal auf jeden Raum, der eine symmetrische Metrik {\bf G}
und einen eindeutigen Ricci Tensor \bm{\cal R} besitzt, auszudehnen:
\begin{equation}
{{\cal R}^{\dot{\mu}}}_\nu = {R^{\sigma\dot{\mu}}}_{\sigma\nu} =
{R^{\dot{\mu}\sigma}}_{\nu\sigma} \ .
\label{Ricci}
\end{equation}
Einen Raum mit diesen Eigenschaften  k"onnte man dann als einen "`fast Riemannschen
Raum"' bezeichnen. Wenden wir dieses verallgemeinerte Kriterium auf unsere Situation
an, so zeigt sich, da\3 die B"undelgeometrie unseres trivialisierbaren Eichfeldes in
der Tat von einer konformal flachen Riemann-Cartan'schen Struktur herr"uhrt! Denn
kontrahieren wir den Kr"ummungstensor \rf{RPP} so erhalten wir f"ur \bm{\cal R}
\alpheqn
\begin{eqnarray}
\label{RicciP}
{{\cal R}^{\dot{\mu}}}_\sigma &=& -2 \scrc^{-2}{\Pdach^\mu}_\sigma\\
\label{calRB}
{\cal R}_{\mu\sigma} &=& -2 B_{\mu\sigma}\ .
\end{eqnarray}
\reseteqn
Damit ergibt sich der Kr"ummungsskalar S zu
\begin{equation}
S \equiv {{\cal R}^\mu}_\mu = -6\scrc^{-2}\ ,
\label{calS}
\end{equation}
welcher nat"urlich aufgrund der Gleichungen \rf{RBB} und \rf{calRB} kovariant
konstant sein mu\3:
\begin{equation}
\nabla_\lambda R_{\dot{\mu}\nu\rho\sigma} = \nabla_\lambda{\cal R}_{\mu\nu} =
\partial_\lambda S = 0\ .
\label{nablaR}
\end{equation}
Dr"ucken wir den Ricci Tensor \bm{{\cal R}} mit Hilfe des Skalars ${\cal S}$ aus,
ergibt sich weiter
\begin{equation}
{\cal R}_{\mu\sigma} = \frac{1}{3}\,S\,\Gdach_{\mu\sigma} ,
\label{calRSGdach}
\end{equation}
was nichts anderes bedeutet, als das wir im wesentlichen einen Einsteinschen Raum mit
drei Dimensionen vorliegen haben. Die Anzahl der Dimensionen ist plausibel, wenn wir
uns vergegenw"artigen, da\3 die B"undel Geometrie auf der 3-dimensionalen
Distribution $\Deltadach$ basiert.

Wir k"onnen nun entweder die gesamte Metrik {\bf G}, zusammem mit der Dimension n =
4, oder die Fasermetrik ${\bf\Gdach}$ mit n = 3 Dimensionen im Weylschen Tensor
\rf{W} verwenden. In beiden F"allen verschwindet {\bf W}, was unsere Definition der
verallgemeinerten konformalen Flachheit rechtfertigt. Die folgende Betrachtung
verdeutlicht dies:

Es existiert ein Theorem in der Riemannschen Geometrie, welches besagt, da\3 jeder
konformal flache Einstein Raum ein Raum mit konstanter Kr"ummung ist. Die
Verallgemeinerung dieses Theorems auf die vorliegende nicht-Riemannsche Struktur ist
in der Tat m"oglich: Unser dreidimensionaler Einsteinsche Raum \rf{calRSGdach}
besitzt sowohl im intrinsischen (n = 3), als auch im extrinsischen (n = 4) Fall einen
verschwindenden Weylschen Tensor, so da\3 die verallgemeinerte Konformalit"at
gesichert scheint. Jedoch taucht nun ein Problem auf: das Verschwinden des Weylschen
Tensors ist nur in n $\ge$ 4 Dimensionen eine hinreichende Bedingung f"ur konformale
Flachheit. Unsere B"undelgeometrie ist jedoch dreidimensional. Aus diesem Grund
m"ussen wir eine weitere Bedingung f"ur konformale Flachheit finden. Diese besteht im
Verschwinden des Tensors \bm{{\cal W}} \cite{SeUr83}:
\begin{equation}
{\cal W}_{\mu\nu\lambda} =
\nabla_\lambda{\cal R}_{\dot{\mu}\nu}- \nabla_\nu{\cal R}_{\dot{\mu}\lambda} +
\frac{1}{4}\,\left( G_{\mu\lambda}\,\partial_\nu{\cal S}-G_{\mu\nu}\,\partial_\lambda
{\cal S}\right)
\end{equation}
Aufgrund der Konstanz des Kr"ummungsskalars \bm{\cal S} und des Ricci Tensors
\bm{{\cal R}} ist dies garantiert.
Der Kr"ummungstensor {\bf R} hat nun, wenn man ihn mit Hilfe der Fasermetrik
${\bf\Gdach}$ ausdr"uckt, in der Tat die Form, die man f"ur einen Raum mit konstanter
Kr"ummung erwartet:
\begin{equation}
R_{\dot{\mu}\nu\sigma\lambda} = -\scrc^{-2}\left(
\Gdach_{\mu\sigma}\Gdach_{\nu\lambda}-\Gdach_{\mu\lambda}\Gdach_{\nu\sigma}\right)\ .
\label{RGG}
\end{equation}
Wir k"onnen also unser oben aufgef"uhrtes Kriterium der verallgemeinerten konformalen
Flachheit auf unseren nicht-Riemannschen Raum anwenden. Die Konsequenz ist, da\3
sich die charakteristischen Fl"achen, falls sie existieren, als dreidimensionale, mit
einer Riemannschen Geometrie konstanter Kr"ummung ausgestattete
Untermannigfaltigkeiten des Basisraums herausstellen. Dies l"a\3t sich z.B. anhand
der Euklidischen Dimeron Konfiguration verdeutlichen: bei dieser bestehen die
charakteristischen Fl"achen aus 3-Sph"aren und sind aufgrund ihrer Einbettung in den
Basisraum ${\bf E_4}$ mit einer konstanten Kr"ummungsgeometrie ausgestattet (s. Bild
1). Jedoch ist diese Geometrie, die durch Projektion aus der Geometrie des
einbettenden Raumes entsteht, nicht identisch mit der charakteristischen
B"undelgeometrie! Trotzdem besitzen beide Geometrien eine konstante Kr"ummung.
Bewegen wir nun eines der Meronzentren ins Unendliche, werden beide Geometrien
identisch (vgl. die Diskussion nach Gleichung \rf{pnablaV}), da die beiden
Distributionen $\Deltaquer$ und $\Deltadach$ ineinander "ubergehen.

\section{Integrabilit"at und Torsion}
\indent

Aufgrund des B"undelisomorphismus k"onnen wir die beiden B"undel $\tauquer$ und
$\taudach$ als identisch annehmen. Auf den ersten Blick scheint dies eine paradoxe
Situation zu sein: Einerseits besagt ein bekanntes Theorem \cite{Che73}, da\3
die Konnexion {\bf A} in $\tauquer$ torsionsfrei sein mu\3, da dessen Geometrie die
Untergeometrie des trivialen einbettenden B"undels $\taukr$ darstellt, welches
nat"urlich keine Torsion besitzt; andererseits besitzt die Standard-Konnexion
${\bf\Gamma}$ in $\taudach$ eine nicht verschwindende Torsion, falls wir dessen
B"undelgeometrie mit globalen kartesischen Koordinaten ausdr"ucken (siehe Abschnitt
\rf{TrivUndTors}). Aus diesem Grund m"ussen wir fragen ob die Torsion {\bf Z} nicht
eliminiert werden kann.

Wir wollen nun untersuchen unter welchen Umst"anden wir die Torsion auch im
kartesischen Koordinatensystem beseitigen k"onnen. Es wird sich zeigen, da\3
die Integrabilit"at der charakteristischen Distribution ein hinreichendes Kriterium
daf"ur darstellt. Das hei\3t, solange die Integrabilit"atsbedingung gilt, k"onnen wir
eine streng Riemannsche Konnexion, die sog. {\em charakteristische Konnexion}
${\bf\Gammastern}$, anstelle der Standard-Konnexion, benutzen. Ferner l"a\3t sich die
korrespondierende Riemannsche Metrik ${\bf\Gstern}$ konstruieren, so da\3
${\bf\Gammastern}$ die Levi-Civita Konnexion von ${\bf\Gstern}$ ist. Die auf diesem
Weg gefundene Riemannsche Struktur ist immer im verallgemeinerten Sinn konformal
flach und symmetrisch.

\subsection{Die Frobenius Bedingung}
\indent

In einer Riemann-Cartan Struktur Struktur gibt es nun einige, die Torsion {\bf Z}
betreffende, Identit"aten. In unserem Fall nehmen diese eine ganz bestimmte Form an,
da wir uns mit einer fast Riemannschen Struktur besch"aftigen. Wenn wir dieser
Struktur die Frobeniuschen Integrabilit"atsbedingungen f"ur die Distribution
$\Deltadach$ hinzuf"ugen, nimmt die Torsion eine Gestalt an, die es uns erm"oglicht
sie vollst"andig zu eliminieren.

Betrachten wir dazu zuerst die zweite Bianchi Identit"at
\begin{equation}
{R^\lambda}_{\sigma\,[\mu\nu;\rho]} = 2 {R^\lambda}_{\sigma\kappa\,[\rho}
{Z^\kappa}_{\mu\nu]}\ .
\label{Bianchi2_R}
\end{equation}
Diese Identit"at ist identisch erf"ullt, da die linke Seite aufgrund der kovarianten
Konstanz der Kr"ummung verschwindet (s. \rf{nablaR}), w"ahrend auf der rechten Seite
die Kr"ummung {\bf R} den charakteristischen Vektor ${\bf\pdach}$ in {\bf Z}
annihiliert. (Vgl. \rf{Rp}, \rf{pR} und \rf{RPB}). Deshalb erhalten wir aus dieser
Gleichung keine Einschr"ankung f"ur die Torsion.

Ganz anders liegt der Fall f"ur die erste Bianchi Identit"at
\begin{equation}
-{R^\rho}_{[\mu\nu\lambda]} = 2\nabla_{[\mu}{Z^\rho}_{\nu\lambda]} +
4{Z^\kappa}_{[\mu\nu}{Z^\rho}_{\lambda]\kappa}\ .
\label{Bianchi1_R}
\end{equation}
Die linke Seite verschwindet, da der Kr"ummungstensor all die Symmetrieeigenschaften
eines gew"ohnlichen Riemannschen Tensors erf"ullt. Aus diesem Grund gehorcht die
Torsion folgender Gleichung
\begin{equation}
\nabla_{[\mu}{Z^\rho}_{\nu\lambda]}+2{Z^\kappa}_{[\mu\nu}{Z^\rho}_{\lambda]\kappa}
= 0\ .
\label{nablaZZZ}
\end{equation}
Falls die charakteristische Distribution integrabel ist, l"a\3t sich dieser Ausdruck
weiter vereinfachen. Die Frobeniussche Integrabilit"atsbedingung f"ur den
Normalenvektor ${\bf\pdach}$ von $\Deltadach$ lautet \cite{We78}
\begin{equation}
\partial_{[\mu}\pdach_{\lambda]} = f_{[\lambda}\pdach_{\mu]} \ .
\label{dpfp}
\end{equation}
Ohne Beschr"ankung der Allgemeinheit k"onnen wir annehmen
\begin{equation}
f^{\mu}\pdach_{\mu} = 0,
\label{fp}
\end{equation}
und demzufolge nimmt die Torsion der Standard-Konnexion ${\bf\Gamma}$, in
"Ubereinstimmung mit \rf{Zpdp}, folgende, spezielle Form an:
\begin{equation}
{Z^\lambda}_{\mu\nu} = - g_{00}\,\pdach^\lambda\pdach_{\,[\mu}f_{\nu\,]} \ .
\label{Zppf}
\end{equation}
Wegen dieser speziellen Form der Torsion {\bf Z} verschwinden die nichtlinearen Terme
in \rf{nablaZZZ} und es verbleibt noch
\begin{equation}
\nabla_{[\mu}{Z^\rho}_{\nu\lambda]} = 0\ .
\label{nablaZ}
\end{equation}
Eine weitere Besonderheit der speziellen Gestalt der Torsion {\bf Z} \rf{Zppf} ist
die Relation
\begin{equation}
{Z^\lambda}_{\mu\nu} + {{Z_{\dot{\nu}}}^{\dot{\lambda}}}_{\mu} +
{Z_{\dot{\mu}\nu}}^{\dot{\lambda}} = 0\ .
\label{ZZZ}
\end{equation}
Dadurch vereinfacht sich der Ausdruck f"ur die Kontorsion {\bf K} \rf{KZZZ} zu
\begin{equation}
{K^\lambda}_{\mu\nu} = 2{Z_{\dot{\nu}\mu}}^{\dot{\lambda}} \ .
\label{KZ}
\end{equation}

Wir sehen, da\3 die Integrabilit"at von $\Deltadach$ eine spezielle Riemann-Cartan
Geometrie erzeugt, welche einer Riemannschen Geometrie ziemlich nahe kommt. Dies
zeigt sich besonders deutlich wenn man die Bewegung eines Testteilchens in dieser
Geometrie untersucht. Hierbei taucht jedoch in einer Riemann-Cartanschen Geometrie
eine gewisse Zweideutigkeit auf \cite{He76}. Wir m"ussen uns n"amlich "uberlegen, ob
das Testteilchen einer autoparallelen oder einer geod"atischen Bewegung unterliegt.
F"ur einen streng Riemannschen Raum stimmen beide Bewegungsformen "uberein und das
Bewegungsproblem ist eindeutig. Die Zweideutigkeit existiert selbstverst"andlich auch
f"ur den vorliegenden Riemann-Cartan Raum, wenn auch in einer etwas schw"acheren Form.
Wir erinnern uns, da\3 der (Einheits) Tangentenvektor {\bf t}, entlang einer
autoparallelen Kurve, bez"uglich der Standard-Konnexion ${\bf\Gamma}$ parallel
verschoben wird:
\begin{equation}
t^\lambda\nabla_\lambda\,t^\mu = 0
\label{tnablat}
\end{equation}
\begin{equation}
\left(t^\lambda\equiv\frac{dx^\lambda}{ds}\right)\ .
\label{deft}
\end{equation}
Das gleiche gilt f"ur eine geod"atische Kurve, jedoch wird hier bez"uglich des
Riemannschen Teils ${\bf\Gammaschl}$ von ${\bf\Gamma}$ parallel verschoben
(s. \rf{GamGamschlK}).
\begin{equation}
t^\lambda\tilde{\nabla}_\lambda\,t^\mu = 0\ .
\label{tnablaschlt}
\end{equation}
Die entsprechenden Differentialgleichungen f"ur die Bewegung des Testteilchens lauten
\alpheqn
\begin{eqnarray}
\label{Bewgl}
\frac{D^2 x^\mu}{ds^2} &\equiv& \frac{d^2 x^\mu}{ds^2} + {\Gamma^\mu}_{\nu\lambda}
\frac{dx^\nu}{ds}\frac{dx^\lambda}{ds} = 0\\
\label{Bewglschl}
\;\frac{\tilde{D}^2x^\mu}{ds^2} &\equiv& \frac{d^2 x^\mu}{ds^2}
+{\Gammaschl^\mu}_{\nu\lambda}\frac{dx^\nu}{ds}\frac{dx^\lambda}{ds} = 0 \ .
\end{eqnarray}
\reseteqn
Die L"osungen dieser Gleichungen werden sich nun im allgemeinen Fall unterscheiden
(s. Abschnitt \rf{AutoUndGeod}).

Im vorliegenden Fall, der durch die Gleichungen \rf{Zppf} und \rf{KZ}
charakterisiert wird, unterscheiden sich die zwei Differentialoperatoren in den
Gleichungen \rf{Bewgl} und \rf{Bewglschl} durch einen Torsionsterm.
\begin{equation}
\frac{D^2x^\mu}{ds^2}=\frac{\tilde{D}^2x^\mu}{ds^2}+2{Z_{\dot{\lambda}\nu}}
^{\dot{\mu}}\;\frac{dx^\nu}{ds}\frac{dx^\lambda}{ds} = 0 \ .
\label{DDschlZ}
\end{equation}
Aus diesem Grund stimmen die beiden Bewegungstypen "uberein falls folgende Bedingung
erf"ullt ist
\begin{equation}
{Z_{\dot{\lambda}\nu}}^{\dot{\mu}}\,t^\nu t^\lambda = 0 \ .
\label{Ztt}
\end{equation}
Das hei\3t falls die Bewegung des Testteilchens auf die charakteristische Fl"ache
beschr"ankt bleibt. Aufgrund des Frobeniusschen Theorems wird der gesamte Raum aus
charakteristischen Fl"achen aufgebaut, so da\3 die oben erw"ahnte Zweideutigkeit
bez"uglich der Teilchenbewegung beseitigt ist, sofern die Anfangsgeschwindigkeit
{\bf t} des Teilchens im Startpunkt orthogonal zum charakteristischen Vektor
${\bf\pdach}$ ist. Mit anderen Worten, die autoparallelen und geod"atischen
Teilchenbahnen sind auf der charakteristischen Fl"ache identisch. Man beachte, da\3
auch der Paralleltransport bez"uglich der Standard-Konnexion ${\bf\Gamma}$ die
Bewegung nicht von der charakteristischen Fl"ache wegf"uhrt (s. Gl. \rf{pnablaV}).

Wir sehen nun, da\3 wir mit Hilfe der Integrabilit"atsbedingung die Zweideutigkeit
bei der Bestimmung der Teilchenbewegung in der Riemann-Cartanschen Struktur teilweise
beseitigen k"onnen. Im folgenden Abschnitt werden wir die M"oglichkeit untersuchen,
wie sich dies vollst"andig bewerkstelligen l"a\3t, indem man eine streng Riemannsche
Struktur einf"uhrt.

\subsection{"Ubergang zu einer Riemannschen Struktur}
\label{RieStr}
\indent

Die Frobenius Bedingung \rf{dpfp} erlaubt wesentlich weitgehendere
Schlu\3folgerungen als das aus den vorhergehenden Betrachtungen zu entnehmen ist.
Sie erm"oglicht es uns nun in der Tat die Torsion vollst"andig zu eliminieren, so
da\3 die Geometrie streng Riemannsch wird. "Uberraschenderweise zeigt es sich dann,
da\3 die entsprechende Konnexion nicht mit dem Riemannschen Teil ${\bf\Gammaschl}$
der Standard-Konnexion ${\bf\Gamma}$ \rf{GamGamschlK} identisch ist.

Um diese neue Konnexion ${\bf\Gammastern}$, die sog. {\em charakteristische
Konnexion}, zu finden, betrachten wir zuerst die Standard-Konnexion ${\bf\Gamma}$
\rf{GamGamschlK} f"ur den Fall, da\3 die Integrabilit"atsbedingung gilt. Mit Hilfe
der Gleichungen \rf{KZ}, \rf{Zpdp} und \rf{dpfp} erhalten wir
\begin{equation}
{\Gamma^\lambda}_{\mu\nu}={\Gammaschl^\lambda}_{\mu\nu} - g_{00}\,\pdach_\nu\left(
\pdach_\mu f^{\dot{\lambda}} - \pdach^\lambda f_\mu\right) \ .
\label{GamGamschlpf}
\end{equation}
Die Integrabilit"atsbedingung impliziert ferner die Existenz einer Funktion $\Phi$,
welche wir dazu ben"utzen wollen, folgenderma\3en einen Gradientenvektor {\bf C} zu
definieren:
\alpheqn
\begin{equation}
C_\mu = f_\mu + \Phi\pdach_\mu
\label{Cfphipdach}
\end{equation}
\begin{equation}
\partial_{[\mu}C_{\nu]} = 0\ .
\label{dC}
\end{equation}
\reseteqn
Mit Hilfe dieses Gradientenfeldes l"a\3t sich die Standard-Konnexion ${\bf\Gamma}$
\rf{GamGamschlpf} folgenderma\3en ausdr"ucken
\begin{equation}
{\Gamma^\lambda}_{\mu\nu}={\Gammaschl^\lambda}_{\;\;\mu\nu}-g_{00}\,\pdach_\nu\left(
\pdach_\mu C^{\dot{\lambda}} - \pdach^\lambda C_\mu\right) \ .
\label{GamGamschlpC}
\end{equation}
Um zur gesuchten charakteristischen Konnexion ${\bf\Gammastern}$ zu gelangen, spalten
wir ${\bf\Gamma}$ in folgender Weise auf:
\begin{equation}
{\Gamma^\lambda}_{\mu\nu}={\Gammastern^\lambda}_{\mu\nu}+{\Kstern^\lambda}_{\mu\nu}
\ ,
\label{GamGamsternKstern}
\end{equation}
mit
\alpheqn
\begin{equation}
{\Gammastern^\lambda}_{\mu\nu}={\Gammaschl^\lambda}_{\mu\nu} - g_{00}\,
C^{\dot{\lambda}} \pdach_\mu\pdach_\nu + g_{00}\,
\pdach^\lambda\left(\pdach_\nu C_\mu-\pdach_\mu C_\nu\right)
\label{GamsternGamschlCpp}
\end{equation}
und
\begin{equation}
{\Kstern^\lambda}_{\mu\nu} = -g_{00}\,\pdach^\lambda\pdach_\mu C_\nu\ .
\label{KsternppC}
\end{equation}
\reseteqn
Bevor wir jetzt anstelle von ${\bf\Gamma}$ mit ${\bf\Gammastern}$ arbeiten k"onnen,
m"ussen wir uns davon "uberzeugen, da\3 die in den Abschnitten \rf{Eig1} und
\rf{Eig2} gefundenen Eigenschaften von ${\bf\Gamma}$ nicht durch ${\bf\Gammastern}$
zerst"ort werden. Als erstes bemerken wir dazu, da\3 die charakteristische
Konnexion ${\bf\Gammastern}$ den gleichen wesentlichen Teil \bm{\gamma} \rf{gamPGam}
wie die Standard-Konnexion ${\bf\Gamma}$ besitzt. Um weiter die "Anderung der
Kr"ummung \rf{defR} zu untersuchen, geben wir deren, durch \rf{GamGamsternKstern}
induzierte, Aufspaltung an:
\begin{equation}
{R^\lambda}_{\nu\mu\sigma}={\Rstern^\lambda}_{\nu\mu\sigma} +
\nabla_\mu{\Kstern^\lambda}_{\nu\sigma}-\nabla_\sigma{\Kstern^\lambda}_{\nu\mu} +
{\Kstern^\lambda}_{\rho\sigma}{\Kstern^\rho}_{\nu\mu} -
{\Kstern^\lambda}_{\rho\mu}{\Kstern^\rho}_{\nu\sigma} +
2{Z^\rho}_{\sigma\mu}{\Kstern^\lambda}_{\nu\rho}\ .
\label{RRstern}
\end{equation}
Die besondere Struktur der Kontorsion ${\bf\Kstern}$ \rf{KsternppC} impliziert nun
das Verschwinden der quadratischen Terme auf der rechten Seite von Gleichung
\rf{RRstern}. Ben"utzt man weiter die kovariante Konstanz von ${\bf\pdach}$
\rf{nablap1} zusammen mit Gleichung \rf{DCZC}, so gilt:
\begin{equation}
\nabla_\mu{\Kstern^\lambda}_{\nu\sigma}-\nabla_\sigma{\Kstern^\lambda}_{\nu\mu} =
2{Z^\rho}_{\mu\sigma}{\Kstern^\lambda}_{\nu\rho} \ .
\end{equation}
Wir sehen also aus Gleichung \rf{RRstern}, da\3 die Kr"ummung unver"andert bleibt
\begin{equation}
{R^\lambda}_{\nu\mu\sigma}={\Rstern^\lambda}_{\nu\mu\sigma}\ .
\end{equation}
Als n"achstes wollen wir zeigen, da\3 eine Metrik ${\bf\Gstern}$ existiert, so da\3
die neue Konnexion ${\bf\Gammastern}$ gerade die Levi-Civita Konnexion von
${\bf\Gstern}$ ist; das hei\3t es mu\3 gelten
\begin{equation}
\nablastern_\lambda \Gstern_{\mu\nu} = 0 \ .
\label{nablasternGstern}
\end{equation}
Solch eine Riemannsche Metrik, die sog. {\em charakteristische} Metrik, l"a\3t sich
leicht finden, indem wir den Gradienten {\bf C} folgenderma\3en ausdr"ucken
\begin{equation}
C_\mu = \psi^{-1}\partial_\mu\psi = \partial_\mu\ln\psi \ .
\label{Cpsi}
\end{equation}
Ber"ucksichtigen wir weiter, da\3 der umskalierte charakteristische Vektor {\bf p}
\begin{equation}
p_\mu = \psi\pdach_\mu
\label{ppsipdach}
\end{equation}
bez"uglich der charakteristischen Konnexion ${\bf\Gammastern}$ konstant ist
\begin{equation}
\nablastern_\lambda p_\mu = 0 \ ,
\label{nablasternp}
\end{equation}
so impliziert dies, da\3 {\bf p} ein Gradientenvektor und $\psi$ demzufolge ein
integrierender Faktor f"ur ${\bf\pdach}$ ist. Gehen wir analog Gleichung
\rf{GdachGpp} vor, so ergibt sich die neue Metrik ${\bf\Gstern}$ zu
\begin{equation}
\Gstern_{\mu\nu} = \scrc^2 B_{\mu\nu} + g_{00}\;p_\mu p_\nu \ ,
\label{GsternBpp}
\end{equation}
was mit der Forderung nach kovarianter Konstanz \rf{nablasternGstern} konsistent
ist. Die charakteristische Metrik ${\bf\Gstern}$ definiert nun die von uns gesuchte
Riemannsche Struktur! Man beachte, da\3 der reskalierte charakteristische Vektor
{\bf p} bez"uglich der Metrik ${\bf\Gstern}$ die Einheitsl"ange besitzt:
\alpheqn
\begin{equation}
p^{\stern{\mu}} p_\mu = \Gstern_{\mu\nu} p^{\stern{\mu}} p^{\stern{\nu}} = g_{00}
\label{psternp}
\end{equation}
\begin{equation}
p^{\stern{\mu}} = \psi^{-1}\pdach^\mu\ .
\label{psternpsip}
\end{equation}
\reseteqn
Es taucht die Frage auf, ob die in Abschnitt \rf{VerallgKonf} definierte
verallgemeinerte Konformalit"at der Riemann-Cartan Struktur in die gew"ohnliche
konformal flache Struktur eines rein Riemannschen Raumes "ubergeht. Eine Berechnung
des Weylschen Tensors \rf{W} unter Verwendung der charakteristischen Metrik
${\bf\Gstern}$ zeigt, da\3 dieser in der Tat, sowohl im intrinsischen
Fall (n = 3), als auch im extrinsischen Fall (n = 4), verschwindet. Man beachte, da\3
der Kr"ummungstensor {\bf R}, der Ricci Tensor \bm{\cal R}, der Skalar ${\cal S}$,
und die Fasermetrik ${\bf\Gdach}$ die gleichen Objekte sowohl im Riemannschen,
als auch im nicht-Riemannschen Fall sind, wogegen die Konnexionen ${\bf\Gamma}$,
${\bf\Gammastern}$ und die Metriktensoren {\bf G}, ${\bf\Gstern}$ sich in den beiden
F"allen, entlang der charakteristischen Richtung unterscheiden. Auf jeden Fall ist
die Riemannsche Struktur konformal flach, obwohl die Riemannsche Metrik
${\bf\Gstern}$ im allgemeinen der Minkowskischen- bzw. Euklidischen Metrik {\bf g}
nicht proportional ist. Aus diesem Grunde mu\3 es eine Koordinatentransformation
geben, welche ${\bf\Gstern}$ proportional zu {\bf g} macht. In speziellen F"allen
(z.B. der Euklidischen Dimeron L"osung) ist diese Proportionalit"at schon f"ur das
kartesische Koordinatensystem erreicht! Die charakteristischen Fl"achen sind
3-dimensionale Riemannsche Untermannigfaltigkeiten, welche eine konstante Kr"ummung
aufweisen.

Der "Ubergang zu einer streng Riemannschen Struktur beseitigt also die Zweideutigkeit
in der Riemann-Cartan Geometrie was die Bewegung von Testteilchen betrifft. Um diesen
mathematischen Sachverhalt auf eine physikalische Situation anzuwenden, ist es
sinnvoll die charakteristische Konnexion ${\bf\Gammastern}$ gegen"uber der
Standard-Konnexion ${\bf\Gamma}$ zu verwenden, was aufgrund der
Integrabilit"atsbedingung hier ja m"oglich ist. Auf den Integralfl"achen von
$\Deltadach$ finden wir folgende Bewegungsgleichung f"ur die Geod"aten:
\begin{equation}
\frac{\Dstern^2 x^\mu}{ds^2}=\frac{D^2 x^\mu}{ds^2} - {\Kstern^\mu}_{\nu\lambda}
\;\frac{dx^\nu}{ds}\frac{dx^\lambda}{ds} = 0\ .
\label{DsternDK}
\end{equation}
Wir sehen also, da\3 die Geod"aten der Metrik ${\bf\Gstern}$ auf den
charakteristischen Fl"achen sowohl mit den Geod"aten der Riemann-Cartan Metrik
{\bf G} als auch mit den Autoparallelen der Standard-Konnexion ${\bf\Gamma}$
"ubereinstimmen. Letztere k"onnen wir deshalb als die Fortsetzung der Geod"aten in
den einbettenden Raum betrachten. Diese Schlu\3folgerung gilt, wie schon erw"ahnt,
nur f"ur den Fall, da\3 die charakteristische Distribution $\Deltadach$ integrabel
ist. Andernfalls existieren keine charakteristische Fl"achen und demzufolge auch kein
${\bf\Gammastern}$ !

\section{Die Yang-Mills Gleichungen}
\indent

Bis jetzt haben wir uns haupts"achlich mit allgemeinen, trivialisierbaren
Feld{\em konfigu\-rationen} besch"aftigt, welche die Yang-Mills Gleichungen im
allgemeinen nicht erf"ullen. Wir wollen uns nun den trivialisierbaren {\em L"osungen}
der homogenen Yang-Mills Gleichung zuwenden, welche eine Untermenge der allgemeinen
Konfigurationen bilden. Es wird sich zeigen, da\3 die Feldgleichungen eine
hinreichende, aber nicht notwendige Bedingung f"ur die Existenz von charakteristischen
Fl"achen darstellen. Ferner implizieren die Ergebnisse des vorherigen Abschnittes die
Existenz einer streng Riemannschen Struktur zus"atzlich zu der allgemeineren und immer
vorhandenen Riemann-Cartan Struktur, vorausgesetzt die charakteristische Distribution
ist integrabel. Die Kombination dieser zwei Argumente besagt nun nichts anderes, als
da\3 die G"ultigkeit der Yang-Mills Gleichungen die Existenz einer Riemannschen
Struktur sichert. Die allgemeinen Eigenschaften solch einer
"`Riemann-Yang-Mills"'-Struktur sollen in diesem Abschnitt untersucht werden.

\subsection{Die geometrische Bedeutung der Feldgleichungen}
\indent

Im folgenden wollen wir annehmen, da\3 die trivialisierbaren Eichfelder die homogenen
Yang-Mills Gleichungen erf"ullen
\begin{equation}
D^\mu F_{i\mu\nu} = 0 \ .
\label{DF}
\end{equation}
Um die geometrischen Eigenschaften dieser Feldgleichungen zu untersuchen, beachten
wir da\3 die intrinsische Kr"ummung {\bf F} f"ur jedes trivialisierbare Eichfeld
{\bf A} nach Gleichung \rf{triva} durch die extrinsische Kr"ummung {\bf B}
ausgedr"uckt werden kann. Aus diesem Grund l"a\3 sich die Yang-Mills Gleichung
\rf{DF} mit Hilfe der {\bf B}-Felder formulieren
\begin{equation}
D^\mu F_{i\mu\nu} = g_{00}\,{\epsilon_i}^{jk}B_{k\mu}\left(D^\mu B_{j\nu} -
{\Pdach^\mu}_\nu\left(D^\lambda B_{j\lambda}\right)\right)\ = 0\ .
\label{DFBP}
\end{equation}
Definieren wir nun die Tensorgr"o\3e {\bf T} durch
\begin{equation}
D_\mu B_{j\nu} - \Pdach_{\mu\nu}\left(D^\lambda B_{j\lambda}\right) =
{T^\lambda}_{\nu\mu} B_{j\lambda}\ ,
\label{DBPT}
\end{equation}
so gilt folgende Beziehung zwischen {\bf T} und dem wesentlichen Teil \bm{\gamma}
\rf{gamPGam}
\begin{equation}
{T^\lambda}_{\nu\mu} = {\gamma^\lambda}_{\nu\mu} - \Pdach_{\mu\nu}
{\gamma^{\lambda\sigma}}_\sigma\ .
\label{TgamPgam}
\end{equation}
Aufgrund von Gleichung \rf{gamPGam} annihiliert ${\bf\pdach}$ das Objekt {\bf T}
\begin{equation}
\pdach_\lambda{T^\lambda}_{\nu\mu} = 0\ ;
\label{pT}
\end{equation}
ferner ist {\bf T} symmetrisch in den letzten zwei Indizes
\begin{equation}
{T^\lambda}_{\mu\nu} = {T^\lambda}_{\nu\mu}\ .
\label{TT}
\end{equation}
Die homogene Yang-Mills Gleichung l"a\3t sich nun mit Hilfe von {\bf T}
folgenderma\3en ausdr"ucken
\begin{equation}
D^\mu F_{i\mu\nu}\equiv g_{00}\,{F_i}^{\lambda\mu}\,T_{\lambda\mu\nu}
\stackrel{!}{=}0\ ,
\label{DFFT}
\end{equation}
In dieser Form l"a\3t sich f"ur das Yang-Mills Problem sofort eine allgemeine L"osung
angeben: Vergegenw"artigen wir uns, da\3 die Kr"ummungskoeffizienten ${\bf F}_i$ f"ur
jeden nicht entarteten Punkt {\bf x} des Basisraumes drei linear unabh"angige
2-Formen aus $\Deltadach_x$ darstellen, so lautet die allgemeinste Form f"ur {\bf T}
\begin{equation}
T_{\lambda\mu\nu}=S_{\lambda\mu\nu}+S_{\lambda\mu}\pdach_\nu+S_{\lambda\nu}\pdach_\mu
+S_\lambda\pdach_\mu\pdach_\nu\ .
\label{TSp}
\end{equation}
Die hier eingef"uhrten Objekte {\bf S} liegen vollst"andig in dem zu $\taudach$
korrespondierenden assoziertem Tensorb"undel, das hei\3t es gilt z.B.
\begin{equation}
\pdach^\lambda S_{\lambda\mu\nu} = \pdach^\lambda S_{\lambda\mu} =
\pdach^\lambda S_\lambda = 0\ .
\label{Sp}
\end{equation}
Damit die Yang-Mills Gleichungen erf"ullt werden, m"ussen die Gr"o\3en {\bf S} total
symmetrisch sein
\alpheqn
\begin{equation}
S_{\lambda\mu\nu} = S_{\mu\lambda\nu} = S_{\mu\nu\lambda}\ ,
\end{equation}
\begin{equation}
S_{\mu\lambda} = S_{\lambda\mu}\ .
\end{equation}
\reseteqn
Da jetzt die allgemeine Gestalt von {\bf T} bekannt ist, k"onnen wir aus Gleichung
\rf{TgamPgam} den wesentlichen Teil \bm{\gamma} bestimmen
\begin{equation}
{\gamma^\lambda}_{\nu\mu} = {S^\lambda}_{\nu\mu} + {S^\lambda}_\mu\pdach_\nu +
{S^\lambda}_\nu\,\pdach_\mu + S^\lambda\pdach_\nu\pdach_\mu - \frac{1}{2}
\Pdach_{\nu\mu}\left( {S^{\lambda\sigma}}_\sigma + g_{00}\,S^\lambda\right)\ .
\label{gammaSpP}
\end{equation}
Damit ergibt sich die Standard-Konnexion ${\bf\Gamma}$ zu
\begin{equation}
{\Gamma^\lambda}_{\nu\mu} = {\gamma^\lambda}_{\nu\mu} - \pdach^\lambda
\left(S_{\nu\mu} + S_\nu\pdach_\mu\right)\ .
\label{GamgampS}
\end{equation}
Die geometrische Bedeutung der Yang-Mills Gleichung wird nun deutlich: Betrachten
wir die Ableitung des charakteristischen Vektors, so erhalten wir mit \rf{GamgampS}
\begin{equation}
\partial_\mu\pdach_\nu = {\Gamma^\lambda}_{\nu\mu}\pdach_\lambda =
- g_{00}\left(S_{\nu\mu} + S_\nu\pdach_\mu\right)\ ,
\label{dpSSp}
\end{equation}
was nichts anderes bedeutet, als da\3 die Integrabilit"at der charakteristischen
Distribution $\Deltadach$ aufgrund der Yang-Mills Gleichung nun gew"ahrleistet ist.
Denn jetzt gilt (vgl. mit \rf{dpfp})
\begin{equation}
\partial_{[\mu}\pdach_{\nu\,]} = - g_{00}\,S_{[\nu}\pdach_{\mu\,]}\ .
\label{dpSp}
\end{equation}
Die Gestalt der charakteristischen Fl"achen l"a\3t sich nach Gleichung \rf{dpSSp}
durch $S_{\mu\nu}$ und dem {\em Torsionsvektor} $S_\mu$ bestimmen
\alpheqn
\begin{equation}
{\Pdach^\lambda}_\mu\partial_\lambda\pdach_\nu = - g_{00}\,S_{\mu\nu}
\label{PdpS}
\end{equation}
\begin{equation}
S_\nu \equiv - \pdach^\mu\partial_\mu\pdach_\nu
\label{Spdp}
\end{equation}
\reseteqn
Man beachte, da\3 der Torsionsvektor nur dann ungleich Null ist, falls die
charakteristischen Linien, bez"uglich der kanonischen Konnexion \bm{\omegakr} "uber
${\bf E_4}$ bzw ${\bf M_4}$, keine Geraden sind. Aus diesem Grund besitzt die
Standard-Konnexion ${\bf\Gamma}$ eine nicht-verschwindende Torsion {\bf Z}
\begin{equation}
{Z^\lambda}_{\mu\nu} = \pdach^\lambda\pdach_{[\mu}S_{\nu]}\ ,
\label{ZppS}
\end{equation}
welche die Abweichung der charakteristischen Linien von euklidischen bzw.
minkowskischen Geod"aten angibt. (Die charakteristischen Linien sind nat"urlich
Geod"aten bez"uglich der charakteristischen Metrik ${\bf\Gstern}$)

Aus Gleichung \rf{dpSSp} k"onnen wir nun noch eine weitere Konsequenz ziehen.
Betrachten wir zwei trivialisierbare Eichfelder {\bf A}, welche identische
charakteristische Fl"achen aufweisen sollen und die sich nicht durch eine
Eichtransformation ineinander "uberf"uhren lassen, so basieren ihre
Standard-Konnexionen ${\bf\Gamma}$ auf den gleichen Objekten $S_\mu$ und
$S_{\mu\nu}$; sie unterscheiden sich jedoch bez"uglich der $S_{\lambda\mu\nu}$.

\subsection{Die Riemann-Yang-Mills Struktur}
\indent

Im vorigen Abschnitt haben wir gesehen, da\3 die Yang-Mills Gleichung eine
hinreichende Bedingung f"ur die Integrabilit"at der charakteristischen Distribution
$\Deltadach$ darstellt. Vom physikalischen Standpunkt aus gesehen ist es eine
befriedigende Situation die ziemlich formale Bedingung der Integrabilit"at durch
eine dynamische Feldgleichung ausdr"ucken zu k"onnen. Wir erhalten als Ergebnis
eine Art Riemann-Yang-Mills Raum-Zeit Struktur, welche auf einer flachen
Hintergrundgeometrie aufgebaut ist.

Der "Ubergang zu einer Riemannschen Konnexion ${\bf\Gammastern}$ wird durch die
Argumente von Abschnitt \rf{RieStr} begr"undet. Ben"utzen wir die Standard-Konnexion
${\bf\Gamma}$ \rf{GamgampS}, die wir aus der homogenen Yang-Mills Gleichung
hergeleitet hatten, so erhalten wir mit Hilfe von \rf{GamGamsternKstern},
\rf{GamsternGamschlCpp} und \rf{KsternppC} sofort einen Ausdruck f"ur die
charakteristische Konnexion ${\bf\Gammastern}$
\begin{equation}
{\Gammastern^\lambda}_{\mu\nu}={\Gamma^\lambda}_{\mu\nu} + g_{00}\,
\pdach^{\stern{\lambda}}\pdach_\mu C_\nu\ .
\label{GamsternGamppC}
\end{equation}
Um die Rechnung zu vereinfachen werden wir in diesem Ausdruck nicht das Ergebnis
\rf{GamgampS} f"ur die Standard-Konnexion ${\bf\Gamma}$ einsetzen, sondern wir
werden ${\bf\Gammastern}$ durch eine konformal flache Konnexion ${\bf\Gammaquer}$
ausdr"ucken, welche folgenderma\3en definiert ist:
\begin{equation}
{\Gammaquer^\lambda}_{\mu\nu} = {g^\lambda}_\mu\,C_\nu + {g^\lambda}_\nu\,C_\mu -
C^\lambda g_{\mu\nu}\ .
\label{defGammaquer}
\end{equation}
${\bf\Gammaquer}$ ist die Levi-Civita Konnexion einer konformal flachen Metrik
${\bf\Gquer}$
\begin{equation}
\Gquer_{\mu\nu} = \psi^2 g_{\mu\nu}\ .
\label{defGquer}
\end{equation}
Die konformal flache Konnexion ${\bf\Gammaquer}$ l"a\3t sich nun leicht in Gleichung
\rf{GamsternGamppC} einf"ugen, wenn wir die Beziehung zwischen dem Gradienten
{\bf C} \rf{Cfphipdach} und dem Torsionsvektor {\bf S} \rf{dpSp} ben"utzen
\alpheqn
\begin{equation}
S_\mu = -g_{00}\,{\Pdach^\nu}_\mu\,C_\nu
\label{SPdachC}
\end{equation}
\begin{equation}
\Phi = g_{00}\left(\pdach^\nu C_\nu\right)\ .
\label{PhipC}
\end{equation}
\reseteqn
Definieren wir ferner die Abweichungen $\left(s_{\lambda\mu\nu}, s_{\lambda\mu}
\right)$ der symmetrischen Tensoren $\left(S_{\lambda\mu\nu}, S_{\lambda\mu}\right)$
von ihren konformal flachen Werten $\left(\Squer_{\lambda\mu\nu}, \Squer_{\lambda\mu}
\right)$, welche durch
\alpheqn
\begin{equation}
\Squer_{\lambda\mu\nu} = \Pdach_{\lambda\mu}S_\nu + \Pdach_{\mu\nu}S_\lambda +
\Pdach_{\nu\lambda}S_\mu
\label{SquerPS}
\end{equation}
und
\begin{equation}
\Squer_{\mu\nu} = \Phi\Pdach_{\mu\nu}
\label{SquerPhiPdach}
\end{equation}
\reseteqn
gegeben sind, so gilt:
\alpheqn
\begin{equation}
S_{\lambda\mu\nu} = \Squer_{\lambda\mu\nu} + s_{\lambda\mu\nu}
\label{SSquers1}
\end{equation}
\begin{equation}
S_{\lambda\nu} = \Squer_{\lambda\nu} + s_{\lambda\nu}
\label{SSquers2}
\end{equation}
\begin{equation}
S_\lambda = \Squer_\lambda\ .
\label{SSquer}
\end{equation}
\reseteqn
Damit finden wir f"ur die charakteristische Konnexion ${\bf\Gammastern}$ folgenden
Ausdruck
\begin{equation}
{\Gammastern^\lambda}_{\mu\nu} = {\Gammaquer^\lambda}_{\mu\nu} +
{\Qquer^\lambda}_{\mu\nu}\ .
\label{GastGaqQq}
\end{equation}
Hierbei setzt sich die Abweichung ${\bf\Qquer}$ der charakteristischen Konnexion
${\bf\Gammastern}$ von ${\bf\Gammaquer}$ aus den Abweichungen {\bf s} von
${\bf\Squer}$ wie folgt zusammen:
\begin{equation}
{\Qquer^\lambda}_{\mu\nu} = {s^\lambda}_{\mu\nu} + {s^\lambda}_\mu\,\pdach_\nu+
{s^\lambda}_\nu\,\pdach_\mu - \frac{1}{2}\Pdach_{\mu\nu}\,{s^{\lambda\sigma}}_\sigma
\label{Qquer}
\end{equation}

Die Zerlegung von ${\bf\Gammastern}$ in den konformal flachen Teil ${\bf\Gammaquer}$
und den Rest ${\bf\Qquer}$ ber"ucksichtigt die Tatsache, da\3 es im Riemannschen
Fall keine Torsion {\bf Z} gibt, obwohl ein nicht verschwindender Torsionsvektor
$S_\mu$ in der Theorie existiert. Der Grund hierf"ur besteht darin, da\3 die
charakteristischen Linien im allgemeinen Fall gekr"ummt sind (vgl. \rf{Spdp}).
Dieser anscheinende Widerspruch wird nun gerade durch die Aufspaltung \rf{GastGaqQq}
beseitigt, indem der Torsionsvektor in ${\bf\Gammaquer}$ absorbiert wird. Der
verbleibende Teil ${\bf\Qquer}$ wird dann vollst"andig torsionsfrei, sieht aber sonst
wie die Standard-Konnexion ${\bf\Gamma}$ aus.

${\bf\Qquer}$ definiert eine zus"atzliche Struktur "uber der konformal flachen
Hintergrundsmetrik. Diese Metrik ist nun durch Gleichung \rf{defGquer} gegeben und
gehorcht der Bedingung
\begin{equation}
\nablastern_\lambda\Gquer_{\mu\nu} = 2Q_{\mu\nu\lambda}\ .
\label{nablasternGquerQquer}
\end{equation}
L"ost man diese Gleichung nach der charakteristischen Konnexion auf, so ergibt sich
die Auf\-spaltung \rf{GastGaqQq}, wobei ${\bf\Qquer}$ durch die Gr"o\3e {\bf Q}
\footnote{Indizes mit einem Querbalken werden mit Hilfe der Hintergrundmetrik
\rf{defGquer} bewegt.} ausgedr"uckt wird.
\begin{equation}
{\Qquer^\lambda}_{\mu\nu} = {Q_{\mu\nu}}^{\bar{\lambda}} -
\left({Q^{\bar{\lambda}}}_{\nu\mu} + {Q^{\bar{\lambda}}}_{\mu\nu}\right)
\label{QquerQ}
\end{equation}
Hierbei ist {\bf Q} der symmetrische Teil ${\bf\Sigma '}$ von ${\bf\Qquer}$:
\alpheqn
\begin{eqnarray}
\label{QSigma'}
Q_{\mu\nu\lambda} &=& - {\Sigma '}_{{\bar{\mu}}\nu\lambda} \equiv -
\frac{1}{2}\left(\Qquer_{\bar{\mu}\nu\lambda}+\Qquer_{\bar{\nu}\mu\lambda}\right)\\
\label{Sigma'}
{{\Sigma '}^\lambda}_{\mu\nu} &=& {s^\lambda}_{\mu\nu} + {s^\lambda}_\mu\,\pdach_\nu
-\frac{1}{4}\left(\Pdach_{\mu\nu}\,{s^{\lambda\sigma}}_\sigma +
{\Pdach^\lambda}_\nu\, {s_{\mu\sigma}}^\sigma\right)
\end{eqnarray}
\reseteqn

Die einfachsten L"osungen sind nun vom {\em konformal flachen} Typ. Aufgrund ihrer
Definition besteht diese L"osungsklasse aus reinen Hintergrund Konfigurationen,
das hei\3t, es gilt ${\bf\Qquer}\equiv 0$ und deshalb ${\bf\Gammastern} \equiv
{\bf\Gammaquer}$. Die allgemeinste, nicht-triviale L"osung dieses Typs ist die schon
in \cite{BrSo86,BrSo87} untersuchte euklidische Dimeron Konfiguration. Deshalb
soll hier eine kurze, auf zwei neuen Aspekten, basierende Betrachtung gen"ugen.

Man beachte, da\3 der Tensor $S_{\mu\nu}$ aus Gleichung \rf{PdpS} beim
euklidischen Di-Meron proportional dem Projektor ${\bf\Pdach}$ ist. Das bedeutet,
die charakteristischen Fl"achen sind 3-Sph"aren mit dem Radius $\Phi^{-1}$. Da die
Funktion $\Phi(x)$ "uber jeder dieser 3-Sph"aren konstant ist, mu\3 ihr Gradient
den charakteristischen Vektor ${\bf\pdach}$ bestimmen. In der Tat gilt:
\begin{equation}
\partial_\mu\Phi=-\frac{1}{2}\left(S^\lambda S_\lambda-\Phi^2-\frac{\psi^2}{c^2}
\right)\pdach_\mu\ ,
\label{dPhi}
\end{equation}
wobei der Torsionsvektor $S_\lambda$ und der inverse Radius $\Phi$ mit dem
konformalen Gradienten ${\bf C}$ \rf{Cpsi} "uber die Gleichungen \rf{PhipC} und
\rf{SPdachC} zusammenh"angen.

Ferner wird die Fasermetrik ${\bf\Gdach}$ f"ur die reinen Hintergrundl"osungen
isotrop
\begin{equation}
\Gdach_{\mu\nu} = \Gstern_{\mu\nu} - g_{00} p_\mu p_\nu = \psi^2\Pdach_{\mu\nu}\ .
\label{GdachpsiP}
\end{equation}
Umgekehrt ersieht man daraus, da\3 jede isotrope L"osung notwendigerweise konformal
flach ist, und deshalb mit der Dimeron L"osung "ubereinstimmen mu\3. Wir k"onnen
deshalb die Schlu\3folgerung ziehen, da\3 die nicht-triviale Hintergrundl"osung immer
konformal flach und isotrop sein mu\3. In diesem Sinne besitzt die Hintergrundl"osung
die h"ochste Symmetrie, die in der gesamten L"osungsmenge gefunden werden kann.
\begin{figure}[b]
\vspace{-20cm}
\epsfig{file=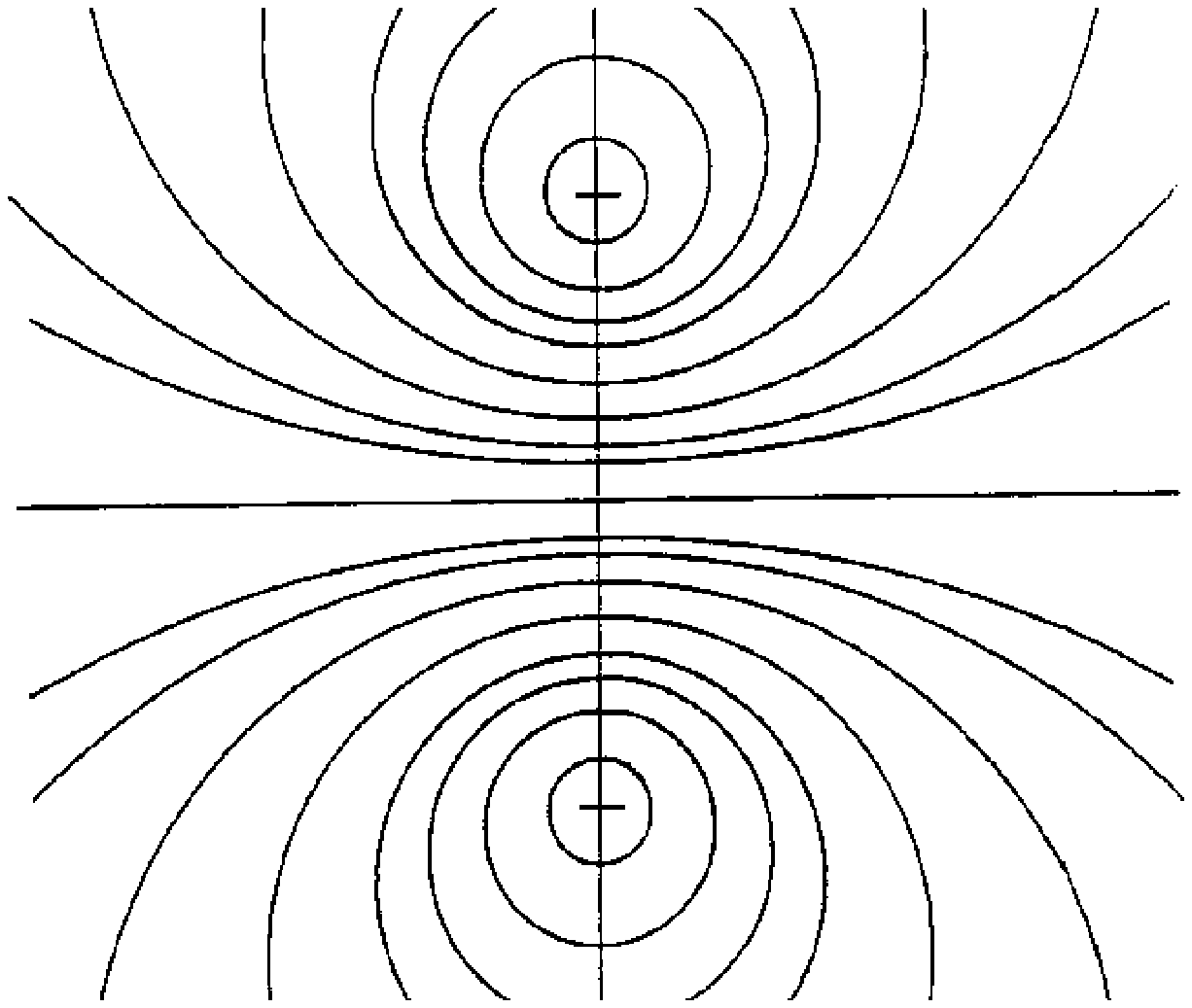}
\vspace{2cm}
\caption[Charakteristische Fl\"achen der Dimeron-L\"osung]{Die charakteristischen Fl"achen der
Dimeron-L"osung sind 3-Sph"aren $\Phi({\bf x}) = \mbox{konst.}$ (vgl. \rf{dPhi}).
Wenn die Meron-Zentren sich bei $x^0=a,b$ befinden, lautet die Gleichung f"ur die
charakteristischen Fl"achen mit dem Radius $\Phi^{-1}$}
\[
x^0 = \frac{a+b}{2}\plmi\frac{a-b}{2}\left(1+\left[\frac{a-b}{2}\,\Phi\right]^{-2}
\right)^{\frac{1}{2}}\nonumber
\]
\end{figure}

\section{Geod"atische und autoparallele Kurven}
\label{AutoUndGeod}
\indent

In einer nicht-Riemannschen Struktur k"onnen geod"atische und autoparallele Kurven
betr"achtliche Unterschiede aufweisen. Dies gilt ebenfalls f"ur die vorliegende
fast Riemannsche Struktur, welche auf den charakteristischen Fl"achen streng
Riemannsch wird. Wir wollen nun beide Typen von Kurven anhand der euklidischen
Dimeron-L"osung untersuchen.

Betrachten wir zuerst die Geod"atengleichung \rf{DsternDK} f"ur den Riemannschen
Fall
\begin{equation}
\frac{d^2 x^\mu}{ds^2} + {\Gammaquer^\mu}_{\nu\lambda}\frac{dx^\nu}{ds}
\frac{dx^\lambda}{ds} = 0\ ,
\label{dGammaquerdd}
\end{equation}
wobei die konformal flache Konnexion ${\bf\Gammaquer}$ durch Gleichung
\rf{defGammaquer} gegeben ist, mit folgendem konformalen Skalenfaktor\cite{BrSo86}
\begin{equation}
\psi = \frac{{\left|{\bf a} - {\bf b}\right|}^2}
{{\left|{\bf x} - {\bf a}\right|} {\left|{\bf x} - {\bf b}\right|}}\ .
\label{psiDimeron}
\end{equation}
Das in \rf{dGammaquerdd} ben"utzte Linienelement ds wird durch die konformal flache
Metrik ${\bf\Gquer}$ aus Gleichung \rf{defGquer} bestimmt
\begin{equation}
ds^2 = \Gquer_{\mu\nu}dx^\mu dx^\nu\ .
\end{equation}
Da wir in einem euklidischen Raum rechnen, erweist es sich als g"unstiger wenn wir
das euklidische Linienelement ben"utzen
\begin{equation}
d\sigma^2 = g_{\mu\nu}dx^\mu dx^\nu\ ,
\label{euklLinel}
\end{equation}
mit welchem die Bewegungsgleichung \rf{dGammaquerdd} folgende Form annimmt
\begin{equation}
\frac{d^2 x^\mu}{d\sigma^2} + {\hdach^\mu}_\rho\;{\Gammaquer^\mu}_{\nu\lambda}
\frac{dx^\nu}{d\sigma}\frac{dx^\lambda}{d\sigma} = 0\ .
\label{dhdachGammaquerdd}
\end{equation}
Der hierbei auftauchende Projektor ${\bf\hdach}$ annihiliert die
Vierergeschwindigkeit $dx^\lambda / d\sigma$ und sieht daher folgenderma\3en aus:
\begin{equation}
{\hdach^\mu}_\rho = {g^\mu}_\rho + \frac{dx^\mu}{d\sigma}\frac{dx_\rho}{d\sigma}\ .
\label{defhdach}
\end{equation}
Setzen wir die konformal flache Konnexion ${\bf\Gammaquer}$ in \rf{dhdachGammaquerdd}
ein, erhalten wir eine Bewegungsgleichung, wie man sie aus der Speziellen
Relativit"atstheorie kennt:
\begin{equation}
\frac{d^2 x^\mu}{d\sigma^2} = F^\mu\ .
\label{d2F}
\end{equation}
Hierbei steht die Viererkraft {\bf F} senkrecht auf der Vierergeschwindigkeit, d.h.
\begin{equation}
F^\mu = - {\hdach^\mu}_\rho C^\rho\ ,
\label{FhdachC}
\end{equation}
wie es f"ur die "`relativistische"' Konsistenz notwendig ist.

F"ur die numerische Auswertung erweist es sich jedoch als vorteilhaft, wenn wir mit
einer Gleichung vom Newtonschen Typ rechnen. Diese erhalten wir aus der
relativistischen Form \rf{FhdachC} durch eine Parametertransformation, welche die
"`Eigenzeit"' $\sigma$ folgenderma\3en in die "`Newtonsche Zeit"' $\tau$ "uberf"uhrt:
\begin{equation}
\frac{d\sigma}{d\tau} = \psi
\label{defNew}
\end{equation}
Somit lautet die Newtonsche Form der Bewegungsgleichung
\begin{equation}
\frac{d^2 x^\mu}{d\tau^2} = - \partial_\mu{\cal V}\ ,
\label{dddcalV}
\end{equation}
mit dem Potential ${\cal V}$
\begin{equation}
{\cal V} = -\frac{1}{2}\psi^2\ .
\label{calVpsi}
\end{equation}
Diese Bewegunggsgleichung beschreibt die Bewegung eines Testteilchens in Gegenwart
zweier anziehender Zentren die an den Meron Orten ${\bf a}$ und ${\bf b}$ lokalisiert
sind (s. Bild 2). Die "`nicht-relativistische"' Form der Bewegungsgleichung
\rf{dddcalV} und ihre "`relativistische"' Form erm"oglichen es uns nun, die
geod"atischen Linien mit der Terminologie der Newtonschen Mechanik zu untersuchen.
Kontrahieren wir \rf{dddcalV} mit der Geschwindigkeit $dx_\mu / d\tau$, so ergibt
sich ein Energieerhaltungsgesetz in der Form
\begin{equation}
\frac{d{\cal E}}{d\tau} \equiv \frac{d}{d\tau}\left({\cal T} + {\cal V}\right)=0\ .
\label{Eerh}
\end{equation}
Ferner zeigt sich, da\3 die gesamte Energie ${\cal E} = {\cal T} + {\cal V}$ Null
ist, da die Geschwindigkeit des Testteilchens mit der konformalen Skalenfunktion
$\psi$ "ubereinstimmt (vgl \rf{defNew}), so da\3 wir f"ur die kinetische Energie
${\cal T}$ erhalten
\begin{equation}
{\cal T} = -\frac{1}{2}\frac{dx^\mu}{d\tau}\frac{dx_\mu}{d\tau} =
+\frac{1}{2}\psi^2 = -{\cal V}\ .
\label{defcalT}
\end{equation}
Das bedeutet, da\3 sich das Testteilchen, welches sich auf einer Geod"aten bewegt,
in einem gebundenen Zustand befindet und nicht ins Unendliche gelangen kann.

Diese Aussage k"onnen wir noch weiter pr"azisieren, indem wir die urspr"ungliche,
"`relativistische"' Form der Bewegungsgleichung \rf{tnablaschlt} betrachten. Da die
zwei Einheitsvektoren {\bf t} und {\bf p} entlang einer Geod"aten parallel
verschoben werden, bleibt der euklidische Winkel $\alpha$, welcher von beiden
eingeschlossen wird, konstant. Das hei\3t, da\3 der Winkel $\alpha$, unter dem eine
Geod"ate die charakteristische Fl"achen schneidet, entlang dieser Kurve eine
konstante Gr"o\3e ist. Da jede charakteristische Fl"ache gerade eines der beiden
Meronzentren umschlie\3t (s. Bild 1), mu\3 jede Geod"ate entweder in einem
Meronzentrum endigen, oder sie ist eine, auf der charakteristischen Fl"ache liegende,
geschlossene Kurve. Die Verallgemeinerung dieses Ergebnisses auf den Multi-Meron Fall
ist offenkundig; vorausgesetzt er existiert. Bild 3 zeigt den Verlauf der Geod"aten,
die sich aus der "`Newtonschen"' Bewegungsgleichung \rf{dddcalV} in zwei Dimensionen
ergeben (Beschr"ankung der geod"atischen Bewegung auf eine 2-Ebene durch die
SO(3)-Symmetrieachse der Dimeron-Konfiguration)~.\footnote{Inzwischen haben sich
durch eine neuere Untersuchung\cite{HeSo89} weitere Hinweise ergeben, da\3 die
geod"atischen Linien in einer Riemannschen Struktur vom trivialisierbaren Typ stets
auf die topologischen Punktdefekte zulaufen und dort endigen, falls sie keine auf
einer charakteristischen Fl"ache liegende geschlossene Kurve bilden.}

Als n"achstes wollen wir, ausgehend von Gleichung \rf{Bewgl}, die autoparallelen
Trajektorien des Testteilchens unter dem Einflu\3 der Riemann-Cartan Geometrie
untersuchen. Die autoparallele Bewegungsgleichung unterscheidet sich von der
Geod"atischen durch den Einflu\3 der Kontorsion ${\bf\Kstern}$ \rf{KsternppC}. Dieser
Term wird nun die Newtonsche Bewegungsgleichung \rf{dddcalV} ab"andern. Nach einigen
einfachen Rechnungen ergibt sich die autoparallele Version der Geod"atengleichung zu
\begin{equation}
\frac{d^2 x^\mu}{d\tau^2} = -\partial_\mu{\cal V} - \left({\hdach^\mu}_\rho
\pdach^\rho\right)\left(p_\nu\frac{dx^\nu}{d\tau}\right)
\left(C_\lambda\frac{dx^\lambda}{d\tau}\right)\ .
\label{dddcalVhpC}
\end{equation}
Der zus"atzliche Term auf der rechten Seite verschwindet auf einer charakteristischen
Linie und bei einer auf die charakteristischen Fl"achen  eingeschr"ankten Bewegung.
Aufgrund der Projektoreigenschaft von ${\bf\hdach}$ \rf{defhdach} verletzt die
Autoparallelen-Gleichung \rf{dddcalVhpC} nicht das Energieerhaltungsgesetz \rf{Eerh}.

Ein qualitatives Bild der Autoparallelen als L"osung von \rf{dddcalVhpC} erhalten
wir wieder durch Untersuchung des durch die Tangente {\bf t} und dem umskalierten
charakteristischen Vektor {\bf p} eingeschlossenen Winkels $\alpha$. Da $\alpha$
entlang einer Autoparallelen sich nicht "andert, erhalten wir eine Konstante der
Bewegung durch
\begin{equation}
\cos\alpha = G_{\mu\nu}\pdach^\mu t^\nu\ .
\label{cosalpha}
\end{equation}
Wir wollen hier zweckm"a\3igerweise den Riemannschen Winkel $\alpha$ durch
dessen euklidisches Analogon ersetzen
\begin{eqnarray}
\cos\alpha_e = g_{\mu\nu}\pdach^\mu t^\nu\ , \\
\left(\tdach^\nu = t^\nu\frac{ds}{d\sigma}\right)\ .\nonumber
\end{eqnarray}
Damit erhalten wir
\begin{equation}
\cos^2\alpha_e={\psi^2\cos^2\alpha\left(\sin^2\alpha+\psi^2\cos^2\alpha\right)}^{-1}
\label{cos2alpha}
\end{equation}
Beide Winkel sind identisch f"ur $\alpha = 0, \pi/2$. Starten wir nun aber das
Testteilchen unter einem Winkel $0<\alpha<\pi/2$, das hei\3t unter
abnehmendem Potential ${\cal V}$, so bewegt sich das Testteilchen auf ein
Meronzentrum zu. In diesem Fall strebt $\alpha_e$ gegen Null was nichts anderes
bedeutet, als da\3 sich die autoparallele Kurve schneller und direkter dem
Meronzentrum n"ahert als die geod"atische Kurve. Schie\3t man das Teilchen dagegen
von einem Meronzentrum weg, so nimmt ${\cal V}$ zu, und $\alpha_e$ n"ahert sich einem
rechten Winkel an, das bedeutet das Testteilchen st"urzt nicht mehr in das andere
Meronzentrum, sondern bewegt sich letztendlich auf einem station"aren Kreis um das
erste Meronzentrum (s. Bild 4).

\begin{figure}[t]
\vspace{-20cm}
\epsfig{file=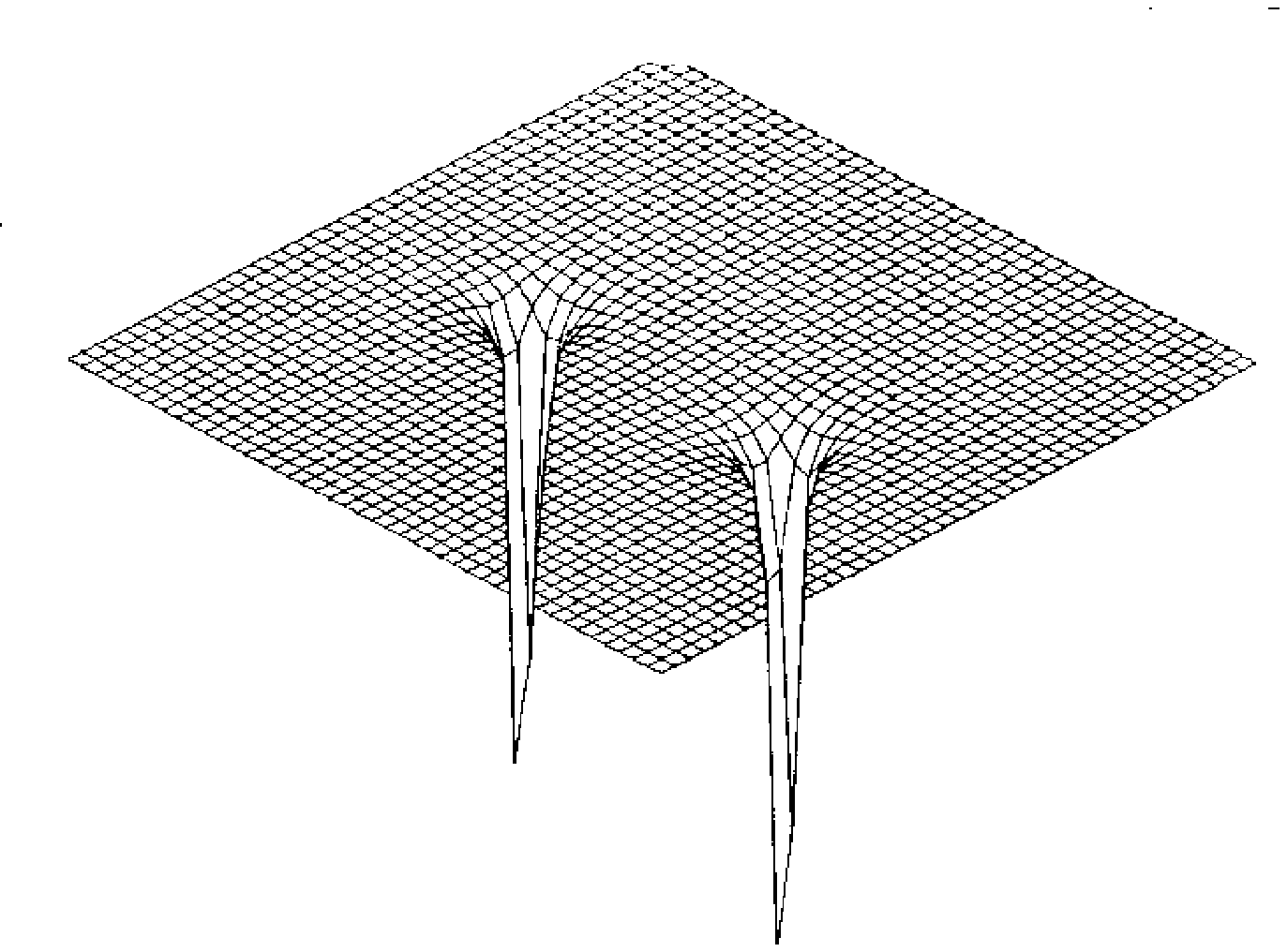}
\vspace{2cm}
\caption[Das Dimeron-Potential]{Das Potential ${\cal V}$ wird von
zwei anziehenden Zentren erzeugt, welche sich an den Meronzentren {\bf a} und {\bf b}
befinden. Die L"osungen der geod"atischen Gleichung \rf{dGammaquerdd} werden durch
Newtonsche Teilchen-Trajektorien in diesem Potential dargestellt. Da die
Gesamtenergie verschwindet \rf{defcalT}, bilden Testteilchen und  Dimeron einen
gebundenen Zustand.}
\end{figure}

\begin{figure}[t]
\vspace{-20cm}
\epsfig{file=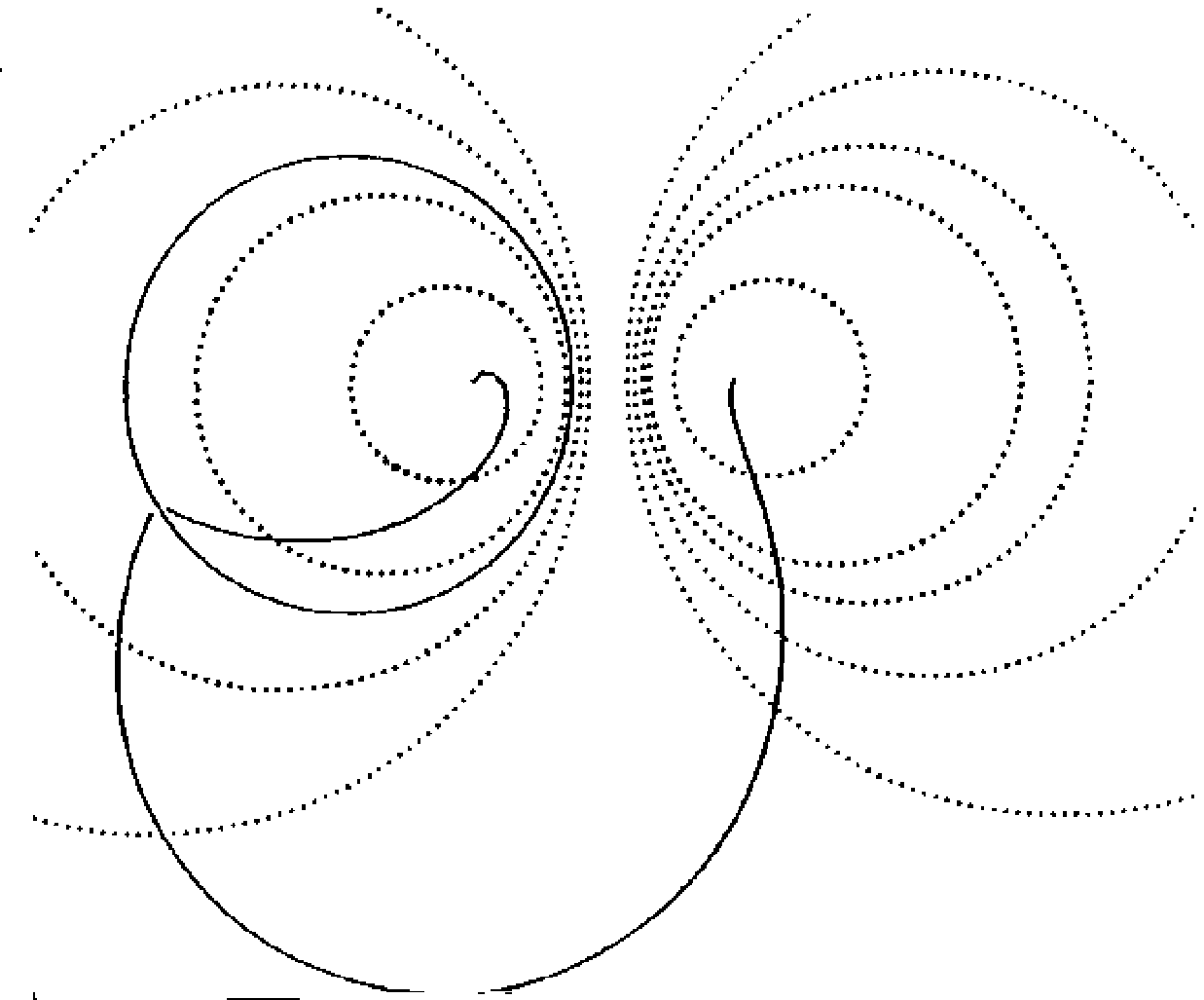}
\vspace{2cm}
\caption[Newtonsche Geod\"aten]{Das 2-dimensionale Newtonsche
Problem \rf{dddcalV}, \rf{calVpsi} erlaubt kreisf"ormige Trajektorien (gepunktete
Kurve). Die Kreise repr"asentieren, geometrisch gesehen, die charakteristischen
Fl"achen, welche unter konstantem Winkel von einer beliebigen Teilchen-Trajektorie
geschnitten werden.}
\end{figure}

\begin{figure}[t]
\vspace{-20cm}
\epsfig{file=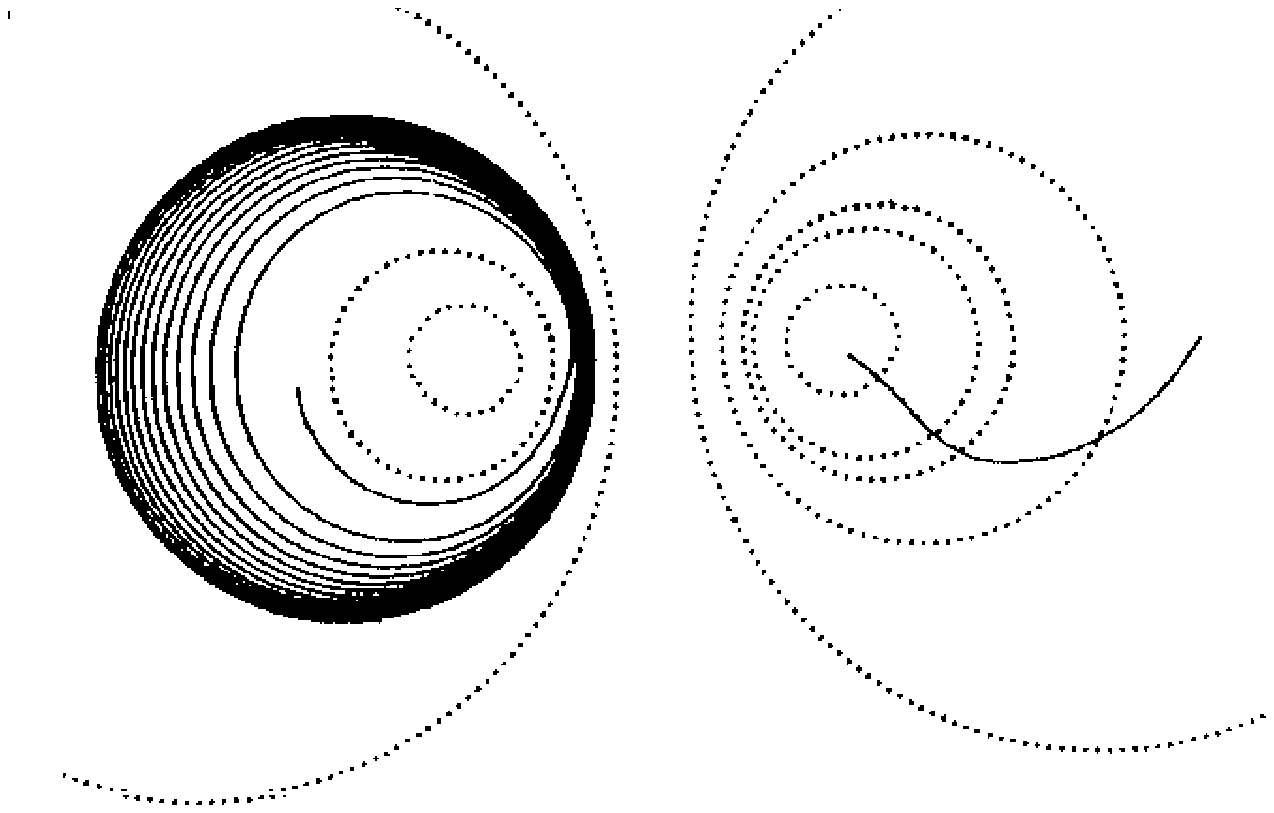}
\vspace{2cm}
\caption[Autoparallele Trajektorien]{Die autoparallelen Trajektorien
schneiden, wenn sie sich dem anziehenden Zentrum n"ahern, die charakteristischen
Fl"achen zunehmend orthogonal (oben). Im Falle einer vom Zentrum wegf"uhrenden
Bewegung n"ahert sich die Teilchenbahn einem, auf einer charakteristischen Fl"ache,
liegenden Kreis an (unten).}
\end{figure}


\chapter{\ \ Die Expansion des leeren Universum}

Unsere Wahl \rf{Te2C4fG} f"ur eine allgemeine Energie-Impuls-Tensordichte ${\bf\Te}$
des Vakuums (bzw. f"ur das "`Gravitationsfeld"') erm"oglicht es uns allgemeinere, auf
dem Ansatz \rf{VerallgMikroGl1} - \rf{VerallgFeldst} beruhende L"osungen der
Einsteingleichungen \rf{ELpTG} zu finden. Der interessante Punkt hierbei besteht in
der Frage, ob die allgemeineren L"osungen die de Sitter Konfigurationen letztendlich
approximieren, so da\3 man letztere bez"uglich dieser Klasse von St"orungen als stabil
bezeichnen kann. Der korrespondierende Energie-Impuls-Tensor ${\bf\Te}$ der
zugrundeliegenden Klasse von St"orungen wurde schon in \rf{TotEnImp} gefunden und der
Einstein-Tensor ist durch \rf{VerallgE} gegeben, so da\3 die Einsteingleichungen
\rf{ELpTG} das folgende gekoppelte System von Differentialgleichungen f"ur die
intrinsischen und extrinsischen Variablen $\varphi$ und $\phi$ erzeugen
\alpheqn
\begin{eqnarray}
\label{Dgl_a}
\frac{\dot{\calR}^2-\sigma}{\calR^2} &=& L^2_p\left[\left(\phi\varphi-\dot{\phi}
\right)^2 + \phi^4\right]\\[0.5cm]
\label{Dgl_b}
\frac{\sigma-\dot{\calR}^2}{\calR^2} - 2\,\frac{\ddot{\calR}}{\calR} &=& L^2_p
\left[\left(\phi\varphi-\dot{\phi}\right)^2+\phi^4-\left(\frac{2\phi}{\scrc}\right)^2
\right]\ .
\end{eqnarray}
\reseteqn
F"ur die nachfolgenden Untersuchungen dieser Bewegungsgleichungen f"uhren wir die
neue Variable $\calS$ ein:
\begin{equation}
\calS := \calR\phi\ .
\label{DefcalS}
\end{equation}
Eine Kombination der Gleichungen \rf{Dgl_a}, \rf{Dgl_b} ergibt nun
\alpheqn
\begin{eqnarray}
\label{Einstein1}
\dot{\calR}^2 - \sigma + \calR\ddot{\calR} &=& 2\frac{L^2_p}{\scrc^2}\,\calS^2\\
\label{Einstein2}
\dot{\calR}^2 - \sigma &=& L^2_p\left(\dot{\calS}^2 + \frac{\calS^4}{\calR^2}\right)\ .
\end{eqnarray}
\reseteqn

Dieses Gleichungssystem l"a\3t nun schon einige interessante Eigenschaften f"ur das
vorliegende kosmologische Modell erkennen: F"ur das offene Universum $(\sigma = +1)$
kann der Radius ${\calR}$ des Weltalls keinen extremalen Wert $(\dot{\calR} = 0)$
annehmen. Das bedeutet, da\3 das Universum in diesem Fall f"ur immer kontrahiert oder
expandiert. F"ur $\sigma=0$ ($\leadsto$ Euklidische Foliation) oder $\sigma = -1$
($\leadsto$ geschlossenes Universum) kann dagegen eine Phase der Expansion auf eine
Kontraktion folgen (oder umgekehrt). Man beachte auch, da\3 die Gleichungen
\rf{Einstein1}, \rf{Einstein2} zeitumkehrinvariant sind, d.h. der zeitlich umgekehrte
Ablauf ist ebenfalls eine L"osung der Gleichungen. Aus diesem Grund mu\3 f"ur jede
expansive L"osung die korrespondierende kontraktive L"osung existieren. Diese
Tatsache f"uhrt nun zu der interessanten Frage welche der beiden L"osungen stabil
ist. Aufgrund der Zeitumkehrinvarianz {\em kann nur eine} der beiden L"osungstypen
stabil sein! Hierbei bezieht sich die Stabilit"atsfrage prim"ar auf den
Vakuumgrundzustand (de Sitter-Universum und dessen Euklidische Foliation) wie in
Abschnitt \rf{EuklFol} beschrieben. Uns interessiert also, welche der beiden
Euklidischen Foliationen des de Sitter-Universums durch die allgemeine L"osung der
Bewegungsgleichungen \rf{Einstein1}, \rf{Einstein2} f"ur $t\to +\infty$
approximiert wird.

Man beachte hier, da\3 der Begriff des Grundzustandes den globalen 4-Raum {\em und}
seine Foliation einschlie\3t. Einen anderen Foliationstyp des {\em gleichen} 4-Raumes
sehen wir aufgrund der absoluten Bedeutung von Raum und Zeit als ein
unterschiedliches Universum an, welches wir aus diesem Grund von der
Stabilit"atsanalyse ausschlie\3en. Um dies an einem Beispiel aufzuzeigen, wollen wir
die verschiedenen Foliationen der de Sitter-Raum-Zeit (d.h. dem Raum konstanter
Kr"ummung) untersuchen, welche von der dynamischen (1+3)-Aufspaltung
\rf{calHvarphiphicalB}, \rf{HvarphicalB} herr"uhren. Da der Kr"ummungstensor {\bf R}
\rf{VerallgR} einen Raum konstanter Kr"ummung beschreibt, mu\3 er die Form \rf{Rnull}
annehmen. Dies ergibt die folgenden Gleichungen f"ur den Radius $\calR$
\alpheqn
\begin{eqnarray}
\label{calR1}
\varphi^2 - \frac{\sigma}{\calR^2} &=& \frac{1}{l^2}\\
\label{calR2}
\frac{\sigma-\dot{\calR}^2}{\calR^2} + \frac{\ddot{\calR}}{\calR} &=& 0\ .
\end{eqnarray}
\reseteqn
Mit Hilfe der "`Hubble Relation"' \rf{Hvarphi} erhalten wir f"ur die erste
Gleichung \rf{calR1}:
\begin{equation}
\frac{\sigma-\dot{\calR}^2}{\calR^2} + \frac{1}{l^2} = 0\ .
\label{calR1_a}
\end{equation}
Die Konsistenzforderung mit der zweiten Gleichung \rf{calR2} ergibt dann
\begin{equation}
\ddot{\calR} - \frac{1}{l^2}\,\calR = 0\ .
\label{calR2_a}
\end{equation}
Die allgemeine L"osung von \rf{calR2_a} l"a\3t sich leicht finden:
\begin{equation}
\calR(\theta) = \calR_+\exp\left[-\frac{\theta}{l}\right] + \calR_-\exp\left[
\frac{\theta}{l}\right]\ .
\label{AllgLsg}
\end{equation}
Setzen wir dieses Ergebnis in \rf{calR2} ein, so erhalten wir eine Bedingung f"ur die
Konstante~$\sigma$
\begin{equation}
\sigma = \dot{\calR}^2 - \calR\ddot{\calR} = - 4\,\frac{\calR_+\cdot\calR_-}{l^2}\ .
\label{sigma}
\end{equation}
Aufgrund der Argumentation nach Gleichung \rf{fparasigmacalR} kann die Konstante
$\sigma$ jedoch nur die Werte  $\sigma = 0,\plmi 1$ annehmen.

Der erste Wert $(\sigma = 0)$ impliziert, da\3 zumindest eine der zwei
Integrationskonstanten $\calR_+$, $\calR_-$, welche in der allgemeinen L"osung
\rf{AllgLsg} enthalten sind, verschwinden m"ussen. Dadurch erhalten wieder die beiden
urspr"unglichen de Sitter Foliationen \rf{calRexp}, welche einer Euklidischen
Foliation des 4-Raumes entsprechen. W"ahlen wir dagegen den Wert $\sigma = -1$
(geschlossenes sph"arisches Universum), so m"ussen die zwei Amplituden
$\calR_{\stackrel{+}{-}}$ positiv sein:
\begin{equation}
\calR_+ = \calR_- = +\frac{l}{2}\ .
\label{Ampli_geschl_Univ}
\end{equation}
Ohne Beschr"ankung der Allgemeinheit erhalten wir f"ur die entsprechende L"osung
\begin{equation}
{}^{(-1)}\calR(\theta) = l\cosh\frac{\theta}{l}\ .
\label{Lsg_geschl_Univ}
\end{equation}
Der Radius des Universums nimmt im Gegensatz zur urspr"unglichen de Sitter Foliation
$(\sigma=0)$ einen minimalen Wert an. Das offene Universum $(\sigma = +1)$ ist
charakterisiert durch
\begin{equation}
\calR_+ = -\calR_- = l
\label{Ampli_off_Univ}
\end{equation}
und entsteht aus einer Singularit"at ("`big bang"')
\begin{eqnarray}
\label{Lsg_off_Univ}
{}^{(+1)}\calR(\theta) = l\sinh\frac{\theta}{l}\\
\left(\theta \ge 0\right)\ .\nonumber
\end{eqnarray}
Unseres Wissens ist diese Foliation $(\sigma = +1)$ in der Literatur noch nicht
untersucht worden \cite{HaEll73}.

Der interessante Punkt hierbei ist, da\3 sich beide Zeit-Funktionen ${}^{(\plmi
1)}\calR(\theta)$ \rf{Lsg_geschl_Univ}, \rf{Lsg_off_Univ} vom exponentiellen Gesetz
$\Rnull(\theta)$ \rf{calRexp} nur f"ur endliche Werte von $\theta$ unterscheiden und
sich mit anwachsender Zeit $\theta$ dem Wert der Hubble-Konstanten
${\dis\frac{1}{l}}$ der de Sitter-Foliation schnell ann"ahern. Abgesehen von diesem
Merkmal m"ussen wir uns aber vor Augen halten, da\3 sich die 3-Geometrien und
3-Topologien beider L"osungen f"ur alle Zeiten $\theta$ unterscheiden! Der Grund
hierf"ur ist, da\3 die Geometrie auf den charakteristischen Fl"achen von Hand
eingef"uhrt und f"ur alle Zeiten konstant gehalten wird. (vgl. die Argumentation nach
Gleichung \rf{fparasigmacalR}). Dies zeigt uns die Bedeutung der SO(3)-Kr"ummung
\bm{\calF}$_i$ \rf{Dgl4} f"ur die Dynamik der (1+3)-Aufspaltung.

Wir kommen nun zu dem wichtigsten Punkt, n"amlich der Frage, ob die drei
Foliationstypen die Einsteinschen Feldgleichungen \rf{Einstein1}, \rf{Einstein2}
erf"ullen. Auf den ersten Blick scheint dies nahezuliegen, da die Einsteinschen
Gleichungen nur die 4-dimensionale Raum-Zeit Geometrie, welche von der Art der
Foliation unabh"angig ist, {\em lokal} mit der Energie-Impulsdichte der Materie
verbinden. Die {\em globale} Topologie der entsprechenden 4-R"aume und 3-R"aume wird
jedoch nicht eindeutig fixiert! Wir erwarten deshalb nicht, da\3 die Einsteinschen
Gleichungen eine bestimmte Topologie, die zu einer L"osung der mikroskopischen
Gleichungen f"ur die (1+3)-Aufspaltung geh"ort, von vornherein auschlie\3en. Das
Gegenteil ist jedoch der Fall: Da die rechte Seite der Einsteinschen Gleichungen
nicht nur die 4-Geometrie der Raum-Zeit enth"alt, sondern auch von deren Foliation
abh"angt (vgl. \rf{ETeTm}), besteht die M"oglichkeit, da\3 die eine oder andere
Topologie ausgeschlossen wird. Um dies zu zeigen, setzen wir Gleichung \rf{sigma} in
die Bewegungsgleichung \rf{Einstein1} ein und erhalten
\begin{equation}
\calR\ddot{\calR} = \frac{L^2_p}{\scrc^2}\,\calS^2\ .
\label{calRcalRpp1}
\end{equation}
Aufgrund von \rf{calR2_a} ist dies in "Ubereinstimmung mit der urspr"unglichen
Definition der Variablen $\calS$ \rf{DefcalS}. Ferner schlie\3en wir aus der
Bewegungsgleichung \rf{Einstein2}
\begin{equation}
\calR\ddot{\calR} = L^2_p\left(\dot{\calS}^2 + \frac{\calS^4}{\calR^2} \right)\ ,
\label{calRcalRpp2}
\end{equation}
bzw. mit Hilfe von \rf{calRcalRpp1}
\begin{equation}
\calS^2 = \scrc^2\dot{\calS}^2 + \frac{\scrc^2}{\calR^2}\,\calS^4\ .
\label{calS_in_calS}
\end{equation}
Die Kombination von \rf{DefcalS} mit \rf{calS_in_calS} ergibt
\begin{equation}
\dot{\calR}^2 = \frac{1}{l^2}\,\calR^2\ ,
\label{calRpplcalR}
\end{equation}
was mit Gleichung \rf{calR1_a} nur f"ur $\sigma = 0$ konsistent ist! Die
Einsteinschen Gleichungen selektieren also die flache Foliation aus der "ubrigen
Menge der Foliationen (sph"arisch und hyperbolisch) aus. Das bedeutet, da\3 die
L"osungen der Einsteinschen Gleichungen \rf{Einstein1}, \rf{Einstein2} f"ur
$\sigma\ne 0$, falls sie existieren, nicht das de Sitter-Universum als den
Vakuumgrundzustand beschreiben. Aus diesem Grund m"ussen wir uns auf die flache
Foliation $\sigma = 0$ beschr"anken, wenn wir die Stabilit"at der expandierenden und
kontrahierenden Phase untersuchen.

Im folgenden werden wir das Stabilit"atsproblem f"ur $\sigma = 0$ numerisch
untersuchen, das mit dem Problem der kosmischen Zeitrichtung eng verbunden ist. F"ur
die numerischen Rechnungen ist es n"utzlich, alle Gr"o\3en in Einheiten der
Planck-L"ange $L_p$ zu messen, d.h. wir setzen
\alpheqn
\begin{eqnarray}
t &:=& \frac{\theta}{L_p}\\
r &:=& \frac{\calR}{L_p}\\
s &\equiv& \calS\\
\frac{dr}{dt} \equiv \dot{r} &:=& \dot{\calR} \equiv \frac{d\calR}{d\theta}\\
\frac{ds}{dt} \equiv \dot{s} &:=& L_p\dot{\calS} \equiv L_p
\frac{d\calS}{d\theta}\\
\Lambda_p &:=& 2\frac{L_p}{\scrc}\ .
\end{eqnarray}
\reseteqn
Damit vereinfachen sich die Gleichungen \rf{Einstein1}, \rf{Einstein2} zu
\alpheqn
\begin{eqnarray}
\label{Einst1}
\ddot{r} = \frac{\sigma-\dot{r}^2+{\dis\frac{1}{2}}\,\Lambda_p^2\,s^2}{r}\\[0.5cm]
\label{Einst2}
\dot{r}^2 = \sigma + \dot{s}^2 + \frac{s^4}{r^2}\ .
\end{eqnarray}
\reseteqn
Dieses Gleichungssystem vereinfacht sich f"ur die flache Foliation zu
\alpheqn
\begin{eqnarray}
\label{Flach1}
\ddot{r} &=& \frac{{\dis\frac{1}{2}}\,\Lambda_p^2\,s^2-\dot{r}^2}{r}\\[0.5cm]
\label{Flach2}
\dot{r}^2 &=& \dot{s}^2 + \frac{s^4}{r^2}\ .
\end{eqnarray}
\reseteqn

Wir wollen uns nat"urlich davon "uberzeugen, da\3 der Vakuumgrundzustand von
Abschnitt \rf{EuklFol} wirklich eine L"osung von \rf{Flach1} und \rf{Flach2} ist.
Dieser Grundzustand sieht bei Verwendung der oben eingef"uhrten Gr"o\3en
folgenderma\3en aus:
\alpheqn
\begin{eqnarray}
\label{Grund1}
r &\to& r_*(t) = \rho_*\exp\left[\frac{t}{\tau_*}\right]\\[0.5cm]
\label{Grund2}
s &\to& s_*(t) = \zeta_*\exp\left[\frac{t}{\tau_*}\right]\ .
\end{eqnarray}
\reseteqn
Dr"ucken wir die Konstanten $\tau_*$ und $\Lambda_p$ mit Hilfe der L"angenparameter
$\left\{l,L,L_p\right\}$ aus, so ergibt sich
\alpheqn
\begin{eqnarray}
\label{Deftaustern}
\tau_* &=& -\frac{l}{L_p}\\
\label{DefLambdap}
\Lambda_p &=& \frac{2L}{l}\ .
\end{eqnarray}
\reseteqn
Setzen wir nun die Konfiguration \rf{Grund1}, \rf{Grund2} in die Bewegungsgleichung
\rf{Flach1}, \rf{Flach2} ein, so ergibt sich eine L"osung, falls (i) zwischen der
Zeit-Konstanten $\tau_*$ und dem L"angenparameter $\Lambda_p$ die folgende Beziehung
herrscht
\begin{equation}
\tau^2_* = \frac{4}{\Lambda_p^2}\left(1 + \frac{4}{\Lambda_p^2} \right)
\label{Deftaustern2}
\end{equation}
und (ii), falls das Verh"altnis der Amplituden folgenden Wert annimmt:
\begin{equation}
\left(\frac{\rho_*}{\zeta_*}\right)^2 = 1 + \left(\frac{2}{\Lambda_p}\right)^2\ .
\label{VerhAmpli1}
\end{equation}
Aufgrund der Gleichung \rf{LpLl} ist die Beziehung \rf{Deftaustern2}
konsistent mit \rf{Deftaustern}, \rf{DefLambdap}. Falls das de Sitter-Universum durch
die Euklidische Foliation aufgebaut wird, gen"ugt es also in der Tat den
Vakuum-Bewegungsleichungen \rf{Einstein1}, \rf{Einstein2} und kann wegen seiner hohen
Symmetrie als Vakuum-Grundzustand betrachtet werden!

Nach diesen Vorbemerkungen k"onnen wir uns nun unserer urspr"unglichen Absicht
zuwenden, die Stabilit"at der expansiven $(\tau_* > 0)$ und kontraktiven $(\tau_* <
0)$ Phase des Grundzustandes zu untersuchen. Dazu benutzen wir in \rf{Flach1},
\rf{Flach2} einen linearisierenden Ansatz
\alpheqn
\begin{eqnarray}
\label{rschl}
r(t) &=& r_*(t) + \tilde{r}(t)\\
\label{sschl}
s(t) &=& s_*(t) + \tilde{s}(t)
\end{eqnarray}
\reseteqn
und erhalten
\alpheqn
\begin{eqnarray}
\label{Linear1}
\tau_*\dot{\tilde{r}} - \frac{\zeta_*}{\rho_*}\,\tau_*\dot{\tilde{s}} +
\left(\frac{4}{\tau_*\Lambda^2_p}\right)^2 \tilde{r} - \frac{8}{\Lambda^2_p}
\frac{\zeta_*}{\rho_*}\,\tilde{s} = 0\\[0.5cm]
\label{Linear2}
\ddot{\tilde{r}} + \frac{2}{\tau_*}\dot{\tilde{r}} + \frac{1}{\tau^2_*}\tilde{r} =
\frac{\zeta_*}{\rho_*}\,\Lambda^2_p\,\tilde{s}\ .
\end{eqnarray}
\reseteqn
Um einen qualitativen "Uberblick "uber den Unterschied im Stabilit"atsverhalten der
zwei Grundzustands-Konfigurationen zu erhalten, bringen wir \rf{Linear2} in eine Form
wie sie bei der Beschreibung eines ged"ampften harmonischen Oszillators unter dem
Einflu\3 einer "au\3eren Kraft auftritt:
\begin{equation}
\frac{d}{dt}\left(\frac{1}{2}\,\dot{\tilde{r}}^2 + \frac{1}{2}\,
\frac{\tilde{r}^2}{\tau^2_*}\right) = -\frac{2}{\tau_*}\,\dot{\tilde{r}}^2 +
\frac{\zeta_*}{\rho_*}\,\Lambda^2_p\,\tilde{s}\,\dot{\tilde{r}}\ .
\label{HarmOsz}
\end{equation}
F"ur $\tau_* > 0$ (expandierendes Universum) ist der Energieverlust durch Reibung
positiv und die St"orung $\tilde{r}(t)$ der de Sitter-Konfiguration wird f"ur $t
\to\infty$ verschwinden (vorausgesetzt die Arbeit, welche die "au\3ere Kraft
verrichtet, kompensiert nicht den Energieverlust). Der expandierende Grundzustand
scheint deshalb stabil zu sein, wogegen die kontrahierende Phase $(\tau_* < 0)$ sich
als instabil erweist.

Um diese Vermutung genauer zu untersuchen machen wir f"ur $\tilde{r}$ und $\tilde{s}$
einen Exponential-Ansatz
\alpheqn
\begin{eqnarray}
\label{ExpAns_r}
\tilde{r}(t) &=& \tilde{\rho}\exp\left[\frac{t}{\tilde{\tau}}\right]\\[0.5cm]
\label{ExpAns_s}
\tilde{s}(t) &=& \tilde{\zeta}\exp\left[\frac{t}{\tilde{\tau}}\right]\ .
\end{eqnarray}
\reseteqn
Damit ergibt sich f"ur die Amplituden $\tilde{\rho}$ und $\tilde{\zeta}$ das folgende
homogene Gleichungssystem:
\alpheqn
\begin{eqnarray}
\label{HomGl1}
\left[\frac{\tau_*}{\tilde{\tau}} + \left(\frac{4}{\tau_*\Lambda^2_p}\right)^2\right]
\tilde{\rho} - \frac{\zeta_*}{\rho_*}\left(\frac{\tau_*}{\tilde{\tau}} +
\frac{8}{\Lambda^2_p}\right)\tilde{\zeta} &=& 0\\[0.5cm]
\label{HomGl2}
\left[1+2\frac{\tau_*}{\tilde{\tau}}+\left(\frac{\tau_*}{\tilde{\tau}}\right)^2
\right]\tilde{\rho} - \frac{\zeta_*}{\rho_*}\left(\tau_*\Lambda_p\right)^2
\tilde{\zeta} &=& 0\ .
\end{eqnarray}
\reseteqn

Der D"ampfungs-Parameter $\tilde{\tau}$ der kleinen Abweichungen
$\left\{\tilde{r},\tilde{s} \right\}$ \rf{ExpAns_r}, \rf{ExpAns_s} ergibt sich wie
gew"ohnlich aus der Forderung einer verschwindenden Koeffizienten-Determinante f"ur
\rf{HomGl1}, \rf{HomGl2}, d.h.
\begin{equation}
\left(\frac{\tau_*}{\tilde{\tau}}\right)^3 + \left(2+\frac{8}{\Lambda^2_p}\right)
\left(\frac{\tau_*}{\tilde{\tau}}\right)^2 - 3\,\frac{\tau_*}{\tilde{\tau}} -
\frac{8}{\Lambda^2_p} = 0\ .
\label{Poly3}
\end{equation}
Die {\em exakten} Wurzeln dieser Gleichung dritten Grades sind
\alpheqn
\begin{eqnarray}
\label{Wurzel1}
\left(\frac{\tau_*}{\tilde{\tau}}\right)_1 &=& 1\\[0.5cm]
\left(\frac{\tau_*}{\tilde{\tau}}\right)_{2,3} &=& -\frac{1}{2}\left(3 +
\frac{8}{\Lambda^2_p}\right)\plmi\frac{1}{2}\,\sqrt{\left(3+\frac{8}{\Lambda^2_p}
\right)^2 - \frac{32}{\Lambda^2_p}}\nonumber\\[0.5cm]
\label{Wurzel2}
&=& -1 -\frac{1}{2}\sqrt{1+\left(2\tau_*\right)^2} \plmi \frac{1}{2}
\sqrt{9+\left(2\tau_*\right)^2}\ .
\end{eqnarray}
\reseteqn
Aus dem Ansatz \rf{ExpAns_r}, \rf{ExpAns_s} ersehen wir, da\3 die beiden negativen
Wurzeln \rf{Wurzel2} die {\em Stabilit"at} des {\em expandierenden} $(\tau_* > 0)$
und die {\em Instabilit"at} des {\em kontrahierenden} Grundzustandes $(\tau_* < 0)$
implizieren. Die Wurzel \rf{Wurzel1} dagegen bedeutet, da\3 die {\em expandierende
und kontrahierende} Phase stabil ist, da die St"orungen \rf{ExpAns_r}, \rf{ExpAns_s}
das gleiche zeitliche Verhalten aufweisen wie der ungest"orte Grundzustand
\rf{Grund1}, \rf{Grund2}. Aus diesem Grund bewirken die St"orungen mit der Amplitude
\rf{Wurzel1} nur eine "Anderung der Grundzustandsamplituden $\rho_*$, $\zeta_*$.
Diese Tatsache erkennt man auch, wenn man das Verh"altnis dieser St"orungsamplituden
aus dem linearen Gleichungssystem \rf{HomGl1}, \rf{HomGl2} berechnet:
\begin{equation}
\frac{\tilde{\rho}}{\tilde{\zeta}} = \left(1+\frac{4}{\Lambda^2_p}\right)
\frac{\zeta_*}{\rho_*}
\label{VerhAmpli2}
\end{equation}
bzw. mit Hilfe von \rf{VerhAmpli1}
\begin{equation}
\left|\frac{\tilde{\rho}}{\tilde{\zeta}}\right| = \left|\frac{\tilde{\rho_*}}
{\tilde{\zeta_*}}\right|\ .
\label{VerhAmpli3}
\end{equation}
Solch eine Beziehung f"ur kleine Abweichungen $\tilde{\rho}$, $\tilde{\zeta}$
ben"otigen wir aufgrund der Konstanz des Verh"altnisses von $\rho_*$ zu $\zeta_*$ wie
in Gleichung \rf{VerhAmpli1} aufgef"uhrt.

Das Ergebnis unserer linearen Stabilit"atsanalyse besteht also darin, da\3 der
expandierende Grundzustand stabil ist, w"ahren der kontrahierende sich als unstabil
erweist. Diese Effekte k"onnen wir anhand der numerischen Integration (Fig.VII.1) der
urspr"unglichen Bewegungsgleichungen \rf{Flach1}, \rf{Flach2} deutlich erkennen.

(Fig.VII.1) zeigt die zeitliche Entwicklung des Radius des Universums, wenn die de
Sitter-Werte f"ur den Radius $\calR$ (nicht jedoch f"ur $\calS$) als
Anfangsbedingungen genommen werden. Das Universum weicht dann von der de
Sitter-Entwicklung ab, kehrt aber bald wieder zu ihr zur"uck und entwickelt sich dann
de Sitter gem"a\3. Integriert man jedoch in die Vergangenheit (f"ur die zeitlich
gespiegelte L"osung ist dies die Zukunft), so endet die Entwicklung des Universums
nach einigen Planck-Zeiten. Es gibt dabei zwei Arten von Singularit"aten, welche die
r"uckw"artige Entwicklung beenden: (i) der Radius des Universums geht gegen Null (wie
in der Einsteinschen Theorie) und (ii) $\dot{s}^2$ wird bei endlichem Radius $r$
negativ. Diese Art von Singularit"at taucht in der Einsteinschen Theorie nicht auf
und zeigt einen Zusammenbruch der klassischen Theorie schon bei einer endlichen
Ausdehnung des Universums. Auf jeden Fall "`"lebt"' das kontrahierende Universum
nicht l"anger als ein paar Planck-Zeiten. Aus diesem Grund ist die "`positive"'
Zeitrichtung durch die anhaltende Expansion des Universums bevorzugt! Als n"achstes
m"u\3te man untersuchen, wie dieses Ergebnis durch die Gegenwart von Materie
$({\bf\TM})\ne 0)$ in Gleichung \rf{ETeTm} modifiziert wird.

Wir wollen nun am Schlu\3 noch einige Bemerkungen "uber das wirkliche Universum
machen. Falls die strenge de Sitter-Konfiguration \rf{dscalRdxdydz} in der Natur
realisiert ist, tr"agt die beobachtbare Hubble-Konstante $H$ \rf{Hnumerisch} einen
Wert von ungef"ahr $10^{-56}\,{\dis\frac{1}{cm^2}}$ zur kosmologischen Konstanten
$\lambda_c$ in den Einstein-Gleichungen \rf{ETMLambdaG} bei:
\begin{equation}
E_{\mu\nu} + \lambda_c\,G_{\mu\nu} = 0\ .
\label{ElambdacG}
\end{equation}
Solch ein kleiner Wert l"a\3t sich sich gegenw"artig experimentell nicht feststellen.
Um zum Beispiel einen zus"atzlichen Effekt in der Merkur-Periheldrehung in der
Gr"o\3enordnung von einer Bogensekunde pro Jahrhundert hervorzubringen (dies w"urde
experimentell feststellbar sein \cite{Ri77}), m"u\3te die kosmologische Konstante
$\lambda_c$ einen Wert von ungef"ahr $10^{-42}\,{\dis\frac{1}{cm^2}}$ besitzen.
Vergleichen wir dies mit dem obigen Wert von $\lambda_c$, so scheint jeglicher Versuch
einer experimentellen Verifikation sinnlos zu sein.

Die Untersuchungen in dieser Arbeit zeigen jedoch eine weitere alternative Sicht
dieses Problems. Wie wir aus Fig.IV.1 ersehen, ben"otigt das Universum eine gewisse
Zeit, um  die strenge de Sitter-Konfiguration zu erreichen. Die entsprechende
Relaxationszeit $\tilde{\tau}$ wird durch die zweite und dritte Wurzel
von Gleichung \rf{Wurzel2} angegeben. F"ur eine grobe Absch"atzung dieser Werte
ben"utzen wir Gleichung \rf{WertL} f"ur $L$. Damit erhalten wir f"ur die
dimensionslose Konstante $\Lambda_p$ in \rf{DefLambdap} einen Wert von $10^{-30}$.
Aufgrund dieser kleinen Gr"o\3e liegt die eine Wurzel $\tilde{\tau}$ in der
Gr"o\3enordnung der Planck-Zeit $T_p\approx 10^{-42}$ sec und die andere in der
Gr"o\3enordnung der Hubble-Zeit $T_H\approx 10^{10}$ Jahre . Da die letztere
Zeitskala in der Gr"o\3enordnung des Alters des Weltalls liegt, scheint uns die
Annahme realistisch, da\3 das Universum die de Sitter-Konfiguration gegenw"artig noch
nicht erreicht hat. Aus diesem Grund mu\3 die beobachtbare Hubble-Rate $\varphi$
\rf{Hvarphi} nicht unbedingt mit dem de Sitter-Gleichgewichtswert $l$ \rf{varphitol}
"ubereinstimmen. Diese Tatsache sollte aber gen"ugen, um die modifizierten
Einstein-Gleichungen \rf{ETeTm} anstelle von \rf{ETMLambdaG} zur Beschreibung des
realen Universums zu benutzen, da der Vakuum-Tensor ${\bf\Te}$ des gegenw"artigen
Universums noch nicht seine Grundzustandsform (proportional zu {\bf G}) erreicht hat.
Zuk"unftige Untersuchungen werden zeigen ob dies in unserem Sonnensystem zu
experimentell beobachtbaren Effekten f"uhrt.\footnotemark
\footnotetext{Dies f\"uhrte nach Beendigung der vorliegenden Arbeit zu
  einer Reihe weiterer Ver\"offentlichungen, 
  siehe~\cite{MaSo91, MaSo92, MaSo96}}
\begin{figure}
\epsfig{file=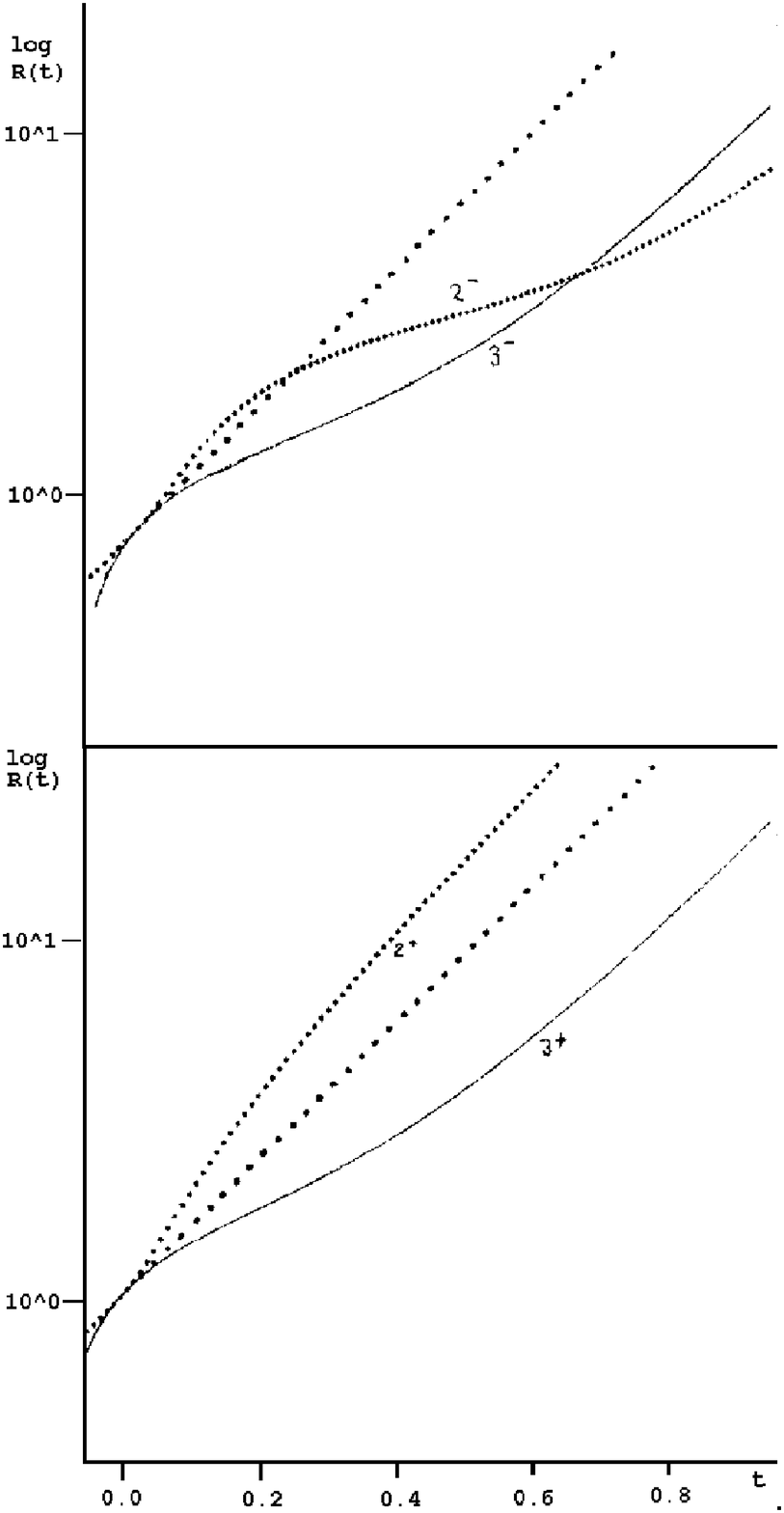,clip=,height=20cm}
\caption{Der Radius $\calR(t)$ des deSitter Universums f"ur $\sigma = 0$}
\end{figure}
\clearpage
Die de Sitter-L"osung \rf{Grund1} erscheint hier als Gerade (gep"unktelt). Alle
gefundenen L"osungen sind stabil, d.h. sie beschreiben f"ur $t\to\infty$ ein de
Sitter-Universum, was sich an der asymptotischen Parallelit"at zur de Sitter-Geraden
ablesen l"a\3t. Als Anfangswerte wurden f"ur den Radius ${\cal R}(0)$ und seine
Zeitableitung $\dot{{\cal R}}(0)$ die de Sitter-Werte zugrundegelegt, w"ahrend die
Anfangswerte f"ur die Variable ${\cal S}$ "`verstimmt"' wurden; und zwar wurde f"ur
die L"osungen $(2^+)$ und $(2^-)$ der Anfangswert ${\cal S}(0)$ "uber dem de
Sitter-Wert ${\cal S}_*$ hinaus erh"oht, w"ahrend f"ur die L"osungen $(3^+)$ und
$(3^-)$ ${\cal S}(0)$ gegen"uber ${\cal S}_*(0)$ abgesenkt wurde. Die Indizes
$\stackrel{+}{-}$ f"ur die einzelnen L"osungen beziehen sich auf die Wahl des
relativen Vorzeichens von ${\cal S}(0)$ und $\dot{{\cal S}}(0)$. Offensichtlich
bedeutet eine {\em Erh"ohung} ({\em Erniedrigung}) von $|{\cal S}(0)|$, da\3 die
zugeh"orige L"osung zun"achst {\em "uber} ({\em unter}) der de Sitter-Konfiguration
liegt, w"ahrend das relative Vorzeichen dar"uber entscheidet, welche
Gr"o\3enverh"altnisse sich im asymptotischen Bereich $(t\to\infty)$ einstellen.


\chapter*{Glossar}
\addcontentsline{toc}{chapter}{Glossar}
\indent

$\otimes$ \dotfill Tensorprodukt

$\circ$ \dotfill Verkn"upfung

$\bullet$ \dotfill "Uberschiebung

Sp \dotfill Spurbildung

\vspace{1cm}


{\bf A} = $\{A_{i\mu}\}$ \dotfill Konnexion in $\tauschl$

${\bf\Aquer} = \{\Aquer_{i\mu}\}$ \dotfill Konnexion in $\tauquer$

$[\Bquer]$ \dotfill Abbildung $\taudach\to\tauquer$

$[\Bdach]$ \dotfill Abbildung $\tauquer\to\taudach$

$[\Bschl]$ \dotfill Abbildung $\tauasymp\to\tauschl$

$[\Basymp]$ \dotfill Abbildung $\tauschl\to\tauasymp$

\bm{\calB} = $\{\calB_{i\mu}\}$ \dotfill Soldering-Form der B"undel $\tauquer$ und
$\tauschl$

{\bf B} \dotfill reskaliertes \bm{\calB}

{\bf C} = $\{C_{i\mu}\}$ \dotfill dynamischer Teil von {\bf A}

{\bf C} = $\{C_\mu\}$ \dotfill Gradientenvektor, $C_\mu = \partial_\mu C$

{\bf D} = $\{D_\mu\}$ \dotfill eichkovariante Ableitung in $\tauschl$

${\bf\Dquer}$ = $\{\Dquer_\mu\}$ \dotfill eichkovariante Ableitung in $\tauquer$

\bm{\calD} = $\{\calD_\mu\}$ \dotfill allgemein-kovariante Ableitung bzgl.
${\bf\Gamma}$ und {\bf A}.

\bm{\calDhalb} = $\{\calDhalb_\mu\}$ \dotfill allgemein-kovariante Ableitung
bzgl. ${\bf\Gammadach}$ und ${\bf\Aquer}$.

$\calDquerF$ = $\{\calDquer_\mu\}$ \dotfill allgemein-kovariante Ableitung bzgl.
${\bf\Gamma}$ und ${\bf\Aquer}$

\bm{\calDasymp} = $\{\calDasymp_\mu\}$ \dotfill allgemein-kovariante Ableitung bzgl.
${\bf\Gammaasymp}$ und {\bf A}

{\bf E} = $\{E_{\mu\nu}\}$ \dotfill Einstein-Tensor

\bm{\calE} \dotfill spezielles Bezugssystem

$\calE$ \dotfill Energiedichte

{\bf F} = $\{F_{i\mu\nu}\}$ \dotfill SO(3) Kr"ummung von {\bf A}

${\bf\Fquer} = \{\Fquer_{i\mu\nu}\}$ \dotfill SO(3) Kr"ummung von ${\bf\Aquer}$

\bm{\calF} = $\{\calF_{i\mu\nu}\}$ \dotfill SO(3) Eichtensor

{\bf G} = $\{G_{\mu\nu}\}$ \dotfill Riemannsche Metrik

${\bf\Gquer}$ = $\{\Gquer_{\mu\nu}\}$ \dotfill konformal flache Metrik

${\bf\Gstern} = \{\Gstern_{\mu\nu}\}$ \dotfill charakteristische Metrik f"ur ${\bf H}
= 0$

G \dotfill Determinante von {\bf G}

${\bf\Gdach} = \{G_{\mu\nu}\}$ \dotfill Fasermetrik von $\taudach$

{\bf H} = $\{H_{\mu\nu}\}$ \dotfill Hubble-Tensor

H \dotfill Hubble-Konstante

\bm{\calH} = $\{\calH_{i\mu\nu}\}$ \dotfill SO(3) Eichtensor

${\bf M}_4$, ${\bf E}_4$ \dotfill Minkowski- bzw. Euklidischer Raum

${\bf K} = \{{K^\mu}_{\nu\lambda}\}$ \dotfill Kontorsion

${\bf\Kdach} = \{{\Kdach^\mu}_{\nu\lambda}\}$ \dotfill charakteristische Kontorsion

${\bf\Kds} = \{{\Kds^\mu}_{\nu\lambda}\}$ \dotfill orthogonaler Anteil der
charakteristischen Kontorsion

${\bf\Kdp} = \{{\Kdp^\mu}_{\nu\lambda}\}$ \dotfill paralleler Anteil der
charakteristischen Kontorsion

${\bf\Kcup} = \{{\Kcup^\mu}_{\nu\lambda}\}$ \dotfill relative Kontorsion von
${\bf\Gammaasymp}$ bzgl. ${\bf\Gammadach}$

${\bf\Kasymp} = \{{\Kasymp^\mu}_{\nu\lambda}\}$ \dotfill totale Kontorsion

${\bf\Kstern} = \{{\Kstern^\mu}_{\nu\lambda}\}$ \dotfill charakteristische Kontorsion
f"ur ${\bf H} = 0$

$L_P$ \dotfill Planck-L"ange

L, $\calL$ \dotfill L"angenparameter

$L^i$ \dotfill ${\cal SO}$(3)-Generator

${\bf\Mex}$, ${\bf\Min}$ \dotfill "au\3ere bzw. innere Variable

$\calM$ \dotfill Energiedichte

$\calP$ \dotfill Druck

${}^{(\scr W)}{\calM}$ \dotfill Energiedichte des {\bf W}-Feldes

${}^{(\scr W )}{\calP}$ \dotfill Druck des {\bf W}-Feldes

${}^{(\scr F)}{\calM}$ \dotfill Energiedichte des Eichfeldes

${}^{(\scr F )}{\calP}$ \dotfill Druck des Eichfeldes

${\bf\Pdach} = \{\Pdach_{\mu\nu}\}$ \dotfill Projektor auf $\Deltaquer$

${\bf\Pcap}$ \dotfill Projektionsoperator auf $\Deltadach$

${\bf Q} = \{Q_{\mu\nu\lambda}\}$ \dotfill Nichtmetrizit"atstensor

$\bar{\bf Q} = \{\bar{Q}_{\mu\nu\lambda}\}$ \dotfill Abweichung ${\bf\Gammastern}$
von ${\bf\Gammaquer}$

{\bf R} = $\{R_{\mu\nu\lambda\sigma}\}$ \dotfill Kr"ummung von ${\bf\Gamma}$

${\bf\Rstern} = \{\Rstern_{\mu\nu\lambda\sigma}\}$ \dotfill charakteristische
Kr"ummung $({\bf H} = 0)$

\bm{\calR} = $\{{\calR^{\dot{\mu}}}_\nu \} \equiv \{{R^{\dot{\mu}}}_\nu\}$ \dotfill
Ricci-Tensor

R bzw. ${\cal S}$ \dotfill Kr"ummungsskalar

${\bf\Rpara} = \{\Rpara_{\mu\nu\lambda\sigma}\}$ \dotfill paralleler Teil von {\bf R}

${\bf\Rperp} = \{\Rperp_{\mu\nu\lambda\sigma}\}$ \dotfill orthogonaler Teil von
{\bf R}

${\bf\Rdach} = \{\Rdach_{\mu\nu\lambda\sigma}\}$ \dotfill charakteristische Kr"ummung
(bzgl. ${\bf\Gammadach}$)

{\bf S} = $\{{S^i}_j\}$ \dotfill Element aus SO(3)

{\bf S} = $\{S_\mu\},\{S_{\mu\nu}\},\{S_{\mu\nu\lambda}\}$ \dotfill Hilfsgr"o\3en aus
$\taudach$

$\bar{\bf S}$ \dotfill konformal flache {\bf S}

{\bf T} = $\{T_{\mu\nu}\}$ \dotfill gesamter Energie-Impulstensor

{\bf T} = $\{T_{\mu\nu\lambda}\}$ \dotfill Tensor

${\calT}$ \dotfill kinetische Energie eines Testteilchens

\bm{\calT} = $\{\calT_{\mu\nu}\}$ \dotfill metrischer Energie-Impulstensor

${\bf\Tnull} = \{\Tnull_{\mu\nu}\}$ \dotfill Energie-Impulstensor des Vakuums

${\bf\TB} = \{\TB_{\mu\nu}\}$ \dotfill kanonischer Energie-Impulstensor des {\bf B}
Feldes

${\bf\Te} = \{\Te_{\mu\nu}\}$ \dotfill Energie-Impulstensor von ${\bf\Mex}$

${\bf\TF} = \{\TF_{\mu\nu}\}$ \dotfill kanonischer Energie-Impulstensor des
Eichfeldes

\bm{\calTF} = $\{\calTF_{\mu\nu}\}$ \dotfill metrischer Energie-Impulstensor des
Eichfeldes

\bm{\calTB} = $\{\calTB_{\mu\nu}\}$ \dotfill metrischer Energie-Impulstensor des
{\bf B} Feldes

${\bf\Tid} = \{\Tid_{\mu\nu}\}$ \dotfill Energie-Impulstensor einer idealen
Fl"ussigkeit

${\bf\TM} = \{\TM_{\mu\nu}\}$ \dotfill Energie-Impulstensor der Materie

${\bf\TW} = \{\TW_{\mu\nu}\}$ \dotfill kanonischer Energie-Impulstensor des
{\bf W}Feldes

{\bf U} = $\{U_\mu\}$ \dotfill Tangentenvektor

{\bf V} = $\{V_\lambda\}$ \dotfill Schnitt in $\taudach$

${\cal V}$ \dotfill Potential

{\bf W} = $\{W_\lambda\}$ \dotfill Kompensationsfeld

{\bf W} = $\{W_{\mu\nu\rho\lambda}\}$ \dotfill Weylscher-Tensor

{\bf Z} = $\{{Z^\mu}_{\nu\lambda}\}$ \dotfill Torsion

${\bf\Zdach}$ = $\{{Z^\mu}_{\nu\lambda}\}$ \dotfill charakteristische Torsion

${\bf\Zds} = \{{\Zds^\mu}_{\nu\lambda}\}$ \dotfill orthogonaler Anteil der
charakteristischen Torsion

${\bf\Zdp} = \{{\Zdp^\mu}_{\nu\lambda}\}$ \dotfill paralleler Anteil der
charakteristischen Torsion

${\bf\Zcup}$ = $\{{\Zcup^\mu}_{\nu\lambda}\}$ \dotfill relative Torsion

${\bf\Zasymp}$ = $\{{\Zasymp^\mu}_{\nu\lambda}\}$ \dotfill totale Torsion

\vspace{1cm}


$\scrc$ \dotfill Konstante

$\hat{\bf e} = \{\ndach_\mu,{\hat{e}_i}{}^\mu\}$ \dotfill Vierbein, welches
$\Deltaquer$ aufspannt

{\bf f} = $\{f_\mu\}$ \dotfill Frobeniusvektor

$f_{||}$, $f_{\perp}$ \dotfill Parameter

{\bf g} $ =\mbox{diag}(\plmi 1,-1,-1,-1)$ \dotfill Minkowski bzw. Euklidische Metrik

{\bf g} $ = \{g_{ij}\}$ \dotfill Fasermetrik von $\tauquer$

{\bf h} = $\{h_{\mu\nu}\}$ \dotfill reduzierter Hubble-Tensor

${\bf\hdach} = \{\hdach_{\mu\nu}\}$ \dotfill Projektor senkrecht zur
Vierergeschwindigkeit

${\bf\hplus} = \{\hplus_{\mu\nu}\}$ \dotfill symmetrischer Teil von {\bf h}

${\bf\hminus} = \{\hminus_{\mu\nu}\}$ \dotfill antisymmetrischer Teil von {\bf h}

{\bf k} = $\{k_{\mu\nu}\}$ \dotfill symmetrisches $\taudach$-Objekt

{\bf l} = $\{l_\mu\}$ \dotfill Schnitt aus $\taudach$

m \dotfill Gravonmasse

l \dotfill L"angenparameter

$l^i$ \dotfill ${\cal SO}$(1,3) bzw. ${\cal SO}$(4) Boost-Generator

{\bf p} = $\{p_\mu\}$ \dotfill charakteristisches Vektorfeld, normal zu $\Deltadach$

${\bf\pdach}$ = \dotfill normiertes char. Vektorfeld "uber {\bf E}$_4$

${\bf\pdual}$ = $\{\pdual_{\mu\nu\lambda}\}$ \dotfill Poincar\'e-Dual von {\bf p}

{\bf s} \dotfill Abweichung $\bar{\bf S}$ von {\bf S}

{\bf t} = $\{t_{\mu\nu}\}$ \dotfill totaler Korrekturterm

${\bf\tB} = \{\tB_{\mu\nu}\}$ \dotfill Korrekturterm des {\bf B}-Feldes

${\bf\tW} = \{\tW_{\mu\nu}\}$ \dotfill Korrekturterm des {\bf W}-Feldes

{\bf u} = $\{u_i\}$ \dotfill SO(3)-Vektoranteil von {\bf U}

u \dotfill SO(3)-Skalaranteil von {\bf U}

{\bf w} = $\{w_\mu\}$ \dotfill Schnitt von $\taudach$

\vspace{1cm}


$\Deltadach$ \dotfill charakteristische Distribution

$\Deltaquer$ \dotfill repr"asentative Distribution

${\bf\Gamma} = \{{\Gamma^\mu}_{\nu\lambda}\}$ \dotfill ${\cal GL}(4,{\bf R})$ Konnexion

${\bf\Gammaquer} = \{{\Gammaquer^\mu}_{\nu\lambda}\}$ \dotfill Levi-Civita Konnexion
von ${\bf\Gquer}$

${\bf\Gammakr} = \{{\Gammakr^\mu}_{\nu\lambda}\}$ \dotfill kanonisch flache
Konnexion

${\bf\Gammastern} = \{{\Gammastern^\mu}_{\nu\lambda}\}$ \dotfill charakteristische
Konnexion $({\bf H} = 0)$

${\bf\Gammadach} = \{{\Gammadach^\mu}_{\nu\lambda}\}$ \dotfill charakteristische
Konnexion

${\bf\Gammaschl} = \{{\Gammaschl^\mu}_{\nu\lambda}\}$ \dotfill Levi-Civita Konnexion
bzgl. {\bf G}

${\bf\Gammaasymp} = \{{\Gammaasymp^\mu}_{\nu\lambda}\}$ \dotfill Konnexion in
$\tauasymp$

$\Lambda = \{{\Lambda^\mu}_\nu\}$ \dotfill Element aus SO(1,3)

$\Lambda$  \dotfill totale Lagrangedichte

$\Lambda_B$ \dotfill Lagrangedichte von {\bf B}

$\Lambda_F$ \dotfill Lagrangedichte von {\bf F}

$\Lambda_W$ \dotfill Lagrangedichte von {\bf W}

\bm{\nabla} = $\{\nabla_\mu\}$ \dotfill koordianten-kovariante Ableitung bzgl.
${\bf\Gamma}$

\mbox{\bm{\nabladach}} = $\{\nabladach_\mu\}$ \dotfill charakteristische Ableitung

\mbox{\bm{\nds}} = $\{\nds_\mu\}$ \dotfill orthogonaler Teil der charakteristischen
Ableitung

\mbox{\bm{\ndp}} = $\{\ndp_\mu\}$ \dotfill paralleler Teil der charakteristischen
Ableitung

\bm{\nablaasymp} = $\{\nablaasymp_\mu\}$ \dotfill koordinaten-kovariante
Ableitung in $\tauasymp$

${\bf\Omega} = \{\Omega_{\mu\nu}\}$ \dotfill Kr"ummung von \bm{\omega}

${\bf\Omegakr} = \{\Omegakr_{\mu\nu}\}$ \dotfill $\lambdakr$ B"undelkr"ummung

$\Phi$ \dotfill Hilfsfunktion

${\bf\Sigma} = \{{\Sigma^\mu}_{\nu\lambda}\}$ \dotfill totale Spindichte

${\bf\SigmacalB}, {\bf\SigmaB} = \{\SigmacalB_{\mu\nu\lambda}\},
\{\SigmaB_{\mu\nu\lambda}\}$ \dotfill Spindichte des \bm{\calB}-Feldes bzw. {\bf
B}-Feldes

${\bf\Sigmap} = \{{\Sigmap^\mu}_{\nu\lambda}\}$ \dotfill Spindichte des {\bf p}-Feldes

${\bf\SigmaW} = \{{\Sigma^\mu}_{\nu\lambda}\}$ \dotfill Spindichte des {\bf W}-Feldes

${\bf\Sigma}^\prime = \{\Sigma^\prime_{\mu\nu\lambda}\}$ \dotfill symmetrischer
Teil von $\bar{\bf Q}$

\vspace{1cm}


$\alpha$ \dotfill Winkel zwischen den Geod"aten und den charakteristischen Linien

$\alpha_e$ \dotfill Euklidisches Analogon von $\alpha$

\bm{\gamma} = ${\gamma^\mu}_{\nu\lambda}$ \dotfill wesentlicher Teil von ${\bf\Gamma}$

$\lambdakr$ \dotfill Triviales Prinzipalb"undel mit ${\bf M}_4$, bzw. ${\bf E}_4$ als
Basisraum

\bm{\nu} = $\{\nu_i\}$ \dotfill Schnitt von $\tauquer$

\bm{\omegakr} = $\{\omegakr_{\mu\nu\lambda}\}$ \dotfill Flache Konnexion in
$\lambdakr$

$\{\wp_{\mu\nu}\}$, $\{\wp_\mu\}$ \dotfill Hilfsgr"o\3en

\bm{\partial} = $\{\partial_\mu\}$ \dotfill partielle Ableitung

$\psi$ \dotfill integrierender Faktor f"ur {\bf p}

$\sigma$ \dotfill Eigenzeit

$\tau$ \dotfill Newtonsche Zeit

$\tau_4$ \dotfill Tangentialb"undel der Raum-Zeit

$\tauschl$, $\tauasymp_4$ \dotfill isomorphe B"undel

$\taukr$ \dotfill triviales Tangentialb"undel der flachen Raum-Zeit

$\taudach$ \dotfill charakteristisches B"undel

$\tauquer$ \dotfill repr"asentatives B"undel

$\theta$ \dotfill Universalzeit

\begin{appendix}
\chapter{Nachbemerkung}

Die Wahl \rf{Fquer0} f"ur die innere Kr"ummung ${\bf\Fquer}$ ist nicht so speziell
wie es vielleicht auf den ersten Blick erscheinen k"onnte. Man kann n"amlich zeigen
(vgl.~\rf{FquersigmacalR2calB}, \rf{RdachsigmacalR2calB}), da\3 der obige
Speziallfall \rf{Fquer0} nur einen $(\sigma = 0)$ von drei m"oglichen F"allen
\mbox{$(\sigma = 1,0,-1)$} der charakteristischen Untergeometrie darstellt. Die
charakteristischen Fl"achen tragen in jedem Fall eine Geometrie konstanter Kr"ummung
(offen, geschlossen oder flach).

Das Ergebnis \rf{FquersigmacalR2calB}, \rf{RdachsigmacalR2calB} ist allgemeing"ultig
wenn man auf der linken Seite statt der gesamten SO(3)-Kr"ummung ${\bf\Fquer}$
bzw.~statt des gesamten charakteristischen Kr"ummungstensor ${\bf\Rdach}$ nur deren
Einschr"ankungen ${\bf\Fquer}\Big|_{\Deltadach}$ und ${\bf\Rdach}\Big|_{\Deltadach}$
auf die charakteristischen Fl"achen verwendet, d.h es gilt stets
\begin{eqnarray}
\label{FquerImmer}
\Fquer_{i\rho\sigma}{\calB^\rho}_\mu{\calB^\sigma}_\nu &=& \frac{\sigma}{\calR^2(
\theta)}{\epsilon_i}^{jk}\calB_{j\mu}\calB_{k\nu}\\
\label{RdachImmer}
\Rdach_{\mu\sigma\alpha\beta}{\calB^\alpha}_\nu{\calB^\beta}_\lambda &=&
\frac{\sigma}{\calR^2(\theta)}\left[\calB_{\mu\nu}\calB_{\sigma\lambda} -
\calB_{\mu\lambda}\calB_{\sigma\nu}\right]\ .
\end{eqnarray}
Demzufolge tragen also die charakteristischen Fl"achen stets eine innere 3-Geometrie
konstanter Kr"ummung, {\em unabh"angig von der Symmetrie der Massenverteilung im
Universum.} Dieser Sachverhalt k"onnte eine Erkl"arung daf"ur abgeben, warum die
3-Geometrie des Weltalls auf kosmischer Skala so extrem homogen und isotrop
erscheint, obwohl j"ungste experimentelle Beobachtungen auf eine ziemlich starke
Inhomogenit"at der Massenverteilung hinweisen (kosmische Leerr"aume $\leadsto$
"`Hubble bubbles"'~\cite{Tre90}). "Ubrigens wurde dieser Sachverhalt f"ur den
Spezialfall der Einbettung der charakteristischen 3-Fl"achen in einen Euklidischen
4-Raum schon in einer fr"uheren Arbeit nachgewiesen~\cite{MaSo89_b}.

\end{appendix}



\clearpage
\thispagestyle{empty}

Herrn Dr.~M.Sorg m"ochte ich an dieser Stelle f"ur die Themenstellung und f"ur
seine st"andige Gespr"achsbereitschaft meinen besonderen Dank ausprechen. Viele
anregende Diskussionen mit ihm trugen wesentlich zum Gelingen dieser Arbeit bei.

Herrn Prof.~Dr.~Dr.~h.c W.Weidlich danke ich f"ur die "Ubernahme des Hauptberichts
und Herrn Prof.~Dr.~H.R.Trebin f"ur seine Bem"uhungen als Mitberichter.

\end{document}